\documentclass{article}

\usepackage{arxiv}
\usepackage[utf8]{inputenc} 
\usepackage[T1]{fontenc}    
\usepackage{hyperref}       
\usepackage{url}            
\usepackage{booktabs}       
\usepackage{amsfonts}       
\usepackage{nicefrac}       
\usepackage{microtype}      
\usepackage{lipsum}		
\usepackage{amsthm}
\usepackage{graphicx}
\usepackage{natbib}
\usepackage{doi}
\usepackage{etoc}
\usepackage{amsmath,amssymb,amsthm}
\usepackage{mathtools}
\usepackage{bm}
\usepackage{enumitem}

\usepackage[T1]{fontenc}
\usepackage{soul} 
\usepackage{phfqit}
\usepackage{proba} 
\usepackage{amssymb}
\usepackage{amsmath}
\usepackage{amsthm}
\usepackage{amsfonts}
\usepackage{mathtools}
\usepackage{xspace}
\usepackage{bm}
\usepackage{nicefrac}
\usepackage{commath}
\usepackage{bbm}
\usepackage{boxedminipage}
\usepackage{xparse}
\usepackage{xcolor}
\usepackage{float}
\usepackage{multirow}
\usepackage{graphicx}
\usepackage{caption}
\usepackage{subcaption}
\usepackage{ifthen}
\usepackage{algpseudocode}
\usepackage{colortbl} 
\usepackage{hyperref}
\usepackage[capitalize,noabbrev]{cleveref}
\usepackage{soul}
\usepackage{fancyhdr}   
\usepackage{tikz}
\usetikzlibrary{positioning}
\usetikzlibrary{fit}
\usetikzlibrary{calc}
\usetikzlibrary{backgrounds}
\usetikzlibrary{shapes}
\usetikzlibrary{patterns}
\usetikzlibrary{matrix}

\crefname{assumption}{Assumption}{Assumptions}
\Crefname{assumption}{Assumption}{Assumptions}

\crefname{fact}{Fact}{Fact}
\Crefname{fact}{Fact}{Fact}

\crefname{claim}{Claim}{Claim}
\Crefname{claim}{Claim}{Claim}

\crefname{observation}{Observation}{Observation}
\Crefname{observation}{Observation}{Observation}

\allowdisplaybreaks

\newcommand{\namedref}[2]{\hyperref[#2]{#1~\ref*{#2}}\xspace}

\renewcommand{\poly}[1]{\ensuremath{\mathrm{poly}}\xspace\left(#1\right)}

\renewcommand{\Tilde}[1]{\widetilde{#1}}

  \newcommand{\ceil}[1]{\ensuremath{\left\lceil{#1}\right\rceil}\xspace}

  \newcommand{\iabs}[1]{\ensuremath{\vert{#1}\vert}\xspace}

  \newcommand{\eps}[0]{\varepsilon}
  \let\epsilon\eps
  
  \newcommand{\cA}{\ensuremath{{\mathcal A}}\xspace}

  \newcommand{\cD}{\ensuremath{{\mathcal D}}\xspace}
  
  \newcommand{\cF}{\ensuremath{{\mathcal F}}\xspace}
  \newcommand{\cG}{\ensuremath{{\mathcal G}}\xspace}
  \newcommand{\cH}{\ensuremath{{\mathcal H}}\xspace}
  \newcommand{\cI}{\ensuremath{{\mathcal I}}\xspace}
  \newcommand{\cJ}{\ensuremath{{\mathcal J}}\xspace}

  \newcommand{\cM}{\ensuremath{{\mathcal M}}\xspace}
  \newcommand{\cN}{\ensuremath{{\mathcal N}}\xspace}
  \newcommand{\cO}{\ensuremath{{\mathcal O}}\xspace}
  \newcommand{\cP}{\ensuremath{{\mathcal P}}\xspace}

  \newcommand{\cT}{\ensuremath{{\mathcal T}}\xspace}
  \newcommand{\cU}{\ensuremath{{\mathcal U}}\xspace}
  \newcommand{\cV}{\ensuremath{{\mathcal V}}\xspace}

  \newcommand{\bbC}{\ensuremath{{\mathbb C}}\xspace}
  
  \newcommand{\bbE}{\ensuremath{{\mathbb E}}\xspace}

  \newcommand{\bbI}{\ensuremath{{\mathbb I}}\xspace}
  
  \newcommand{\bbK}{\ensuremath{{\mathbb K}}\xspace}
  \newcommand{\bbN}{\ensuremath{{\mathbb N}}\xspace}
  \newcommand{\bbO}{\ensuremath{{\mathbb O}}\xspace}

  \newcommand{\bbR}{\ensuremath{{\mathbb R}}\xspace}

  \newcommand{\bbU}{\ensuremath{{\mathbb U}}\xspace}
  \newcommand{\bbV}{\ensuremath{{\mathbb V}}\xspace}

  \newcommand{\bbZ}{\ensuremath{{\mathbb Z}}\xspace}
  
  \newcommand{\fU}{\ensuremath{{\mathfrak U}}\xspace}

  \newcommand{\fc}{\ensuremath{{\mathfrak c}}\xspace}
  \newcommand{\fg}{\ensuremath{{\mathfrak g}}\xspace}
  \newcommand{\fh}{\ensuremath{{\mathfrak h}}\xspace}

  \newcommand{\fu}{\ensuremath{{\mathfrak u}}\xspace}
  \newcommand{\fm}{\ensuremath{{\mathfrak m}}\xspace}

  \newcommand{\tfg}{\ensuremath{\widetilde{\mathfrak g}}\xspace}
  
  \newcommand{\defeq}[0]{\ensuremath{{\;\vcentcolon=\;}}\xspace}

  \newcommand{\var}[0]{\mathop{\mathrm{Var}}\xspace}
  
  \let\leq\leqslant
  \let\le\leqslant
  \let\geq\geqslant
  \let\ge\geqslant
  
\usepackage[titlenumbered,ruled, lined, linesnumbered, commentsnumbered, longend,algoruled,boxed]{algorithm2e}
\makeatletter 
\g@addto@macro{\@algocf@init}{\SetKwInOut{Input}{Input}}
\makeatother
\makeatletter 
\g@addto@macro{\@algocf@init}{\SetKwInOut{Output}{Output}}
\makeatother
\makeatletter 
\g@addto@macro{\@algocf@init}{\SetKwInOut{Parameter}{Parameter}}
\makeatother

\newtheorem{theorem}{Theorem}[section]
\newtheorem{lemma}{Lemma}[section]
\newtheorem{claim}{Claim}[section]
\newtheorem{proposition}{Proposition}[section]
\newtheorem{corollary}{Corollary}[section]
\newtheorem{fact}{Fact}[section]
\newtheorem{example}[theorem]{Example}

\newtheorem{observation}{Observation}[section]

\newtheorem{definition}{Definition}[section]

\theoremstyle{remark}
\newtheorem{remark}[theorem]{Remark}

\theoremstyle{plain}

\newcommand{\CurB}[1]{\left\{#1\right\}}

\newcommand{\commutator}[2]{\left[#1, #2\right]}
\newcommand{\icommutator}[2]{[#1, #2]}

\newcommand{\EXP}[1]{\mathbb{E}\left[#1\right]}
\newcommand{\EXPD}[2]{\underset{#1}{\mathbb{E}}\left[#2\right]}
\newcommand{\VARD}[2]{\underset{#1}{\var}\left[#2\right]}

\newcommand{\Tr}[1]{\tr\CurB{#1}}
\newcommand{\iTr}[1]{\tr\{#1\}}

\newcommand{\absU}[0]{U\left(\VECTOR{\theta}\right)}
\newcommand{\absUd}[0]{U^\dagger\left(\VECTOR{\theta}\right)}

\newcommand{\state}[1]{\left|#1\right>}
\newcommand{\cstate}[1]{\left<#1\right|}

\newcommand{\Inner}[2]{\left<#1\middle|#2\right>}
\newcommand{\Outer}[2]{\left|#1\right>\left<#2\right|}

\newcommand{\lag}[0]{\mathfrak{g}}

\newcommand{\SPAN}[1]{\text{span}\CurB{#1}}

\newcommand{\VECTOR}[1]{\boldsymbol{#1}}
\newcommand{\vtheta}{\VECTOR{\theta}}

\newcommand{\Prob}[1]{\Pr\left[#1\right]}

\renewcommand{\su}[1]{\mathfrak{su}\left(#1\right)}

\renewcommand{\Var}[1]{\var\left[#1\right]}

\renewcommand{\SU}[1]{\mathrm{SU}\left(#1\right)}

\newcommand{\Spin}[1]{\mathrm{Spin}\left(#1\right)}

\newcommand{\sgn}[0]{\operatorname{sgn}}

\renewcommand{\GL}[1]{\mathrm{GL}\left(#1\right)}
\newcommand{\gl}[1]{\mathfrak{gl}\left(#1\right)}

\newcommand{\Ad}{\text{Ad}}
\newcommand{\ad}{\text{ad}}

\newcommand{\End}[1]{\mathrm{End}\left(#1\right)}
\newcommand{\iEnd}[1]{\mathrm{End}(\!#1)}

\newcommand{\Hom}[3]{\mathrm{Hom}_{#1}\left(#2, #3\right)}

\newcommand{\Id}{\text{id}}

\newcommand{\tx}{\widetilde{x}}

\newcommand{\Lie}[1]{\text{Lie}\left(#1\right)}

\newcommand{\Symt}[1]{\mathrm{Sym}^2\left(#1\right)}
\newcommand{\Altt}[1]{\wedge^2\left(#1\right)}

\newcommand{\iSymt}[1]{\mathrm{Sym}^2(\!#1)}
\newcommand{\iAltt}[1]{\wedge^2(\!#1)}

\newcommand{\SWAP}{\mathrm{SWAP}}

\newcommand{\Comm}[1]{\mathrm{Comm}\left(#1\right)}
\newcommand{\iComm}[1]{\mathrm{Comm}(#1)}

\newcommand{\expm}[1]{\exp\left(#1\right)}
\newcommand{\iexpm}[1]{\exp(#1)}
\newcommand{\snorm}[1]{\norm{#1}^S}
\newcommand{\inorm}[1]{\!\|#1\!\|}
\newcommand{\isnorm}[1]{\!\|#1\!\|^S}

\newcommand{\infsnorm}[1]{\snorm{#1}_\infty}
\newcommand{\onesnorm}[1]{\snorm{#1}_1}

\newcommand{\iinfsnorm}[1]{\snorm{#1}_\infty}
\newcommand{\ionesnorm}[1]{\snorm{#1}_1}

\newcommand{\compactG}[0]{\widetilde{G}}

\newcommand{\Bernoulli}[1]{\mathrm{Bernoulli}(#1)}

\newcommand{\trho}[0]{\widetilde{\rho}}
\newcommand{\tcM}[0]{\widetilde{\cM}}
\newcommand{\teps}[0]{\widetilde{\varepsilon}}

\newcommand{\ispan}[1]{\operatorname{span}\{#1\}}

\newcommand{\opl}[0]{\lambda^*}
\newcommand{\opc}[0]{c_{\opl}}

\newcommand{\Mproj}{\cM^{(\lambda^\prime,a^\prime)\otimes(\nu^\prime,b^\prime)}_{\lambda,a,1\to\lambda,b,2}}

\newcommand{\JJ}{\mathcal{J}}
\newcommand{\dimV}{\dim V_\lambda}

\crefname{algocf}{algorithm}{algorithms}
\Crefname{algocf}{Algorithm}{Algorithms}

\allowdisplaybreaks

\title{Structure makes quantum circuits trainable}

\author{Wenbo Xie \\
   Department of Computer Science\\
   Purdue University\\
   West Lafayette, USA \\
   \texttt{xie401@purdue.edu} \\
   \And
   Wojciech Szpankowski \\
   Department of Computer Science\\
   Purdue University\\
   West Lafayette, USA \\
   \texttt{szpan@purdue.edu} \\
   \And
   Ananth Grama \\
   Department of Computer Science\\
   Purdue University\\
   West Lafayette, USA \\
   \texttt{ayg@cs.purdue.edu} \\}

\hypersetup{
pdftitle={Structure makes quantum circuits trainable},
pdfsubject={quant-ph},
pdfauthor={Wenbo Xie, Wojciech Szpankowski, Ananth Grama},
}

\begin{document}
\maketitle

\begin{abstract}
The scalability of variational quantum algorithms is limited by barren plateaus, where gradients vanish exponentially in system size. This {\em worst-case} result raises fundamental questions about the trainability of quantum AI models. We present a major advance towards resolving this trainability question by answering whether a {\em given} quantum circuit has a barren plateau. We do this using an analytical framework that bounds the gradient variance for arbitrary input states $\rho$ and observables $\cM$ by decomposing the unitary representation of the associated compact Lie group. When the circuit segments on both sides of a gate form $2$-designs, we obtain the variance exactly and prove a sufficient condition for trainability: if $\rho^{\otimes 2}$ and $\cM^{\otimes 2}$ are sufficiently concentrated on the same irreducible invariant subspace of polynomial dimension, and the differentiated gate moves this subspace by an inverse-polynomial amount, then the gradient variance is at least inverse-polynomial. We further present a sampling algorithm that estimates the gradient variance from runs of the ansatz, without constructing the representation-theoretic projectors. Our algorithm decides whether the variance exceeds a tolerance $\tau$ with the number of runs polynomial in $1/\tau$, including cases in which the bounds are inconclusive. This resolves the question of the trainability of the circuit before expensive training procedures.
\end{abstract}

Corresponding Author: Wojciech Szpankowski

\section{Introduction}
\label{sec:intro}

Parameterized quantum circuits (PQCs) are essential substrates of variational quantum algorithms (VQA)~\cite{cerezo2021variational}, offering a platform for useful computation on noisy intermediate-scale quantum (NISQ) devices~\cite{peruzzo2014variational,anschuetz2026arbitrary,dalzell2023quantum}. A PQC prepares a quantum state characterized by parameters $\theta$, which a classical optimizer updates to minimize a cost function. This hybrid loop underlies the variational quantum eigensolver~\cite{you2022convergence,anschuetz2021critical}, the quantum approximate optimization algorithm~\cite{farhi2014quantum,yoshioka2023fermionic}, and a range of quantum machine learning models~\cite{shaydulin2024evidence,mitarai2018quantum,farhi2018classification,havlivcek2019supervised,larocca2022group,herman2023expressivity,you2021exponentially}. The scalability of PQCs is, however, limited by their trainability.
A central obstacle to trainability of PQCs is the presence of barren plateaus (BP): a regime in which the gradient of the cost function concentrates exponentially around zero as the number of qubits grows. The variance of gradient estimators then vanishes, and the signal available to the optimizer becomes indistinguishable from statistical noise, rendering gradient-based optimization ineffective at scale~\cite{mcclean2018barren,anschuetz2022quantum,cerezo2021cost,wang2021noise,martin2023barren}. This worst-case phenomenon raises the fundamental question of the trainability of PQCs. It opens the possibility of costly training procedures that may eventually fail because the optimizer does not make meaningful progress towards desired parameterizations of PQCs for given training data. In this paper, we present a significant advance towards answering the question of whether a given PQC has a barren plateau (as opposed to the corresponding worst case assertion), thereby addressing its trainability question before expensive PQC optimization.

A natural question is whether a given ansatz necessarily induces a barren plateau. Existing theory ties this to the expressivity and algebraic structure of the circuit: highly expressive ans\"atze that form approximate $2$-designs provably exhibit exponentially vanishing gradients~\cite{mcclean2018barren,arrasmith2022equivalence}. A complementary line of work decomposes the Lie algebra $\mathfrak{g}$ generated by the ansatz to bound the variance of partial derivatives~\cite{fontana2024characterizing,ragone2024lie}. This algebra-centric view is powerful, but its exact variance formulas typically assume that the observable is generated by $\mathfrak{g}$. For a general observable, the bounds acquire covariance terms with competing contributions, and verifying that the variance is non-vanishing requires case-by-case analysis.

In this work, we bound the variance of gradients for \emph{arbitrary} input states $\rho$ and observables $\cM$. Rather than decomposing the Lie algebra, we decompose the unitary representation $(\mathcal{H}^{\otimes 2}, \Pi^{\otimes 2})$ of the associated compact, connected Lie group. When the circuit segments surrounding a target parameter are sufficiently deep, the variance separates into contributions from the input state, the observable, and the algebraic generators, each expressed through the invariant subspaces of the representation. This shifts the diagnosis of trainability away from the algebra towards the structure of the subspaces that $\rho$ and $\cM$ actually occupy. This viewpoint yields a concrete criterion: the presence or absence of a barren plateau is governed not by the ansatz alone, but by how the tensor squares $\rho^{\otimes 2}$ and $\cM^{\otimes 2}$ are distributed across the irreducible invariant subspaces of $\cH^{\otimes 2}$, and by how the differentiated generator couples these subspaces. When both segments are $2$-designs we evaluate the variance exactly through this decomposition (\cref{main:text:thm:exact}), and we prove a sufficient condition for trainability (\cref{main:text:thm:sufficient,main:text:cor:leakage}): if $\rho^{\otimes 2}$ and $\cM^{\otimes 2}$ are sufficiently supported on \emph{the same irreducible invariant subspace, whose dimension scales polynomially with the system size $n$}, and the differentiated generator moves this subspace by an inverse-polynomial amount, then the gradient variance has an inverse-polynomial lower bound, ruling out a barren plateau. Polynomial dimension alone is not enough: a state and an observable that occupy different invariant subspaces of polynomial dimension can give exactly zero variance. Our upper bound further shows that the gradient variance can be exponentially suppressed when their support spreads over exponentially large invariant subspaces. 

Unlike Lie-algebraic frameworks \cite{fontana2024characterizing,ragone2024lie}, our proposed condition focuses on the state and the observable, applying even to observables that are not necessarily generated by $\fg$. When the observable is generated by $\fg$, the variance is also given by the Lie-algebraic formula, and several of our worked examples in the paper are of this type. For number-conserving fermionic circuits with the observable $X^{(1)}X^{(2)}$, which is not generated by the algebra, the exact formula of \cref{main:text:thm:exact} gives the variance in closed form, and the component of the observable inside the algebra accounts for only about half of it (\cref{eq:compound-var}). Recent work has explored similar state-dependent concentrations, but only within specific constraints: \cite{diaz2023showcasingbarrenplateautheory} showed that concentration in matchgate circuits is determined by how arbitrary inputs and observables distribute across invariant subspaces of the $\Spin{2n}$ representation, while \cite{Schatzki_2024} demonstrated state-dependent trainability through invariant subspaces restricted to blockwise-universal equivariant QNNs and symmetry-compatible observables. Our work generalizes these principles to arbitrary compact connected ansatz groups, as well as arbitrary states and observables, by resolving the complete tensor-square representation.

Our analytical bounds are generally not tight and evaluating the exact formula requires irreducible decomposition of $\cH^{\otimes 2}$, so they cannot always determine whether a specific PQC induces a barren plateau. To decide this directly, we introduce a sampling algorithm that estimates the gradient variance from runs of the given ansatz, without building the representation-theoretic projectors~\cite{collins2022weingarten}. Each run executes the circuit twice, at a cost linear in the circuit depth and the number of generators. It returns a bounded estimate whose bias is either set by a finite-difference step, or is exactly unbiased when the differentiated gate admits a parameter-shift rule~\cite{wierichs2022general}. As a result, deciding whether the variance exceeds a tolerance $\tau$ requires a number of runs polynomial in $1/\tau$, so trainability at any polynomial tolerance is decided with polynomial resources. Only resolving an exponentially small variance to relative precision requires exponentially many runs, as it would for any sampling method.

While our sufficient condition ties the absence of a barren plateau to an invariant subspace of polynomial dimension, this does not automatically guarantee that the system is easy to simulate classically. The presence of such a subspace should not be viewed as a standalone rule for efficient classical simulation. Our condition focuses only on the second-moment representation that controls the gradient variance; it does not imply that the cost function $\Tr{\cM\Pi(g)\rho\Pi(g)^\dagger}$ can be evaluated efficiently for an arbitrary circuit parameter. A true classical simulation would also require that the relevant invariant subspaces, their projectors or bases, the restricted group action, and the coordinates of the state and observable can all be built and handled efficiently. None of these properties come from the subspace dimensions alone. Furthermore, trainability is not limited to this situation: in the example of \cref{eq:compound-var} the variance decays only polynomially although the state is spread over all particle-number sectors, most of which have exponentially large dimension. Therefore, without extra structural and computational assumptions, the absence of a barren plateau in our framework does not mean the system is classically simulable. If those extra assumptions do hold, however, the representation-theory breakdown might offer a practical path to an efficient simplified model. The examples in this paper without a barren plateau can nevertheless be simulated classically: in the collective-spin example of \cref{eq:su2-lower} the state stays in an invariant subspace of dimension at most $n+1$, and in the free-fermion example (Supplementary \cref{app:example:ff:local}) and the example of \cref{eq:compound-var} the gates and the observable are quadratic in the fermionic operators of the Jordan--Wigner transformation. Structure that provably prevents barren plateaus often also allows classical simulation~\cite{cerezo2025does,goh2023lie}; we therefore make no claim of quantum advantage.

The remainder of the paper is organized as follows. \cref{main:text:subsec:prelim} sets out the definitions for the ansatz defined in \cref{main:ansatz}. \cref{main:text:sec:expectation-variance} gives the exact gradient variance for $2$-designs and analytical upper and lower bounds on it. \cref{main:text:theory:sec:discussion} explores the implications of these results and states the sufficient condition for trainability. Next, \cref{main:text:sec:algorithm} presents the sampling algorithm and its complexity. \cref{main:text:sec:discussion} concludes with limitations and extensions. Finally, the Methods analyze the expressibility of the ansatz (\cref{main:text:subsec:express}) and give the derivations of the results (\crefrange{main:text:subsec:exact}{main:text:subsec:error}).

\section{Results}
\subsection{Preliminary}
\label{main:text:subsec:prelim}

We consider a compact and connected Lie group $G$ with an associated Lie algebra $\fg$. Let $(\cH, \Pi)$ be a finite-dimensional unitary representation of $G$; that is, $\Pi : G \to \mathrm{U}(\cH)$ is a Lie group homomorphism assigning to each group element a unitary operator on $\cH$. Furthermore, let $\{g_1, \dots, g_k\} \subseteq \fg$ be a spanning set of $\fg$.

\begin{remark}
 Although we primarily utilize a spanning set, the following results also hold assuming access to a generating set of $\fg$. We present a discussion of this extension in Supplementary Material~\cref{sec:generating_set}.
\end{remark}

To construct a parameterization of group elements, we analyze the local structure of $G$ induced by its Lie algebra. The exponential map $\exp: \fg \to G$ is a local diffeomorphism from a neighborhood $\fU_0$ of the origin in $\fg$ onto a neighborhood $U_e$ of the identity element in $G$~\cite{kirillov2008introduction}. In particular, every element in $U_e$ can be uniquely written in the form $\exp(X)$ for some $X \in \fU_0$. This yields a local parameterization of elements in $U_e$ in terms of $\fg$, where $X$ admits a decomposition $X = \sum_{j=1}^k \theta_j g_j$ with respect to the spanning set $\{g_j\}$.
Moreover, since $G$ is connected, any neighborhood of the identity generates the entire group. That is, every element of $G$ can be written as a finite product of elements from $U_e$ (see Supplementary S2 and S3 for details, and Lemma~1.18 of~\cite{ziller2012liegroups}). This observation provides a natural parameterization of elements in $G$ via compositions of exponentials.

Motivated by this structure, we define a parameterized quantum circuit (PQC) with a periodic layered architecture. Related constructions have been studied by others~\cite{mcclean2018barren, larocca2024review, fontana2024characterizing}. Specifically, we define

\begin{align}
\absU  = \prod_{l=1}^L \prod_{q = 1}^k \Pi\!\left(\exp(\theta_{l,q} g_q)\!\right), \label{main:ansatz}
\end{align}
where $\theta_{l,q} \in [-\pi,\pi]$ for all $l\in[L]$ and $q\in[k]$. Here, $\Pi$ denotes the unitary representation of $G$ introduced above: each factor $\exp(\theta_{l,q} g_q) \in G$ is a group element, and $\Pi$ maps it to a unitary operator on $\cH$. The outer symbols $\prod$ denote the ordered operator product over layers $l$ and generators $q$, and should not be confused with the representation map $\Pi$.

\begin{example}
    A classic example involves the group $G = \mathrm{SU}(2^n)$ under the standard representation $(\mathbb{C}^{2^n}, \mathrm{id})$, where $\mathrm{id}$ denotes the identity map. The Lie algebra associated with $\mathrm{SU}(2^n)$ is $\mathfrak{su}(2^n)$, which is spanned by the \emph{traceless} Pauli strings
    \begin{align*}
        \mathcal{P} = \Bigl\{P_q = i \bigotimes_{j=1}^n \sigma_{q_j} : \sigma_{q_j} \in \{I, X, Y, Z\}\Bigr\} \setminus \{i\, I^{\otimes n}\},
    \end{align*}
    the identity string being excluded since $\mathfrak{su}(2^n)$ consists of the traceless anti-Hermitian matrices. Since the representation is the identity map, $\Pi(\exp(\theta_{l,q} P_q)) = \exp(\theta_{l,q} P_q)$, and the ansatz reduces to
    \begin{align*}
        \absU = \prod_{l=1}^L \prod_{q=1}^{|\mathcal{P}|} \exp(\theta_{l,q} P_q).
    \end{align*}
\end{example}

Having specified the parameterized circuit, we now turn to the objective function. Given a quantum state $\rho \in \End{\cH}$ and a measurement operator $\cM$, we define
\begin{align}
    C(\vtheta) = \Tr{\cM \absU \rho \absUd}. \label{main:objective}
\end{align}
This represents the expected measurement outcome with respect to the quantum state parameterized by $\vtheta$.

To analyze the optimization landscape induced by \eqref{main:ansatz}, we examine the behavior of the partial derivatives (PD) of the objective function $C(\vtheta)$. In particular, for each $\ell \in [L]$ and $j \in [k]$, we study the quantity $\partial_{\theta_{\ell,j}} C(\vtheta)$ and its statistical properties under random initialization of the parameters $\vtheta$.

The partial derivatives are not immediately accessible in the abstract group formulation. Nevertheless, they can be derived by exploiting the structural properties of Lie groups. Since $G$ is connected, the fundamental theorem of Lie theory ensures the existence of a unique Lie algebra homomorphism $\pi_* : \fg \to \mathfrak{u}(\cH)$ such that
\begin{align}
    \Pi(\exp(X)) = \exp\bigl(\pi_*(X)\bigr), \quad \forall X \in \fg. \label{main:text:commuting:diagram}
\end{align}
Consequently, the ansatz defined in \cref{main:ansatz} admits the equivalent matrix representation
\begin{align}
    U(\vtheta) = \prod_{l=1}^L \prod_{q = 1}^k \exp\!\bigl(\theta_{l,q}\,\pi_*(g_q)\bigr). \label{main:reduced:ansatz}
\end{align}
In this form, each generator $\pi_*(g_q)$ lies in $\mathfrak{u}(\cH)$, and thus the circuit is expressed entirely in terms of matrix exponentials. This allows for explicit differentiation with respect to the parameters.

Let $U_-$ and $U_+$ denote the portions of the circuit preceding and following the gate $\exp(\theta_{\ell,j}\pi_*(g_j))$, respectively, in the written order of the product in \cref{main:reduced:ansatz}, so that $U_+$ acts on $\rho$ first; $U_-$ includes the gate itself, which commutes with $\pi_*(g_j)$ and could equally be included in $U_+$. Then, differentiating the objective function with respect to $\theta_{\ell,j}$ yields:
\begin{align}
    \partial_{\theta_{\ell,j}} C
    = \Tr{\cM \, U_- \commutator{\pi_*(g_j)}{\,U_+ \rho U_+^\dagger}U_-^\dagger}. \label{main:text:PD}
\end{align}

A key phenomenon governing the behavior of $\partial_{\theta_{\ell,j}} C(\vtheta)$ is the \emph{barren plateau} (BP). Informally, a PQC exhibits a barren plateau if the PDs of the objective function concentrate around zero with vanishing variance as the system size increases, thereby rendering gradient-based optimization ineffective. More formally (see, e.g., \cite{mcclean2018barren}), the ansatz \eqref{main:ansatz} is said to exhibit a barren plateau if
\begin{align}
    \bbE_{\vtheta}\!\left[\partial_{\theta_{\ell,j}} C(\vtheta)\right] = 0
    \quad \text{and} \quad
    \Var{\partial_{\theta_{\ell,j}} C(\vtheta)} \in \cO(b^{-n}) \quad \text{for some } b >1 , \label{main:def:BP}
\end{align}
where $n$ denotes the system size. When this condition holds, the PQC is considered untrainable in the sense that gradient-based optimization does not make progress.

\subsection{Variance bounds from the representation decomposition}
\label{main:text:sec:expectation-variance}

We present analytical upper and lower bounds on the variance of the partial derivatives of the objective function, together with an exact expression that holds when the circuit segments on both sides of the differentiated gate are $2$-designs (\cref{main:text:thm:exact}). Prior algebraic treatments, most notably by \cite{fontana2024characterizing}, express these bounds using the dimensions of the simple ideals in a decomposition of the dynamical Lie algebra $\fg$. Instead, our analysis starts from the decomposition of the associated unitary representation $(\cH^{\otimes 2}, \Pi^{\otimes 2})$. Using representation theory rather than Lie algebraic methods, we can assess trainability directly from the invariant subspaces that support the state $\rho$ and the observable $\cM$. This approach avoids the need to explicitly build the Lie-algebra decomposition and removes the assumption that $\rho$ and $\cM$ must be in the Lie algebra $\fg$.

Since $(\cH,\Pi)$ is a finite-dimensional unitary representation of a
compact Lie group, both it and its tensor-product representation
$(\cH^{\otimes 2},\Pi^{\otimes 2})$ are completely reducible. The Hilbert
space and its tensor square therefore admit the decompositions
$\cH=\bigoplus_\lambda m^\prime_\lambda \cH_\lambda$ and
\begin{equation}
  \cH^{\otimes 2} = \bigoplus_{\lambda} m_{\lambda}V_\lambda.
  \label{eq:tensor-decomp}
\end{equation}
Here $\lambda$ indexes the distinct, pairwise non-isomorphic irreducible
representations of the group; $\cH_\lambda$ and $V_\lambda$ are the
corresponding irreducible invariant subspaces of $\cH$ and
$\cH^{\otimes 2}$, with multiplicities $m^\prime_\lambda$ and $m_\lambda$. When a
multiplicity exceeds one, the decomposition contains several isomorphic
copies of the same subspace; by unitarity all of these invariant
subspaces, isomorphic copies included, are mutually orthogonal.
Consequently, for any operator $A\in\End{\cH^{\otimes 2}}$, the
second-moment (Haar-twirl) operator can be written as
\begin{equation}
   \int_G \Pi(g)^{\otimes 2} A (\Pi(g)^\dagger)^{\otimes 2} d\mu_G(g) = \sum_{\lambda} \sum_{a,b=1}^{m_\lambda}\frac{\Tr{A\cJ^\dagger_{\lambda,ab}}}{\dim V_\lambda}\cJ_{\lambda, ab},
  \label{eq:second-moment}
\end{equation}
where $\mu_G$ is the Haar measure on $G$ and $\cJ_{\lambda,ab}=\sum_{m=1}^{\dim V_\lambda} \lvert\phi_{\lambda,b,m}\rangle\langle\phi_{\lambda,a,m}\rvert$ maps the $a$-th copy of $V_\lambda$ onto its $b$-th copy, with $\{\lvert\phi_{\lambda,a,m}\rangle\}$ an orthonormal basis of the $a$-th copy (and analogously for $b$); the bases are chosen such that $\Pi(g)^{\otimes 2}$ has the same matrix on every copy of $V_\lambda$, so that the intertwiner $\cJ_{\lambda,ab}$ commutes with $\Pi(g)^{\otimes 2}$ (Supplementary \cref{sec:prelim}). The orthogonal projectors onto these irreducible components may be constructed explicitly through Weingarten calculus \cite{collins2022weingarten}.

Our analysis rests on the convergence of the parameterized circuits of \cref{main:ansatz}. Let $\mathbb{U}_\pm$ denote the distributions induced by the circuit segments $U_\pm$ when their parameters are drawn uniformly from $[-\pi,\pi]^{\mathrm{length}(U_\pm)}$. By \cref{main:express:thm} (\cref{main:text:subsec:express}), for any $\varepsilon\in(0,1)$, there exists a threshold depth $L^{\ast}$ such that any circuit segment of at least $L^{\ast}$ layers induces an ensemble whose second-moment operator is $\varepsilon$-close to that of the Haar measure on $\Pi(G)$. Choosing the total depth $L\geq 2L^{\ast}$ guarantees that, for every partition of the ansatz into $U_-$ and $U_+$, at least one of ${U}_\pm$ forms an $\varepsilon$-approximate $2$-design. We assume $L\geq 2L^{\ast}$ throughout.

The expectation of the partial derivative vanishes at any depth when $\exp(2\pi\,\pi_*(g_j))$ is a multiple of the identity, as for the gates of \cref{fig:variance-bounds,fig:placeholder} and for Pauli generators $\pi_*(g_j)=-\tfrac{i}{2}P$, and when $\bbU_-$ or $\bbU_+$ is a $2$-design (\cref{lemma:expectation_main}); its variance is then the second moment. The results presented below are derived for exact $2$-designs, and we assume that $\varepsilon$ is small enough for them to apply. If $\varepsilon$ is measured in the norm induced by the Schatten-$1$ norm on $\End{\cH^{\otimes 2}}$, H\"older's inequality shows that replacing an $\varepsilon$-approximate segment by the Haar measure on $\Pi(G)$ changes the second moment by at most $4\varepsilon(\iinfsnorm{\pi_*(g_j)})^2(\iinfsnorm{\cM})^2$; the lower bounds of \cref{main:text:theory:sec:discussion} therefore remain informative only when $\varepsilon$ is well below them, that is, below an inverse polynomial in $n$. Depending on the convergence of the two ensembles, the bounds fall into three operational regimes, according to which of the ensembles $\bbU_\pm$ forms an approximate $2$-design: $\bbU_-$ only (Case~1), $\bbU_+$ only (Case~2), or both (Case~3). Below, we state the exact variance and the bounds of Case~3, with $\isnorm{\cdot}_p$ the Schatten-$p$ norm; \crefrange{main:text:subsec:exact}{main:text:subsec:lower} give the derivations, the bounds of Cases~1 and~2, and upper bounds for a rank-one state and observable.

When both ensembles are $2$-designs the variance can be evaluated in closed form, and the bounds below for that regime are consequences of this identity. We write $\cT_G(A)\defeq\int_G \Pi(g)^{\otimes 2} A\, (\Pi(g)^\dagger)^{\otimes 2} d\mu_G(g)$ for the Haar twirl of \cref{eq:second-moment}.

\begin{proposition}[Exact variance]
\label{main:text:thm:exact}
    Suppose that both $\bbU_-$ and $\bbU_+$ are $2$-designs with respect to $\Pi(G)$, and let $\Theta\defeq\cT_G(\cM^{\otimes 2})$ and $R\defeq\cT_G(\rho^{\otimes 2})$. Then
    \begin{align}
        \var&=\Tr{\Theta\,\commutator{\pi_*(g_j)\otimes\bbI}{\commutator{\bbI\otimes\pi_*(g_j)}{R}}} \nonumber\\
        &=\sum_{\lambda,\nu}\sum_{a,b}^{m_\lambda}\sum_{c,d}^{m_\nu}\frac{\Tr{\cJ_{\lambda,ab}^\dagger\cM^{\otimes 2}}\cdot\Tr{\cJ_{\nu,cd}^\dagger\rho^{\otimes 2}}}{\dim V_\lambda\cdot\dim V_\nu}\Tr{\cJ_{\lambda, ab}\commutator{\pi_*(g_j)\otimes \bbI}{\commutator{\bbI\otimes \pi_*(g_j)}{\cJ_{\nu, cd}}}} \label{main:text:eq:exact}\\
        &=-\frac14\Tr{\Theta\,\commutator{\Delta_j}{\commutator{\Delta_j}{R}}},\qquad \Delta_j\defeq\pi_*(g_j)\otimes\bbI-\bbI\otimes\pi_*(g_j). \label{main:text:eq:exact:delta}
    \end{align}
\end{proposition}
Its proof is given in \cref{main:text:subsec:exact}. The first line of \cref{main:text:thm:exact} is a standard identity of Weingarten type~\cite{collins2022weingarten}; the contribution of the proposition is its evaluation through the decomposition of \cref{eq:tensor-decomp} in the second line, on which the bounds below and the trainability criteria of \cref{main:text:theory:sec:discussion} are built. When $i\cM\in\pi_*(\fg)$, the variance is also given by the Lie-algebraic formula of \cite{fontana2024characterizing} (Supplementary \cref{sec:comparison_prior_works}); the corresponding formula for the variance of the loss function itself is given in \cite{ragone2024lie}.

\subsubsection{Upper bound}
\label{sec:upper-bound}
This section gives the upper bound in the general case, in which both $U_-$ and $U_+$ converge to $2$-designs (Case~3); this is so for every gate with at least $L^\ast$ layers on each side. The edge cases, in which only one of the two segments converges (Cases~1 and~2, for a gate with fewer than $L^\ast$ layers on one side), and the bounds for a rank-one state and observable are given, with their derivations, in \cref{main:text:subsec:upper}. When both segments converge, the Haar integrals close on both sides, and the
bound decouples into independent state and observable contributions:
\begin{align}
   \var\leq  4\left(\sum_{\lambda^\prime}\sum_{c,d} \frac{\abs{\Tr{\rho^{\otimes 2} \cJ_{\lambda^\prime, cd} }}}{\sqrt{\dim V_{\lambda^\prime}}}\right)\cdot \left(\sum_{\lambda}\sum_{a,b}\frac{\abs{\Tr{\cM^{\otimes 2} \cJ_{\lambda, ab}} }}{\sqrt{\dim V_\lambda}}\right)\cdot \left(\norm{\pi_*(g_j)}^S_{\infty}\right)^2. \label{main:text:eq:upperbound_case3}
\end{align}

\begin{remark}[Observables with the same derivative]
\label{main:text:rem:reduced}
    The observable $\cM$ is arbitrary. The derivative of \cref{main:text:PD} is unchanged, for every value of the parameters, (i) when an operator that commutes with $\Pi(G)$, for instance, a multiple of the identity, is added to $\cM$, and (ii) when $\cM$ is replaced by its compression $P_A\cM P_A$ to an invariant subspace $\cH_A\subseteq\cH$, with orthogonal projector $P_A$, that supports the state, $P_A\rho P_A=\rho$. Neither operation changes the variance or the expressions of \cref{main:text:thm:exact}, but both change the right-hand sides of the upper and lower bounds, and each of these observables gives a valid bound on the variance of $\cM$. Unless otherwise stated, we therefore evaluate the upper bounds for the \emph{reduced observable} of $\cM$, $\cM-\int_G\Pi(g)\cM\Pi(g)^\dagger d\mu_G(g)$, which has no component in the commutant of $\Pi(G)$ and equals $\cM-\iTr{\cM}\,\bbI/\dim\cH$ for irreducible $\Pi$.
\end{remark}

When a multiplicity $m_\lambda$ exceeds one, the decomposition into copies of $V_\lambda$ is not unique, and the right-hand sides also depend on which copies are chosen; every choice gives a valid bound. Moreover, since $\abs{\partial_{\theta_{\ell,j}}C}\leq2\,\snorm{\pi_*(g_j)}_\infty\snorm{\cM}_\infty$ for all parameter values, the variance never exceeds the trivial bound $4(\snorm{\pi_*(g_j)}_\infty)^2(\snorm{\cM}_\infty)^2$, and the upper bounds are informative only where they lie below it.

\subsubsection{Lower bound}
\label{sec:lower-bound}

For the projected operators of the lower bound, we write
\begin{align*}
    \rho_{\lambda, a, 1 \to \lambda, b, 2}^{\,(\lambda^\prime, a^\prime) \otimes (\nu^\prime, b^\prime)} \defeq P_{\lambda^\prime_{\lambda, a, 1}, a^\prime_{\lambda, a, 1}}\otimes P_{\nu^\prime_{\lambda, a, 1}, b^\prime_{\lambda, a, 1}} \rho^{\otimes 2} P_{\lambda^\prime_{\lambda, b, 2}, a^\prime_{\lambda, b, 2}}\otimes P_{\nu^\prime_{\lambda, b, 2}, b^\prime_{\lambda, b, 2}},
\end{align*}
where the
$P_{\lambda^\prime, a^\prime}\otimes P_{\nu^\prime, b^\prime}$ are the orthogonal projectors onto the corresponding tensor-product spaces $\cH_{\lambda^\prime, a^\prime}\otimes \cH_{\nu^\prime, b^\prime}$, the subscripts $(\lambda,a,1)$ and $(\lambda,b,2)$ label the tensor-product spaces that contain the $a$-th and the $b$-th copy of $V_\lambda$, respectively, and $\Mproj$ is defined analogously with $\cM^{\otimes 2}$ in place of $\rho^{\otimes 2}$. Here $\cH_{\lambda^\prime, a^\prime}$ is the $a^\prime$-th copy of $\cH_{\lambda^\prime}$ in $\cH$, and the copies of each $V_\lambda$ are those of the Clebsch--Gordan decomposition of these tensor-product spaces, so that every copy lies in exactly one of them.

Again, with simultaneous
convergence of $U_\pm$ the variance is given by \cref{main:text:eq:exact}; with $H\defeq\pi_*(g_j)$, its lower bound satisfying (\cref{main:text:subsec:lower})
\begin{align}
    \begin{split}
        \var \geq&\abs{\sum_{\lambda}\sum_{a,b,c,d}^{m_\lambda}\frac{\Tr{\cJ_{\lambda,ab}^\dagger\cM^{\otimes 2}}\cdot \Tr{\cJ_{\lambda,cd}^\dagger\rho^{\otimes 2}}}{(\dim V_\lambda)^2}\Tr{\cJ_{\lambda, ab}\commutator{H\otimes \bbI}{\commutator{\bbI\otimes H}{\cJ_{\lambda, cd}}}}} \\
        &\hspace{0.1cm}-2\sum_{\nu}\sum_{c,d}^{m_\nu}\frac{\snorm{\cJ_{\nu, dd}\sqrt{{\rho^{\otimes 2}}}}_2\snorm{\cJ_{\nu, cc}\sqrt{{\rho^{\otimes 2}}}}_2}{\dim V_\nu}\\
        &\hspace{0.3cm}\times\abs{\Tr{\left(\sum_{\lambda\neq \nu}\sum_{a,b}^{m_\lambda}\frac{\Tr{\cJ^\dagger_{\lambda, ab} \cM_{\nu, c, 1 \to \nu, d, 2}^{\,(\lambda^\prime, a^\prime) \otimes (\nu^\prime, b^\prime)}}}{\dim V_\lambda}\cJ_{\lambda, ab}^{(\lambda^\prime, a^\prime, \nu^\prime, b^\prime)_{\nu, c, 1}, \; (\lambda^\prime, a^\prime, \nu^\prime, b^\prime)_{\nu, d, 2}}\right) (H \otimes \bbI) \cJ_{\nu, cd} (H\otimes \bbI)}},
    \end{split} 
    \label{main:text:eq:lowerbound_case3}
\end{align}
where 
\begin{align*}
    &\cJ_{\lambda, ab}^{(\lambda^\prime, a^\prime, \nu^\prime, b^\prime)_{\nu, c, 1}, \; (\lambda^\prime, a^\prime, \nu^\prime, b^\prime)_{\nu, d, 2}}
    \defeq P_{\lambda^\prime_{\nu, c, 1}, a^\prime_{\nu, c, 1}}\otimes P_{\nu^\prime_{\nu, c, 1}, b^\prime_{\nu, c, 1}} \cJ_{\lambda, ab} P_{\lambda^\prime_{\nu, d, 2}, a^\prime_{\nu, d, 2}}\otimes P_{\nu^\prime_{\nu, d, 2}, b^\prime_{\nu, d, 2}} \\
    =& \begin{cases}
        J_{\lambda, ab}  & V_{\lambda, a} \subseteq  \cH_{\lambda^\prime_{\nu, d, 2}, a^\prime_{\nu, d, 2}}\otimes \cH_{\nu^\prime_{\nu, d, 2}, b^\prime_{\nu, d, 2}} \text{ and } V_{\lambda, b} \subseteq \cH_{\lambda^\prime_{\nu, c, 1}, a^\prime_{\nu, c, 1}}\otimes \cH_{\nu^\prime_{\nu, c, 1}, b^\prime_{\nu, c, 1}} \\
        0 & o.w.
    \end{cases}
\end{align*}denotes the restricted intertwiner obtained by sandwiching $\cJ_{\lambda,ab}$ between the projectors onto the Clebsch--Gordan factors $\cH_{\lambda^\prime_{\nu, c, 1}, a^\prime_{\nu, c, 1}}\otimes \cH_{\nu^\prime_{\nu, c, 1}, b^\prime_{\nu, c, 1}}$ and $\cH_{\lambda^\prime_{\nu, d, 2}, a^\prime_{\nu, d, 2}}\otimes \cH_{\nu^\prime_{\nu, d, 2}, b^\prime_{\nu, d, 2}}$. These factors carry the $c$-th and $d$-th copies of $V_\nu$, respectively. The terms $V_{\lambda, a}$ and $V_{\lambda, b}$ denote the $a$-th and $b$-th copies of $V_{\lambda}$. Finally, $\cM^{(\lambda',a')\otimes(\nu',b')}_{\nu,c,1\to\nu,d,2}$ is the correspondingly restricted image of $\cM^{\otimes 2}$. Full derivations of the upper bounds in \cref{main:text:eq:upperbound_case1,main:text:eq:upperbound_case2,main:text:eq:upperbound_case3} and the lower bounds in \cref{main:text:eq:lowerbound_case1,main:text:eq:lowerbound_case2,main:text:eq:lowerbound_case3} are given in the Supplementary Information.

\subsection{Trainability criteria and representative regimes}
\label{main:text:theory:sec:discussion}

We now examine the physical implications of these results. Throughout this discussion, we assume that both the observable $\cM$ and the generator $\pi_*(g_j)$ have constant Schatten-infinity norm, i.e., $\iinfsnorm{\cM} = \cO(1)$ and $\iinfsnorm{\pi_*(g_j)}=\cO(1)$; this matters because several of the objects in the bounds above carry an implicit $n$-dependent scale, and a variance that grows with $n$ under one convention may decay under another. In the upper bound, we take $\cM$ to be reduced in the sense of \cref{main:text:rem:reduced}; traceless when $\Pi$ is irreducible, unless stated otherwise, because the derivative does not depend on the component of $\cM$ in the commutant, whereas the bound does. For simplicity, we further restrict ourselves to the regime in which both $U_\pm$ form $2$-designs with respect to $\Pi(G)$, so that \cref{main:text:thm:exact} and the bounds of \cref{main:text:eq:upperbound_case3,main:text:eq:lowerbound_case3} apply.

\begin{figure}[t]
  \centering
  \includegraphics[scale=0.4]{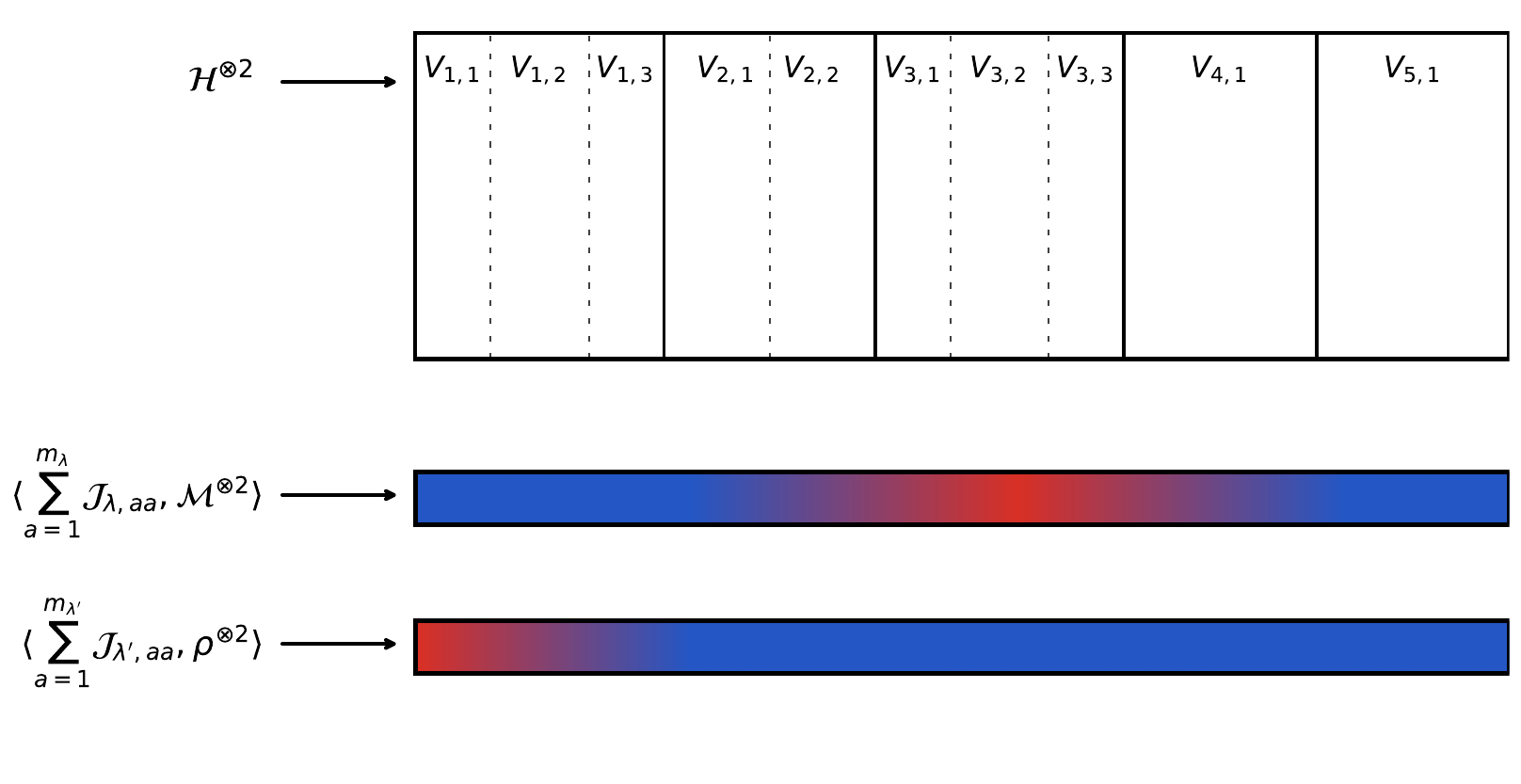}
  \caption{An illustrative example of the conditions for a non-vanishing upper bound. The tensor product space $\cH^{\otimes2}$ decomposes into five irreducible representations $V_j$, where $V_{j,k}$ denotes the $k$-th isomorphic copy of $V_j$; the dimensions of $V_1$ and $V_3$ are assumed to scale polynomially with system size. The bars encode the normalized magnitude of the inner product $\bigl\langle\sum_{a=1}^{m_{\lambda^\ast}}[\dots],\cdot\bigr\rangle$, with deep red indicating strictly dominant values and blue indicating negligible ones.}
  \label{fig:upper-bound}
\end{figure}

\paragraph{When the upper bound certifies a plateau}
When the commutant of $\Pi(G)^{\otimes2}$ has polynomial dimension, ensuring that \cref{main:text:eq:upperbound_case3} is $\Omega(1/\poly{n})$ requires the product of three quantities to remain at least \emph{inverse-polynomially} large in the number of qubits $n$: the state overlap $\isnorm{\sum_{\lambda}\sum_{a,b}^{m_\lambda} \iTr{\cJ_{\lambda,ab}^{\dagger}\rho^{\otimes2}}\cJ_{\lambda,ab}/\dim V_\lambda}_2$, the corresponding observable overlap, and the Schatten-infinity norm of the generator $\iinfsnorm{\pi_*(g_j)}$ (Supplementary \cref{sec:criteria:upper}). A sufficient condition is the existence of irreducible sub-representations $V_\lambda,V_{\lambda'}$ of polynomial dimension, $\dim V_\lambda,\dim V_{\lambda'}=\cO(\poly{n})$, on which the Hilbert--Schmidt overlaps $\bigl\langle\sum_{a=1}^{m_\lambda}\cJ_{\lambda,aa},\cM^{\otimes2}\bigr\rangle_{HS}$ and $\bigl\langle\sum_{a=1}^{m_{\lambda'}}\cJ_{\lambda',aa},\rho^{\otimes2}\bigr\rangle_{HS}$ are dominant. \Cref{fig:upper-bound} illustrates this condition schematically.

In the absence of such polynomially scaling subspaces for the state, that is, when $\rho^{\otimes2}$ has only exponentially small weight on the irreducible representations below some exponentially large dimension, the state overlap is exponentially small, and the upper bound suppresses the variance to exponentially small values unless the observable overlap or the dimension of the commutant of $\Pi(G)^{\otimes2}$ grows exponentially (Supplementary \cref{sec:criteria:upper}). Under the standard representation of $\SU{2^n}$, for instance, the tensor square splits into symmetric and alternating parts, both of dimension $\cO(2^{2n})$; the state and observable trace contributions then scale as $\cO(1/2^n)$ and $\cO(1)$, and \cref{main:text:eq:upperbound_case3} yields
\begin{align}
  \mathrm{Var}\bigl[\partial_{\theta_{\ell,j}}C\bigr]\;=\;\cO\!\left(2^{-n}\right).
  \label{eq:sun-upper}
\end{align}A similar example regarding the unitary representation of $\Spin{2n}$ (also known as free fermions) with non-local observables $\Outer{0^{\otimes n}}{0^{\otimes n}}$ can be found in \cref{app:example:ff:non-local}.

It is worth stating how \cref{eq:sun-upper} compares with what is already known. For this ansatz, the variance is known exactly. For a pure state and a traceless observable, \cref{main:text:thm:exact} or, equivalently, the Lie-algebraic formula of \cite{fontana2024characterizing}, since every traceless observable lies in $i\,\su{2^n}$, gives, with $d=2^n$ (Supplementary \cref{app:example:SUN}),
\begin{align}
  \mathrm{Var}\bigl[\partial_{\theta_{\ell,j}}C\bigr]\;=\;\frac{2\,(\snorm{\pi_*(g_j)}_2)^2\,(\snorm{\cM}_2)^2}{(d^2-1)(d+1)},
  \label{eq:sun-exact}
\end{align}
which, for a Pauli observable and a Pauli generator ($\iinfsnorm{\cM}=\iinfsnorm{\pi_*(g_j)}=1$), equals $2d^2/((d^2-1)(d+1))\approx2\cdot2^{-n}$. For the same choice, \cref{main:text:eq:upperbound_case3} evaluates to $4/(d+1)+4/\sqrt{d^2-1}\approx8\cdot2^{-n}$. For the canonical example, our bound therefore has the correct exponent and overestimates the variance by a factor that tends to four; it is not exponentially weaker than the Lie-algebraic result. The remaining loss has two sources: \cref{main:text:eq:upperbound_case3} separates the state and observable contributions by triangle inequality \emph{before} either is evaluated, so it cannot register cancellations between them; and each factor is then controlled by a Schatten norm rather than by its exact projection onto the commutant. On the other hand, \cref{main:text:eq:upperbound_case3} requires no relationship whatsoever between $\cM$ and $\fg$, whereas the exact compact-group formula of \cite[Thm.~2.9]{fontana2024characterizing} presumes $i\cM=d\phi(\fm)$ for some $\fm\in\fg$. The bound is this tight only because two irreducible subspaces carry everything here. When many do, the sum of absolute values in \cref{main:text:eq:upperbound_case3} can exceed even the trivial bound, as the collective-spin example below shows.

\paragraph{Why the upper bound cannot be inverted}
An exponentially vanishing upper bound certifies a barren plateau, but an inverse-polynomially large upper bound does \emph{not} by itself rule one out. Because the derivation evaluates the state and measurement contributions independently, the bound overestimates the variance whenever $\rho$ and $\cM$ possess disjoint geometries. The failure is not marginal, and it admits an exact witness. Consider the direct-product group $\SU{d}\times\SU{d'}$ acting on $\bbC^{d}\oplus\bbC^{d'}\oplus\bbC^{2^n-d-d'}$, trivially on the last summand, with a pure state $\rho$ and a rank-one projector $\cM$ supported on mutually orthogonal summands $\bbC^{d}$ and $\bbC^{d'}$. Every trace connecting the two summands vanishes
identically, by orthogonality of the corresponding projectors, so that
\begin{align}
  \mathrm{Var}\bigl[\partial_{\theta_{\ell,j}}C\bigr]\;=\;0,
  \qquad\text{while}\qquad
  \text{\cref{main:text:eq:upperbound_case3}}\;=\;\frac{8\,(\snorm{\pi_*(g_j)}_\infty)^2}{\sqrt{d(d+1)\,d'(d'+1)}}.
  \label{eq:separation-witness}
\end{align}
The bound is of order $1/(dd')$. Choosing $d,d'=\poly{n}$ makes it inverse-polynomially large while the true variance is exactly zero; both $d$ and $d'$ must be polynomial in $n$, since the bound is exponentially small as soon as either of them is exponentially large. \Cref{eq:separation-witness} is evaluated for the projector itself; for the reduced observable, the bound is of the same order. This particular overestimate disappears when $\cM$ is compressed to the invariant subspace $\bbC^{d}$ that supports $\rho$ (\cref{main:text:rem:reduced}(ii)), but an inverse-polynomially large upper bound for a variance that is exactly zero also occurs for an observable that is reduced and cannot be compressed. For $\SU{d_1}\times\SU{d_2}$, which acts irreducibly on $\bbC^{d_1}\otimes\bbC^{d_2}$, with $d_1=\poly{n}$, a product state, an observable $\cM_1\otimes\bbI$ on the first factor with $\cM_1$ traceless and $\cM_1^2=\bbI$, and a differentiated gate generated by $\bbI\otimes H_2$ on the second, the derivative vanishes identically, and yet \cref{main:text:eq:upperbound_case3} is of order $1/d_1$: the bound sees the generator only through its norm, and no sharpening of the constants repairs it. A detailed derivation of both examples is given in Supplement~\ref{sec:S72}.

\paragraph{What the lower bound certifies}
For a rigorous guarantee \emph{against} barren plateaus, one must therefore turn to \cref{main:text:eq:lowerbound_case3}. A sufficient condition for non-vanishing variance is the existence of an invariant subspace $V_\lambda$ with $\dim V_\lambda=\poly{n}$ on which both the state and the observable carry large trace overlaps, $\iTr{\cJ^{\dagger}_{\lambda,aa}\cM^{\otimes2}}$ and $\iTr{\cJ^{\dagger}_{\lambda,aa}\rho^{\otimes2}}$, provided that the differentiated generator moves this subspace by an inverse-polynomial amount, that is, $\bigl(\isnorm{(\bbI-\cJ_{\lambda,aa})(\pi_*(g_j)\otimes\bbI)\cJ_{\lambda,aa}}_2\bigr)^2\geq(\dim V_\lambda)^2/\poly{n}$, and that the subtractive cross terms vanish. The isolated subspace contribution, the first term of \cref{main:text:eq:lowerbound_case3}, then scales inverse-polynomially. Neither proviso holds automatically: the cross terms need not vanish, and the first term of \cref{main:text:eq:lowerbound_case3} contains the traces $\iTr{\cJ_{\lambda,ab}\icommutator{\pi_*(g_j)\otimes\bbI}{\icommutator{\bbI\otimes\pi_*(g_j)}{\cJ_{\lambda,cd}}}}$; for $a=b=c=d$ such a trace equals $2\bigl(\isnorm{(\bbI-\cJ_{\lambda,aa})(\pi_*(g_j)\otimes\bbI)\cJ_{\lambda,aa}}_2\bigr)^2$, which vanishes when $\pi_*(g_j)\otimes\bbI$ maps the copy into itself.

\begin{figure}[t]
  \centering
  \includegraphics[scale=0.4]{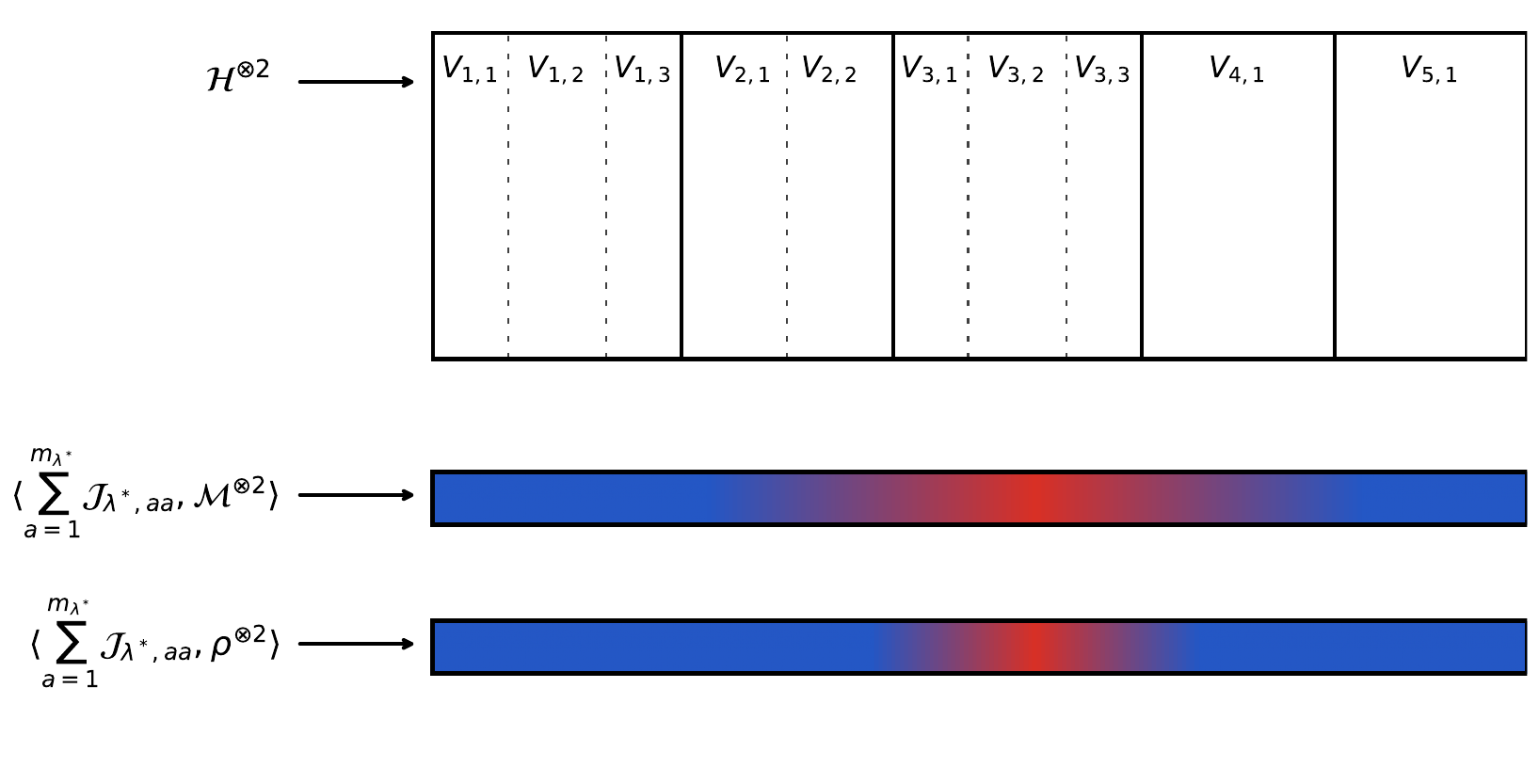}
  \caption{Conditions for a non-vanishing \emph{lower} bound under the
  concentrated support assumptions. The supports of $\rho^{\otimes2}$ and
  $\cM^{\otimes2}$ are concentrated on a common polynomial-dimensional sector
  $m_{\lambda^\ast}V_{\lambda^\ast}$, so the isolated contribution dominates and
  the cross-sector interference is negligible.}
  \label{fig:lower-bound}
\end{figure}

This is not the only route to a variance that is at least inverse-polynomial, and the more interesting route is the one in which the cross terms do \emph{not} vanish but cancel. When the supports of $\rho^{\otimes2}$ and $\cM^{\otimes2}$ are concentrated on a common polynomial-dimensional sector $m_{\lambda^\ast}V_{\lambda^\ast}$, the isolated contribution dominates and the cross terms are small (\cref{main:text:subsec:concentration}). \Cref{fig:lower-bound} illustrates this concentration, and \cref{main:text:thm:sufficient,main:text:cor:leakage} below states the conditions precisely.

\begin{theorem}[Lower bound under concentration]
\label{main:text:thm:sufficient}
  Let $\bbU_\pm$ be $2$-designs with respect to $\Pi(G)$, let the copies of each $V_\lambda$ be those of the Clebsch--Gordan decomposition, and let $\lambda^\ast$ be a label such that $\rho^{\otimes2}$ is supported on $m_{\lambda^\ast}V_{\lambda^\ast}$. Write $H\defeq\pi_*(g_j)$, $\Theta\defeq\cT_G(\cM^{\otimes2})$ and $P_{\lambda^\ast}\defeq\sum_{a=1}^{m_{\lambda^\ast}}\cJ_{\lambda^\ast,aa}$ for the orthogonal projector onto $m_{\lambda^\ast}V_{\lambda^\ast}$, let
  \begin{align*}
    I_{\lambda^\ast}\defeq\sum_{a,b,c,d=1}^{m_{\lambda^\ast}}\frac{\Tr{\cJ_{\lambda^\ast,ab}^{\dagger}\cM^{\otimes2}}\cdot\Tr{\cJ_{\lambda^\ast,cd}^{\dagger}\rho^{\otimes2}}}{(\dim V_{\lambda^\ast})^2}\Tr{\cJ_{\lambda^\ast,ab}\commutator{H\otimes\bbI}{\commutator{\bbI\otimes H}{\cJ_{\lambda^\ast,cd}}}}
  \end{align*}
  be the term $\lambda=\lambda^\ast$ of the sum in \cref{main:text:theory:isolated:terms}, and let $s_c\defeq\isnorm{\cJ_{\lambda^\ast,cc}\sqrt{\rho^{\otimes2}}}_2$. Then
  \begin{align}
    \var\geq\abs{I_{\lambda^\ast}}-\frac{2}{\dim V_{\lambda^\ast}}\sum_{c,d=1}^{m_{\lambda^\ast}}s_cs_d\,\abs{\Tr{(\bbI-P_{\lambda^\ast})\Theta\,(H\otimes\bbI)\,\cJ_{\lambda^\ast,cd}\,(H\otimes\bbI)}},
    \label{main:text:eq:concentration}
  \end{align}
  where the second term is at most $2(\iinfsnorm{H})^2m_{\lambda^\ast}\iinfsnorm{(\bbI-P_{\lambda^\ast})\cM^{\otimes2}(\bbI-P_{\lambda^\ast})}$ and vanishes, with $\var=I_{\lambda^\ast}$, when $\cM^{\otimes2}$ is supported on $m_{\lambda^\ast}V_{\lambda^\ast}$ as well. In particular, if the second term is at most $\abs{I_{\lambda^\ast}}/2$ and $\abs{I_{\lambda^\ast}}=\Omega(1/\poly{n})$, then $\var\geq\abs{I_{\lambda^\ast}}/2=\Omega(1/\poly{n})$, and there is no barren plateau.
\end{theorem}

\begin{corollary}[Lower bound with leakage]
\label{main:text:cor:leakage}
  Let $\bbU_\pm$ form $2$-designs with respect to $\Pi(G)$, let the copies of each $V_\lambda$ be those of the Clebsch--Gordan decomposition, let $\lambda^\ast$ be an arbitrary label, and let $H$, $\Theta$, $P_{\lambda^\ast}$, $I_{\lambda^\ast}$ and $s_c$ be defined as in \cref{main:text:thm:sufficient}. Then
  \begin{align}
    \begin{split}
      \var\geq{}&\abs{I_{\lambda^\ast}}-\frac{2}{\dim V_{\lambda^\ast}}\sum_{c,d=1}^{m_{\lambda^\ast}}s_cs_d\,\abs{\Tr{(\bbI-P_{\lambda^\ast})\Theta\,(H\otimes\bbI)\,\cJ_{\lambda^\ast,cd}\,(H\otimes\bbI)}}\\
      &-(\iinfsnorm{H})^2(\iinfsnorm{\cM})^2\sum_{\nu\neq\lambda^\ast}(4+2m_\nu)\,\iTr{\textstyle\sum_{c=1}^{m_\nu}\cJ_{\nu,cc}\,\rho^{\otimes2}},
    \end{split}
    \label{main:text:eq:leakage}
  \end{align}
  where $\iTr{\sum_{c}\cJ_{\nu,cc}\,\rho^{\otimes2}}$ is the weight of $\rho^{\otimes2}$ on $m_\nu V_\nu$. In particular, if the two subtracted terms in \cref{main:text:eq:leakage} sum to at most $\iabs{I_{\lambda^\ast}}/2$ and $\iabs{I_{\lambda^\ast}}=\Omega(1/\poly{n})$, then $\var\geq\iabs{I_{\lambda^\ast}}/2=\Omega(1/\poly{n})$, and there is no barren plateau.
\end{corollary}

The proofs of \cref{main:text:thm:sufficient,main:text:cor:leakage} are given in Supplementary \cref{sec:criteria:lower}, where each term of \cref{main:text:eq:concentration} is related to a step of the argument in \cref{main:text:subsec:concentration}. The isolated contribution $I_{\lambda^\ast}$ is built from the trace overlaps of $\rho^{\otimes2}$ and $\cM^{\otimes2}$ with the copies of $V_{\lambda^\ast}$. For $m_{\lambda^\ast}=1$ it equals $2\,\iTr{\cJ_{\lambda^\ast,11}\cM^{\otimes2}}\iTr{\cJ_{\lambda^\ast,11}\rho^{\otimes2}}\bigl(\isnorm{(\bbI-\cJ_{\lambda^\ast,11})(H\otimes\bbI)\cJ_{\lambda^\ast,11}}_2\bigr)^2/(\dim V_{\lambda^\ast})^2$, the product of the two trace overlaps and of a factor that measures how far $H\otimes\bbI$ moves $V_{\lambda^\ast}$; it vanishes when $H\otimes\bbI$ maps $V_{\lambda^\ast}$ into itself, however large the overlaps are. Moreover $\abs{I_{\lambda^\ast}}\leq4(\iinfsnorm{H})^2(\ionesnorm{\cM})^2/\dim V_{\lambda^\ast}$, so that, for an observable whose trace norm is polynomially bounded, such as a projector onto a pure state, the isolated contribution can be inverse-polynomial only if $\dim V_{\lambda^\ast}$ is polynomial in $n$. The second term is the residual cross term built from \cref{eq:conc-crossnorm} and \cref{eq:conc-crosstrace}, with \cref{eq:conc-crosstrace} written without the restricted operators. Since $P_{\lambda^\ast}$ commutes with $\Pi(g)^{\otimes2}$ and with $\Theta$, the operator $(\bbI-P_{\lambda^\ast})\Theta=\cT_G\bigl((\bbI-P_{\lambda^\ast})\cM^{\otimes2}(\bbI-P_{\lambda^\ast})\bigr)$ depends only on the part of $\cM^{\otimes2}$ outside $m_{\lambda^\ast}V_{\lambda^\ast}$; for a rank-one projector $\cM=\Outer{\Psi}{\Psi}$, the operator norm of this part is $1-\iTr{P_{\lambda^\ast}\cM^{\otimes2}}$, which is small when the trace overlap of $\cM^{\otimes2}$ with $m_{\lambda^\ast}V_{\lambda^\ast}$ is close to one. If $\rho^{\otimes2}$ also has weight outside $m_{\lambda^\ast}V_{\lambda^\ast}$, the other sectors contribute as well, and \cref{main:text:cor:leakage} bounds their contribution by \cref{main:text:eq:leakage}. Since the multiplicities $m_\nu$ can grow exponentially with $n$ (in the collective-spin representation below, the one-dimensional irreducible representation occurs $\binom{2n}{n}/(n+1)\geq2^n/(n+1)$ times in $\cH^{\otimes2}$), the condition of \cref{main:text:cor:leakage} can require this weight to be exponentially small.

For a pure state $\rho=\Outer{\psi}{\psi}$ with $\state{\psi}$ in an irreducible invariant subspace $\cH_{\lambda^\prime,a^\prime}$ of $\cH$, $\rho^{\otimes2}$ is supported on the irreducible invariant subspace of $\cH_{\lambda^\prime,a^\prime}\otimes\cH_{\lambda^\prime,a^\prime}$ whose highest weight is twice that of $\cH_{\lambda^\prime}$ if and only if $\state{\psi}$ is a generalized coherent state~\cite{perelomov1986generalized}, that is, up to a phase, the image under some $\Pi(g)$ of a highest-weight vector of $\cH_{\lambda^\prime,a^\prime}$~\cite{lichtenstein1982system}. The assumption of \cref{main:text:thm:sufficient} on the support of $\rho^{\otimes2}$ is therefore restrictive for pure states; \cref{main:text:cor:leakage} applies to every state.

\paragraph{A worked instance: collective spin}
As a concrete demonstration of this concentration effect, we consider the $n$-qubit collective-spin representation of $\SU{2}$ in which the generators act as $\pi_*(X)=\sum_{j=1}^{n}X^{(j)}$ and analogously for $Y,Z$, so that $\cH=(\bbC^{2})^{\otimes n}$ decomposes into total-spin sectors of maximal weight $n/2$.

For this representation, one can derive the exact variance and compare it with the lower bound. Take $\rho=|\lambda^\ast\rangle\langle\lambda^\ast|$, $\lambda^\ast\geq\frac12$, a highest-weight vector, so that $\rho^{\otimes2}$ is supported on a single copy of $V_{2\lambda^\ast}$ with $\dim V_{2\lambda^\ast}=4\lambda^\ast+1=\cO(n)$, and set $\cM=S_X+S_Y+S_Z$ with $H=-iS_Z$. Spatial symmetries and swap-operator identities reduce the isolated contribution \cref{main:text:theory:isolated:terms} to exactly $(\lambda^\ast)^3/3$, while the residual cross terms \cref{eq:conc-crosstrace}, evaluated through total-spin eigenvalues, contribute $(\lambda^\ast)^3/3-2(\lambda^\ast)^2/3$ with their sign, and the variance is the difference of the two (Supplementary \cref{app:example:SU2}). The cubic leading-order terms cancel identically, leaving
\begin{align}
  \mathrm{Var}\bigl[\partial_{\theta_{\ell,j}}C\bigr]
  \;=\;\tfrac{2}{3}(\lambda^\ast)^2
  \label{eq:su2-lower}
\end{align}
for every $\lambda^\ast\geq\frac12$; the lower bound \cref{main:text:eq:lowerbound_case3} takes this value for $\lambda^\ast\geq2$ and is smaller for $\lambda^\ast<2$, where $(\lambda^\ast)^3/3-2(\lambda^\ast)^2/3$ is negative. Two remarks are in order. First, \cref{eq:su2-lower} is stated in the unnormalized convention in which $\iinfsnorm{\cM}=\Theta(n)$ and $\iinfsnorm{H}=\Theta(n)$; the apparent \emph{growth} of the variance with system size is an artifact of that convention. Restoring the normalization fixed at the beginning of this section, and using $\lambda^\ast=\Theta(n)$, the dimensionless variance is
\begin{align}
  \frac{\Var{\partial_{\theta_{\ell,j}}C}}
       {(\iinfsnorm{\cM})^{2}\,(\infsnorm{H})^{2}}
  \;=\;\Omega\!\left(n^{-2}\right),
  \label{eq:su2-lower-normalized}
\end{align}
which is inverse-polynomial, not constant, and therefore still rules out a barren plateau in the sense of \cref{main:def:BP}. Second, and more importantly, the exact cancellation of the cubic terms is what distinguishes this example: the isolated contribution and the cross terms are individually of order $(\lambda^\ast)^3$ and the surviving signal is a subleading $(\lambda^\ast)^2$. Discarding the cross terms would give $(\lambda^\ast)^3/3$, and bounding the variance by the sum of the absolute values of the two contributions would give $(\lambda^\ast)^3/3+\abs{(\lambda^\ast)^3/3-2(\lambda^\ast)^2/3}$; for $\lambda^\ast>2$, both exceed the variance $2(\lambda^\ast)^2/3$, by the factors $\lambda^\ast/2$ and $\lambda^\ast-1$, respectively, so that the cancellation between the two contributions must be kept. This example illustrates the mechanism of the lower bound; it does not go beyond the algebraic theory, because $i\cM\in\pi_*(\su{2})$ and the Lie-algebraic formula of \cite{fontana2024characterizing} yields the same value (Supplementary \cref{app:example:SU2}). In addition, we observe a similar phenomenon when applying our bounds to the unitary representation of $\Spin{2n}$ with a local observable $Z^{(1)}$, which also lies in the algebra; full details are provided in \cref{app:example:ff:local}.

\begin{figure}[t]
    \centering
    \begin{subfigure}[t]{0.49\textwidth}
        \centering
        \includegraphics[width=\linewidth]{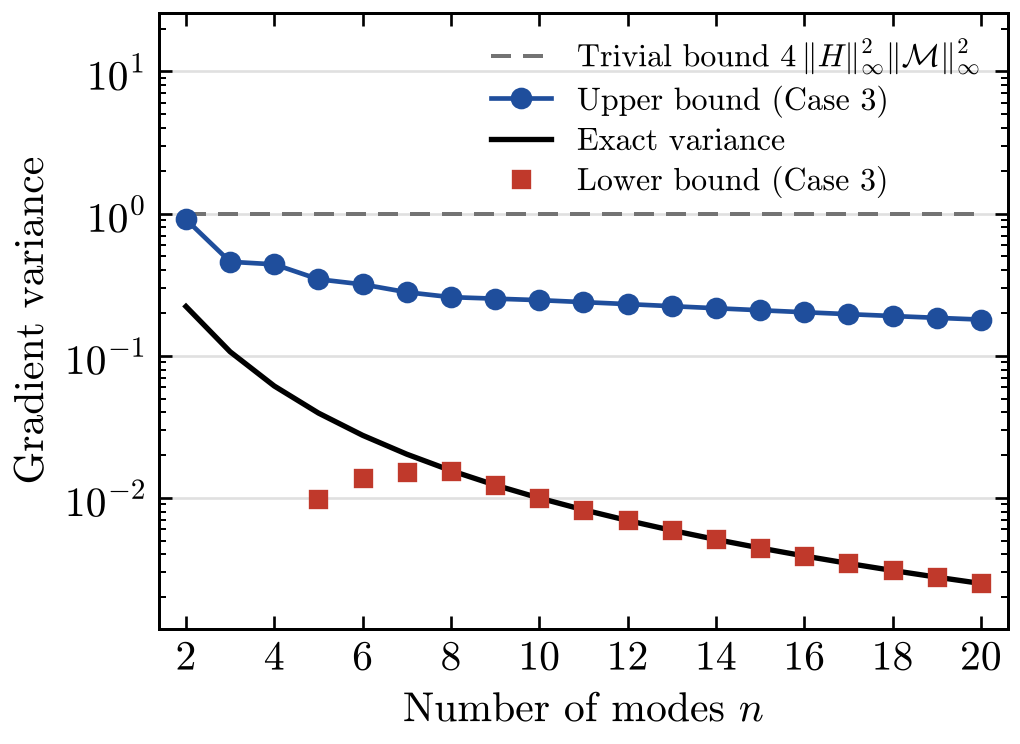}
        \caption{}
        \label{fig:free:fermion:bounds}
    \end{subfigure}\hfill
    \begin{subfigure}[t]{0.49\textwidth}
        \centering
        \includegraphics[width=\linewidth]{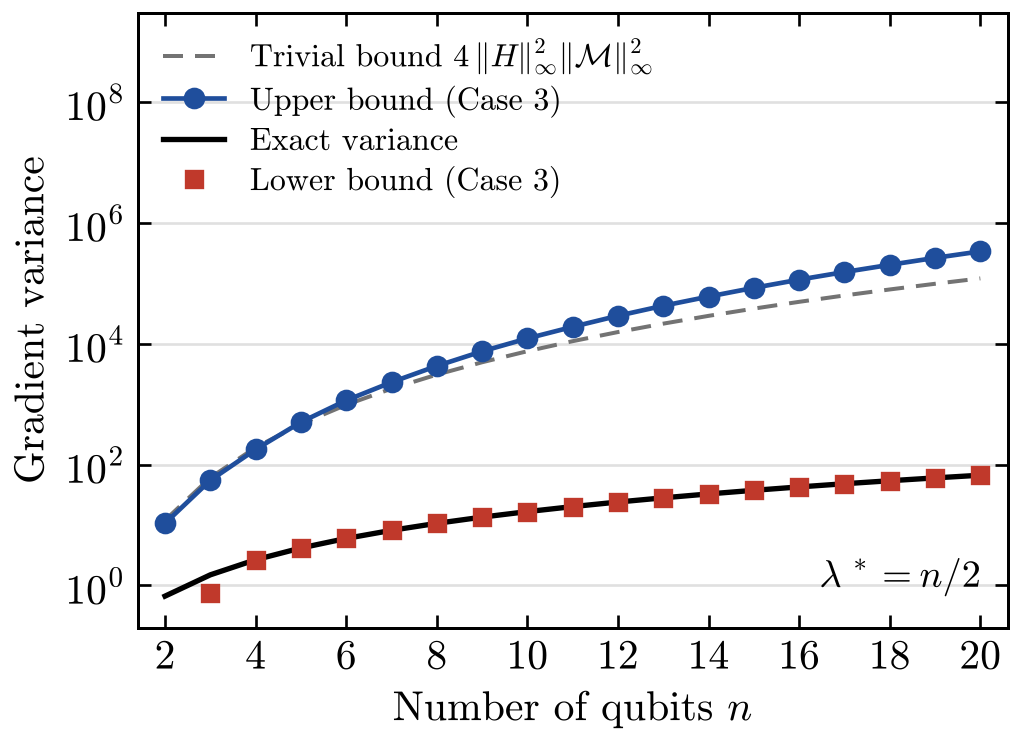}
        \caption{}
        \label{fig:su2-collective-spin-bounds}
    \end{subfigure}
    \caption[Exact variance and bounds for the free-fermion and collective-spin examples]{%
        Exact variance (solid line), lower bound of \cref{main:text:eq:lowerbound_case3} (squares), upper bound of \cref{main:text:eq:upperbound_case3} (circles) and trivial bound $4(\norm{H}^S_{\infty})^2(\norm{\cM}^S_{\infty})^2$ (dashed) on a logarithmic scale, for $2\leq n\leq 20$. Both bounds are evaluated after compression to the irreducible sub-representation that carries $\rho$ (\cref{main:text:rem:reduced}(ii)); the formulas below are derived in Supplementary \cref{app:example:SU2,app:example:ff:local}.
        (a)~Free-fermion example with $\rho=\Outer{0^{\otimes n}}{0^{\otimes n}}$,
        $\cM=Z^{(1)}=-i\gamma_1\gamma_2$ and $H=\frac{1}{2}\gamma_1\gamma_2$, so that $\norm{H}^S_{\infty}=\frac{1}{2}$. The exact variance is $\frac{4(n-1)}{n(2n-1)^2}$. The lower bound matches it for $n\geq8$, is $\frac{(n-1)(n-4)}{n(2n-1)^2}$ for $n<8$, and is not positive for $n\leq4$. The upper bound, evaluated on the even-parity sector, is $\frac{1}{2n-1}\bigl[1+\frac{\sqrt{2}}{n}\sum_{j=1}^{\lfloor n/2\rfloor}
        \abs{8j^2-n}\sqrt{\binom{2n}{n-2j}\big/\binom{2n}{n}}\,\bigr]$; on the full Fock space, the same bound is larger ($1.54$ instead of $0.44$ at $n=4$). Variance and upper bound decay
        polynomially, as $n^{-2}$ and $n^{-1/2}$, respectively.
        (b)~$\SU{2}$ collective-spin example with $\rho=\Outer{\opl}{\opl}$, $\opl=n/2$,
        $\cM=S_X+S_Y+S_Z$, and $H=-iS_Z$, so that
        $\norm{H}^S_{\infty}=n/2$. The exact variance is $\frac23(\opl)^2=\frac{n^2}{6}$. The lower bound equals it for $n\geq4$, is $\frac34$ at $n=3$ and vanishes at $n=2$. The upper bound, evaluated on the spin-$n/2$ sector, is $\frac{n^2}{2\sqrt{2n+1}}\sum_{\lambda=0}^{n}\sqrt{2\lambda+1}\,
        \abs{\lambda(\lambda+1)-\frac{n(n+2)}{2}}$; it lies above the trivial bound $\frac34n^4$ for all $n\geq5$.}
    \label{fig:variance-bounds}
\end{figure}

\Cref{fig:variance-bounds} compares the exact variances of these two examples with the bounds. For free fermions (panel a) the lower bound is tight for $n\geq8$, and the upper bound is informative but loose: it decays as $n^{-1/2}$, against $n^{-2}$ for the variance. For the collective spin (panel b) the lower bound is tight for $n\geq4$, whereas the upper bound exceeds the trivial bound $4(\infsnorm{H})^2(\iinfsnorm{\cM})^2=3n^4/4$ for every $n\geq5$ and carries no information. The reason is that \cref{main:text:eq:upperbound_case3} adds the absolute values of the contributions of all $n+1$ sectors $V_\lambda$, each weighted by $\sqrt{\dim V_\lambda}$, and thereby loses the cancellations between them.

\paragraph{An observable outside the algebra}
In the two examples above, as for the standard representation of $\SU{2^n}$, the observable is generated by the algebra, and the Lie-algebraic theory already gives the variance. \Cref{main:text:thm:exact} needs neither this nor a state confined to one irreducible subspace, and we work out one such case in its entirety. Let $G=\SU{n}$ act on $n$ qubits through the number-conserving fermionic (compound) representation $\pi_*(X)=\sum_{a,b}X_{ab}\,a_a^\dagger a_b$, with Jordan--Wigner operators $a_a$, and take $\rho=\Outer{+}{+}^{\otimes n}$, $\cM=X^{(1)}X^{(2)}$ and $\pi_*(g_j)=-\frac{i}{2}(X^{(1)}X^{(2)}+Y^{(1)}Y^{(2)})$, which is the setting of \cref{fig:placeholder}. The observable splits into a hopping and a pairing part,
\begin{align}
  X^{(1)}X^{(2)}=\underbrace{a_1^\dagger a_2+a_2^\dagger a_1}_{\cM_h}+\underbrace{a_1^\dagger a_2^\dagger+a_2a_1}_{\cM_p},
  \label{eq:compound-split}
\end{align}
of which $\cM_h$ lies in $i\pi_*(\su{n})$, whereas $\cM_p$ changes the particle number and lies outside the algebra. Under conjugation by $\Pi(G)$ the two parts transform in representations without a common irreducible constituent, so that $\cT_G(\cM^{\otimes2})=\cT_G(\cM_h^{\otimes2})+\cT_G(\cM_p^{\otimes2})$ and, by \cref{main:text:thm:exact}, the variance is the sum of the variances of the two parts. Evaluating each with the Schur orthogonality relations (Supplementary \cref{app:example:compound}) gives, for every $n\geq2$,
\begin{align}
  \mathrm{Var}\bigl[\partial_{\theta_{\ell,j}}C\bigr]
  \;=\;\underbrace{\frac{n}{(n-1)(n+1)^2}}_{\text{hopping}}+\underbrace{\frac{n-2}{n^2(n-1)}}_{\text{pairing}}
  \;=\;\frac{2n^3-3n-2}{n^2(n-1)(n+1)^2}.
  \label{eq:compound-var}
\end{align}
The variance decays as $2/n^2$, so the ansatz is trainable, although $\rho^{\otimes2}$ is spread over many irreducible subspaces and the observable is not generated by the algebra. The Lie-algebraic formula applied to the component of $\cM$ inside the algebra accounts for the first term only, which is $56\%$ of the variance at $n=4$ and one half of it asymptotically. The lower bound of \cref{main:text:eq:lowerbound_case3}, shown in \cref{fig:placeholder}, is positive there but captures only part of the variance, about $8\%$ at $n=4$. Closed forms such as \cref{eq:compound-var} rely on the structure of this example. For circuits without such structure, neither \cref{main:text:thm:exact} nor the bounds can be evaluated in practice, and this is the situation for which the estimator of \cref{main:text:sec:algorithm} is intended.

\subsection{Algorithms}
\label{main:text:sec:algorithm}
In \cref{main:text:sec:expectation-variance,main:text:theory:sec:discussion}, we established upper and lower bounds that outline the conditions on the input quantum state and measurement under which the objective function avoids barren plateaus. These bounds, however, do not exhaustively determine the barren plateau landscape. To characterize the presence of a barren plateau definitively, one must directly estimate the variance of the partial derivative $\partial_{\theta_{\ell,j}}C$. As discussed
above, a direct analytical evaluation conventionally requires decomposing the unitary representation $\cH^{\otimes 2}$ and explicitly constructing the associated projectors $\cJ_{\lambda, ab}$. Although this construction is theoretically accessible through standard Lie-algebraic methods~\cite{hall2013lie} and Weingarten calculus~\cite{collins2022weingarten}, it becomes computationally intractable for large system sizes $n$. In this section, we introduce a sampling-based algorithm that circumvents this representation-theoretic bottleneck: it runs the ansatz itself, one shot at a time, without constructing the projectors $\cJ_{\lambda, ab}$ or decomposing $\cH^{\otimes 2}$. Because every output is bounded, it decides whether the variance exceeds a tolerance $\tau$ with a number of runs polynomial in $1/\tau$ (\cref{main:text:prop:tolerance}).

\subsubsection{Finite-difference formulation}
\label{main:text:sec:finite-difference}
We first restructure the variance into a form that accommodates random sampling. Rather than relying on an explicit algebraic decomposition, we proceed through a finite-difference approximation.
Following \cref{lemma:expectation}, we assume that the first moment of the
partial derivative strictly vanishes, so that the true variance can be written
as
\begin{align}
  \begin{split}
    \var = \EXPD{U_{-} \sim \bbU_{-}}
    {\Tr{\cM^{\otimes 2}(U_{-})^{\otimes 2}
      \commutator{\pi_*(g_j) \otimes I}
        {\commutator{I \otimes \pi_*(g_j)}
          {\EXPD{U_{+}\sim \bbU_{+}}{(U_{+})^{\otimes 2}\rho^{\otimes 2}(U_{+}^\dagger)^{\otimes 2}}}}
      (U_{-}^\dagger)^{\otimes 2}}}.
  \end{split}
  \label{main:text:eq:var-fd-start}
\end{align}
Recalling the identity
$\frac{d}{dt}\!\left(\expm{t \pi_*(g_j)} A \expm{-t \pi_*(g_j)}\right)\big|_{t=0}
= [\pi_*(g_j), A]$ for any operator $A$, we approximate the nested commutators
with a central finite difference of sufficiently small step size $\varepsilon$.
For notational simplicity, we define the finite-difference superoperators $f$
and $g$,
\begin{align*}
  f(\varepsilon;A,\pi_*(g_j)) &\defeq
    \frac{\Ad_{\expm{\varepsilon \pi_*(g_j)}\otimes \bbI}(A)
          - \Ad_{\expm{-\varepsilon \pi_*(g_j)}\otimes \bbI}(A)}{2\varepsilon}, \\
  g(\varepsilon;A,\pi_*(g_j)) &\defeq
    \frac{\Ad_{\bbI\otimes \expm{\varepsilon \pi_*(g_j)}}(A)
          - \Ad_{\bbI\otimes \expm{-\varepsilon \pi_*(g_j)}}(A)}{2\varepsilon},
\end{align*}
where $\Ad_{M}(\cdot) \defeq M(\cdot)M^{-1}$ denotes the adjoint action of any
invertible linear operator $M$. Applying this approximation, we define the
approximate variance $\bbV_{\mathrm{FD}}$ through the composition $f \circ g$,
\begin{align*}
  \bbV_{\mathrm{FD}} \defeq \EXPD{U_{-} \sim \bbU_{-}}
  {\Tr{\cM^{\otimes 2}(U_{-})^{\otimes 2}\, f\circ g
    \!\left(\EXPD{U_{+}\sim \bbU_{+}}{(U_{+})^{\otimes 2}\rho^{\otimes 2}(U_{+}^\dagger)^{\otimes 2}}\right)
    (U_{-}^\dagger)^{\otimes 2}}}.
\end{align*}
With the shifted unitary operators
$U_{\pm \varepsilon} \defeq \expm{\pm \varepsilon \pi_*(g_j)} \otimes \bbI
= \Pi(\expm{\pm\varepsilon g_j})\otimes \bbI$ and
$V_{\pm \varepsilon} \defeq \bbI \otimes \expm{\pm \varepsilon \pi_*(g_j)}
= \bbI \otimes \Pi(\expm{\pm\varepsilon g_j})$ (using
\cref{main:text:commuting:diagram}) and the cost along the differentiated gate,
\begin{align*}
  C_{U_\pm}(s) \defeq \Tr{\cM\, U_{-}\expm{s\pi_*(g_j)}U_{+}\,\rho\,U_{+}^\dagger\expm{-s\pi_*(g_j)}U_{-}^\dagger},
\end{align*}
the approximate variance is (\cref{main:text:subsec:fd})
\begin{align}
  \bbV_{\mathrm{FD}} = \EXPD{U_{\pm} \sim \bbU_{\pm}}{\left(\frac{C_{U_\pm}(\varepsilon) - C_{U_\pm}(-\varepsilon)}{2\varepsilon}\right)^{2}},
  \label{main:text:eq:vfd-factorized}
\end{align}
the mean square of a central difference quotient. In the same way, \cref{main:text:eq:var-fd-start} reads $\var = \EXPD{U_{\pm} \sim \bbU_{\pm}}{\big(\partial_s C_{U_\pm}(0)\big)^{2}}$. Both quantities can therefore be sampled with ordinary runs of the $n$-qubit circuit.

\subsubsection{Variance estimation algorithm}
\label{main:text:sec:variance-estimation}
Building on the finite-difference approximation derived above, we now formalize a sampling procedure to evaluate the approximate variance $\bbV_{\mathrm{FD}}$.

\begin{algorithm}[H]
\caption{Sampling-Based Variance Estimator}\label{main:text:new:alg}
\Input{input quantum state $\rho$, measurement $\cM$, step size $\varepsilon\in (0,1)$, unitary distributions $\bbU_{+}$ and $\bbU_{-}$}
Prepare two copies of the initial state, $\rho^{\otimes 2}$\;
Sample binary variables $b_1, b_2 \sim \Bernoulli{1/2}$\;
Sample unitaries $U_{+} \sim \bbU_{+}$ and $U_{-} \sim \bbU_{-}$\;
Apply the parameter-shifted unitary sequence
$\rho_{\mathrm{out}} \leftarrow (U_{-})^{\otimes 2}\, U_{(-1)^{b_2}\cdot\varepsilon}\,V_{(-1)^{b_1}\cdot\varepsilon}\,(U_{+})^{\otimes 2}\, \rho^{\otimes 2}\, (U_{+}^\dagger)^{\otimes 2}\, V_{(-1)^{b_1}\cdot\varepsilon}^\dagger\, U_{(-1)^{b_2}\cdot\varepsilon}^\dagger\, (U_{-}^\dagger)^{\otimes 2}$\;
Measure $\rho_{\mathrm{out}}$ with respect to the observable $\cM^{\otimes 2}$ and store the scalar outcome as $O$\;
\Return{the single-shot estimate $\dfrac{1}{\varepsilon^2}\cdot (-1)^{b_1 + b_2}\cdot O$}
\end{algorithm}

Every operator in the step that defines $\rho_{\mathrm{out}}$ acts on one copy at a time, so that $\rho_{\mathrm{out}}$ is the product of the two states $U_{-}\expm{s\pi_*(g_j)}U_{+}\,\rho\,U_{+}^\dagger\expm{-s\pi_*(g_j)}U_{-}^\dagger$ with $s=(-1)^{b_2}\varepsilon$ and $s=(-1)^{b_1}\varepsilon$, and the outcome of $\cM^{\otimes 2}$ is distributed as the product of the outcomes of $\cM$ measured on these two states. \Cref{main:text:new:alg} can therefore be run as two executions of the $n$-qubit circuit whose outcomes are multiplied.

The following lemma establishes that this sampling procedure provides an
unbiased estimator for the approximate variance $\bbV_{\mathrm{FD}}$.

\begin{lemma}
  \Cref{main:text:new:alg} provides an unbiased estimate of the approximate
  variance $\bbV_{\mathrm{FD}}$.
  \label{main:text:tester:unbiased}
\end{lemma}

A detailed proof of \cref{main:text:tester:unbiased} is provided in
\cref{sec:tester} of the Supplementary Information.

\begin{figure}
    \centering
    \includegraphics[scale=0.9]{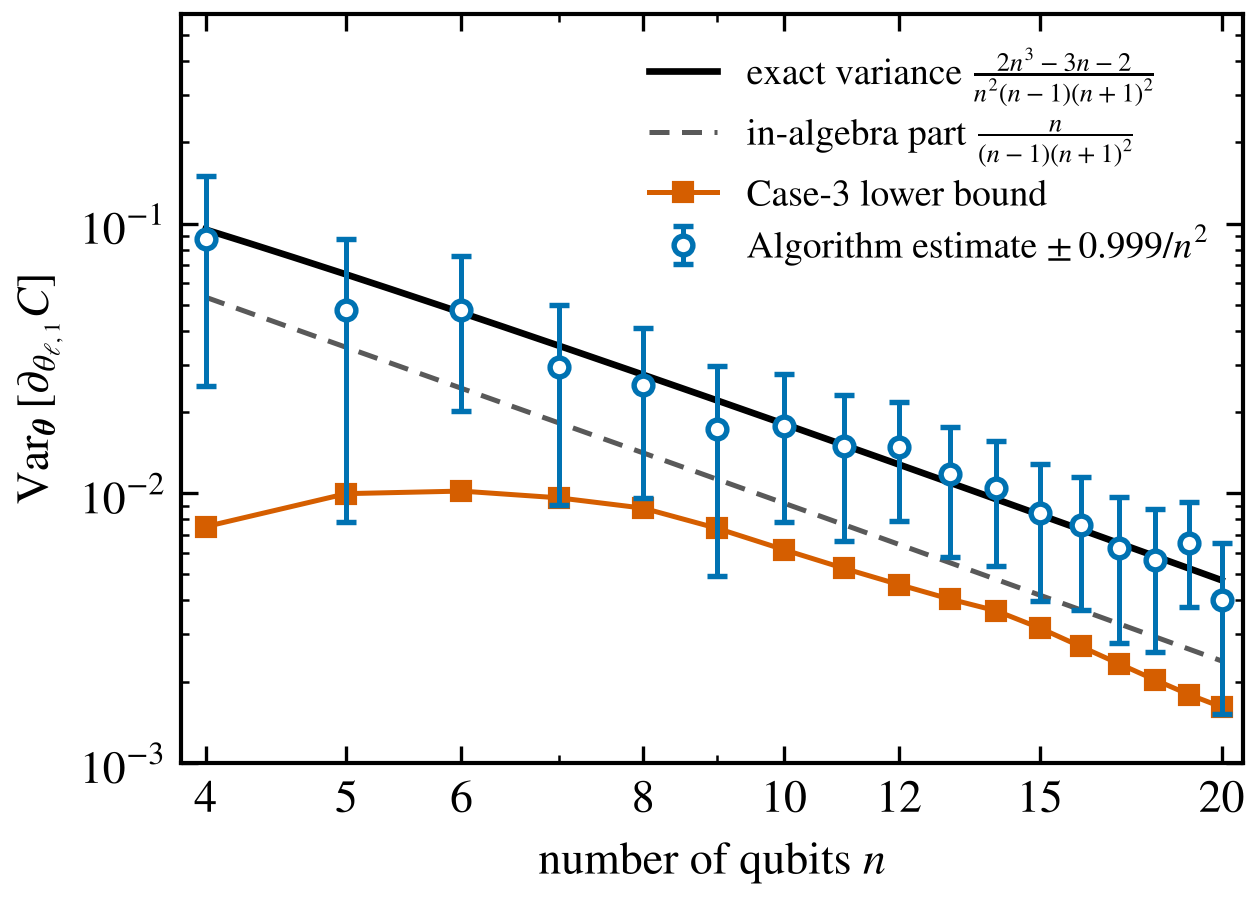}
    \caption{%
        Variance of $\partial_{\theta_{\ell,1}}C$ for the $\mathrm{SU}(n)$ compound ansatz, $n=4,\dots,20$. \textit{Setting.} $G=\mathrm{SU}(n)$ acts on $\cH=(\mathbb{C}^2)^{\otimes n}\cong\bigoplus_{m=0}^{n}\Lambda^m(\mathbb{C}^n)$ through the compound representation $\Pi(V)=\bigoplus_{m=0}^{n}C_m(V)$, where the $m$-th compound matrix $C_m(V)$ acts on the Hamming-weight-$m$ subspace; equivalently $\pi_*(X)=\sum_{a,b}X_{ab}\,a_a^\dagger a_b$ with Jordan--Wigner operators $a_a=Z_1\cdots Z_{a-1}\,|0\rangle\langle 1|_a$. The ansatz~\eqref{main:reduced:ansatz} is $U(\vtheta)=\prod_{l=1}^{L}\prod_{j=1}^{n^2-1}\exp\!\big(\theta_{l,j}\,\pi_*(g_j)\big)$ with $\theta_{l,j}\sim\mathrm{Unif}[-\pi,\pi]$ and the basis $\{g_j\}=\{-i(E_{pq}+E_{qp}),\,E_{pq}-E_{qp}\}_{p<q}\cup\{i(E_{rr}-E_{r+1,r+1})\}_{r=1}^{n-1}$ of $\mathfrak{su}(n)$, i.e.\ the gates are generated by $-\tfrac{i}{2}(X_p\mathcal{Z}_{pq}X_q+Y_p\mathcal{Z}_{pq}Y_q)$, $-\tfrac{i}{2}(Y_p\mathcal{Z}_{pq}X_q-X_p\mathcal{Z}_{pq}Y_q)$ and $-\tfrac{i}{2}(Z_r-Z_{r+1})$, where $\mathcal{Z}_{pq}=\prod_{p<k<q}Z_k$. We take $\rho=|+\rangle\langle+|^{\otimes n}$, $\cM=X_1X_2$, and the generator of the differentiated gate $H=\pi_*(g_1)=-\tfrac{i}{2}(X_1X_2+Y_1Y_2)$ ($\|H\|_\infty=\|\cM\|_\infty=1$), with $U_\pm$ Haar-distributed on $\Pi(\mathrm{SU}(n))$ (Case~3). \textit{Curves.} Solid line: exact variance $\mathrm{Var}(n)=\frac{2n^3-3n-2}{n^2(n-1)(n+1)^2}\approx 2/n^2$ of \cref{eq:compound-var}, derived in Supplementary \cref{app:example:compound}. Dashed line: its first term $\frac{n}{(n-1)(n+1)^2}$, the contribution of the component of $\cM$ inside the algebra, which is all that the Lie-algebraic formula accounts for. Squares: Case-3 lower bound \cref{main:text:eq:lowerbound_case3}. Circles: output of \Cref{main:text:new:alg} in its exact form, with the four-term shift rule for $H$ ($\gamma=2$), $\delta=10^{-3}$ and $\teps=0.999/n^2$, using $\mathcal{N}=\lceil 32\ln(2/\delta)/\teps^{2}\rceil$ runs ($6.2\times10^{4}$ at $n=4$ to $3.9\times10^{7}$ at $n=20$; the finite-difference form needs $4.6\times10^{8}$ to $1.8\times10^{14}$ runs for the same guarantee). Every run was simulated, with the value of the cost function at each shift computed in a time polynomial in $n$ as described in Supplementary \cref{app:example:compound}. Error bars: the certified accuracy $\pm\teps=\pm0.999/n^2$, guaranteed with probability at least $1-\delta=0.999$.}
    \label{fig:placeholder}
\end{figure}

\paragraph{Exact shift rules.}
When the spectrum of $\pi_*(g_j)$ is known, the finite difference can be replaced by an exact parameter-shift rule~\cite{wierichs2022general}. If $\Omega_1,\dots,\Omega_R$ are the distinct positive differences between eigenvalues of $i\pi_*(g_j)$, then $s\mapsto C_{U_\pm}(s)$ is a trigonometric polynomial with these frequencies, and there are shifts $x_1,\dots,x_{2R}$ and coefficients $c_1,\dots,c_{2R}$, independent of $U_{\pm}$, $\rho$ and $\cM$, such that $\partial_s C_{U_\pm}(0)=\sum_{k}c_k\,C_{U_\pm}(x_k)$. In the exact form of \Cref{main:text:new:alg}, $k_1$ and $k_2$ are drawn independently, each $k$ with probability $\abs{c_k}/\gamma$, where $\gamma\defeq\sum_k\abs{c_k}$, the shifts $x_{k_2}$ and $x_{k_1}$ are used in place of $(-1)^{b_2}\varepsilon$ and $(-1)^{b_1}\varepsilon$, and the run returns $\gamma^2\,\mathrm{sgn}(c_{k_1})\,\mathrm{sgn}(c_{k_2})\,O$. This output is an unbiased estimate of $\var$ itself, and it is bounded by $\gamma^2(\infsnorm{\cM})^2$ irrespective of the shifts. For a Pauli generator $\pi_*(g_j)=-\tfrac{i}{2}P$, this is the usual two-term rule, with shifts $\pm\pi/2$ and $\gamma=1$. For a generator with spectrum $\{0,\pm i\}$, such as the gates of \cref{fig:placeholder}, a four-term rule is exact: shifts $\pm\pi/4$ and $\pm3\pi/4$ with coefficients $\pm1/(8\sin^2\tfrac{\pi}{8})$ and $\mp1/(8\sin^2\tfrac{3\pi}{8})$, and $\gamma=2$.

\subsubsection{Error and complexity analysis}
\label{main:text:sec:error-complexity}
The error introduced by the finite-difference approximation and the statistical uncertainty inherent to the sampling procedure are bounded separately in \cref{main:text:subsec:error} (\cref{main:text:tester:lemma:1,main:text:tester:lemma:2}), which also gives the number of runs for the exact form. These bounds turn the comparison of the estimate with a tolerance into a test with a guaranteed outcome.

\begin{proposition}[Deciding trainability]
  \label{main:text:prop:tolerance}
  Let $\tau>0$ and $\delta\in(0,1)$, and let $\bbO$ be the mean of $\cN$
  independent outputs of \Cref{main:text:new:alg}, where either:
  (i) the exact form is used and
  $\cN\geq 8\gamma^4(\infsnorm{\cM})^4\ln(2/\delta)/\tau^2$, or
  (ii) the finite-difference form is used with $\varepsilon$ small enough
  that $\abs{\bbV_{\mathrm{FD}}-\var}\leq\tau/4$, and $\cN$ satisfies
  \cref{main:text:tester:lemma:2} with $\teps=\tau/4$.
  Then $\abs{\bbO-\var}\leq\tau/2$ with probability at least $1-\delta$. In
  particular, the rule ``declare the ansatz trainable at tolerance $\tau$ if
  $\bbO>\tau$'' is correct with probability at least $1-\delta$ whenever
  $\var>3\tau/2$ or $\var<\tau/2$.
\end{proposition}

By \cref{main:text:tester:lemma:1}, condition (ii) holds for some $\varepsilon^2=\Theta(\tau)$. For $\gamma$, $\infsnorm{\cM}$ and $\infsnorm{\pi_*(g_j)}$ of order one, the number of runs is therefore $\cO(\tau^{-2}\log\delta^{-1})$ in case (i) and $\cO(\tau^{-4}\log\delta^{-1})$ in case (ii). For any tolerance $\tau=1/\poly{n}$, trainability is decided with polynomially many runs of the circuit, including in cases where the bounds of \cref{main:text:sec:expectation-variance,main:text:theory:sec:discussion} are inconclusive. The proof is given in \cref{sec:tester} of the Supplementary Information.

\subsubsection{Complexity, generality, and comparison with gradient sampling}
\label{main:text:sec:algorithm-discussion}
A single run of \Cref{main:text:new:alg} executes the circuit on each of the two copies, at a cost of $\cO(L \cdot k)$ gates per copy for a circuit with $L$ parameterized layers and $k$ Lie-algebra generators (or basis elements). The number of runs is determined by the tolerance: by \cref{main:text:prop:tolerance}, deciding whether $\var$ exceeds $\tau$ takes $\cO(\tau^{-2}\log\delta^{-1})$ runs with an exact shift rule and $\cO(\tau^{-4}\log\delta^{-1})$ with finite differences, which is polynomial whenever $\tau=1/\poly{n}$. Resolving an exponentially small variance to relative precision would instead require exponentially many runs, as it would for any sampling method, but deciding trainability does not require it.

The most direct alternative is to sample circuits and estimate the gradient of each. Squaring a single noisy gradient estimate biases the result upward by its shot-noise variance, and the standard unbiased remedy multiplies two independent gradient estimates obtained for the same sampled circuit, $Z = \hat{X}_1 \hat{X}_2$. Run as two executions of the $n$-qubit circuit, \Cref{main:text:new:alg} is an estimator of this cross-product type in its leanest form: each factor is a single shot at a randomly signed shift, so a run costs two circuit executions, whereas the deterministic version, in which each factor is a full difference $(O_{+\varepsilon}-O_{-\varepsilon})/(2\varepsilon)$, costs four. Both are unbiased for $\bbV_{\mathrm{FD}}$ and bounded by $(\infsnorm{\cM})^2/\varepsilon^2$, so for the same Hoeffding guarantee \Cref{main:text:new:alg} needs half as many circuit executions. The deterministic version has the smaller variance: for the same number of circuit executions, its estimate has about $2.5$ times less variance in the setting of \cref{fig:placeholder} at $n=4$, so it is preferable when sampling stops on the observed spread rather than on the worst-case bound; \cref{main:text:prop:tolerance} holds for both.

A different route~\cite{PRXQuantum.4.040335} maps the gradient variance of randomly initialized circuits exactly onto averages over Clifford circuits, which a classical computer can simulate efficiently. \Cref{main:text:new:alg} instead runs on the quantum device and uses no structure of the generators beyond, in its exact form, their spectrum.

The exact form removes both the bias and the $\varepsilon^{-4}$ factor of the finite-difference form. For the gates of \cref{fig:placeholder}, it reduces the number of runs needed for the same guarantee by a factor of about $7\times10^{3}$ at $n=4$ and $5\times10^{6}$ at $n=20$.

Notably, our method applies to any parameterized quantum circuit $\prod_{l=1}^{L}\expm{i\theta_l H_l}$ and does not require the underlying unitary ensembles to converge to a $2$-design. For gates with $\expm{2\pi iH_l}$ a multiple of the identity, the expected gradient vanishes at any depth (\cref{lemma:expectation_main}) and need not be estimated; for other gates, our method can also be deployed to determine whether the expected gradient vanishes, which is a critical prerequisite for barren plateau analysis. This can be done by applying the same finite-difference estimation to the adjoint action of the Hamiltonian generators, thereby avoiding algebraically demanding state-space decompositions. When the expected gradient is instead found to be non-zero, the optimization landscape already possesses sufficient global features to drive parameter updates through standard first-order techniques, which renders the higher-order moment algorithm unnecessary. Complete derivations of the finite-difference estimator, together with rigorous finite-sample error bounds and probability guarantees, are provided in the Supplementary Information.

\section{Discussion}
\label{main:text:sec:discussion}

In this work, we establish a Lie-theoretic framework for characterizing the trainability of parameterized quantum circuits. The exact variance of \cref{main:text:thm:exact}, and the bounds derived in \cref{main:text:eq:upperbound_case1,main:text:eq:upperbound_case2,main:text:eq:upperbound_case3} and \cref{main:text:eq:lowerbound_case1,main:text:eq:lowerbound_case2,main:text:eq:lowerbound_case3} establish that the presence of barren plateaus is not solely a property of the ansätze, but also depends heavily on how the input state $\rho$ and the measurement operator $\cM$ act on the Hilbert space $\cH$. Additionally, the proposed sample-based estimators offer a practical approach for estimating the variance of the partial derivatives $\partial_{\theta_{\ell, j}} C$ without needing direct access to the structure of the compact and connected Lie group $G$ or its associated Lie algebra $\fg$.

Beyond these contributions, several open questions remain. First, the derived bounds are generally not tight, meaning they cannot be used to definitively determine whether a given ansatz induces a barren plateau. As discussed in \cref{main:text:theory:sec:discussion}, there are specific choices of $\cM$, $\rho$, and unitary representations $\Pi$ where the upper bound scales as $\Omega(1/\poly{n})$, while the true variance remains exponentially small (in the example given there, exactly zero). Additionally, the lower bound may not reliably show whether the objective function has a barren plateau when both $\cM^{\otimes 2}$ and $\rho^{\otimes 2}$ concentrate on a reducible, polynomial-dimensional invariant subspace of $\cH^{\otimes 2}$. Such a concentration does not by itself exclude a barren plateau: in the example of \cref{eq:separation-witness} both lie in irreducible subspaces of polynomial dimension, but in different ones, whose direct sum is such a reducible subspace, and the variance is exactly zero. This gap highlights an inherent limitation in using these bounds for definitive characterization.

\Cref{main:text:thm:exact} characterizes the variance exactly when both circuit segments are $2$-designs, but evaluating it requires the irreducible decomposition of $\cH^{\otimes 2}$, which is available only for structured ans\"atze. Outside that regime, to reliably characterize whether a given ansatz induces a barren plateau, one would need upper and lower variance bounds that are close to the true variance. Future work must also focus on clarifying the exact relationship between the irreducible representations, the measurement operator, and the input quantum state. One possible approach to achieve this is to analyze the variance of the curvature of the objective function $C$, which involves evaluating its second-order derivatives.

Second, the estimation algorithm decides trainability only up to a tolerance. With polynomially many runs it separates variances above $3\tau/2$ from variances below $\tau/2$ for any $\tau = 1/\poly{n}$ (\cref{main:text:prop:tolerance}), but resolving an exponentially small variance to relative precision requires exponentially many runs, as it does for any sampling method. Upper and lower bounds that are provably close to the true variance would remove this limitation where they apply; whether such bounds can be evaluated efficiently for general ans\"atze remains open.

Finally, we address the extension of this framework to non-compact Lie groups. As a representative example, consider the affine group of the real line (the "$ax + b$" group). It can be faithfully represented by the group of $2 \times 2$ real upper-triangular matrices:
\begin{align}
    G = \CurB{ \begin{pmatrix} a & b \\ 0 & 1 \end{pmatrix} : a > 0, b \in \mathbb{R} }.
\end{align}
Unlike the compact subgroups typically generated by parameterized quantum circuits, this group is strictly non-compact and non-unimodular. 

In the non-compact setting, the Haar measure on the entire group $G$ is no longer finite (see Proposition 9.3.3 of \cite{Cohn2013}). Although such groups remain locally compact and admit left- and right-invariant Haar measures, these measures are generally not identical. Consequently, the projection onto the commutant cannot be realized via group averaging in the manner used in Supplementary \cref{subsec:commutant}. This prevents a direct extension of the expressibility analysis developed for compact groups and requires alternative mathematical tools from the representation theory of non-compact groups.

A practical workaround is to consider the closure $\overline{\Pi(G)}$ of $\Pi(G)$ in the unitary group $U(\cH)$, which is a compact Lie group, and to work with its associated Lie algebra. In this setting, the procedure proposed in \cite{larocca2022diagnosing} can be used to identify a spanning set for the Lie algebra of $\overline{\Pi(G)}$. One can then construct the ansatz using this spanning set, restoring compatibility with the compact-group framework and enabling the application of the expressibility analysis developed in this work.

\section{Methods}
\label{main:text:sec:problem:statement}

\subsection{Expressibility}
\label{main:text:subsec:express}

To analyze gradient variance, we must establish when the circuit \eqref{main:ansatz} behaves like a random unitary drawn from its reachable group. This is not merely a question of independent interest: the variance bounds derived in \cref{main:text:sec:expectation-variance} rely on replacing averages over the circuit parameters with Haar averages over the group $\Pi(G)$, and this replacement is justified only when the ansatz converges to a $2$-design. Expressibility is therefore the bridge that makes our variance analysis well defined. In this section we quantify this convergence.

To do so, we first generalize the notion of an $\varepsilon$-approximate $t$-design, a standard tool for quantifying expressibility. Prior studies \cite{PhysRevA.80.012304, Mele_2024, larocca2022diagnosing, haah2025efficient} typically evaluate $\varepsilon$-approximate $t$-designs relative to the Haar measure on $SU(2^n)$ or $SO(2^n)$. However, this is inapplicable to structured ans\"atze whose reachable set is a \emph{proper} compact subgroup: when a circuit is generated by a Lie algebra whose associated Lie group is a proper compact subgroup of $SU(2^n)$, the Haar measure on $SU(2^n)$ no longer serves as a meaningful reference. To address this, we define $\varepsilon$-approximate $t$-designs relative to arbitrary compact, connected matrix Lie groups.

\begin{definition}[$\varepsilon$-Approximate $t$-Design]
Let $\bbU$ be a distribution over a set of unitary matrices, and let $\compactG$ be a compact, connected matrix Lie group containing the support of $\bbU$. For $t \geq 0$ and $\varepsilon > 0$, we say $\bbU$ is an $\varepsilon$-approximate $t$-design with respect to $\compactG$ (equivalently, $\bbU$ $(\varepsilon\text{-})$converges to a $t$-design with respect to $\compactG$) if
    \begin{align}
        \norm{\EXPD{V\sim \bbU}{(V)^{\otimes t} \otimes (V^*)^{\otimes t}} - \EXPD{U\sim \mu_{\compactG}}{(U)^{\otimes t} \otimes (U^*)^{\otimes t}}} \leq \varepsilon,
    \end{align}
where $\mu_{\compactG}$ denotes the Haar measure on $\compactG$ and $\norm{\cdot}$ is any sub-multiplicative matrix norm.
\end{definition}

Because the parameters are sampled from $[-\pi, \pi]^{Lk}$, the reduced ansatz \eqref{main:reduced:ansatz} induces a distribution $\bbU_U$ supported on the compact, connected matrix Lie group $\Pi(G) \subseteq \mathrm{U}(\cH)$ associated with the induced Lie algebra $\tilde{\mathfrak{g}} = \pi_*(\fg)$. Its expressibility is therefore naturally measured against the Haar measure $\mu_{\Pi(G)}$. The following theorem shows that this convergence is achieved in the large-depth limit.

\begin{theorem}
    Let $\compactG$ be the compact, connected matrix Lie group associated with the Lie algebra $\tfg$ spanned or generated by anti-Hermitian matrices $H_1, \cdots, H_\cN$. Let $t\in\bbN_{\geq 1}$. Define
    \begin{align}
        \cV_l(\vtheta_l) := \prod_{j=1}^\cN \exp\!\bigl((\vtheta_l)_j H_j\bigr),\qquad\vtheta_l\in[-\pi,\pi]^\cN. \label{main:text:eq:single:layer}
    \end{align}
    Then,
    \begin{align}
        \lim_{L \to \infty} \norm{\EXPD{\vtheta_1,\cdots,\vtheta_L \sim [-\pi, \pi]^\cN}{\prod_{l=1}^L\left((\cV_l)^{\otimes t}\otimes (\cV_l^*)^{\otimes t}\right)} - \EXPD{U \sim \mu_{\compactG}}{(U)^{\otimes t}\otimes (U^*)^{\otimes t}}} = 0
    \end{align}
    for any sub-multiplicative matrix norm.
    \label{main:express:thm}
\end{theorem}

The proof of \cref{main:express:thm} is provided in Supplementary \cref{sec:express}. Together with the observations above, \cref{main:express:thm} implies that $\bbU_U$ converges to an $\varepsilon$-approximate $t$-design with respect to $\Pi(G)$ as the depth $L$ grows; equivalently, the ansatz $U(\vtheta)$ of \eqref{main:ansatz} forms an $\varepsilon$-approximate $t$-design with respect to $\Pi(G)$ in the large-depth limit. Consequently, the $t$-th moment operator of the circuit converges to the corresponding moment of the Haar measure on $\Pi(G)$, and the statistical properties of the PQC can be analyzed through Haar averages over $\Pi(G)$, consistent with established methodologies. The variance-estimation algorithm of \cref{main:text:sec:algorithm} does not rely on this convergence: it samples the circuit at the given depth.

Unlike previous studies, however, our result does not yield an explicit bound on the depth $L$ needed to reach a prescribed error $\varepsilon \in (0,1)$. The proof proceeds through the difference operator $\mathcal{A}^{(t)} \defeq M^{(t)}_{\nu} - M^{(t)}_{\mu}$, where $M^{(t)}_\nu \defeq \bbE_{\vtheta}[U_1^{\otimes t}\otimes (U_1^*)^{\otimes t}]$ is the single-layer $t$-th moment operator and $M^{(t)}_\mu \defeq \bbE_{U\sim \mu_G}[U^{\otimes t}\otimes (U^*)^{\otimes t}]$, with $U_1(\vtheta) = \prod_{q=1}^k \exp(\theta_{1,q}\pi_*(g_q))$ a single layer of the ansatz. We can guarantee that the spectral radius of $\cA^{(t)}$ is strictly less than 1, but because $\cA^{(t)}$ is not necessarily normal, this does not by itself force its sub-multiplicative norm below one.

Nevertheless, whenever $\inorm{\cA^{(t)}} \leq \lambda < 1$, the depth needed for an $\varepsilon$-approximate $t$-design can be made explicit.

\begin{proposition}
    Suppose that $\inorm{\cA^{(t)}} \leq \lambda \in (0,1)$. Then
    \begin{align}
        \norm{
        \EXPD{\vtheta}{U(\vtheta)^{\otimes t} \otimes (U(\vtheta)^*)^{\otimes t}} 
        - 
        \EXPD{U \sim \mu_G}{\Pi(U)^{\otimes t} \otimes (\Pi(U)^*)^{\otimes t}}
        }
        \leq \varepsilon
    \end{align}
    for all $\varepsilon\in (0,1)$, provided that $L \ge \dfrac{\log(\varepsilon)}{\log(\lambda)}$.
    \label{main:text:express:eps:approx:k:design}
\end{proposition}

\subsection{Exact variance}
\label{main:text:subsec:exact}
With the convergence guarantees of \cref{main:text:sec:expectation-variance} in place, we evaluate the statistical moments of the partial derivatives, beginning with the expectation.
\begin{lemma}
    If $\exp(2\pi\,\pi_*(g_j))$ is a multiple of the identity, or if $\bbU_-$ or $\bbU_+$ is a $2$-design with respect to $\Pi(G)$, the first moment satisfies:
    \begin{align}
        \EXPD{U_\pm\sim\bbU_\pm}{\partial_{\theta_{\ell,j}} C} = 0.
    \end{align}\label{lemma:expectation_main}
\end{lemma}
Because the expectation vanishes, the variance reduces exactly to the second
moment,
\begin{align}
    \VARD{\vtheta}{\partial_{\theta_{\ell, j}}C} = \EXPD{U_\pm\sim \bbU_\pm}{\Tr{\cM U_- \commutator{\pi_*(g_j)}{U_+ \rho U_+^\dagger} U_-^\dagger}^2}. \label{main:text:true:var}
\end{align}
It therefore suffices to bound this second moment, which we do in turn from
above and from below. The four traces of \cref{eq:compact} reassemble into a single double commutator. Writing $\var$ for $\var_{\vtheta}\bigl[\partial_{\theta_{\ell, j}}C\bigr]$ and using the identity $[\pi_*(g_j),X]^{\otimes2}=[\,\pi_*(g_j)\otimes\bbI,\,[\,\bbI\otimes \pi_*(g_j),\,X^{\otimes2}\,]\,]$, the variance takes the compact form
\begin{align}
    \var = \Tr{\cT_-(\cM)\commutator{\pi_*(g_j)\otimes\bbI}{\commutator{\bbI\otimes \pi_*(g_j)}{\cT_+(\rho)}}}, \label{eq:compact}
\end{align}
where the twirl super-operators $\cT_\pm$ are defined below. To verify this, expand \cref{main:text:true:var} using
$\Tr{X}^{2}=\tr\{X\otimes X\}$ and $(AB)^{\otimes2}=A^{\otimes2}B^{\otimes2}$:
\begin{align}
\label{eq:var1}
    \begin{split}
         \VARD{\vtheta}{\partial_{\theta_{\ell, j}}C}=& \Tr{\cT_-(\cM)\cdot \pi_*(g_j)^{\otimes 2}\cdot \cT_+(\rho)} -\Tr{(\bbI\otimes \pi_*(g_j))\cdot \cT_-(\cM) \cdot(\pi_*(g_j)\otimes \bbI)\cdot \cT_+(\rho)} \\
        &\hspace{0.5cm}-\Tr{(\pi_*(g_j)\otimes \bbI)\cdot \cT_-(\cM) \cdot(\bbI \otimes \pi_*(g_j))\cdot \cT_+(\rho)} +\Tr{\pi_*(g_j)^{\otimes 2}\cdot \cT_-(\cM) \cdot \cT_+(\rho)}
    \end{split}
\end{align}
where the twirl super-operators are
\begin{align*}
  \cT_-(\cdot)  \defeq \EXPD{U_- \sim \bbU_-}{ (U_{-}^\dagger)^{\otimes 2}(\cdot)^{\otimes 2}(U_{-})^{\otimes 2} },\qquad \cT_+(\cdot)  \defeq \EXPD{U_+ \sim \bbU_+}{ (U_{+})^{\otimes 2}(\cdot)^{\otimes 2}(U_{+}^\dagger)^{\otimes 2} }
\end{align*}
Expanding the nested commutator in \cref{eq:compact} reproduces, by cyclicity of the trace, the four terms of \cref{eq:var1}.

The first two lines of \cref{main:text:thm:exact} are \cref{eq:compact} with both twirls evaluated by \cref{eq:second-moment}. The third uses that $\Theta$ and $R$ commute with $\pi_*(g_j)\otimes\bbI+\bbI\otimes\pi_*(g_j)$; with $\cT_-(\cM)$ and $\cT_+(\rho)$ in place of $\Theta$ and $R$ it already holds when only one of the two ensembles is a $2$-design (proof of Supplementary \cref{app:lem:rank1}). The operator $\Delta_j$ changes sign when the two copies are exchanged, so it maps the symmetric square $\Symt{\cH}$ into the alternating square $\Altt{\cH}$ and conversely.

To organize the resulting estimates, write the anti-Hermitian generators as $H_{l,q}:=\pi_*(g_q)$, and introduce the one-parameter moment operator
\begin{align}
    P^{t}_{H}(\cdot) = \frac{1}{2\pi}\int_{-\pi}^{\pi}\exp(\theta H)^{\otimes t}\,(\cdot)\,\exp(-\theta H)^{\otimes t}d\theta,
  \label{eq:moment-operator}
\end{align}
which averages over the one-parameter subgroup generated by $H$.

\subsection{Upper bounds}
\label{main:text:subsec:upper}

We now bound the exact variance~\eqref{eq:compact} from above in the three regimes of \cref{main:text:sec:expectation-variance}. We provide detailed derivations for Case~1; Cases~2 and~3 follow the same steps.

\paragraph{Case 1: $U_-$ converges to a $2$-design while $U_+$ does not.}
Because $\bbU_-$ is a $2$-design, its twirl reduces to the Haar twirl~\eqref{eq:second-moment}:
\begin{align}
\cT_-(\cM)=\sum_\lambda\sum_{a,b=1}^{m_\lambda}
  \frac{\tr\{\cM^{\otimes2}\cJ^\dagger_{\lambda,ab}\}}{\dim V_\lambda}\,\cJ_{\lambda,ab}.
\label{eq:Tminus}
\end{align}
Because $U_+$ is \emph{not} a $2$-design, its twirl cannot be collapsed
to~(\ref{eq:second-moment}). Instead we retain the exact finite-depth channel: averaging each
post-gate parameter $\theta_{l,q}\in[-\pi,\pi]$ independently turns $\cT_+$
into the ordered composition of the one-parameter moment operators,
\begin{align}
    \cT_+(\rho)=\bigl(\cP^2_{H_{\ell,j+1}}\circ\cdots\circ\cP^2_{H_{L,k}}\bigr)(\rho^{\otimes2}),
    \qquad
    \cP^t_H(\cdot)=\frac{1}{2\pi}\int_{-\pi}^{\pi}\expm{\theta H}^{\otimes t}(\cdot)\,\expm{-\theta H}^{\otimes t}\,d\theta, \label{eq:Tplus}
\end{align} with $H_{l,q}=\pi_*(g_q)$, the composition running over every gate of $U_+$, from the gate $(L,k)$ next to $\rho$, which acts first, to the gate $(\ell,j{+}1)$ just after the differentiated gate. Identity~\eqref{eq:Tplus} is exact at any depth; Case~1 only means we may \emph{not} replace it by the Haar twirl. Now we establish the first inequality. Inserting~\eqref{eq:Tminus} into~\eqref{eq:compact} and applying the triangle inequality over the $(\lambda,a,b)$ sum gives
\begin{align}
    \var \leq \sum_{\lambda,a,b} \frac{\abs{\Tr{\cM^{\otimes2}\cJ^\dagger_{\lambda,ab}}}}{\dim V_\lambda}\, \abs{\Tr{\cJ_{\lambda,ab}\,X}}, \qquad X:=\bigl[\,\pi_*(g_j)\otimes\bbI,\,[\,\bbI\otimes \pi_*(g_j),\,\cT_+(\rho)\,]\,\bigr].
\end{align}
We then use two standard Schatten-norm estimates.

\emph{(a) Cauchy--Schwarz for the trace pairing}, with $\snorm{\cJ_{\lambda,ab}}_{2}=\sqrt{\dim V_\lambda}$, we find
\begin{align}
    \abs{\Tr{\cJ_{\lambda,ab}\,X}} \leq\snorm{\cJ_{\lambda,ab}}_{2}\,\snorm{X}_{2} =\sqrt{\dim V_\lambda}\;\snorm{X}_{2}.
\end{align}

\emph{(b) The double-commutator bound}, using
$\snorm{[A,Y]}_{2}\leq 2\snorm{A}_{\infty}\snorm{Y}_{2}$ twice together with $\snorm{\pi_*(g_j)\otimes\bbI}_{\infty}=\snorm{\bbI\otimes \pi_*(g_j)}_{\infty}=\snorm{\pi_*(g_j)}_{\infty}$:
\begin{align}
    \snorm{X}_{2}\leq 2\snorm{\pi_*(g_j)\otimes\bbI}_{\infty}\cdot 2\snorm{\bbI\otimes \pi_*(g_j)}_{\infty}\, \snorm{\cT_+(\rho)}_{2} =4\,\bigl(\snorm{\pi_*(g_j)}_{\infty}\bigr)^2\,\snorm{\cT_+(\rho)}_{2}.
\end{align}
The factor $\sqrt{\dim V_\lambda}$ from~(a) cancels one power of the $1/\dim V_\lambda$ prefactor, leaving $1/\sqrt{\dim V_\lambda}$; the factor $4=2^2$ is the price of the two commutators. Collecting, and writing $\cT_+(\rho)$ via~\eqref{eq:Tplus},
\begin{align}
    \var\leq 4\sum_\lambda\sum_{a,b}
     \frac{\abs{\Tr{\cJ^\dagger_{\lambda,ab}\cM^{\otimes2}}}}{\sqrt{\dim V_\lambda}}\,
     \bigl(\snorm{\pi_*(g_j)}_{\infty}\bigr)^2\,
     \snorm{\cP^2_{H_{\ell,j+1}}\!\circ\cdots\circ\cP^2_{H_{L,k}}(\rho^{\otimes2})}_2.
    \label{main:text:eq:upperbound_case1}
\end{align}The bound is a product of three independent factors: the \emph{observable}'s overlap with the representation basis ($\abs{\iTr{\cJ^\dagger\cM^{\otimes2}}}/\sqrt{\dim V_\lambda}$), the \emph{generator} strength ($(\snorm{\pi_{*}(g_j)}_{\infty})^2$), and how the finite-depth tail $U_+$ \emph{spreads the state} across the two copies (the Schatten-$2$ norm of the moment-operator composition acting on $\rho^{\otimes2}$).

\paragraph{Case 2: $U_+$ converges to a $2$-design while $U_-$ does not.}
Symmetrically, convergence of $U_+$ closes the Haar integral over $\rho^{\otimes 2}$, and the observable is instead controlled by the finite-depth moment operators:
\begin{align}
    \var\leq  4\sum_{\lambda}\sum_{a,b} \frac{\abs{\Tr{\cJ^\dagger_{\lambda,ab}(\rho^{\otimes 2})}}}{\sqrt{\dim V_{\lambda}}}\cdot \left(\snorm{\pi_*(g_j)}_\infty\right)^2\cdot \snorm{\cP^2_{H_{\ell, j}}\circ \cdots \circ \cP^2_{H_{1,1}}(\cM^{\otimes 2})}_2. \label{main:text:eq:upperbound_case2}
\end{align}

\Cref{main:text:eq:upperbound_case3} follows from \cref{main:text:thm:exact} by the triangle inequality and H\"older's inequality.

These general bounds remain loose when the state and observable are supported on disjoint invariant subspaces. A sharper estimate is available in the practically important regime where both $\rho=\lvert\psi\rangle\langle\psi\rvert$ and $\cM=\lvert\Psi\rangle\langle\Psi\rvert$ are rank one. There, the tensor square decomposes canonically into symmetric and alternating parts, $\cH^{\otimes 2}=\Symt{\cH}\oplus\Altt{\cH}$, and a rank-one operator $\lvert\varphi\rangle\langle\varphi\rvert^{\otimes 2}$ has strictly zero support on $\Altt{\cH}$. The twirls $\cT_-(\cM)$ and $\cT_+(\rho)$ are therefore both supported on $\Symt{\cH}$, the term $\frac12\iTr{\cT_-(\cM)\,\Delta_j\,\cT_+(\rho)\,\Delta_j^\dagger}$ of the expanded double commutator vanishes because $\Delta_j$ exchanges the symmetric and the alternating square, and, whenever at least one of the two ensembles is a $2$-design,
\begin{align}
    \var=\frac12\,\mathrm{Re}\,\Tr{\cT_-(\cM)\,\Delta_j^\dagger\Delta_j\,\cT_+(\rho)}. \label{main:text:eq:rank1:identity}
\end{align}
This identity leads to the bounds below; in Case~3 the result is at most half of \cref{main:text:eq:upperbound_case3}, whereas in Cases~1 and~2 the bounds are of a different form and need not be smaller than \cref{main:text:eq:upperbound_case1,main:text:eq:upperbound_case2}.
The copies of the Clebsch--Gordan decomposition (\cref{sec:lower-bound}) need not lie in $\Symt{\cH}$ or in $\Altt{\cH}$, since for two qubits under the collective action of $\SU{2}$, the tensor product of the triplet and the singlet subspaces is an example. So the sums below run over all copies (Supplementary \cref{sec:upper:rank1}).
Below we bound the inner trace by H\"older with the
operator/trace-norm pair instead of Cauchy--Schwarz, using
$\snorm{\JJ_{\lambda,ab}}_{\infty}=1$ we find
\begin{equation}
|\tr\{\JJ_{\lambda,ab}\,X\}|
\le\snorm{\JJ_{\lambda,ab}}_{\infty}\,\snorm{X}_{1}
=\snorm{X}_{1}.
\end{equation}
No $\sqrt{\dimV}$ is produced this time, so the full $1/\dimV$ stays in the
denominator; the state enters through its restricted trace norm
$\snorm{\rho^{(\lambda',a')\otimes(\nu',b')}_{\lambda,a,1\to\lambda,b,2}}_{1}$,
while $\frac12\Delta_j^\dagger\Delta_j$ contributes at most $2(\snorm{H}_{\infty})^2$, where $H\defeq\pi_*(g_j)$.

\paragraph{Case 1 (rank-one).} With $U_-$ converged and $U_+$ not,
bounding the non-converged action on the state by its restricted $1$-norm gives
\begin{align}
     \var\leq    2\sum_{\lambda} \sum_{a,b}\frac{\abs{\Tr{\cJ^\dagger_{\lambda, ab}\cM^{\otimes 2}}}\cdot (\infsnorm{H})^2 \cdot \onesnorm{\rho_{\lambda, a, 1 \to \lambda, b, 2}^{(\lambda^\prime, a^\prime) \otimes (\nu^\prime, b^\prime)}}}{\dim V_\lambda}. \label{main:text:eq:rank1_upperbound_case1}
\end{align}

\paragraph{Case 2 (rank-one).} Symmetrically, with $U_+$ converged,
\begin{align}
    \var\leq     2\sum_{\lambda} \sum_{a,b}\frac{\abs{\Tr{\cJ^\dagger_{\lambda, ab}\rho^{\otimes 2}}} \cdot (\infsnorm{H})^2 \cdot \onesnorm{\cM_{\lambda, a, 1 \to \lambda, b, 2}^{(\lambda^\prime, a^\prime) \otimes (\nu^\prime, b^\prime)}}}{\dim V_\lambda}. \label{main:text:eq:rank1_upperbound_case2}
\end{align}

\paragraph{Case 3 (rank-one).} When both segments converge, \cref{main:text:eq:rank1:identity} gives $\var\leq2(\infsnorm{H})^2\,\onesnorm{\cT_G(\rho^{\otimes 2})\,\cT_G(\cM^{\otimes 2})}$. In terms of the $m_\lambda\times m_\lambda$ matrices $(x^\lambda)_{ab}\defeq\Tr{\cJ^\dagger_{\lambda, ab} \rho^{\otimes 2} }$ and $(y^\lambda)_{ab}\defeq\Tr{\cJ^\dagger_{\lambda, ab} \cM^{\otimes 2}}$, the bound reads
\begin{align}
     \var \leq    2\,(\infsnorm{H})^2\sum_{\lambda}\frac{\onesnorm{x^\lambda\,y^\lambda}}{\dim V_{\lambda}}, \label{main:text:eq:rank1_upperbound_case3}
\end{align}
and it does not depend on how the copies are chosen. When $x^\lambda$ and $y^\lambda$ commute for every $\lambda$, for instance, when every $V_\lambda$ in which $\rho^{\otimes 2}$ has a component occurs with multiplicity one, the right-hand side equals
\begin{align}
     2\,(\infsnorm{H})^2\sum_{\lambda}\sum_{a,b}\frac{\Tr{\cJ^\dagger_{\lambda, ab} \rho^{\otimes 2} }\cdot \Tr{\cJ_{\lambda, ab}\, \cM^{\otimes 2}}}{\dim V_{\lambda}}=2\,(\infsnorm{H})^2\,\EXPD{g\sim\mu_G}{\Tr{\cM\,\Pi(g)\rho\,\Pi(g)^\dagger}^2}, \label{main:text:eq:rank1_upperbound_case3:commuting}
\end{align}
that is, $2(\infsnorm{H})^2$ times the second moment of the cost function itself. The second factor of each term carries $\cJ_{\lambda,ab}$, not $\cJ_{\lambda,ab}^\dagger$; with $\cJ_{\lambda,ab}^\dagger$ in both factors the sum would depend on the choice of the copies.

\subsection{Lower bounds}
\label{main:text:subsec:lower}

We now establish rigorous lower bounds. Exploiting the cyclic property of the trace, the exact variance can be written in terms of expectations over the two ensembles,
\begin{align*}
    \var = \EXPD{U_{-}\sim \bbU_{-}, U_{+}\sim \bbU_{+}}{ \Tr{\tcM^{\otimes 2} \commutator{H}{\trho}^{\otimes 2}} }    
\end{align*}where the rotated operators are defined as $\tcM \defeq U_{-}^\dagger \cM U_{-}$ and $\trho \defeq U_{+}\rho U_{+}^\dagger$.
Applying the nested-commutator identity $[H,X]^{\otimes 2}=[H\otimes\bbI,[\bbI\otimes H,X^{\otimes 2}]]$ and the twirl maps $\cT_\pm$ collapses this to the compact operator form (\ref{eq:compact}) which we write as
\begin{align}
  \var = \Tr{\cT_{-}(\cM) \commutator{H \otimes \bbI} { \commutator{ \bbI \otimes H} {\cT_{+}(\rho)}} }.
  \label{eq:var-commutator}
\end{align}
As before, we evaluate the bound according to the convergence of the two ensembles.

\paragraph{Case 1: $U_-$ converges while $U_+$ does not.} Separating the diagonal ($a=b$) from the off-diagonal ($a\neq b$) projector contributions with the reverse triangle inequality, and bounding the residual dependence on $U_+\sim\mathbb{U}_+$ through Jensen's inequality, gives
\begin{align}
    \begin{split}
       \var \geq & \abs{\sum_{\lambda} \sum_{a}^{m_\lambda} \frac{\Tr{\cJ_{\lambda,aa}^\dagger\cM^{\otimes 2}}}{\dim V_\lambda} \Tr{\cJ_{\lambda, aa} \commutator{H\otimes \bbI}{\commutator{\bbI\otimes H}{\cT_{+}(\rho)}}}} \\
        &\hspace{0.5cm} - \EXPD{U \sim \bbU_{+}}{\abs{\sum_{\lambda} \sum_{a\neq b}^{m_\lambda} \frac{\Tr{\cJ_{\lambda,ab}^\dagger\cM^{\otimes 2}}}{\dim V_\lambda} \Tr{\cJ_{\lambda, ab}\commutator{H\otimes \bbI}{\commutator{\bbI\otimes H}{U^{\otimes 2} \rho_{\lambda, a, 1 \to \lambda, b, 2}^{\,(\lambda^\prime, a^\prime) \otimes (\nu^\prime, b^\prime)} (U^\dagger)^{\otimes 2}}}} }}.
    \end{split} \label{main:text:eq:lowerbound_case1}
\end{align}

\paragraph{Case 2: $U_+$ converges while $U_-$ does not.} An analogous argument isolates the non-converged measurement dynamics and controls the cross terms with Schatten-$2$ norms and the intermediate operator $\Mproj$:
\begin{align}
    \begin{split}
       \var \geq  & \abs{\sum_{\lambda} \sum_{a}^{m_\lambda} \frac{\Tr{\cJ_{\lambda,aa}^\dagger\rho^{\otimes 2}}}{\dim V_\lambda} \Tr{\cT_{-}(\cM) \commutator{H\otimes \bbI}{\commutator{\bbI\otimes H}{\cJ_{\lambda, aa}}}}} \\
        &\hspace{0.5cm} -\sum_{\lambda} \sum_{a\neq b}^{m_\lambda} \frac{\snorm{\cJ_{\lambda, bb}\sqrt{{\rho^{\otimes 2}}}}_2\cdot \snorm{\cJ_{\lambda, aa}\sqrt{{\rho^{\otimes 2}}}}_2}{(\dim V_\lambda)}\\
        &\hspace{1cm}\times\EXPD{U\sim \bbU_{-}}{\abs{\Tr{(U^\dagger)^{\otimes 2} \cM_{\lambda, a, 1 \to \lambda, b, 2}^{\,(\lambda^\prime, a^\prime) \otimes (\nu^\prime, b^\prime)} U^{\otimes 2} \commutator{H\otimes \bbI}{\commutator{\bbI\otimes H}{\cJ_{\lambda, ab}}}}}}.
    \end{split} \label{main:text:eq:lowerbound_case2}
\end{align}

\paragraph{Case 3: both $U_-$ and $U_+$ converge.} Applying the reverse triangle inequality to \cref{main:text:eq:exact} systematically separates the intra-representation contributions ($\lambda = \nu$) from the inter-representation interference ($\lambda \neq \nu$). To evaluate the absolute value of the trace for these complex cross-terms, we introduce the following inequality for any Hermitian matrix $\cM$:
\begin{align*}
    \abs{\Tr{\cJ_{\lambda, ab}^\dagger M^{\otimes 2}}} \leq \snorm{\cJ_{\lambda, bb}\sqrt{\abs{M^{\otimes 2}}}}_2\cdot \snorm{\cJ_{\lambda, aa}\sqrt{\abs{M^{\otimes 2}}}}_2
\end{align*}
together with the projector orthogonality $\cJ_{\lambda, ab}\cJ_{\nu, cd} = 0$ for $\lambda \neq \nu$ yields \cref{main:text:eq:lowerbound_case3}.

\subsection{Lower bound under concentration}
\label{main:text:subsec:concentration}

We now derive \cref{main:text:thm:sufficient} from \cref{main:text:eq:lowerbound_case3}. Let $\lambda^\ast$ be a label such that $\rho^{\otimes2}$ is supported on $m_{\lambda^\ast}V_{\lambda^\ast}$, that is, $\iTr{(\bbI-P_{\lambda^\ast})\rho^{\otimes2}}=0$ with $P_{\lambda^\ast}=\sum_{a}\cJ_{\lambda^\ast,aa}$. Then $\iTr{\cJ^{\dagger}_{\lambda,cd}\rho^{\otimes2}}=0$ and $\cJ_{\lambda,cc}\sqrt{\rho^{\otimes2}}=0$ for every $\lambda\neq\lambda^\ast$, so that all terms of \cref{main:text:eq:lowerbound_case3} with labels other than $\lambda^\ast$ vanish. The first term of \cref{main:text:eq:lowerbound_case3} is the absolute value of the sum of the isolated contributions,
\begin{align}
  \abs{\sum_{\lambda}\sum_{a,b,c,d}^{m_\lambda}
  \frac{\Tr{\cJ_{\lambda,ab}^{\dagger}\cM^{\otimes2}}\cdot
        \Tr{\cJ_{\lambda,cd}^{\dagger}\rho^{\otimes2}}}{(\dim V_\lambda)^2}
  \Tr{\cJ_{\lambda,ab}\commutator{H\otimes\bbI}{\commutator{\bbI\otimes H}{\cJ_{\lambda,cd}}}}},
  \label{main:text:theory:isolated:terms}
\end{align}
and it reduces to the absolute value of its term $\lambda=\lambda^\ast$, in which $H\otimes\bbI$ enters through the traces 
\begin{align*}
    \Tr{\cJ_{\lambda^\ast,ab}\commutator{H\otimes\bbI}{\commutator{\bbI\otimes H}{\cJ_{\lambda^\ast,cd}}}}.
\end{align*}The second term of \cref{main:text:eq:lowerbound_case3} reduces to its term $\nu=\lambda^\ast$, the residual cross term, in which each summand of
\begin{align}
  2\sum_{c,d}^{m_{\lambda^\ast}}
  \frac{\snorm{\cJ_{\lambda^\ast,dd}\sqrt{\rho^{\otimes2}}}_2
        \snorm{\cJ_{\lambda^\ast,cc}\sqrt{\rho^{\otimes2}}}_2}
       {\dim V_{\lambda^\ast}}
  \label{eq:conc-crossnorm}
\end{align}
is multiplied by
\begin{align}
  \abs{\Tr{\left(\sum_{\lambda\neq\lambda^\ast}\sum_{a,b}^{m_\lambda}
  \frac{\Tr{\cJ^{\dagger}_{\lambda,ab}\,
  \cM_{\lambda^\ast,c,1\to\lambda^\ast,d,2}^{\,(\lambda',a')\otimes(\nu',b')}}}{\dim V_\lambda}
  \cJ_{\lambda,ab}^{(\lambda',a',\nu',b')_{\lambda^\ast,c,1};\,
  (\lambda',a',\nu',b')_{\lambda^\ast,d,2}}\right)
  (H\otimes\bbI)\,\cJ_{\lambda^\ast,cd}\,(H\otimes\bbI)}},
  \label{eq:conc-crosstrace}
\end{align}
for the same $c,d$. By Supplementary \cref{sec:criteria:lower}, \cref{eq:conc-crosstrace} is equal to the same trace with $(\bbI-P_{\lambda^\ast})\cT_-(\cM)$ in place of the sum in the large parentheses, its absolute value is at most $(\iinfsnorm{H})^2\dim V_{\lambda^\ast}\iinfsnorm{(\bbI-P_{\lambda^\ast})\cM^{\otimes2}(\bbI-P_{\lambda^\ast})}$, and the residual cross term is therefore at most $2m_{\lambda^\ast}(\iinfsnorm{H})^2\iinfsnorm{(\bbI-P_{\lambda^\ast})\cM^{\otimes2}(\bbI-P_{\lambda^\ast})}$. This gives \cref{main:text:thm:sufficient}. If $\rho^{\otimes2}$ has non-zero weight $\iTr{\sum_c\cJ_{\nu,cc}\rho^{\otimes2}}$ on $m_\nu V_\nu$ for labels $\nu\neq\lambda^\ast$, the terms with these labels decrease the bound by at most $(4+2m_\nu)(\iinfsnorm{H})^2(\iinfsnorm{\cM})^2$ times this weight (Supplementary \cref{sec:criteria:lower}), which gives \cref{main:text:cor:leakage}.

\subsection{Finite-difference expansion}
\label{main:text:subsec:fd}
Expanding the composition $f\circ g$ in the definition of $\bbV_{\mathrm{FD}}$ with the shifted unitary operators $U_{\pm\varepsilon}$ and $V_{\pm\varepsilon}$ of \cref{main:text:sec:finite-difference} yields
\begin{align*}
  \begin{split}
    \bbV_{\mathrm{FD}} =\ & \frac{1}{4\varepsilon^2} \left(
      \EXPD{U_{-} \sim \bbU_{-}}{\Tr{\cM^{\otimes 2}(U_{-})^{\otimes 2} U_\varepsilon V_\varepsilon
        \!\left(\EXPD{U_{+}\sim \bbU_{+}}{(U_{+})^{\otimes 2}\rho^{\otimes 2}(U_{+}^\dagger)^{\otimes 2}}\right)
        V_{-\varepsilon} U_{-\varepsilon} (U_{-}^\dagger)^{\otimes 2}}}\right. \\
    &- \EXPD{U_{-} \sim \bbU_{-}}{\Tr{\cM^{\otimes 2}(U_{-})^{\otimes 2} U_\varepsilon V_{-\varepsilon}
        \!\left(\EXPD{U_{+}\sim \bbU_{+}}{(U_{+})^{\otimes 2}\rho^{\otimes 2}(U_{+}^\dagger)^{\otimes 2}}\right)
        V_\varepsilon U_{-\varepsilon}(U_{-}^\dagger)^{\otimes 2}}} \\
    &- \EXPD{U_{-} \sim \bbU_{-}}{\Tr{\cM^{\otimes 2}(U_{-})^{\otimes 2}U_{-\varepsilon} V_\varepsilon
        \!\left(\EXPD{U_{+}\sim \bbU_{+}}{(U_{+})^{\otimes 2}\rho^{\otimes 2}(U_{+}^\dagger)^{\otimes 2}}\right)
        V_{-\varepsilon} U_\varepsilon(U_{-}^\dagger)^{\otimes 2}}} \\
    &+ \left.\EXPD{U_{-} \sim \bbU_{-}}{\Tr{\cM^{\otimes 2}(U_{-})^{\otimes 2}U_{-\varepsilon} V_{-\varepsilon}
        \!\left(\EXPD{U_{+}\sim \bbU_{+}}{(U_{+})^{\otimes 2}\rho^{\otimes 2}(U_{+}^\dagger)^{\otimes 2}}\right)
        V_\varepsilon U_\varepsilon(U_{-}^\dagger)^{\otimes 2}}}\right).
  \end{split}
\end{align*}
Every operator in this expansion acts on one copy at a time: for $s, t \in \{\pm\varepsilon\}$,
\begin{align*}
  (U_{-})^{\otimes 2}\, U_{s} V_{t}\, (U_{+})^{\otimes 2}
  = \big(U_{-}\expm{s\pi_*(g_j)}U_{+}\big)\otimes\big(U_{-}\expm{t\pi_*(g_j)}U_{+}\big).
\end{align*}
Since the trace of a tensor product is the product of the traces, each term of the expansion is a product of two single-copy expectation values $C_{U_\pm}(s)\,C_{U_\pm}(t)$, which gives \cref{main:text:eq:vfd-factorized}.

\subsection{Error and sample complexity}
\label{main:text:subsec:error}
To quantify the performance of \Cref{main:text:new:alg}, we bound separately the error introduced by the finite-difference approximation and the statistical uncertainty inherent to the sampling procedure. We begin with the deterministic error of the finite-difference composition.

\begin{lemma}
  For the anti-Hermitian generator $\pi_*(g_j)$, any Hermitian operator $A$,
  and any step size $\varepsilon\in (0,1)$, the finite-difference
  approximation error is bounded by
  \begin{align*}
    \onesnorm{f\circ g(A) - \commutator{\pi_*(g_j)\otimes \bbI}{\commutator{\bbI \otimes \pi_*(g_j)}{A}}}
    \leq \frac{16}{3}\, \varepsilon^2\, (\infsnorm{\pi_*(g_j)})^4\, \onesnorm{A} + \mathcal{O}(\varepsilon^4).
  \end{align*}
  \label{main:text:tester:lemma:1}
\end{lemma}

A detailed proof of \cref{main:text:tester:lemma:1} is provided in
\cref{sec:tester} of the Supplementary Information.

Next, we establish the statistical sample complexity required to reach an additive error $\teps$. Writing the empirical estimator over $\cN$ independent runs as $\bbO = \frac{1}{\cN}\sum_{j=1}^{\cN} \bbO^{(j)}$, we apply Hoeffding's concentration inequality to bound its deviation.

\begin{lemma}
  Let $\varepsilon\in (0,1)$. For any desired precision $\teps > 0$ and failure
  probability $\delta \in (0, 1)$, if the total sample count satisfies
  \begin{align*}
    \cN \geq \frac{2 (\infsnorm{\cM})^4}{\teps^2 \varepsilon^4} \ln\!\left(\frac{2}{\delta}\right),
  \end{align*}
  then the empirical estimator $\bbO$ tracks the approximate variance
  $\bbV_{\mathrm{FD}}$ up to an additive error $\teps$ with probability at
  least $1 - \delta$.
  \label{main:text:tester:lemma:2}
\end{lemma}

A detailed proof of this sample-complexity bound is provided in
\cref{sec:tester} of the Supplementary Information.

Finally, combining the triangle inequality with
\cref{main:text:tester:lemma:1,main:text:tester:lemma:2}, we bound the total
absolute error between the empirical estimator $\bbO$ and the true variance
$\var$. With probability at least $1-\delta$,
\begin{align}
  \abs{\bbO - \var} \nonumber
  &\leq \abs{\bbO - \bbV_{\mathrm{FD}}} + \abs{\bbV_{\mathrm{FD}} - \var} \nonumber\\
  &\leq \teps + \frac{16}{3}\, \varepsilon^2\, (\infsnorm{\pi_*(g_j)})^4\, \infsnorm{\cM^{\otimes 2}}\, \onesnorm{\rho^{\otimes 2}} + \mathcal{O}(\varepsilon^4).
\end{align}
In the exact form of \Cref{main:text:new:alg}, the outputs are unbiased for $\var$ and lie in $[-\gamma^2(\infsnorm{\cM})^2,\,\gamma^2(\infsnorm{\cM})^2]$, so the same argument gives $\abs{\bbO-\var}\leq\teps$ with probability at least $1-\delta$ as soon as
\begin{align}
  \cN \geq \frac{2\gamma^4(\infsnorm{\cM})^4}{\teps^2}\ln\!\left(\frac{2}{\delta}\right),
  \label{main:text:eq:N-exact}
\end{align}
with no bias term and no dependence on a step size.

\section*{Acknowledgment}

This work was supported by NSF Center for Science of Information (CSoI) Grant
CCF-0939370 and  by NSF Grant CCF-2211423.

\section*{Data availability}
The numerical values plotted in \cref{fig:variance-bounds,fig:placeholder} are available in the folder \texttt{Source\_Data\_CSV} of the \href{https://github.com/SteveMTse/Structure-makes-quantum-circuits-trainable-}{GitHub repository} (\href{https://doi.org/10.5281/zenodo.24182089}{DOI: 10.5281/zenodo.24182089}), and are provided as Source Data with this paper. The exact variances and the bounds in \cref{fig:variance-bounds}, and the exact variance in \cref{fig:placeholder}, also follow from the closed-form expressions in the figure captions and in Supplementary \cref{app:example:SU2,app:example:ff:local,app:example:compound}.

\section*{Code availability}
The code is available in the folder \texttt{sun\_large} of the \href{https://github.com/SteveMTse/Structure-makes-quantum-circuits-trainable-}{GitHub repository} (\href{https://doi.org/10.5281/zenodo.24182089}{DOI: 10.5281/zenodo.24182089}). It is a CUDA C++ implementation of \Cref{main:text:new:alg} for the $\SU{n}$ compound ansatz of \cref{fig:placeholder}, with input state $\Outer{+}{+}^{\otimes n}$ and observable $X^{(1)}X^{(2)}$, which simulates the runs of the algorithm on graphics processing units.

\section*{Competing interests}
The authors declare no competing interests.

\section*{Author contributions}
W.X. developed the theory and the sampling algorithm, proved the results, carried out the numerical computations and wrote the manuscript. A.G. and W.S. supervised the project, contributed to the analysis and the interpretation of the results, and revised the manuscript. All authors read and approved the final manuscript.

\bibliographystyle{unsrt}
\bibliography{references}

\clearpage
\appendix

\renewcommand{\thesection}{S\arabic{section}}
\renewcommand{\thefigure}{S\arabic{figure}}
\renewcommand{\thetable}{S\arabic{table}}

\setcounter{section}{0}
\setcounter{figure}{0}
\setcounter{table}{0}
\setcounter{theorem}{0}\setcounter{lemma}{0}\setcounter{claim}{0}\setcounter{proposition}{0}
\setcounter{corollary}{0}\setcounter{fact}{0}\setcounter{observation}{0}\setcounter{assumption}{0}
\setcounter{conjecture}{0}\setcounter{definition}{0}
\numberwithin{equation}{section}
\setcounter{algocf}{0}
\renewcommand{\thealgocf}{S\arabic{algocf}}


\pdfbookmark[0]{Supplementary Materials}{supplementary-materials} 
\etoctoccontentsline{part}{Supplementary Materials} 
\etocsettocstyle{\begin{center}{\Large\bfseries Supplementary Materials}\end{center}%
  \markboth{SUPPLEMENTARY MATERIALS}{SUPPLEMENTARY MATERIALS}}{\bigskip} 
\etocsetnexttocdepth{2} 
\localtableofcontents

\section{Setting and Main Results}
\label{sec:supp:setting}

Let $G$ be a compact and connected Lie group with Lie algebra $\fg$. Let $(\cH, \Pi)$ be a finite-dimensional unitary representation of $G$. Suppose $\CurB{g_1, \cdots, g_k} \subseteq \fg$ is a spanning set of $\fg$. Consider the parametrized unitary ans\"atze
\begin{align}
    \absU = \prod_{l=1}^L \prod_{q = 1}^k \Pi\left(\expm{\theta_{l,q} X_{l, q}}\right) \label{ansatz}
\end{align}where $\theta_{l,q}\in [-\pi, \pi]$ and $X_{l,q}\in \fg$ are indexed basis. Here, for the same $q$, we assume $X_{1, q} = X_{2, q} = \cdots = X_{L, q} = g_q$. Let $\rho$ be a quantum state supported on $\cH$, and let $\cM$ be a measurement operator. We define the objective function
\begin{align}
    C(\vtheta) = \Tr{\cM \absU \rho \absUd}. \label{obj}
\end{align}

In this paper, we establish the following two results.

\begin{theorem}
\label{thm:variance_bounds}
Let $C(\vtheta)$ be the objective function of a parameterized quantum circuit acting on an input state $\rho$ with measurement operator $\cM$. Assume the circuit depth satisfies $L \ge 2L^*$ and that the expectation of the partial derivative vanishes (\cref{lemma:expectation}). Let $\bbU_-$ and $\bbU_+$ denote the unitary distributions induced by the circuit segments preceding and following the parameter $\theta_{\ell,j}$ in the written order of the product in \cref{ansatz}. 

Depending on the convergence of $\bbU_-$ and $\bbU_+$ to an $\varepsilon$-approximate 2-design with respect to the group $G$, the variance $\VARD{\vtheta}{\partial_{\theta_{\ell,j}} C}$ satisfies the following bounds:
\begin{enumerate}
    \item If only $\bbU_-$ (resp. $\bbU_+$) converges to a 2-design, the variance is bounded by the Haar integration over the measurement (resp. state) operator and the one-parameter subgroup generators of the non-convergent segment. These asymmetric limits are bounded above by \cref{app:upperbound:case_1} (resp. \cref{app:upperbound:case_2}) and below by \cref{app:lowerbound:case_1} (resp. \cref{app:lowerbound:case_2}).
    
    \item If both $\bbU_-$ and $\bbU_+$ form 2-designs, the variance is fully decoupled and given exactly by \cref{eq:true:var} in terms of the representation-theoretic projections of $\rho$ and $\cM$. The resulting symmetric upper and lower bounds are given in \cref{app:upperbound:case_3} and \cref{app:lowerbound:case_3}. Moreover, with the copies of the Clebsch--Gordan decomposition (\cref{sec:upper:rank1}), \cref{app:lowerbound:case_3} implies the lower bound \cref{app:crit:concentration} of \cref{app:thm:concentration} whenever $\rho^{\otimes 2}$ is supported on the copies of one irreducible representation $V_{\lambda^\ast}$, and, for every state, the same bound with the term \cref{main:text:eq:leakage} of the main text subtracted (\cref{main:text:cor:leakage} of the main text). In particular, there is no barren plateau if the contribution of $V_{\lambda^\ast}$ to the first term of \cref{app:lowerbound:case_3} is at least inverse-polynomial in $n$ in absolute value and the terms subtracted from it are together at most half of its absolute value.
\end{enumerate}
\end{theorem}

\begin{theorem}
    Consider a quantum circuit $C(\vtheta)$ under the aforementioned setting (see \cref{main:text:subsec:prelim}). Let the measurement $\cM$ and $\rho$ be the input quantum state that has support on $\cH$. Then, for any $\ell \in [L]$ and $j \in [k]$, there exists a sampling-based algorithm that estimates the variance of the partial derivative $\partial_{\theta_{\ell,j}} C$ from runs of the circuit and, for any tolerance $\tau > 0$, separates variances above $3\tau/2$ from variances below $\tau/2$ with a number of runs polynomial in $1/\tau$ (\cref{app:tester:prop}).
\end{theorem}

The remainder of the paper is organized as follows. In \cref{sec:prelim}, we review basic concepts from Lie groups, Lie algebras, and representation theory. In \cref{subsec:commutant}, we introduce the notion of the commutant. In \cref{subsec:prop:topo:lin:alg}, we establish several auxiliary results that will be used throughout the paper. Next, in \cref{sec:reduction}, we reduce the ans\"atze defined in \cref{ansatz} to a matrix form. In \cref{sec:express}, we analyze the expressibility of \cref{ansatz}. We then derive upper and lower bounds for the variance of the partial derivative $\partial_{\theta_{\ell,j}} C$ in \cref{sec:upper} and \cref{sec:lower}, respectively. In \cref{sec:criteria} we derive from these bounds the trainability criteria of \cref{main:text:theory:sec:discussion} of the main text, among them a lower bound for states and observables concentrated on the copies of one irreducible representation. Finally, in \cref{sec:tester}, we present a sampling algorithm that estimates the variance directly, show how it decides trainability at a given tolerance, and discuss its cost.

\section{Preliminary}
\label{sec:prelim}
\subsection{Review of Lie Group, Lie Algebra, and Representation Theory}
\subsubsection{Exponential Map}
In this section, we briefly review the general definition of the exponential map. This construction extends the familiar matrix exponential to an abstract Lie algebra, allowing one to define exponentiation even when the Lie algebra is not realized as a matrix algebra.
\begin{definition} [Exponential Map \cite{kirillov2008introduction}]
    Let $G$ be a real or complex Lie group, $\fg = T_eG$. Then the exponential map $\exp:\fg \to G$ is defined by
    \begin{align}
        \expm{X} = \gamma_X(1)
    \end{align}for all $X\in \fg$\label{def:exp:map}
\end{definition}

Here, $T_eG$ denotes the tangent space of $G$ at the identity element $e$. 
The map $\gamma_X : \bbK \to G$ (typically with $\bbK = \bbR$ or $\bbC$) is the 
unique Lie group morphism satisfying
\begin{align*}
    \gamma_X\prime(0) = \frac{d}{dt}\gamma_X(t)\bigg|_{t=0} = X .
\end{align*}This curve is called the \emph{one-parameter subgroup generated by $X$}. 
From this definition it follows that
\begin{align*}
    \gamma_X(t) = \gamma_{tX}(1) = \exp(tX),
\end{align*}which generalizes the usual matrix exponential parameterization of one-parameter subgroups. The exponential map also satisfies several fundamental properties. Below we list the properties that will be needed in this paper.

\begin{proposition} [Properties of Exponential Maps \cite{kirillov2008introduction}]
    Let $G$ be a real or complex Lie group and $\fg = T_eG$.
    \begin{itemize}
        \item The exponential map is a diffeomorphism between some neighborhood of $0$ in $\fg$ and a neighborhood of $e$ in $G$. The local inverse map will be denoted by $\log$.

        \item For any morphism of Lie groups $\varphi:G_1 \to G_2$ and any $X\in \fg_1$, we have $\exp(\varphi_*(X)) = \varphi(\exp(X))$, where $\varphi_* : \fg_1 \to \fg_2$ denotes the differential of $\varphi$ at the identity.
    \end{itemize} \label{prop:exp:map}
\end{proposition}

Throughout this work, unless stated otherwise, the term \emph{exponential map} refers to this general exponential map.

\subsubsection{Lie Group and Lie Algebra}
In this section, we briefly recall several notations from Lie group, Lie algebra and representation theory that will be used throughout this paper. We begin with Lie groups and Lie algebras

\begin{definition} [Lie Group (\cite{hall2013lie})]
    A \textbf{Lie group} is a smooth manifold $G$ that is also a group and such that the group product $G\times G \to G$ and inverse map $G \to G$ are smooth.
    $\label{def:lie:gp}$
\end{definition}

\begin{definition} [Lie Algebra (\cite{hall2013lie})]
    A \textbf{finite-dimensional real or complex Lie algebra} is a finite-dimensional real or complex vector space $\fg$, together with a map $\commutator{\cdot}{\cdot}:\fg\times \fg \to \fg$, with the following properties 
    \begin{enumerate}
        \item $\commutator{\cdot}{\cdot}$ is bilinear
        \item $\commutator{\cdot}{\cdot}$ is skew-symmetric, i.e., $[X,Y] = -[Y,X]$ for all $X,Y\in \lag$
        \item The Jacobi identity holds, i.e., 
        \begin{align}
            \commutator{X}{\commutator{Y}{Z}} + \commutator{Y}{\commutator{Z}{X}} + \commutator{Z}{\commutator{X}{Y}} = 0
        \end{align}for all $X,Y,Z\in \lag$
    \end{enumerate}
    $\label{def:lie:alg}$
\end{definition}

In topology, a space $S$ is called \textbf{compact} if every open cover of $S$ admits a finite subcover, and \textbf{connected} if it cannot be written as the union of two disjoint nonempty open subsets. Since a Lie group $G$ is both a smooth manifold and a topological space, these notions naturally apply to Lie groups. When $G$ is a matrix Lie group, however, compactness and connectedness admit more concrete characterizations and enjoy several useful properties arising from the ambient matrix topology. We therefore recall the notions of compactness and connectedness in the setting of matrix Lie groups.

\begin{definition} [Compactness of Matrix Lie Group (\cite{hall2013lie})]
    A \textbf{matrix Lie group} $G$ is said to be \textbf{compact} if it is compact in the usual topological sense as a subset of $M_n(\bbC)\cong \bbR^{2n^2}$
\end{definition}

\begin{definition} [Connected Matrix Lie Subgroup (\cite{hall2013lie})]
    If G is a matrix Lie group with Lie algebra $\fg$, then $H \subset G$ is a \textbf{connected Lie subgroup} of $G$ if the following conditions are satisfied:
    \begin{enumerate}
        \item $H$ is a subgroup of $G$.
        \item The Lie algebra $\fh$ of $H$ is a Lie subalgebra of $\fg$.
        \item Every element of $H$ can be written in form $e^{X_1}e^{X_2}\cdots e^{X_m}$, with $X_1, X_2,\cdots, X_m\in \fh$.
    \end{enumerate}\label{def:connected:lie:subgp}
\end{definition}

\begin{remark}
    For a compact Lie group $G$, there exists a unique normalized Haar measure $\mu_G$, which is bi-invariant. In particular, for any $h\in G$ and any integrable function $f$,
    \begin{align*}
        \int_G f(gh) d\mu_G(g) = \int_G f(hg) d\mu_G(g) = \int_G f(g) d\mu_G(g)
    \end{align*}(See Section~9 of \cite{Cohn2013}.)
\end{remark}

The following results connects Lie sub-algebras and Lie sub-groups.

\begin{theorem} [Lie's Third Theorem \cite{hall2013lie}]
    Let $G$ be a matrix Lie group with Lie algebra $\fg$ and let $\fh$ be a Lie sub-algebra of $\fg$. Then, there exists a unique connected Lie subgroup $H$ of $G$ with Lie algebra $\fh$. \label{fact:2}
\end{theorem}

\subsubsection{Representation Theory}
We recall basic notions from representation theory that will be used throughout the paper.

\begin{definition} [Lie Group Homomorphism (\cite{hall2013lie})]
    Let $G$ and $H$ be matrix Lie groups. A map $\Phi$ from $G$ to $H$ is called \textbf{Lie group homomorphism} if (1) $\Phi$ is a group homomorphism, which satisfies $\varphi(g_1 g_2) = \varphi(g_1)\varphi(g_2)$ for all $g_1, g_2\in G$, and (2) $\Phi$ is continuous. If, in addition, $\Phi$ is bijective and $\Phi^{-1}$ is continuous, then $\Phi$ is called a \textbf{Lie group isomorphism}.\label{def:group:homo}
\end{definition}

Similarly, we can define the Lie algebra homomorphisms. 

\begin{definition} [Lie Algebra Homomorphism (\cite{hall2013lie})]
    If $\fg$ and $\fh$ are Lie algebras (\cref{def:lie:alg}), then a linear map $\phi:\fg \to \fh$ is called a \textbf{Lie algebra homomorphism} if $\phi(\commutator{X}{Y}) = \commutator{\phi(X)}{\phi(Y)}$ for all $X,Y\in \lag$. If, in addition, $\phi$ is bijective, then $\phi$ is called a \textbf{Lie algebra isomorphism}.
    $\label{def:lie:alg:homo}$
\end{definition}

These notions are connected by the fundamental theorem of Lie theory. For our purposes, we highlight the following result.

\begin{theorem}[Fundamental Theorem of Lie Theory {\cite{kirillov2008introduction}}]
    Every Lie group homomorphism $\varphi: G_1 \to G_2$ induces a Lie algebra homomorphism $\varphi_* : \fg_1 \to \fg_2$. Moreover, if $G_1$ is connected, then the map
    \begin{align*}
        \Hom{}{G_1}{G_2} &\to \Hom{}{\fg_1}{\fg_2} \\
        \varphi &\mapsto \varphi_*
    \end{align*}
    is injective, where $\Hom{}{G_1}{G_2}$ and $\Hom{}{\fg_1}{\fg_2}$ denote the sets of Lie group and Lie algebra homomorphisms, respectively.
    \label{fund:thm:lie}
\end{theorem}

Then, we can define the representation for Lie groups and Lie algebras.

\begin{definition} [Representation]
    Let $G$ be a matrix Lie group. A \textbf{finite-dimensional complex representation} of $G$ is a Lie group homomorphism 
    \begin{align}
        \Pi: G \to \GL{V}
    \end{align}Similarly, let $\fg$ be a real or complex Lie algebra. A \textbf{finite-dimensional complex representation} of $\fg$ is a Lie algebra homomorphism $\pi$ of $\fg$ into $\gl{V}$\label{def:rep}
\end{definition}

A representation may admit invariant subspaces, which lead to the notion of irreducibility. 

\begin{definition} [Irreducible Representation]
    Let $\Pi$ be a finite-dimensional real or complex representation of a matrix Lie group $G$, acting on a space $V$. A space $W$ of $V$ is called \textbf{invariant (or $G$-invariant)} if $\Pi(A)w\in W$ for all $w\in W$ and all $A\in G$. An invariant subspace $W$ is called \textbf{non-trivial} if $W\neq \CurB{0}$ and $W\neq V$. A representation with no non-trivial invariant subspace is called \textbf{irreducible}. \label{def:ir:rep}
\end{definition}

\begin{definition} [Completely Reducible Representation]
    A representation $V$ is called \textbf{completely reducible} or \textbf{semi-simple} if it is isomorphic to a direct sum of irreducible representations, i.e., $V \simeq \bigoplus V_\lambda$, $V_\lambda$ are irreducible.
\end{definition}

If isomorphic irreducible components are grouped together, one may write $V\simeq n_\lambda V_\lambda$, where $n_\lambda \in \bbZ_{+}$ are \textbf{\emph{multiplicities}} and $V_i$ are pairwise non-isomorphic irreducible representations. 

\begin{definition} [Unitary Representation]
    If $V$ is a finite-dimensional inner product space and $G$ is a matrix Lie group, a representation $\Pi: G \to \GL{V}$ is \textbf{unitary} if $\Pi(g)$ is a unitary operator for all $g\in G$. For a Lie algebra $\fg$, we say a representation $\pi: \fg \to \gl{V}$ is \textbf{unitary} if $\pi(A)$ is anti-Hermitian for all $A\in \fg$. 
\end{definition}

As stated in Theorem 4.30 of \cite{kirillov2008introduction}, \emph{every finite-dimensional unitary representation of a compact Lie group is completely reducible}. In particular, for any finite-dimensional unitary representation $(V \subseteq \bbC^{d},\Pi)$ of a compact Lie group $G$, one can naturally induce a representation of $G$ on the tensor product space $V^{\otimes 2}$. Using standard properties of tensor products in linear algebra, one easily verifies that the tensor product representation $(V^{\otimes 2}, \Pi^{\otimes 2})$ is again unitary.

By the complete reducibility of unitary representations, the tensor product representation decomposes into mutually orthogonal, irreducible invariant subspaces (see Theorem 4.30 in \cite{hall2013lie}):
\begin{align}
    V^{\otimes 2} = \bigoplus_{\lambda} m_\lambda V_\lambda,
\end{align}Consequently, for any operator $A \in \End{V^{\otimes 2}}$, the second-moment operator can be expressed as
\begin{align}
    \int_G \Pi(g)^{\otimes 2} A (\Pi(g)^\dagger)^{\otimes 2} d\mu_G(g) = \sum_{\lambda} \sum_{a,b=1}^{m_\lambda}\frac{\Tr{A\cJ^\dagger_{\lambda,ab}}}{\dim V_\lambda}\cJ_{\lambda, ab},\label{eq:35}
\end{align}where $\mu_G$ is the Haar measure over the compact Lie group $G$. Here, $\cJ_{\lambda,ab} = \sum_{m=1}^{\dim V_\lambda} \Outer{\phi_{\lambda, b, m}}{\phi_{\lambda, a, m}}$ maps the $a$-th copy of $V_\lambda$ to its $b$-th copy, and $\CurB{\state{\phi_{\lambda, a, m}}}$ represents an orthonormal basis for the $a$-th copy (with analogous definitions for $b$). The orthonormal bases of the $m_\lambda$ copies of $V_\lambda$ cannot be chosen independently of one another: they are chosen such that $\Pi(g)^{\otimes 2}$ has the same matrix on every copy. That is, for every $g\in G$ there is a unitary matrix $D^{\lambda}(g)$ of size $\dim V_\lambda\times\dim V_\lambda$, which does not depend on the copy index $a$, such that
\begin{align*}
    \Pi(g)^{\otimes 2}\state{\phi_{\lambda, a, m}} = \sum_{m^\prime=1}^{\dim V_\lambda} D^{\lambda}_{m^\prime m}(g)\state{\phi_{\lambda, a, m^\prime}}\quad\text{for all } a=1,\cdots,m_\lambda \text{ and all } m=1,\cdots,\dim V_\lambda.
\end{align*}Such bases exist because the copies of $V_\lambda$ are isomorphic representations. Indeed, let $\CurB{\state{\phi_{\lambda, 1, m}}}$ be an orthonormal basis of the first copy, and for $a\geq 2$ let $S_a$ be an invertible linear map from the first copy onto the $a$-th copy with $S_a\Pi(g)^{\otimes 2}\state{v}=\Pi(g)^{\otimes 2}S_a\state{v}$ for all $g\in G$ and all $\state{v}$ in the first copy. The operator $S_a^\dagger S_a$ commutes with the irreducible action of $G$ on the first copy, so by Schur's lemma $S_a^\dagger S_a=c_a\bbI$ with $c_a>0$. Hence $T_a\defeq S_a/\sqrt{c_a}$ is unitary, and the vectors $\state{\phi_{\lambda, a, m}}\defeq T_a\state{\phi_{\lambda, 1, m}}$ form an orthonormal basis of the $a$-th copy that satisfies the equation above with $D^{\lambda}_{m^\prime m}(g)=\cstate{\phi_{\lambda, 1, m^\prime}}\Pi(g)^{\otimes 2}\state{\phi_{\lambda, 1, m}}$. With this choice of the bases, every $\cJ_{\lambda,ab}$ commutes with $\Pi(g)^{\otimes 2}$:
\begin{align*}
    \Pi(g)^{\otimes 2}\cJ_{\lambda,ab} = \sum_{m, m^\prime=1}^{\dim V_\lambda} D^{\lambda}_{m^\prime m}(g)\Outer{\phi_{\lambda, b, m^\prime}}{\phi_{\lambda, a, m}} = \cJ_{\lambda,ab}\Pi(g)^{\otimes 2}.
\end{align*}The first equality is the equation above for the $b$-th copy. The second equality uses $\cstate{\phi_{\lambda, a, m^\prime}}\Pi(g)^{\otimes 2} = \sum_{m=1}^{\dim V_\lambda}D^{\lambda}_{m^\prime m}(g)\cstate{\phi_{\lambda, a, m}}$, which holds because $\cstate{\phi_{\lambda, a, m^\prime}}\Pi(g)^{\otimes 2}\state{\phi_{\lambda, a, m}}=D^{\lambda}_{m^\prime m}(g)$ and because $\Pi(g)^{\otimes 2}$ maps the orthogonal complement of the $a$-th copy into itself. For this reason we refer to the operators $\cJ_{\lambda,ab}$ as intertwiners. Without this property \cref{eq:35} does not hold: for $A=\cJ_{\lambda,ab}$ its right-hand side equals $\cJ_{\lambda,ab}$, because $\Tr{\cJ_{\lambda,ab}\cJ^\dagger_{\lambda^\prime,cd}}=\delta_{\lambda\lambda^\prime}\delta_{ac}\delta_{bd}\dim V_\lambda$, whereas its left-hand side commutes with $\Pi(h)^{\otimes 2}$ for every $h\in G$ by the invariance of the Haar measure.

\begin{remark}
    The orthogonal projector onto each irreducible representation can be explicitly constructed using Weingarten calculus \cite{collins2022weingarten}. Furthermore, when an irreducible representation appears with multiplicity greater than one, its isomorphic copies are mutually orthogonal, and thus each corresponds to a distinct orthogonal projector.
\end{remark}

\subsection{Commutant}
\label{subsec:commutant}
We begin by recalling the notion of a commutant, which plays a central role in our analysis. Intuitively, the commutant of a set of operators consists of all endomorphisms that commute with every element of that set. More precisely, it is defined as the following 

\begin{definition} [$k$-th fold commutant]
    Let $V$ be a finite-dimensional complex Hilbert space and $A\subseteq \End{V}$ be any set of linear operators. The \textbf{$k$-th fold commutant of $A$} is 
    \begin{align}
        \Comm{k, A} = \CurB{B \in \End{V^{\otimes k}}\enspace:\enspace \commutator{B}{X^{\otimes k}} = 0,\forall X\in A},
    \end{align}where $\End{V^{\otimes k}}$ is is the set of all linear operators from $V^{\otimes k}$ to itself. \label{def:commutant}
\end{definition}

Now, we consider the case $A = \cU(G) \defeq \CurB{\cU(g):\forall g \in G}$, where $G$ be any compact Lie group and $(\cU, V \subseteq \bbC^d)$ a finite-dimensional unitary representation of $G$, with $\cU: G\to \mathrm{U}(V)$ a Lie group homomorphism. By \cref{def:commutant}, the $k$-th fold commutant of $\cU(G)$ is given by
\begin{align}
    \Comm{k, \cU(G)} = \CurB{B \in \End{V^{\otimes k}}\enspace:\enspace \commutator{B}{\cU(g)^{\otimes k}} = 0,\forall g\in G}
\end{align}
The following proposition shows that the Haar-twirling operator over $G$ is an orthogonal projector from a operator that has support on $V^{\otimes k}$ onto $\iComm{k, \cU(G)}$. 

\begin{proposition} Let $\mu$ be a bi-invariant Haar measure defined over $G$. Then,
    \begin{align}
        \cP_k(\cdot) = \int_G (\cU(g))^{\otimes k} \cdot (\cU(g)^\dagger)^{\otimes k} d\mu(g)
    \end{align}is an orthogonal projector onto $\iComm{k, \cU(G)}$ \label{prop:cP:is:orth:proj}
\end{proposition}

This result is well known; for completeness, we provide a brief proof sketch.

\begin{proof} [Sketch Proof of \cref{prop:cP:is:orth:proj}]
    To show the $\cP_k$ is projector, we consider $\cU(g)$ for any $g\in G$. Then, using \cref{def:commutant} and property of Kronecker product, we get
    \begin{align}
        &\cP_k\left((\cU(g))^{\otimes k}\cdot (\cU(g)^\dagger)^{\otimes k}\right) \\
        =& \int_G (\cU(h))^{\otimes k} \left((\cU(g))^{\otimes k}\cdot (\cU(g)^\dagger)^{\otimes k} \right)(\cU(h)^\dagger)^{\otimes k} d\mu(h)  \\
        =& \int_G (\cU(h)\cU(g))^{\otimes k} \cdot (\cU(g)^\dagger\cU(h)^\dagger)^{\otimes k} d\mu(h) \\
        =& \int_G (\cU(hg))^{\otimes k} \cdot (\cU(g^{-1}h^{-1}))^{\otimes k} d\mu(h) & \text{(\cref{def:group:homo})} \\
        =& \int_G (\cU(h))^{\otimes k} \cdot (\cU(h^{-1}))^{\otimes k} d\mu(h) = \cP_k & \text{(Haar measure over $G$ is bi-invariant)}.
    \end{align}Using identity above and linearity of $\cP_k$, we get
    \begin{align}
        \cP_k^2 = \int_G \cP_k\left((\cU(g))^{\otimes k}\cdot (\cU(g)^\dagger)^{\otimes k}\right) d\mu(g) = \int_G \cP_k d\mu(g) = \cP_k.
    \end{align}Moreover, $\cP_k^\dagger = \cP_k$ is trivial. All together, $\cP_k$ is an orthogonal projector onto $\iComm{k, \cU(G)}$. 
\end{proof}

\subsection{Auxiliary Results}
\label{subsec:prop:topo:lin:alg}

In this section, we establish several auxiliary results that will be used throughout the paper.

\begin{theorem} [Thm 5.6.12 in \cite{0521386322}]
    Let $A$ be any square matrix. Then, $\lim_{L\to \infty} A^L = 0$ if and only if the spectral radius of $A$ is strictly smaller than 1, i.e., $\rho(A) < 1$. \label{thm:5.6.12:converge:small:sr}
\end{theorem}

\begin{fact}
    If $H$ is an anti-Hermitian matrix, then 
    \begin{align*}
        \abs{\Tr{\sum_{d}\cJ_{\lambda, ad} H^{\otimes 2}}} \leq \dim V_\lambda \cdot \infsnorm{H^{\otimes 2}}
    \end{align*} \label{fact:abs:trJH}
\end{fact}

\begin{proof}
    Using this decomposition and the definition $\sum_{d}\cJ_{\lambda, ad} = \sum_{d} \sum_{m} \Outer{\phi_{\lambda, d, m}}{\phi_{\lambda, a, m}}$, we evaluate the trace:$$\begin{aligned}
    &\abs{\Tr{\sum_{d}\cJ_{\lambda, ad} H^{\otimes 2}}} \\
    =& \abs{\Tr{\sum_{d} \sum_{m} \Outer{\phi_{\lambda, d, m}}{\phi_{\lambda, a, m}} \left(\sum_{\nu} \nu \Outer{\nu}{\nu}\right)}} \\
    \leq& \left(\max_{\nu} |\nu|\right) \cdot \abs{\Tr{\sum_{d} \sum_{m} \Outer{\phi_{\lambda, d, m}}{\phi_{\lambda, a, m}} \left(\sum_{\nu}\Outer{\nu}{\nu}\right)}} 
\end{aligned}$$Since $\sum_{\nu}\Outer{\nu}{\nu} = \bbI$, and recognizing that the maximum absolute eigenvalue is the spectral norm:$$\begin{aligned}
    \abs{\Tr{\sum_{d}\cJ_{\lambda, ad} H^{\otimes 2}}} \leq \infsnorm{H^{\otimes 2}} \cdot \abs{\sum_{d}\sum_{m}\Inner{\phi_{\lambda, a, m}}{\phi_{\lambda, d, m}}}
\end{aligned}$$Since the basis is orthonormal, $\Inner{\phi_{\lambda, a, m}}{\phi_{\lambda, d, m}} = 1$ if and only if $d = a$, and evaluates to $0$ otherwise. Summing over the $\dim V_\lambda$ internal states $m$:$$\begin{aligned}
    \abs{\Tr{\sum_{d}\cJ_{\lambda, ad} H^{\otimes 2}}} \leq& \dim V_\lambda \cdot \infsnorm{H^{\otimes 2}}
\end{aligned}$$
\end{proof}

\begin{fact} [Holder inequality for Schatten norms (\cite{shebrawi2013trace})]
    Let $A_1, \cdots, A_k\in M_n(\bbC)$ be any matrices. Then, 
    \begin{align}
        \abs{\Tr{A_1\cdots A_k}} \leq \prod_{j=1}^k \norm{A_j}^S_{p_j}
    \end{align}where $\sum_{j=1}^k \frac{1}{p_j} = 1$ for all $p_j \in [1, \infty]$ and $\norm{\cdot}^S_p = \left(\Tr{\abs{A}^p}\right)^{1/p}$ is the Schatten-$p$ norm, $\abs{A} = \sqrt{A^\dagger A}$. \label{fact:holder}
\end{fact}
\begin{observation}
Then, for projector $\cJ_{\lambda, ab}$, we have
\begin{align}
    \snorm{\cJ_{\lambda, ab}}_2 = \left(\tr{|\cJ_{\lambda, ab}|^2}\right)^{1/2} = \left(\Tr{(\sqrt{\cJ_{\lambda, ab}^\dagger \cJ_{\lambda, ab}})^2}\right)^{1/2} = \left(\Tr{\cJ_{\lambda, aa}}\right)^{1/2} = \sqrt{\dim V_\lambda}
\end{align} \label{obs:norm{P}S2}
\end{observation}

\begin{observation}
    For any two complex matrices $A, B\in M_n(\bbC)$, we have
    \begin{align}
        \norm{A\otimes B}^S_{\infty} =& \sigma_{\text{max}}(A\otimes B) \\
        =& \sqrt{\lambda_{\text{max}}\left((A\otimes B)^\dagger(A\otimes B)\right)} = \sqrt{\lambda_{\text{max}}\left((A^\dagger A)\otimes (B^\dagger B)\right)} \\
        =& \sqrt{\lambda_{\text{max}}\left(A^\dagger A\right) \lambda_{\text{max}}\left(B^\dagger B\right)} \\
        =& \norm{A}^S_{\infty}\norm{B}^S_{\infty}
    \end{align}where $\sigma_{\text{max}}$ is the maximum singular value and $\lambda_{\text{max}}$ is the maximum eigenvalue. \label{obs:norm{A otimes B}Sinfty}
\end{observation} 

\begin{fact}
    For any linear operators $H$ and $M$ acting on a Hilbert space, the nested commutator of their tensor products simplifies to the tensor square of their commutator:
    \begin{align*}
        [H, M]^{\otimes 2} = [H \otimes I, [I \otimes H, M^{\otimes 2}]]
    \end{align*} \label{app:prelim:nested:comm}
\end{fact}

\begin{proof} [Sketch Proof of \cref{app:prelim:nested:comm}]
    Using the mixed-product property of the tensor product, $(A \otimes B)(C \otimes D) = AC \otimes BD$, and the standard definition of the commutator, $[X, Y] = XY - YX$, we have
    \begin{align*}
        [H \otimes I, [I \otimes H, M^{\otimes 2}]] &= [H \otimes I, M \otimes HM - M \otimes MH] \\
        &= [H \otimes I, M \otimes [H, M]] \\
        &= HM \otimes [H, M] - MH \otimes [H, M] \\
        &= (HM - MH) \otimes [H, M] \\
        &= [H, M]^{\otimes 2}
    \end{align*}
\end{proof}

\begin{fact}
    Let $H\in \fu(V)$ be anti-Hermitian and $\sum_{j}\lambda_j \Outer{\lambda_j}{\lambda_j}$ be the spectral decomposition of $H$. Then, for any integer $k \geq 1$ and any Hermitian matrix $M \in i\fu(V^{\otimes k})$, 
    \begin{align}
        \cP^k_H(M) =& \frac{1}{2\pi}\int_t (e^{tH})^{\otimes k} M (e^{-tH})^{\otimes k} dt \nonumber \\
        =& \sum_{\substack{(j^{(1)}_1, \cdots, j^{(1)}_k) \\ (j^{(2)}_1, \cdots, j^{(2)}_k)}} \alpha\left(\lambda_{j^{(1)}_1\cdots j^{(1)}_k}, \lambda_{j^{(2)}_1\cdots j^{(2)}_k}\right)\Outer{\lambda_{j_1^{(1)}\cdots j_k^{(1)}}}{\lambda_{j_1^{(1)}\cdots j_k^{(1)}}}M\Outer{\lambda_{j_1^{(2)}\cdots j_k^{(2)}}}{\lambda_{j_1^{(2)}\cdots j_k^{(2)}}}
    \end{align}where $\state{\lambda_{j_1^{(\eta)}\cdots j_k^{(\eta)}}} = \bigotimes_{i=1}^k\state{\lambda_{j_i^{(\eta)}}}$ is an eigenvector of $(e^{tH})^{\otimes k}$ associated with the eigenvalue $\expm{t\lambda_{j_1^{(\eta)}\cdots j_k^{(\eta)}}} \allowbreak =  \expm{t(\sum_{i=1}^k\lambda_{j_i^{(\eta)}})}$ for $\eta = 1,2$ and $\alpha:\bbC^2 \to \bbR$ is defined by 
    \begin{align}
        &\left(\lambda_{j^{(1)}_1\cdots j^{(1)}_k},\enspace \lambda_{j^{(2)}_1\cdots j^{(2)}_k}\right) \mapsto \nonumber \\
        & \hspace{0.5cm}\bbI\left[\lambda_{j^{(1)}_1\cdots j^{(1)}_k} - \lambda_{j^{(2)}_1\cdots j^{(2)}_k} \neq 0\right]\frac{\sin\left(i\pi\left(\lambda_{j^{(2)}_1\cdots j^{(2)}_k} - \lambda_{j^{(1)}_1\cdots j^{(1)}_k}\right)\right)}{i\pi\left(\lambda_{j^{(2)}_1\cdots j^{(2)}_k} - \lambda_{j^{(1)}_1\cdots j^{(1)}_k}\right)} +\bbI\left[\lambda_{j^{(1)}_1\cdots j^{(1)}_k} - \lambda_{j^{(2)}_1\cdots j^{(2)}_k} = 0\right] 
    \end{align} \label{fact:true:bound:important:fact}
\end{fact}

\begin{proof} [Proof of \cref{fact:true:bound:important:fact}]
    Using the property of Kronecker product and matrix exponential, we note that the spectral decomposition of $\expm{tH}^{\otimes k}$ is given by
    \begin{align}
        \expm{tH}^{\otimes k} = \sum_{j_1,\cdots, j_k}\expm{t\sum_{i=1}^k\lambda_{j_i}} \bigotimes_{i=1}^k \Outer{\lambda_{j_i}}{\lambda_{j_i}},
    \end{align}which implies
    \begin{align}
        &\frac{1}{2\pi}(e^{tH})^{\otimes k} M (e^{-tH})^{\otimes k} \\
        =& \frac{1}{2\pi}\sum_{\substack{(j^{(1)}_1, \cdots, j^{(1)}_k) \\ (j^{(2)}_1, \cdots, j^{(2)}_k)}}\expm{t\left(\sum_{i=1}^k\lambda_{j_i^{(1)}} - \sum_{i=1}^k\lambda_{j_i^{(2)}}\right)} \Outer{\lambda_{j_1^{(1)}\cdots j_k^{(1)}}}{\lambda_{j_1^{(1)}\cdots j_k^{(1)}}}M\Outer{\lambda_{j_1^{(2)}\cdots j_k^{(2)}}}{\lambda_{j_1^{(2)}\cdots j_k^{(2)}}}
    \end{align}for all $t\in [-\pi, \pi]$. Integrating the exponential factor over $[-\pi, \pi]$, we get
    \begin{align}
        &\int_{t\in [-\pi, \pi]} \expm{t\left(\sum_{i=1}^k\lambda_{j_i^{(1)}} - \sum_{i=1}^k\lambda_{j_i^{(2)}}\right)} dt \\
        =& \begin{cases}
            2\pi & \sum_{i=1}^k\lambda_{j_i^{(1)}} = \sum_{i=1}^k\lambda_{j_i^{(2)}} \\
            \frac{\expm{\pi\left(\sum_{i=1}^k\lambda_{j_i^{(1)}} - \sum_{i=1}^k\lambda_{j_i^{(2)}}\right)} - \expm{-\pi\left(\sum_{i=1}^k\lambda_{j_i^{(1)}} - \sum_{i=1}^k\lambda_{j_i^{(2)}}\right)}}{2\pi\left(\sum_{i=1}^k\lambda_{j_i^{(1)}} - \sum_{i=1}^k\lambda_{j_i^{(2)}}\right)} & o.w.
        \end{cases}
    \end{align}Using the fact $-i\cdot i = 1$ and $\sin(x) = \frac{e^{ix} - e^{-ix}}{2ix}$, we can write
    \begin{align}
        &\frac{\expm{\pi\left(\sum_{i=1}^k\lambda_{j_i^{(1)}} - \sum_{i=1}^k\lambda_{j_i^{(2)}}\right)} - \expm{-\pi\left(\sum_{i=1}^k\lambda_{j_i^{(1)}} - \sum_{i=1}^k\lambda_{j_i^{(2)}}\right)}}{2\pi\left(\sum_{i=1}^k\lambda_{j_i^{(1)}} - \sum_{i=1}^k\lambda_{j_i^{(2)}}\right)} \\
        =&\frac{\expm{i\pi\left((-i)\sum_{i=1}^k\lambda_{j_i^{(1)}} - (-i)\sum_{i=1}^k\lambda_{j_i^{(2)}}\right)} - \expm{-i\pi\left((-i) \sum_{i=1}^k\lambda_{j_i^{(1)}} - (-i)\sum_{i=1}^k\lambda_{j_i^{(2)}}\right)}}{2i\pi\left((-i)\cdot \sum_{i=1}^k\lambda_{j_i^{(1)}} - (-i)\cdot \sum_{i=1}^k\lambda_{j_i^{(2)}}\right)} \\
        =& \frac{\sin\left(\pi\left((-i)\cdot \sum_{i=1}^k\lambda_{j_i^{(1)}} - (-i)\cdot \sum_{i=1}^k\lambda_{j_i^{(2)}}\right)\right)}{\pi\left((-i)\cdot \sum_{i=1}^k\lambda_{j_i^{(1)}} - (-i)\cdot \sum_{i=1}^k\lambda_{j_i^{(2)}}\right)}
    \end{align}Here, since $H$ is an anti-Hermitian matrix, all of its eigenvalues are pure imaginary numbers, which implies the value $-i\pi\!\left(\sum_{i=1}^k\lambda_{j_i^{(1)}} - \sum_{i=1}^k\lambda_{j_i^{(2)}}\!\right)$ is real.
\end{proof}

\begin{lemma}
\label{useful:fact:no:neg:1:eigval}
    Let $H_1, \cdots, H_\ell\in \fu(V)$ be anti-Hermitian operators, where $V$ is some Hilbert space, and let $k\geq 1$. Then, the operator $\cP^k_{H_\ell}\circ \cdots \circ \cP^k_{H_1}$ has the following properties.
    \begin{enumerate}
        \item[(i)] Every eigenvalue $c\in\bbC$ satisfies $\abs{c}\leq 1$.
        \item[(ii)] If $c$ is an eigenvalue with $\abs{c}=1$, then $c=1$.
        \item[(iii)] If $A$ is an eigenvector with eigenvalue $1$, then $\cP^k_{H_j}(A)=A$ for all $j=1,\cdots,\ell$; equivalently, $A$ commutes with $(e^{tH_j})^{\otimes k}$ for all $t\in\bbR$ and all $j$.
    \end{enumerate}
    In particular, the operator does not have an eigenvalue in $(-\infty, -1]\cup (1, \infty)$.
\end{lemma}

\begin{proof} [Proof of \cref{useful:fact:no:neg:1:eigval}]
    First, using the fact $\rho(AB) \leq \norm{A}_2\norm{B}_2$, the spectral radius of $\cP^k_{H_\ell}\circ \cdots \circ \cP^k_{H_1}$, i.e., $\rho(\cP^k_{H_\ell}\circ \cdots \circ \cP^k_{H_1})$ is upper bounded by $ \prod_{j=1}^{\ell}\inorm{\cP^k_{H_j}}_2$ (assuming $\cP^k_{H_j}$ are represented in matrix form, acting on the operators on $V^{\otimes k}$ written as vectors, so that the two norm of an operator $X$ is $\norm{X}_2=\sqrt{\iTr{X^\dagger X}}$). By \cref{fact:true:bound:important:fact}, we note that $\inorm{\cP^k_{H_j}}_2 \leq 1$ for all $j=1,\cdots, \ell$ (this also follows from the triangle inequality, as in the computation below), which implies $\cP^k_{H_\ell}\circ \cdots \circ \cP^k_{H_1}$ does not have an eigenvalue $c\in\bbC$ with $\abs{c}>1$. This is (i). Then, it suffices to consider the eigenvalues $c$ with $\abs{c}=1$.

    Suppose $\cP^k_{H_\ell}\circ \cdots \circ \cP^k_{H_1}$ has an eigenvalue $c$ with $\abs{c}=1$. Then, there exists an eigenvector $A\in \iEnd{V^{\otimes k}}$, $A\neq 0$, such that $\cP^k_{H_\ell}\circ \cdots \circ \cP^k_{H_1}(A) = cA$. Taking the two norm on both sides, we obtain that
    \begin{align*}
        \norm{\cP^k_{H_\ell}\circ \cdots \circ \cP^k_{H_1}(A)}_2 = \norm{cA}_2 = \norm{A}_2
    \end{align*}Using the triangle inequality and the fact that 2 norm is invariant under unitary transformation, we get
    \begin{align*}
        &\norm{\cP^k_{H_\ell}\circ \cdots \circ \cP^k_{H_1}(A)}_2 \\
        =&\norm{\frac{1}{2\pi}\int_{t_\ell} (e^{t_{\ell}H_\ell})^{\otimes k}\left(\cP^k_{H_{\ell-1}}\circ \cdots \circ \cP^k_{H_1}(A)\right)(e^{-t_{\ell}H_\ell})^{\otimes k}dt_\ell}_2 \\
        \leq& \frac{1}{2\pi}\int_{t_\ell} \norm{(e^{t_{\ell}H_\ell})^{\otimes k}\left(\cP^k_{H_{\ell-1}}\circ \cdots \circ \cP^k_{H_1}(A)\right)(e^{-t_{\ell}H_\ell})^{\otimes k}}_2 dt_\ell \\
        =& \norm{\cP^k_{H_{\ell-1}}\circ \cdots \circ \cP^k_{H_1}(A)}_2 \\
        \leq& \norm{\cP^k_{H_{\ell-2}}\circ \cdots \circ \cP^k_{H_1}(A)}_2 \\
        \leq& \cdots \\
        \leq& \norm{\cP^k_{H_1}(A)}_2 \\
        \leq& \norm{A}_2
    \end{align*}Combining all of the above, we note that
    \begin{align}
        \norm{\cP^k_{H_\ell}\circ \cdots \circ \cP^k_{H_1}(A)}_2 = \norm{\cP^k_{H_{\ell-1}}\circ \cdots \circ \cP^k_{H_1}(A)}_2 = \cdots = \norm{\cP^k_{H_1}(A)}_2 = \norm{A}_2 \label{eq:prelim:equal:norms}
    \end{align}which implies
    \begin{align}
        \cP^k_{H_\ell}\circ \cdots \circ \cP^k_{H_1}(A) = \cP^k_{H_{\ell-1}}\circ \cdots \circ \cP^k_{H_1}(A) = \cdots = \cP^k_{H_1}(A) = A. \label{eq:prelim:fixed}
    \end{align}To see this, fix $j\in [\ell]$ and let $B\defeq\cP^k_{H_{j-1}}\circ \cdots \circ \cP^k_{H_1}(A)$, with $B\defeq A$ for $j=1$, so that $\norm{\cP^k_{H_j}(B)}_2=\norm{B}_2$ by \cref{eq:prelim:equal:norms}. Let $\state{a}$ and $\state{b}$ run over the orthonormal basis of $V^{\otimes k}$ used in \cref{fact:true:bound:important:fact} for $H=H_j$, which consists of tensor products of eigenvectors of $H_j$, and write $(e^{tH_j})^{\otimes k}\state{a}=e^{t\lambda_a}\state{a}$. The identity of \cref{fact:true:bound:important:fact} is linear in $M$, and every operator on $V^{\otimes k}$ is a complex linear combination of two Hermitian operators, so the identity holds for every operator $M$ on $V^{\otimes k}$; for $M=\Outer{a}{b}$ it reads $\cP^k_{H_j}(\Outer{a}{b})=\alpha(\lambda_a,\lambda_b)\Outer{a}{b}$. Hence the operators $\Outer{a}{b}$, which are orthonormal with respect to the inner product $\iTr{X^\dagger Y}$, form a basis of eigenvectors of $\cP^k_{H_j}$. By \cref{fact:true:bound:important:fact}, every eigenvalue of $\cP^k_{H_j}$ is either $1$ or of the form
    \begin{align*}
        \frac{\sin\!\left(i\pi\big(\lambda_{b}-\lambda_{a}\big)\right)}{i\pi\big(\lambda_{b}-\lambda_{a}\big)}\qquad\text{with }\lambda_a\neq\lambda_b,
    \end{align*}where $i(\lambda_b-\lambda_a)$ is a non-zero real number because the eigenvalues of $H_j$ are purely imaginary. Since the equation
    \begin{align*}
        \abs{\frac{\sin(\pi x)}{\pi x}} = 1
    \end{align*}admits no real solution for $x \in \bbR\setminus\CurB{0}$ (indeed, $\abs{\sin(\pi x)}<\abs{\pi x}$ for every real $x\neq 0$), it follows that every eigenvalue of $\cP^k_{H_j}$ other than $1$ has absolute value smaller than $1$. Expanding $B=\sum_{a,b}\beta_{ab}\Outer{a}{b}$, we obtain
    \begin{align*}
        \norm{B}_2^2-\norm{\cP^k_{H_j}(B)}_2^2=\sum_{a,b}\left(1-\alpha(\lambda_a,\lambda_b)^2\right)\abs{\beta_{ab}}^2 .
    \end{align*}The left-hand side is zero and no term on the right-hand side is negative, so $\beta_{ab}=0$ whenever $\alpha(\lambda_a,\lambda_b)\neq 1$. In other words, $B$ is a linear combination of eigenvectors of $\cP^k_{H_j}$ with eigenvalue $1$, which means $\cP^k_{H_j}(B)=B$. Since $j$ was arbitrary, this shows \cref{eq:prelim:fixed}.

    By assumption, we note that $\cP^k_{H_\ell}\circ \cdots \circ \cP^k_{H_1}(A) = cA$, which means $cA=A$ by \cref{eq:prelim:fixed}, i.e., $c=1$ because $A\neq 0$. In other words, $A$ cannot be an eigenvector of operator $\cP^k_{H_{\ell}}\circ \cdots \circ \cP^k_{H_1}$ with an eigenvalue of absolute value $1$ other than $1$. This is (ii).

    To show (iii), let $A$ be an eigenvector with eigenvalue $c=1$. The argument above applies to it, and \cref{eq:prelim:fixed} gives $\cP^k_{H_1}(A)=A$, then $\cP^k_{H_2}(A)=\cP^k_{H_2}\circ\cP^k_{H_1}(A)=\cP^k_{H_1}(A)=A$, and in the same way $\cP^k_{H_j}(A)=A$ for all $j=1,\cdots,\ell$. Moreover, since $\cP^k_{H_j}(\Outer{a}{b})=\alpha(\lambda_a,\lambda_b)\Outer{a}{b}$ and $\alpha(\lambda_a,\lambda_b)=1$ exactly when $\lambda_a=\lambda_b$, the identity $\cP^k_{H_j}(A)=A$ holds if and only if $\cstate{a}A\state{b}=0$ for all $a,b$ with $\lambda_a\neq\lambda_b$. Since $\cstate{a}(e^{tH_j})^{\otimes k}A(e^{-tH_j})^{\otimes k}\state{b}=e^{t(\lambda_a-\lambda_b)}\cstate{a}A\state{b}$, this is the case if and only if $(e^{tH_j})^{\otimes k}A(e^{-tH_j})^{\otimes k}=A$ for all $t\in\bbR$, i.e., if and only if $A$ commutes with $(e^{tH_j})^{\otimes k}$ for all $t\in\bbR$.

    Finally, every real number $c\in(-\infty, -1)\cup (1, \infty)$ has absolute value $\abs{c}>1$, so by (i) it is not an eigenvalue of the operator. The number $c=-1$ has absolute value $1$ and is different from $1$, so by (ii) it is not an eigenvalue either. Hence the operator does not have an eigenvalue in $(-\infty, -1]\cup (1, \infty)$.
\end{proof}

\section{Reduction}
\label{sec:reduction}
When analyzing the variance of the partial derivatives of the objective function in \cref{obj} and the unitary $t$-design properties of the ansatz, it is analytically advantageous to express each ansatz operator in the matrix form $\exp(\theta H)$. Here, $\theta$ denotes a bounded real parameter and $H$ is an anti-Hermitian generator. This representation is preferred over an mapping of abstract operators, such as $\exp(\theta_{l,q} X_{l,q})$, to unitary matrices as it facilitates a more direct application of tools from linear algebra. In the following, we demonstrate that the ansatz in \cref{ansatz} can be reduced to this canonical form by leveraging the properties established in the previous section.

Recall that $G$ is a compact and connected Lie group with Lie algebra $\fg$ spanned by $X_1,\ldots,X_k\in\fg$, and that $(\cH,\Pi)$ is a finite-dimensional unitary representation of $G$. By the fundamental theorem of Lie theory, the representation $\Pi$ induces a Lie algebra homomorphism $\pi_*:\fg \to \mathfrak{u}(\cH)$. Since $G$ is connected, the homomorphism $\pi_*$ is uniquely determined by $\Pi$ (see \cref{fund:thm:lie}). Moreover, by properties of the exponential map (see \cref{prop:exp:map}), it satisfies
\begin{align}
    \Pi(\exp(\theta X)) = \exp(\theta \,\pi_*(X)),
\end{align}
for all $X\in\fg$ and $\theta\in[-\pi,\pi]$.

Now, define
\begin{align}
    \compactG = \Pi(G) = \CurB{\Pi(g):\forall g\in G}
\end{align} and 
\begin{align}
    \tfg \defeq \pi_*(\fg) = \CurB{\pi_*(X): X\in \fg}
\end{align}The following lemma characterizes the structure of $\compactG$ and the relationship between $\compactG$ and $\tfg$.

\begin{lemma}
    The set $\compactG\subseteq \mathrm{U}(\cH)$ is a compact and connected Lie group. Moreover, the Lie algebra associated with $\compactG$ is $\tfg$, i.e., $\Lie{\compactG} = \tfg$. \label{reduction:thm}
\end{lemma}

\begin{proof} [Proof of \cref{reduction:thm}]
Since $\Pi:G\to \mathrm{U}(\cH)$ is a continuous map and $G$ is compact and connected, its image $\compactG=\Pi(G)$ is also compact and connected. Moreover, by \cref{def:group:homo}, $\compactG$ is closed under multiplication and inversion, and therefore forms a subgroup of $\mathrm{U}(\cH)$. Consequently, $\compactG$ is a compact Lie subgroup of $\mathrm{U}(\cH)$. Next, by \cref{def:rep} and \cref{def:lie:alg:homo}, one verifies that $\tfg$ forms a Lie algebra. Since $\fg$ is spanned by $X_1,\ldots,X_k$ and $\pi_*$ is linear, it follows that $\tfg$ is spanned by $\pi_*(X_1),\ldots,\pi_*(X_k)$. It remains to identify the Lie algebra of $\compactG$. In particular, it suffices to show that $\mathrm{Lie}(\compactG)=T_{\Pi(e)}\compactG=\tfg$, where $\Pi(e)$ is the identity element of $\compactG$.

Since $\tfg\subseteq \mathfrak{u}(V)$ is a Lie subalgebra, Lie's third theorem (\cref{fact:2}) guarantees the existence of a unique connected Lie subgroup $H\subseteq \mathrm{U}(V)$ whose Lie algebra satisfies $T_{e_H}H = \Lie{H} = \tfg$, where $e_H$ is the identity of the Lie group $H$. We first show $H\subseteq \compactG$. By the construction in the proof of Lie's third theorem, $H$ is generated by exponentials of elements in $\tfg$, i.e.,
\begin{align}
    H = \left<\expm{\pi_*(X)}: X\in \fg\right>
\end{align}Thus every element $h\in H$ admits the form
\begin{align}
    h = \expm{\pi_*(X_1)}\cdots \expm{\pi_*(X_{\cN_h})}
\end{align}for some positive integer $\cN_h$ and $X_i\in\fg$. Using the defining property of $\pi_*$,
\begin{align}
    \exp(\pi_*(X_i))=\Pi(\exp(X_i)).
\end{align}Since $\exp(X_i)\in G$ (see \cref{def:exp:map}), it follows that $\Pi(\exp(X_i))\in\compactG$. Because $\compactG$ is closed under multiplication, we conclude that $h\in\compactG$. Hence $H\subseteq \compactG$

Next we prove the reverse inclusion. By \cref{prop:exp:map}, the exponential map is a diffeomorphism between a neighborhood $\fU_0\subseteq\fg$ of $0$ and a neighborhood of the identity $e\in G$. Since $G$ is connected, it can be generated by elements in $U_{\Id}$, where $U_{\Id}$ is any neighborhood of the identity (see Lemma~1.18 of \cite{ziller2012liegroups}). Consequently, every element $g \in G$ can be written as
\begin{align}
    g = \expm{Y_1}\cdots\expm{Y_{\Tilde{\cN}_g}}
\end{align}for some positive integer $\Tilde{\cN}_g$ with $Y_i\in\fU_0$. Applying the homomorphism property of $\Pi$ gives
\begin{align}
    \compactG =& \CurB{\Pi\left(\expm{Y_1}\cdots\expm{Y_{\Tilde{\cN}_g}}\right):g\in G, Y_i \in \fU_0} \nonumber \\
    =& \CurB{\Pi\left(\expm{Y_1}\right)\cdots\Pi\left(\expm{Y_{\Tilde{\cN}_g}}\right):g\in G, Y_i \in \fU_0} \nonumber \\
    =& \CurB{\expm{\pi_*(Y_1)}\cdots\expm{\pi_*(Y_{\Tilde{\cN}_g})}:g\in G, Y_i \in \fU_0}
\end{align}Since $\pi_*(Y_i)\in\tfg$, every such product lies in $H$, which implies $\compactG\subseteq H$. Combining both inclusions yields $H=\compactG$. Therefore, $\Lie{\compactG} = T_{\Pi(e)}\compactG = T_{e_H}H = \tfg$.
\end{proof}

\begin{remark}
    One immediate consequence of \cref{reduction:thm} is that the ans\"atze defined in \cref{ansatz} can be viewed as a parameterized quantum circuit generated by the anti-Hermitian matrices ${\pi_*(X_1), \cdots, \pi_*(X_k)}$. In particular, each layer of the circuit consists of exponentials of these generators,
    \begin{align}
        \absU = \prod_{l=1}^L \prod_{q = 1}^k \expm{\theta_{l,q} \pi_*(X_{l, q})}. \label{reduction:reduced:ansatz}
    \end{align}
\end{remark}

\begin{remark}
    Another consequence is that a subspace $W\subseteq\cH$ is invariant under the standard representation $(\cH,\Id)$ of $\compactG$ if and only if it is invariant under the representation $(\cH,\Pi)$ of $G$. This follows directly from the definition of representation (see \cref{def:rep}) and the fact $\compactG=\Pi(G)$.
\end{remark}

To conclude this section, we discuss the relationship between the Haar measure on the original group $G$ (denoted by $\mu_G$) and the Haar measure on the image group $\compactG$ (denoted by $\mu_{\compactG}$). Since $G$ and $\compactG$ are compact Lie groups, they are locally compact Hausdorff spaces and therefore admit Haar measures. Moreover, compact Lie groups can be realized as compact submanifolds of Euclidean spaces; in particular, there exist integers $n_1,n_2$ such that $G \subseteq \mathbb{R}^{n_1}$ and $\compactG \subseteq \mathbb{R}^{n_2}$. Because the representation map $\Pi : G \to \compactG$ is continuous, it is measurable (see Section 2 of \cite{Bog07}). Hence the pushforward measure $\mu_G \circ \Pi^{-1}$ defines a measure on $\compactG$. By the change-of-variables theorem for pushforward measures (Theorem 3.6.1 in \cite{Bog07}), for any measurable function $g$ on $\compactG$ we have
\begin{align}
    \int_{\compactG} g(y)\, d(\mu_G \circ \Pi^{-1})(y) = \int_G g(\Pi(x))\, d\mu_G(x).
\end{align}Furthermore, the pushforward measure $\mu_G \circ \Pi^{-1}$ is left-invariant on $\compactG$, and therefore is a Haar measure on $\compactG$. Since the Haar measure on a compact group is unique up to a constant factor, it follows that $\mu_G \circ \Pi^{-1}$ coincides with $\mu_{\compactG}$ after normalization. Consequently,
\begin{align}
    \int_{\compactG} g(y)\, d\mu_{\compactG}(y) = \int_G g(\Pi(x))\, d\mu_G(x).
\end{align}

\section{Expressibility}
\label{sec:express}
In this section, we analyze the expressibility of the parameterized quantum circuits defined in \cref{ansatz}. Expressibility is formally quantified by the extent to which a distribution of unitaries approximates a unitary $t$-design. To establish this, we first recall the notion of an $\varepsilon$-approximate $t$-design, which provides a rigorous metric for evaluating the distance between the moment operators of a given distribution and those of a reference Haar measure.

Prior studies, including \cite{PhysRevA.80.012304, Mele_2024, larocca2022diagnosing, haah2025efficient}, have introduced various formulations of approximate $t$-designs to characterize PQC expressibility. However, these definitions are typically defined relative to the Haar measure on the special unitary group $\SU{2^n}$ or $SO(2^n)$. In such formulations, a set of unitaries is considered a $t$-design if its $t$-th moment operator closely approximates the $t$-th moment operator of the Haar measure over $\SU{2^n}$. While this assumption is appropriate when an ansatz is expected to explore the full unitary space, it lacks the generality required for structured ansätze. Specifically, when a circuit is generated by elements of a Lie algebra whose associated Lie group is a proper compact subgroup of $\SU{2^n}$, the $\SU{2^n}$ Haar measure no longer serves as a meaningful reference.

Motivated by these considerations, we generalize the standard definition of an $\varepsilon$-approximate $t$-design from $\SU{2^n}$ to arbitrary compact and connected matrix Lie groups, as formulated below.

\begin{definition} [$\varepsilon$-Approximate $t$-Design]
    Let $\bbU$ be a distribution over a set of unitary matrices, and let $\compactG$ be a compact, connected matrix Lie group containing the support of $\bbU$. For $t \geq 0$ and $\varepsilon > 0$, we say $\bbU$ is a $\varepsilon$-approximate $t$-design (with respect to $\compactG$) if and only if
    \begin{align}
        \norm{\EXPD{V\sim \bbU}{(V)^{\otimes t} \otimes (V^*)^{\otimes t}} - \EXPD{U\sim \mu_{\compactG}}{(U)^{\otimes t} \otimes (U^*)^{\otimes t}}} \leq \varepsilon
    \end{align}where $\mu_{\compactG}$ denotes the Haar measure on $\compactG$.
\end{definition}

We now discuss the expressibility of our PQC. Recall from the preceding section that the PQC given in \cref{ansatz} is structured as a periodic architecture generated by exponentiating elements from a spanning set of the induced Lie algebra $\tfg = \pi_*(\fg)$, where $(\cH, \pi_*)$ represents the induced unitary representation of $\fg$ and $\mathrm{Lie}(\compactG) = \tfg$ (see \cref{reduction:reduced:ansatz}). Given that the parameters are drawn uniformly from $[-\pi, \pi]^{Lk}$, the PQC induces a distribution $\bbU_U$ supported on a compact and connected Lie group $\compactG \subseteq \mathrm{U}(\cH)$. Consequently, the expressibility of such an ansatz can be evaluated relative to the Haar measure on $\compactG$ (denoted $\mu_{\compactG}$).

\begin{theorem}
    Let $\compactG$ be the compact, connected matrix Lie group associated with the Lie algebra $\fg$ spanned by anti-Hermitian matrices $\cH_1, \cdots, \cH_\cN$. Let $t\in\mathbb N_{\geq 1}$. Define
    \begin{align}
        \cV_l(\vtheta_l) := \prod_{j=1}^\cN \exp\!\bigl((\vtheta_l)_j H_j\bigr),\qquad\vtheta_l\in[-\pi,\pi]^\cN.
    \end{align}Then, 
    \begin{align}
        \lim_{L \to \infty} \norm{\EXPD{\vtheta_1,\cdots,\vtheta_L \sim [-\pi, \pi]^\cN}{\prod_{l=1}^L\left((\cV_l)^{\otimes t}\otimes (\cV_l^*)^{\otimes t}\right)} - \EXPD{U \sim \mu_{\compactG}}{(U)^{\otimes t}\otimes (U^*)^{\otimes t}}} = 0
    \end{align}for any sub-multiplicative matrix norm.
    \label{express:thm}
\end{theorem}

Combining the observations above, \cref{express:thm} implies that the distribution $\bbU_U$ converges to a $t$-design as the circuit depth $L$ increases. Before presenting the proof of \cref{express:thm}, we first provide the following lemma.

\begin{lemma}
    Let $\compactG$ be the compact, connected Lie group associated with the Lie algebra $\tfg$ spanned by anti-Hermitian matrices $H_1, \cdots, H_\cN$. For $t\in\mathbb N_{\geq 1}$, define
    \begin{align}
        \cV(\theta_1,\ldots,\theta_\cN) := \prod_{j=1}^\cN \exp\!\bigl(\theta_j H_j\bigr),\qquad(\theta_1,\ldots,\theta_\cN)\in[-\pi,\pi]^\cN .
    \end{align}Then, the operator $\EXPD{\theta_1, \cdots, \theta_\cN\sim [-\pi, \pi]}{(\cV)^{\otimes t}\otimes (\cV^*)^{\otimes t}} - \EXPD{U \sim \mu_{\compactG}}{(U)^{\otimes t}\otimes (U^*)^{\otimes t}}$ has spectral radius strictly smaller than 1, i.e., 
         \begin{align}
         \rho\left(\EXPD{\theta_1, \cdots, \theta_\cN\sim [-\pi, \pi]}{(\cV)^{\otimes t}\otimes (\cV^*)^{\otimes t}} - \EXPD{U \sim \mu_{\compactG}}{(U)^{\otimes t}\otimes (U^*)^{\otimes t}}\right) < 1
     \end{align}Here $\mu_{\compactG}$ denotes the Haar probability measure on $\compactG$, and $(\cdot)^*$ denotes complex conjugation.\label{express:small:spectral:radius}
\end{lemma}

The proof of \cref{express:small:spectral:radius} relies on the commutant of the parameterized set $S\defeq \CurB{\cV(\theta_1, \cdots, \theta_\cN): \theta_j \in [-\pi, \pi]}$. This commutant is characterized by the following two claims.

\begin{claim}
    Let $H_1, \cdots, H_\cN$ be some anti-Hermitian matrices. Let $X$ be any linear operator in $\End{V^{\otimes t}}$, where $V$ is a Hilbert space and $t \in \bbZ_{\geq 1}$. If $\icommutator{\prod_{j=1}\iexpm{\theta_j H_j}^{\otimes t}}{X} = 0$ for all $(\theta_1, \cdots, \theta_\cN)\in [-\pi, \pi]^\cN$, then $\icommutator{\iexpm{\sum_j c_j H_j}^{\otimes t}}{X} = 0$ for all $(c_1, \cdots, c_\cN)\in \bbR^{\cN}$. \label{express:key:obs:1} 
\end{claim}

\begin{proof} [Proof of \cref{express:key:obs:1}]
    Let $(c_1, \cdots, c_\cN)$ be any element in $\bbR^{\cN}$. Consider $U_m^{\otimes t} \defeq \left(\prod_{j=1}\left(\iexpm{(c_j/m) H_j}\right)^{\otimes t}\right)^{m}$. Let $m_0 = \max_{j=1,\cdots,\cN} \ceil{\abs{c_j}}$. Clearly, for all $m \geq m_0$ and $j=1,\cdots, \cN$, $c_j/m \in [-\pi, \pi]$. Applying the assumption, we note that $\commutator{U_m^{\otimes t}}{X} = 0$ for all integers $m \geq m_0$, which implies
    \begin{align}
        \lim_{m \to \infty} \commutator{U_m^{\otimes t}}{X} = 0.
    \end{align}Moreover, since $X$ and $U_m^{\otimes t}$ are all bounded, i.e., $\norm{X}_{op}$ and $\norm{U_m^{\otimes t}}_{op}$ are finite, we know that $\lim_{m \to \infty} \commutator{U_m^{\otimes t}}{X} = \commutator{\lim_{m \to \infty} U_m^{\otimes t}}{X}$. Combining all of the above, we conclude that
    \begin{align}
        0 = \commutator{\lim_{m \to \infty} U_m^{\otimes t}}{X} = \commutator{\expm{\sum_j c_j H_j}^{\otimes t}}{X} 
    \end{align}by Trotterization theorem \cite{trotter}. 
\end{proof}

\begin{claim}
    Let $\compactG$ be the compact and connected Lie group associated with Lie algebra $\tfg$ spanned by anti-Hermitian $B_1, \cdots, B_\cN\in \mathfrak{u}(V)$ for some Hilbert space $V$. Consider the set 
    \begin{align}
        S = \CurB{\prod_{j=1}^{\cN}\exp(\theta_j B_j): \theta_j\in [-\pi, \pi]}
    \end{align}Then, 
    \begin{align}
        \Comm{t, S} = \Comm{t, \compactG}
    \end{align}for any $t\in \bbN_{\geq 1}$\label{express:same:commutant}
\end{claim} 

\begin{proof} [Proof of \cref{express:same:commutant}]
    Since $S$ is clearly a subsets of $\compactG$ by definition, it is not hard to see that $t$-th fold commutant of $\compactG$ is a subset of that of $S$, i.e., $\iComm{t, \compactG} \subseteq \iComm{t, S}$. Therefore, it suffices to show $\iComm{t, S} \subseteq \iComm{t, \compactG}$. Let $X \in \iComm{t, S}$ be arbitrary. In order to show whether $X \in \iComm{t, \compactG}$, it suffices to check if $\int_{\compactG} (U)^{\otimes t} X (U^\dagger)^{\otimes t}d\mu(U) = X$. By \cref{def:connected:lie:subgp}, we note that 
    \begin{align}
        U^{\otimes t} = \expm{Y_1}^{\otimes t}\expm{Y_2}^{\otimes t}\cdots\expm{Y_m}^{\otimes t}, \label{express:same:comm:eq:1}
    \end{align}where $Y_1, \cdots, Y_m\in \tfg$. Given the assumption that $\tfg\subseteq \mathfrak{u}(V)$ is spanned by $B_1, \cdots, B_\cN$ and $\mathfrak{u}(V)$ is a real Lie algebra, we note that $Y_j$ can be written as a linear combination of $B_j$'s, i.e., $Y_j = \sum_{l=1}^\cN c_j^{(l)}B_l$ for some real number $c_j^{(l)}$. Thus, we can write \cref{express:same:comm:eq:1} as the following
    \begin{align}
        U^{\otimes t} = \expm{\sum_{l=1}^\cN c_1^{(l)}B_l}^{\otimes t}\expm{\sum_{l=1}^\cN c_2^{(l)}B_l}^{\otimes t}\cdots\expm{\sum_{l=1}^\cN c_m^{(l)}B_l}^{\otimes t}
    \end{align}Since $B_j\in \mathfrak{u}(V)$ are anti-Hermitian and $X\in \iComm{t, S}$, applying \cref{express:key:obs:1}, we conclude that
    \begin{align}
        \int_{\compactG} (U)^{\otimes t} X (U^*)^{\otimes t}d\mu(U) = \int_{\compactG} X (U)^{\otimes t} (U^*)^{\otimes t}d\mu(U) = X
    \end{align}
\end{proof}

With these preparations, we now prove \cref{express:small:spectral:radius}.

\begin{proof} [Proof of \cref{express:small:spectral:radius}]
    For simplicity, we introduce the following notation:
    \begin{align*}
        M^{(t)}_\nu &\defeq \EXPD{\theta_1, \cdots, \theta_\cN\sim [-\pi, \pi]}{(\cV)^{\otimes t}\otimes (\cV^*)^{\otimes t}}, \\
        M^{(t)}_\mu &\defeq \EXPD{U \sim \mu_{\compactG}}{(U)^{\otimes t}\otimes(U^*)^{\otimes t}}, \\
        \cA^{(t)} &\defeq M^{(t)}_\nu - M_\mu^{(t)}.
    \end{align*}

    First, we show that $\cA^{(t)}$ does not have eigenvalue 1. Suppose $\cA^{(t)}$ has eigenvalue 1. By \cref{express:same:commutant}, we observe that $\cA^{(t)}(A) = 0$ for all $A\in \iComm{t, \compactG}$. It therefore suffices to show there exists no non-zero element $A\notin \iComm{t, \compactG}$ such that $\cA^{(t)}(A) = A$. Suppose, to the contrary, that such an element $A$ exists. To proceed, applying $M_\mu^{(t)}$ on both side of $M_\nu^{(t)}(A) - M_\mu^{(t)}(A) = A$ and using the linearity of $M_\mu^{(t)}$, we obtain that 
    \begin{align}
        M_\mu^{(t)}\left(M_\nu^{(t)}(A)\right) - M_\mu^{(t)}\left(M_\mu^{(t)}(A)\right) = M_\mu^{(t)}\left(A\right).
    \end{align}Then, by \cref{prop:cP:is:orth:proj}, we note that $M_\mu^{(t)}$ is an orthogonal projector, and hence $M_\mu^{(t)}\left(M_\mu^{(t)}(A)\right) = M_\mu^{(t)}(A)$. We next expand $M_\mu^{(t)}\left(M_\nu^{(t)}(A)\right)$. Exchanging the order of integration and invoking the bi-invariance of the Haar measure, we can write 
    \begin{align}
        &\int_{\compactG} \int_{(\theta_1, \cdots, \theta_\cN)} (g)^{\otimes t}(\cV)^{\otimes t} A (\cV^\dagger)^{\otimes t}(g^\dagger)^{\otimes t} d(\theta_1, \cdots, d\theta_\pi) d\mu(g) \\
        =& \int_{(\theta_1, \cdots, \theta_\cN)} \int_{\compactG} (g\cV)^{\otimes t} A (\cV^\dagger g^\dagger)^{\otimes t}d\mu(g) d(\theta_1, \cdots, d\theta_\pi) \\
        =& \int_{(\theta_1, \cdots, \theta_\cN)} \int_{\compactG} (g)^{\otimes t} A ( g^\dagger)^{\otimes t}d\mu(g) d(\theta_1, \cdots, d\theta_\pi) & \text{(Haar measure is bi-invariant and $\cV\in \compactG$)} \\
        =& M_\mu^{(t)}\left(A\right).
    \end{align}Using similar arguments, one can show that $M_\nu^{(t)}\left(M_\mu^{(t)}(A)\right) = M_\mu^{(t)}(A)$ as well. Consequently, combining all of above, we note that 
    \begin{align}
        M_\mu^{(t)}\left(M_\nu^{(t)}(A)\right) - M_\mu^{(t)}\left(A\right) = M_\mu^{(t)}\left(A\right) - M_\mu^{(t)}\left(A\right) = 0 = M_\mu^{(t)}\left(A\right)
    \end{align}that is, $A$ is in the kernel of $M_\mu^{(t)}$. Finally, we observe that this leads to a contradiction. Indeed, since $A$ is in the kernel, we have $A = M_\nu^{(t)}(A) - M_\mu^{(t)}(A) = M_\nu^{(t)}(A)$, i.e., $A$ is an eigenvector of $M_\nu^{(t)}$ with eigenvalue $1$. Since the parameters $\theta_1, \cdots, \theta_\cN$ are independent and uniformly distributed on $[-\pi, \pi]$, the average over them can be taken one parameter at a time, starting with $\theta_\cN$:
    \begin{align*}
        M_\nu^{(t)}(A) =& \frac{1}{(2\pi)^{\cN}}\int_{\theta_1}\cdots\int_{\theta_\cN} (e^{\theta_1 H_1})^{\otimes t}\cdots(e^{\theta_\cN H_\cN})^{\otimes t} A (e^{-\theta_\cN H_\cN})^{\otimes t}\cdots(e^{-\theta_1 H_1})^{\otimes t} d\theta_\cN\cdots d\theta_1 \\
        =& \cP^t_{H_1}\circ\cP^t_{H_2}\circ\cdots\circ\cP^t_{H_\cN}(A),
    \end{align*}where $\cP^t_{H_j}(X)=\frac{1}{2\pi}\int_{\theta}(e^{\theta H_j})^{\otimes t}X(e^{-\theta H_j})^{\otimes t}d\theta$ is the map of \cref{fact:true:bound:important:fact}. Hence $\cP^t_{H_1}\circ\cdots\circ\cP^t_{H_\cN}(A)=A$. By \cref{useful:fact:no:neg:1:eigval}(iii), applied with $k=t$ to the anti-Hermitian operators $H_\cN, \cdots, H_1$, each of the maps $\cP^t_{H_j}$ leaves $A$ unchanged, i.e., $\cP^t_{H_j}(A)=A$ for all $j=1,\cdots,\cN$, and $A$ commutes with $(e^{\theta H_j})^{\otimes t}$ for all $\theta\in\bbR$ and all $j$. Since $\cV(\theta_1, \cdots, \theta_{\cN})^{\otimes t}=(e^{\theta_1 H_1})^{\otimes t}\cdots(e^{\theta_\cN H_\cN})^{\otimes t}$ is a product of such operators, this implies $A$ commutes with $\cV(\theta_1, \cdots, \theta_{\cN})^{\otimes t}$ for all $(\theta_1, \cdots, \theta_{\cN}) \in [-\pi, \pi]^{\cN}$, i.e., $A\in \iComm{t, S}$. By \cref{express:same:commutant}, $\iComm{t, S}=\iComm{t, \compactG}$, and hence $A\in \iComm{t, \compactG}$. This contradicts the assumption $A\notin \iComm{t, \compactG}$. We therefore conclude that there exists no non-zero element $A\notin \iComm{t, \compactG}$ such that $\cA^{(t)}(A) = A$, which means that $\cA^{(t)}$ does not have eigenvalue 1.
    
    To complete the proof, we show that $\cA^{(t)}$ does not have an eigenvalue $c\in\bbC$ with $\abs{c}\geq 1$ and $c\neq 1$. Complex numbers $c$ have to be considered here, because the eigenvalues of $\cA^{(t)}$ need not be real: $M_\nu^{(t)}$ is a product of self-adjoint operators that do not commute with one another, and such a product is in general not self-adjoint. Let $c\in\bbC$ with $\abs{c}\geq 1$ and $c\neq 1$ be arbitrary. Suppose that there exists a nonzero operator $B \in \iEnd{V^{\otimes t}}$ such that $\cA^{(t)}(B) = cB$. Again, applying $M_\mu^{(t)}$ to both sides of this equation and the linearity of $M_\mu^{(t)}$ yields
    \begin{align}
        M_\mu^{(t)}\!\left(M_\nu^{(t)}(B)\right)
        - M_\mu^{(t)}(B) = M_\mu^{(t)}(cB) = cM_\mu^{(t)}(B).
    \end{align}
    Using the identity $M_\mu^{(t)} \circ M_\nu^{(t)} = M_\mu^{(t)}$, we obtain
    \begin{align}
        M_\mu^{(t)}(B) - M_\mu^{(t)}(B)
        = 0
        = cM_\mu^{(t)}(B),
    \end{align}
    which implies that $M_\mu^{(t)}(B) = 0$. Hence, $B$ lies in the kernel of
    $M_\mu^{(t)}$. In particular, we have
    \begin{align}
        M_\nu^{(t)}(B) - M_\mu^{(t)}(B)
        = M_\nu^{(t)}(B)
        = cB,
    \end{align} which shows that $B$ must be an eigenvector of $M_\nu^{(t)}$ with eigenvalue $c$. However, $M_\nu^{(t)}=\cP^t_{H_1}\circ\cdots\circ\cP^t_{H_\cN}$ as shown above, and by \cref{useful:fact:no:neg:1:eigval} this operator has no eigenvalue $c$ with $\abs{c}>1$ (item (i)) and no eigenvalue $c\neq 1$ with $\abs{c}=1$ (item (ii)). Hence the operator $M_\nu^{(t)}$ admits no eigenvalue $c$ with $\abs{c}\geq1$ other than $c=1$, which contradicts the choice of $c$.

    Combining all of the above, we know that $\cA^{(t)}$ has neither the eigenvalue $1$ nor an eigenvalue $c\in\bbC$ with $\abs{c}\geq 1$ and $c\neq 1$. Hence every eigenvalue $c$ of $\cA^{(t)}$ satisfies $\abs{c}<1$. Since $\cA^{(t)}$ has only finitely many eigenvalues, the largest of their absolute values is also smaller than $1$, which implies $\rho(\cA^{(t)}) < 1$.
\end{proof}

The preceding lemmas provide the necessary ingredients for the proof of \cref{express:thm}. We now proceed with the proof.

\begin{proof} [Proof of \cref{express:thm}]
    
    First, we note that random variables $(\cV_l)^{\otimes t}\otimes (\cV_l^*)^{\otimes t}$ are mutually independent and identical, which means
    \begin{align}
        \EXPD{\vtheta_1,\cdots,\vtheta_L \sim [-\pi, \pi]^\cN}{\prod_{l=1}^L\left((\cV_l)^{\otimes t}\otimes (\cV_l^*)^{\otimes t}\right)} = \left(\EXPD{\vtheta \sim [-\pi, \pi]^\cN}{(\cV)^{\otimes t}\otimes (\cV^*)^{\otimes t}}\right)^L
    \end{align}Next, we observe that 
    \begin{align}
        \left(M_\nu^{(t)}\right)^L - M_\mu^{(t)} = \left(M_\nu^{(t)} - M_\mu^{(t)}\right)^L.
    \end{align}for all integers $L \geq 1$. The proof follows the same argument as Proposition 40 in \cite{Mele_2024}; for completeness, we outline it here by induction on $L$. Using the identities $M_\mu^{(t)} M_\nu^{(t)} = M_\nu^{(t)} M_\mu^{(t)} = M_\mu^{(t)}$ from \cref{express:small:spectral:radius} together with the inductive hypothesis, we obtain
    \begin{align}
        \left(M_\nu^{(t)} - M_\mu^{(t)}\right)^{L+1} = \left(\left(M_\nu^{(t)}\right)^L - M_\mu^{(t)}\right)\left(M_\nu^{(t)} - M_\mu^{(t)}\right) = \left(M_\nu^{(t)}\right)^{L+1} - M_\mu^{(t)}
    \end{align}Combining the above observations, we have
    \begin{align}
        \norm{\EXPD{\vtheta_1,\cdots,\vtheta_L \sim [-\pi, \pi]^\cN}{\prod_{l=1}^L\left((\cV_l)^{\otimes t}\otimes (\cV_l^*)^{\otimes t}\right)} - \EXPD{U \sim \mu_{\compactG}}{(U)^{\otimes t}\otimes (U^*)^{\otimes t}}} = \norm{\left(M_\nu^{(t)} - M_\mu^{(t)}\right)^{L}}
    \end{align}Consequently, using \cref{express:small:spectral:radius} and \cref{thm:5.6.12:converge:small:sr}, we conclude that
    \begin{align}
        &\lim_{L\to \infty} \norm{\EXPD{\vtheta_1,\cdots,\vtheta_L \sim [-\pi, \pi]^\cN}{\prod_{l=1}^L\left((\cV_l)^{\otimes t}\otimes (\cV_l^*)^{\otimes t}\right)} - \EXPD{U \sim \mu_{\compactG}}{(U)^{\otimes t}\otimes (U^*)^{\otimes t}}} \nonumber\\
        =& \lim_{L\to \infty} \norm{\left(M_\nu^{(t)} - M_\mu^{(t)}\right)^{L}} = 0
    \end{align}
\end{proof}

\cref{express:thm} shows that the distribution induced by the ans\"atze defined in \cref{ansatz} converges to the Haar measure on $\compactG$ as the number of layers $L$ tends to infinity. In particular, for any $\varepsilon > 0$ and $t \in \bbN_{\geq 1}$, there exists $L^* \in \bbN$ such that for all $L \ge L^*$,
\begin{align}
    \norm{\EXPD{\vtheta_1,\ldots,\vtheta_L \sim [-\pi,\pi]^\cN}{\prod_{l=1}^L \left( (\cV_l)^{\otimes t} \otimes (\cV_l^*)^{\otimes t} \right)} -\EXPD{U \sim \mu_{\compactG}}
    {U^{\otimes t} \otimes (U^*)^{\otimes t}}} = \norm{ \left(M_\nu^{(t)} - M_\mu^{(t)}\right)^L }
    \leq \varepsilon .
\end{align}

\begin{remark}
    The spectral radius of $M_\nu^{(t)} - M_\mu^{(t)}$ depends on the integer $t$. Consequently, for any fixed $\varepsilon \in (0,1)$, the threshold $L^*$ varies with $t$, implying that no universal $L^*$ exists that satisfies the condition for all $t$.
\end{remark}

Unfortunately, the operator $M_\nu^{(t)} - M_\mu^{(t)}$ is not normal in general, and thus its norm cannot be characterized directly in terms of its eigenvalues. Consequently, no simple closed-form expression is available to determine how large $L$ must be in order to obtain an $\varepsilon$-approximate $t$-design. Nevertheless, if one is given that $\inorm{M_\nu^{(t)} - M_\mu^{(t)}} < 1,$ then the powers $(M_\nu^{(t)} - M_\mu^{(t)})^L$ decay exponentially in $L$. In this case, an explicit bound on the required circuit depth $L$ can be derived to achieve an $\varepsilon$-approximate $t$-design.

\begin{proposition}
    If $\enspace\inorm{M_\nu^{(t)} - M_\mu^{(t)}} \leq \lambda < 1$, then 
    \begin{align}
        \norm{\EXPD{\vtheta_1,\cdots,\vtheta_L \sim [-\pi, \pi]^\cN}{\prod_{l=1}^L\left((\cV_l)^{\otimes t}\otimes (\cV_l^*)^{\otimes t}\right)} - \EXPD{U \sim \mu_{\compactG}}{(U)^{\otimes t}\otimes (U^*)^{\otimes t}}} \leq \varepsilon
    \end{align}for $L \geq \frac{\log(\varepsilon)}{\log(\lambda)}$. \label{express:eps:approx:k:design}
\end{proposition}

\begin{proof} [Proof of \cref{express:eps:approx:k:design}]
    Again, the proof follows the same argument as Prop 40 in \cite{Mele_2024}. Using the observation from \cref{express:thm}, we note that 
    \begin{align}
        \norm{\EXPD{\vtheta_1,\cdots,\vtheta_L \sim [-\pi, \pi]^\cN}{\prod_{l=1}^L\left((\cV_l)^{\otimes t}\otimes (\cV_l^*)^{\otimes t}\right)} - \EXPD{U \sim \mu_{\compactG}}{(U)^{\otimes t}\otimes (U^*)^{\otimes t}}} = \norm{\left(M_\nu^{(t)} - M_\mu^{(t)}\right)^{L}}
    \end{align}Using the sub-multiplicity, we have 
    \begin{align}
        &\norm{\left(M_\nu^{(t)} - M_\mu^{(t)}\right)^{L}} \leq \norm{M_\nu^{(t)} - M_\mu^{(t)}}^{L} = \lambda^{L}
    \end{align}Solving $\lambda^{L} \leq \varepsilon$ for $L$, we have $L \geq \frac{\log(\varepsilon)}{\log(\lambda)}$
\end{proof}

\begin{remark}
    Note that $\varepsilon$-approximate $t$-designs are downward-compatible; the $t$-design property implies the $(t-1)$-design property with at least the same precision $\varepsilon$, as the latter is a projection of the former onto a smaller tensor space.
\end{remark}

\section{Upper and Lower Bounds on the Variance}

In this section, we derive analytical upper and lower bounds on the variance of the objective function's partial derivatives. While prior works, such as those by \cite{fontana2024characterizing}, rely on the explicit decomposition of the underlying Lie algebra, our approach pivots to the decomposition of the associated unitary representation. By shifting the focus from an algebraic construction to the existence of representation-theoretic subspaces, the bounds established in \cref{sec:upper,sec:lower} provide a state- and measurement-centric perspective, identifying when the variance is constrained by the invariant subspaces occupied by the state $\rho$ and measurement $\mathcal{M}$.

As discussed in \cref{sec:reduction}, the ansatz defined in \cref{ansatz} is generated by the anti-Hermitian matrices $\pi_*(X_j)$, where $(\cH, \pi_*)$ is a unitary representation of a finite-dimensional Lie algebra $\fg$ with spanning set $\{X_1, \ldots, X_k\} \subseteq \fg$. Throughout this work, we assume the tensor product representation space decomposes as $\cH^{\otimes 2} = \bigoplus_\lambda m_\lambda V_\lambda$.

In addition, throughout this section, we consider an arbitrary density operator $\rho$ and an arbitrary Hermitian observable $\cM$. Fix an index pair $(\ell,j)$ corresponding to a parameter $\theta_{\ell,j}$ in the ans\"atze. For convenience, we partition the circuit given in \cref{ansatz} into two parts relative to this parameter. Define
\begin{align}
    U_{-}(\vtheta_-) = \prod_{l=1}^{\ell-1}\prod_{j^\prime = 1}^{k}\expm{\theta_{l, j^\prime}H_{l,j^\prime}}\prod_{j^\prime = 1}^{j}\expm{\theta_{\ell, j^\prime}H_{\ell, j^\prime}},
\end{align}and 
\begin{align}
    U_{+}(\vtheta_+) = \prod_{j^\prime = j+1}^{k}\expm{\theta_{\ell, j^\prime}H_{\ell, j^\prime}}\prod_{l=\ell+1}^{L}\prod_{j^\prime = 1}^{k}\expm{\theta_{l, j^\prime}H_{l,j^\prime}},
\end{align}where $H_{l,j^\prime} = \pi_*(X_{j^\prime})$ are the generators appearing in the ansatz and the corresponding parameter vectors are $\vtheta_- = (\theta_{1,1}, \cdots, \theta_{\ell, j})^T \in [-\pi, \pi]^{k(\ell - 1) + j}$ and $\vtheta_+ = (\theta_{\ell, j + 1}, \cdots, \theta_{L, k})^T \in [-\pi, \pi]^{kL - (k(\ell - 1) + j)}$. When focusing on the derivative with respect to $\theta_{\ell,j}$, we write $H \defeq \pi_*(X_j)$ for notational simplicity. Here $U_+$, which stands on the right in the written order of the product in \cref{ansatz}, acts on $\rho$ first, and $U_-$ contains the differentiated gate $\expm{\theta_{\ell,j}H}$; since this gate commutes with its generator $H$, the expressions below have the same form if it is counted in $U_+$ instead. 

Since the parameter vectors $\vtheta_\pm$ take values in bounded domains, we equip them with the uniform distribution. Let $\bbU_\pm$ denote the induced distributions over the sets $\CurB{U_\pm(\vtheta_\pm):\vtheta_\pm \in [-\pi,\pi]^{\dim(\vtheta_\pm)}}.$ By combining \cref{reduction:thm} and \cref{express:thm}, it follows that for any $\varepsilon \in (0,1)$, there exists a positive integer $L^*$ such that the product distribution $\bbU_-\bbU_+$ $\varepsilon$-converges to the Haar measure $\mu_{\compactG}$. Consequently, if the circuit depth satisfies $L \ge 2L^*$, then at least one of $\bbU_-$ or $\bbU_+$ $\varepsilon$-converges to 2-design with respect to $\compactG$. In what follows, we assume that $L \ge 2L^*$. Under the notation introduced above, the partial derivative of the objective function $C$ with respect to $\theta_{\ell,j}$ can be written as
\begin{align}
    \partial_{\theta_{\ell,j}} C = \Tr{\cM U_-\commutator{H}{U_+ \rho U_+^\dagger}U_-^\dagger}.
\end{align}Therefore, its variance under the parameter distribution is
\begin{align}
    \Var{\partial_{\theta_{\ell, j}}C} = \EXPD{U_\pm\sim \bbU_\pm}{\Tr{\cM U_- \commutator{H}{U_+ \rho U_+^\dagger} U_-^\dagger}^2} - \left(\EXPD{U_\pm\sim \bbU_\pm}{\Tr{\cM U_- \commutator{H}{U_+ \rho U_+^\dagger} U_-^\dagger}}\right)^2
\end{align}

The following lemma gives conditions under which the first moment vanishes.

\begin{lemma}
    If $\expm{2\pi H}$ is a multiple of the identity, or if $\bbU_-$ or $\bbU_+$ is a $2$-design with respect to $\compactG$, then
    \begin{align}
        \EXPD{U_-\sim\bbU_-,\,U_+\sim\bbU_+}{\partial_{\theta_{\ell,j}} C} = 0.
    \end{align}\label{lemma:expectation}
\end{lemma}

\begin{proof} [Proof of \cref{lemma:expectation}]
    If $\expm{2\pi H}$ is a multiple of the identity, then $\expm{\pi H}$ and $\expm{-\pi H}$ differ by a phase factor, so that, for fixed values of the other parameters, the circuits at $\theta_{\ell,j}=\pi$ and at $\theta_{\ell,j}=-\pi$ differ by this phase factor and give the same value of $C$. As $\theta_{\ell,j}$ is uniform on $[-\pi,\pi]$ and independent of the other parameters, the mean of $\partial_{\theta_{\ell,j}}C$ over $\theta_{\ell,j}$ is the difference of these two values divided by $2\pi$, which is zero, at every depth.

    For the other two conditions, expanding the expectation and regrouping terms yields
    \begin{align}
        \EXPD{U_-\sim \bbU_-, U_+\sim\bbU_+}{\partial_{\theta_{\ell, j}}C} = \Tr{\EXPD{U_-\sim \bbU_-}{U_-^\dagger \cM U_-}\cdot \commutator{H}{\EXPD{U_+\sim \bbU_+}{U_+ \rho U_+^\dagger}}} \label{eq:grad:expanded}
    \end{align}We consider the following two cases. 
    \begin{itemize}
        \item \textbf{Case 1: If $\bbU_-$ converges to 2-design with respect to $\compactG$} Applying the identity $[A, BC] = B[A, C] + [A, B]C$, we obtain
        \begin{align}
            &\Tr{\EXPD{U_-\sim \bbU_-}{U_-^\dagger \cM U_-}\cdot \commutator{H}{\EXPD{U_+\sim \bbU_+}{U_+ \rho U_+^\dagger}}} \\
            \begin{split}
                =& \Tr{\commutator{H}{\EXPD{U_-\sim \bbU_-}{U_-^\dagger \cM U_-}\cdot \EXPD{U_+\sim \bbU_+}{U_+ \rho U_+^\dagger}}} \\
                &\qquad- \Tr{\commutator{H}{\EXPD{U_-\sim \bbU_-}{U_-^\dagger \cM U_-}}\cdot \EXPD{U_+\sim \bbU_+}{U_+ \rho U_+^\dagger}} 
            \end{split}
        \end{align}Since the trace of a commutator vanishes, the first term is zero. Hence,
        \begin{align}
            \EXPD{U_-\sim \bbU_-, U_+\sim\bbU_+}{\partial_{\theta_{\ell, j}}C} = - \Tr{\EXPD{U_+\sim \bbU_+}{U_+ \rho U_+^\dagger}\cdot \commutator{H}{\EXPD{U_-\sim \bbU_-}{U_-^\dagger \cM U_-}}}
        \end{align}
        If $\bbU_-$ converges to 2-design with respect to $\compactG$, then
        \begin{align}
            \EXPD{U_-\sim \bbU_-}{U_-^\dagger \cM U_-} = \int_{\compactG}U_-^\dagger \cM U_-d\mu_{\compactG}(U_-)
        \end{align}By Lemma~C.1 of~\cite{fontana2024characterizing}, $H$ commutes with $\int_{\compactG}U_-^\dagger \cM U_-d\mu_{\compactG}(U_-)$. Consequently,
        \begin{align}
            \EXPD{U_-\sim \bbU_-, U_+\sim\bbU_+}{\partial_{\theta_{\ell, j}}C} = 0
        \end{align}

        \item \textbf{Case 2: If $\bbU_+$ converges to the Haar measure.} If $\bbU_+$ converges to the Haar measure on $\compactG$, then
        \begin{align}
            \EXPD{U_+\sim \bbU_+}{U_+ \rho U_+^\dagger} = \int_{\compactG}U_+ \rho U_+^\dagger d\mu_{\compactG}(U_+)
        \end{align}Again, by observation from Lemma~C.1 of~\cite{fontana2024characterizing}, we conclude that 
        \begin{align}
            \commutator{H}{\EXPD{U_+\sim \bbU_+}{U_+ \rho U_+^\dagger}} = 0.
        \end{align}Substituting into \eqref{eq:grad:expanded}, we obtain
        \begin{align}
            \EXPD{U_-\sim \bbU_-, U_+\sim\bbU_+}{\partial_{\theta_{\ell, j}}C} = 0.
        \end{align}
    \end{itemize}
\end{proof}

By \cref{lemma:expectation}, the variance reduces to the second moment:
\begin{align}
    \VARD{\vtheta\sim [-\pi, \pi]^L}{\partial_{\theta_{\ell, j}}C} = \EXPD{U_\pm\sim \bbU_\pm}{\Tr{\cM U_- \commutator{H}{U_+ \rho U_+^\dagger} U_-^\dagger}^2}. \label{var:expression}
\end{align}Therefore, it suffices to derive upper and lower bounds for the second moment. For convenience, throughout this section we introduce the shorthand $\var\defeq \EXPD{U_\pm\sim \bbU_\pm}{\Tr{\cM U_- \commutator{H}{U_+ \rho U_+^\dagger} U_-^\dagger}^2}$. 

\subsection{Upper Bound} \label{sec:upper}
Now, we derive the upper bound on variance. Expanding \cref{var:expression} and using $\Tr{X}^2 = \Tr{X\otimes X}$, we obtain
\begin{align}
    \var 
    =&\EXPD{U_+ \sim \bbU_+}{\EXPD{U_- \sim \bbU_-}{ \Tr{\cM U_{-}\commutator{H}{U_{+}\rho U_{+}^\dagger} U_{-}^\dagger}^2 }} \\
    =&\EXPD{U_+ \sim \bbU_+}{\EXPD{U_- \sim \bbU_-}{ \Tr{\cM U_{-}\commutator{H}{U_{+}\rho U_{+}^\dagger} U_{-}^\dagger \otimes \cM U_{-}\commutator{H}{U_{+}\rho U_{+}^\dagger} U_{-}^\dagger} }} \\
    =&\EXPD{U_+ \sim \bbU_+}{\EXPD{U_- \sim \bbU_-}{ \Tr{\left(\cM\otimes \cM\right)\left(U_{-}\otimes U_{-}\right)\left(\commutator{H}{U_{+}\rho U_{+}^\dagger }\otimes \commutator{H}{U_{+}\rho U_{+}^\dagger }\right)\left(U_{-}^\dagger\otimes U_{-}^\dagger\right)} }} \\
    =&\EXPD{U_+ \sim \bbU_+}{\Tr{\EXPD{U_- \sim \bbU_-}{ (U_{-}^\dagger)^{\otimes 2}\cM^{\otimes 2}(U_{-})^{\otimes 2} }\cdot  \commutator{H}{U_{+}\rho U_{+}^\dagger}^{\otimes 2}}} \label{var:upperbound:eq:0}
\end{align}For convenience, define
\begin{align*}
    \cT_-(\cdot)  \defeq \EXPD{U_- \sim \bbU_-}{ (U_{-}^\dagger)^{\otimes 2}(\cdot)^{\otimes 2}(U_{-})^{\otimes 2} },\qquad \cT_+(\cdot)  \defeq \EXPD{U_+ \sim \bbU_+}{ (U_{+})^{\otimes 2}(\cdot)^{\otimes 2}(U_{+}^\dagger)^{\otimes 2} }
\end{align*}To proceed, we first rewrite $\commutator{H}{U_{+}\rho U_{+}^\dagger}^{\otimes 2}$. Using the identity $AB\otimes BA = (A\otimes I)(B \otimes B)(I \otimes A)$, we can re-express it as the following
\begin{align}
    \begin{split}
        \commutator{H}{U_{+}\rho U_{+}^\dagger}^{\otimes 2} =& HU_{+}\rho U_{+}^\dagger\otimes HU_{+}\rho U_{+}^\dagger - HU_{+}\rho U_{+}^\dagger\otimes U_{+}\rho U_{+}^\dagger H \\
        &\hspace{1cm}- U_{+}\rho U_{+}^\dagger H\otimes HU_{+}\rho U_{+}^\dagger + U_{+}\rho U_{+}^\dagger H\otimes U_{+}\rho U_{+}^\dagger H  
    \end{split} \nonumber\\
    \begin{split}
        =& H^{\otimes 2}(U_{+}\rho U_{+}^\dagger)^{\otimes 2} - (H\otimes \bbI)(U_{+}\rho U_{+}^\dagger)^{\otimes 2}(\bbI\otimes H) \\
        &\hspace{1cm}- (\bbI\otimes H)(U_{+}\rho U_{+}^\dagger)^{\otimes 2}(H\otimes \bbI) + (U_{+}\rho U_{+}^\dagger)^{\otimes 2} H^{\otimes 2}
    \end{split} \label{var:upperbound:eq:1}
\end{align}Substituting \cref{var:upperbound:eq:1} into \cref{var:upperbound:eq:0} and rearranging terms, we have 
\begin{align}
    \begin{split}
        \var=&\underset{U_+\sim \bbU_+}{\bbE} \left[\Tr{\cT_{-}(\cM)\cdot H^{\otimes 2}(U_{+}\rho U_{+}^\dagger)^{\otimes 2}} -\Tr{(\bbI\otimes H)\cdot \cT_{-}(\cM)\cdot (H\otimes \bbI)(U_{+}\rho U_{+}^\dagger)^{\otimes 2}} \right.\\
        &\left.\hspace{0.5cm}-\Tr{(H\otimes \bbI)\cdot \cT_{-}(\cM)\cdot (\bbI\otimes H)(U_{+}\rho U_{+}^\dagger)^{\otimes 2}} + \Tr{H^{\otimes 2}\cdot \cT_{-}(\cM)\cdot (U_{+}\rho U_{+}^\dagger)^{\otimes 2} }
        \right]
    \end{split} \\
    \begin{split}
        =& \Tr{\cT_-(\cM)\cdot H^{\otimes 2}\cdot \cT_+(\rho)} -\Tr{(\bbI\otimes H)\cdot \cT_-(\cM) \cdot(H\otimes \bbI)\cdot \cT_+(\rho)} \\
        &\hspace{0.5cm}-\Tr{(H\otimes \bbI)\cdot \cT_-(\cM) \cdot(\bbI\otimes H)\cdot \cT_+(\rho)} +\Tr{H^{\otimes 2}\cdot \cT_-(\cM) \cdot \cT_+(\rho)}
    \end{split} \label{true:var}
\end{align}

Now, based on the convergence of $U_+$ and $U_-$, we consider the following three cases:
\paragraph{Case 1:}Assume $\bbU_-$ converges to 2-design with respect to $\compactG$ but $\bbU_+$ does not. Then, 
\begin{align}
    \cT_-(\cM) =\int_{\compactG} \left(U_{-}^\dagger\otimes U_{-}^\dagger\right)\left(\cM\otimes \cM\right)\left(U_{-}\otimes U_{-}\right)d\mu(U_{-}) = \sum_{\lambda}\sum_{a,b} \frac{\Tr{\cJ^\dagger_{\lambda, ab}\cM^{\otimes 2}}}{\dim V_{\lambda}} \cJ_{\lambda, ab}
\end{align}Plugging it into \cref{true:var} yields
\begin{align}
        \var =&\sum_{\lambda}\sum_{a,b} \frac{\Tr{\cJ^\dagger_{\lambda, ab}\cM^{\otimes 2}}}{\dim V_{\lambda}}\left(\Tr{\cJ_{\lambda, ab} H^{\otimes 2}\cT_+(\rho)} -\Tr{(\bbI\otimes H) \cJ_{\lambda, ab} (H\otimes \bbI)\cT_+(\rho)} \right. \nonumber\\ 
        &\left.\hspace{1cm}-\Tr{(H\otimes \bbI) \cJ_{\lambda, ab} (\bbI\otimes H)\cT_+(\rho)} +\Tr{H^{\otimes 2} \cJ_{\lambda, ab} \cT_+(\rho)}\right). \label{true:var:expended}
\end{align}Applying the triangle inequality, we obtain the upper bounded
\begin{align}
    \var \leq &\sum_{\lambda} \frac{\abs{\Tr{\cJ^\dagger_{\lambda, ab}\cM^{\otimes 2}}}}{\dim V_{\lambda}}\left(\abs{\Tr{\cJ_{\lambda, ab} H^{\otimes 2}\cT_+(\rho)}} + \abs{\Tr{(\bbI\otimes H) \cJ_{\lambda, ab} (H\otimes \bbI)\cT_+(\rho)}} \right. \nonumber\\ 
    &\left.\hspace{1cm}+ \abs{\Tr{(H\otimes \bbI) \cJ_{\lambda, ab} (\bbI\otimes H)\cT_+(\rho)}} +\abs{\Tr{H^{\otimes 2} \cJ_{\lambda, ab} \cT_+(\rho)}}\right)
\end{align}By \cref{fact:holder}, we have
\begin{align}
    \abs{\Tr{\cJ_{\lambda, ab} H^{\otimes 2}\cT_+(\rho)}} \leq& \snorm{H^{\otimes 2}}_\infty \snorm{\cT_+(\rho)}_2\snorm{\cJ_{\lambda, ab}}_2, \\
    \abs{\Tr{(\bbI\otimes H) \cJ_{\lambda, ab} (H\otimes \bbI)\cT_+(\rho)}} \leq& \snorm{(\bbI\otimes H)}_\infty \snorm{\cJ_{\lambda, ab}}_2 \snorm{( H\otimes I)}_\infty \snorm{\cT_+(\rho)}_2,
\end{align}and similarly for the remaining two terms. Combining these bounds with \cref{obs:norm{P}S2} and
$\snorm{H^{\otimes 2}}_\infty=(\snorm{H}_\infty)^2$,
$\snorm{H\otimes \bbI}_\infty=\snorm{\bbI\otimes H}_\infty=\snorm{H}_\infty$,
we obtain
\begin{align}
    \var \leq& \sum_{\lambda} \frac{\abs{\Tr{\cJ^\dagger_{\lambda, ab}\cM^{\otimes 2}}}}{\dim V_{\lambda}}\cdot \sqrt{\dim V_\lambda}\left(2\snorm{H^{\otimes 2}}_\infty \snorm{\cT_{+}(\rho)}_2 + 2\snorm{H\otimes \bbI}_\infty \snorm{\bbI\otimes H}_\infty \snorm{\cT_{+}(\rho)}_2\right) \\
    =& \sum_{\lambda} \frac{\abs{\Tr{\cJ^\dagger_{\lambda, ab}\cM^{\otimes 2}}}}{\sqrt{\dim V_{\lambda}}}\cdot 4\left(\snorm{H}_\infty\right)^2\cdot\snorm{\cT_{+}(\rho)}_2
\end{align}Using the notation from \cref{fact:true:bound:important:fact} and the definition of $U_+$, we note that
\begin{align}
    \cT_+(\rho) = &\int_{\vtheta_+} (U_{+})^{\otimes 2}\rho^{\otimes 2} (U_{+}^\dagger )^{\otimes 2}p(\vtheta_+)d\vtheta_{+} 
    = \cP^2_{H_{\ell,j+1}}\circ \cdots \circ \cP^2_{H_{L,k}}(\rho^{\otimes 2}),
\end{align}the gate $(L,k)$, which acts on $\rho$ first, being innermost. Together, we obtain 
\begin{align}
    \var \leq& 4\sum_{\lambda}\sum_{a,b} \frac{\abs{\Tr{\cJ^\dagger_{\lambda, ab}\cM^{\otimes 2}}}}{\sqrt{\dim V_{\lambda}}}\cdot \left(\snorm{H}_\infty\right)^2\cdot\snorm{\cP^2_{H_{\ell,j+1}}\circ \cdots \circ \cP^2_{H_{L,k}}(\rho^{\otimes 2})}_2 \label{app:upperbound:case_1}
\end{align}

\paragraph{Case 2:} Assume $U_+$ converges to 2-design with respect to $\compactG$ but $U_-$ does not. Applying similar argument as in Case 1 yields
\begin{align}
    \var \leq& 4\sum_{\lambda}\sum_{a,b} \frac{\abs{\Tr{\cJ^\dagger_{\lambda, ab}(\rho^{\otimes 2})}}}{\sqrt{\dim V_{\lambda}}}\cdot \left(\snorm{H}_\infty\right)^2\cdot \snorm{\cP^2_{H_{\ell, j}}\circ \cdots \circ \cP^2_{H_{1,1}}(\cM^{\otimes 2})}_2 \label{app:upperbound:case_2}
\end{align}

\paragraph{Case 3:} We assume both $U_-$ and $U_+$ converge to 2-design with respect to $\compactG$. The assumption implies
\begin{align}
    \cT_{-}(\cM) =&\int_{G} \left(U_{-}^\dagger\right)^{\otimes 2} \cM^{\otimes 2}\left(U_{-}\right)^{\otimes 2} d\mu(U_{-}) = \sum_{\lambda}\sum_{a,b} \frac{\Tr{\cJ^\dagger_{\lambda, ab} \cM^{ \otimes 2}}}{\dim V_{\lambda}} \cJ_{\lambda, ab}, \label{eq:upper:1} \\
    \cT_+(\rho) =& \int_{G} \left(U_{+}^\dagger\right)^{\otimes 2}\rho^{\otimes 2} \left(U_{+}\right)^{\otimes 2} d\mu(U_{+}) = \sum_{\lambda}\sum_{a,b} \frac{\Tr{\cJ^\dagger_{\lambda, ab}\rho^{\otimes 2}}}{\dim V_{\lambda}} \cJ_{\lambda, ab}. \label{eq:upper:2}
\end{align}
Substituting \cref{eq:upper:1,eq:upper:2} into \cref{true:var} yields the expansion
\begin{align}
    \begin{split}
        \var =& \sum_{\lambda, \lambda^\prime}\sum_{a,b}^{m_\lambda}\sum_{c,d}^{m_{\lambda^\prime}} \frac{\Tr{\cJ^\dagger_{\lambda, ab} \cM^{\otimes 2}}}{\dim V_{\lambda}} \frac{\Tr{\cJ^\dagger_{\lambda^\prime, cd}\rho^{\otimes 2}}}{\dim V_{\lambda^\prime}} \\
        &\hspace{1cm}\cdot\tr\left\{\cJ_{\lambda^\prime, cd}\left[\cJ_{\lambda, ab}H^{\otimes 2} - (H \otimes \bbI)\cJ_{\lambda, ab}(\bbI \otimes H) \right.\right. \left.\left. - (\bbI \otimes H)\cJ_{\lambda, ab}(H \otimes \bbI) + H^{\otimes 2}\cJ_{\lambda, ab}\right]\right\}
    \end{split} \label{eq:true:var}
\end{align}Applying the triangle inequality, we obtain
\begin{align}
    \begin{split}
        \var\leq& \sum_{\lambda}\sum_{a,b}^{m_\lambda} \frac{\abs{\Tr{\cJ^\dagger_{\lambda, ab}\cM^{\otimes 2}}}}{\dim V_{\lambda}}\sum_{\lambda^\prime}\sum_{c,d}^{m_{\lambda^\prime}} \frac{\abs{\Tr{\cJ^\dagger_{\lambda^\prime, cd}\rho^{\otimes 2}}}}{\dim V_{\lambda^\prime}}\left[\abs{\Tr{\cJ_{\lambda^\prime,cd}\cJ_{\lambda,ab}H^{\otimes 2}}} \right. \\ &\hspace{0.1cm}\left. + \abs{\Tr{\cJ_{\lambda^\prime,cd}(H \otimes \bbI)\cJ_{\lambda,ab}(\bbI \otimes H)}} + \abs{\Tr{\cJ_{\lambda^\prime,cd}(\bbI \otimes H)\cJ_{\lambda,ab}(H \otimes \bbI)}} + \abs{\Tr{\cJ_{\lambda^\prime,cd}H^{\otimes 2}\cJ_{\lambda, ab}}}\right] 
    \end{split} \label{var:upperbound:eq:2}
\end{align}
Using \cref{obs:norm{P}S2}, we note that 
\begin{align*}
    \abs{\Tr{\cJ_{\lambda^\prime,cd} \cJ_{\lambda, ab} H^{\otimes 2}}} \leq& \norm{\cJ_{\lambda^\prime,cd}}^S_2\norm{\cJ_{\lambda,ab}}^S_2\norm{H^{\otimes 2}}^S_{\infty} \\
    =& \sqrt{\rank{\cJ_{\lambda^\prime,cc}}}\sqrt{\rank{\cJ_{\lambda,aa}}} \left(\norm{H}^S_{\infty}\right)^2 \\
    =& \sqrt{\dim {V_{\lambda^\prime}}}\sqrt{\dim{V_{\lambda}}} \left(\norm{H}^S_{\infty}\right)^2, \\
    \abs{\Tr{\cJ_{\lambda^\prime,cd}(H \otimes \bbI)\cJ_{\lambda,ab}(\bbI \otimes H)}} \leq& \norm{ \cJ_{\lambda^\prime, cd} }^S_2 \norm{H\otimes \bbI}^S_{\infty} \norm{ \cJ_{\lambda, ab} }^S_2 \norm{\bbI\otimes H}^S_{\infty} \\
    \leq& \sqrt{\dim {V_{\lambda^\prime}}}\sqrt{\dim{V_{\lambda}}} \left(\norm{H}^S_{\infty}\right)^2,
\end{align*}and similarly for the remaining two terms. Substituting observation above into \cref{var:upperbound:eq:2}, we further bound
\begin{align}
    \var \leq& \sum_{\lambda}\sum_{a,b} \frac{\abs{\Tr{\cJ^\dagger_{\lambda,ab}\cM^{\otimes 2}}}}{\dim V_{\lambda}}\sum_{\lambda^\prime}\sum_{c,d} \frac{\abs{\Tr{\cJ^\dagger_{\lambda^\prime, cd}\rho^{\otimes 2}}}}{\dim V_{\lambda^\prime}}\left[\abs{\Tr{\cJ_{\lambda^\prime,cd}\cJ_{\lambda,ab}H^{\otimes 2}}} \right. \nonumber\\ &\hspace{0.1cm}\left. + \abs{\Tr{\cJ_{\lambda^\prime,cd}(H \otimes \bbI)\cJ_{\lambda,ab}(\bbI \otimes H)}}  + \abs{\Tr{\cJ_{\lambda^\prime,cd}(\bbI \otimes H)\cJ_{\lambda,ab}(H \otimes \bbI)}} + \abs{\Tr{\cJ_{\lambda^\prime,cd}H^{\otimes 2}\cJ_{\lambda,ab}}}\right]  \nonumber\\
    \leq& \sum_{\lambda}\sum_{a,b} \frac{\abs{\Tr{\cM^{\otimes 2} \cJ_{\lambda, ab} }}}{\dim V_{\lambda}}\sum_{\lambda^\prime}\sum_{c,d} \frac{\abs{\Tr{\rho^{\otimes 2} \cJ_{\lambda^\prime, cd} }}}{\dim V_{\lambda^\prime}}\cdot 4\sqrt{\dim {V_{\lambda^\prime}}}\sqrt{\dim{V_{\lambda}}} \left(\norm{H}^S_{\infty}\right)^2 \nonumber\\
    \leq& \sum_{\lambda^\prime}\sum_{c,d} \frac{\abs{\Tr{\rho^{\otimes 2} \cJ_{\lambda^\prime, cd} }}}{\sqrt{\dim V_{\lambda^\prime}}}\left(\sum_{\lambda}\sum_{a,b}\frac{4\abs{\Tr{\cM^{\otimes 2} \cJ_{\lambda, ab} }}}{\sqrt{\dim V_\lambda}}\right)\cdot \left(\norm{H}^S_{\infty}\right)^2 \nonumber\\
    =& 4\left(\sum_{\lambda^\prime}\sum_{c,d} \frac{\abs{\Tr{\rho^{\otimes 2} \cJ_{\lambda^\prime, cd} }}}{\sqrt{\dim V_{\lambda^\prime}}}\right)\cdot \left(\sum_{\lambda}\sum_{a,b}\frac{\abs{\Tr{\cM^{\otimes 2} \cJ_{\lambda, ab}} }}{\sqrt{\dim V_\lambda}}\right)\cdot \left(\norm{H}^S_{\infty}\right)^2 \label{app:upperbound:case_3}
\end{align}
Combining all of the above, the variance is upper bounded by
\begin{align}
    \begin{cases}
        4\sum_{\lambda}\sum_{a,b} \frac{\abs{\Tr{\cJ^\dagger_{\lambda, ab}\cM^{\otimes 2}}}}{\sqrt{\dim V_{\lambda}}}\cdot \left(\snorm{H}_\infty\right)^2\cdot\snorm{\cP^2_{H_{\ell,j+1}}\circ \cdots \circ \cP^2_{H_{L,k}}(\rho^{\otimes 2})}_2 & \bbU_- \text{ converges but }\bbU_{+}\text{ does not}\\
        4\sum_{\lambda}\sum_{a,b} \frac{\abs{\Tr{\cJ^\dagger_{\lambda,ab}(\rho^{\otimes 2})}}}{\sqrt{\dim V_{\lambda}}}\cdot \left(\snorm{H}_\infty\right)^2\cdot \snorm{\cP^2_{H_{\ell, j}}\circ \cdots \circ \cP^2_{H_{1,1}}(\cM^{\otimes 2})}_2 & \bbU_{+}\text{ converges but }\bbU_{-}\text{ does not}\\
        4\left(\sum_{\lambda^\prime}\sum_{c,d} \frac{\abs{\Tr{\rho^{\otimes 2} \cJ_{\lambda^\prime, cd} }}}{\sqrt{\dim V_{\lambda^\prime}}}\right)\cdot \left(\sum_{\lambda}\sum_{a,b}\frac{\abs{\Tr{\cM^{\otimes 2} \cJ_{\lambda, ab}} }}{\sqrt{\dim V_\lambda}}\right)\cdot \left(\norm{H}^S_{\infty}\right)^2 & \text{Both $\bbU_-$ and $\bbU_+$ converge}
    \end{cases}\label{var:upper:bound}
\end{align}

The derivative does not change when an element of $\iComm{1,\compactG}$ is added to $\cM$, whereas the right-hand sides of \cref{var:upper:bound} do (\cref{main:text:rem:reduced} of the main text); unless stated otherwise we evaluate them for the reduced observable. When a multiplicity $m_\lambda$ exceeds one, $\cH^{\otimes 2}$ can be decomposed into copies of $V_\lambda$ in more than one way (see \cref{sec:upper:rank1}); the right-hand sides then also depend on which copies are chosen, and every choice gives a valid bound. Moreover, since $\abs{\partial_{\theta_{\ell,j}}C}\leq2\,\infsnorm{H}\,\infsnorm{\cM}$ for all parameter values, the variance never exceeds $4(\infsnorm{H})^2(\infsnorm{\cM})^2$, and the bounds in \cref{var:upper:bound} are informative only where they lie below this value.

\subsubsection{Rank-1 Measurement and Quantum State}
\label{sec:upper:rank1}
In this section, we show that a refinement is possible in the rank-one setting. Specifically, when the input quantum state and measurement are both rank-one, we obtain an upper bound that captures this behavior. Now consider the case where the input quantum state and measurement are both rank one. In particular,
\begin{align}
    \rho = \Outer{\psi}{\psi}, \quad \cM = \Outer{\Psi}{\Psi},
\end{align}where $\state{\psi},\state{\Psi}\in \cH$ are arbitrary. Before deriving upper bounds on the variance under this assumption, we first recall two standard subspaces of $\cH^{\otimes 2}$.
\begin{definition} [Symmetry and Alternating Subspace]Let $V$ be any vector space. Then,
    \begin{align}
        \Symt{V} = \CurB{x\in V\otimes V: \SWAP(x) = x}
    \end{align}and
    \begin{align}
        \Altt{V} = \CurB{x\in V\otimes V: \SWAP(x) = -x}
    \end{align}where $\SWAP:V\otimes V\to V\otimes V$ is the linear map given by $\SWAP(v\otimes w) = w \otimes v$
\end{definition}

Then, we proceed to derive an upper bound for the true variance given in \cref{true:var}. Before presenting the derivation, we recall that the tensor product space $\cH^{\otimes 2}$ admits a canonical decomposition into symmetric and alternating subspaces. In particular,
\begin{align}
    \cH^{\otimes 2} = \Symt{\cH} \oplus \Altt{\cH}. \label{rank:1:upper:obs1}
\end{align}
Both summands are invariant subspaces, because $\SWAP$ commutes with $U^{\otimes 2}$ for every $U\in\compactG$. An arbitrary irreducible invariant subspace of $\cH^{\otimes 2}$, however, need not be contained in either of them. For the two-qubit collective-spin representation of $\SU{2}$, for instance, $\cH=\cH_1\oplus\cH_0$ is the sum of the triplet $\cH_1$ and the singlet $\cH_0$, the subspace $\cH_1\otimes\cH_0$ of $\cH^{\otimes 2}$ is irreducible, and $\SWAP$ maps it onto $\cH_0\otimes\cH_1$; it lies neither in $\Symt{\cH}$ nor in $\Altt{\cH}$. For this reason the bounds below do not restrict the sums to copies inside $\Symt{\cH}$.

On the other hand, a rank-one operator of the form $\Outer{\varphi}{\varphi}^{\otimes 2}$ has no support on $\Altt{\cH}$. Indeed,
\begin{align}
    \Tr{P_{\mathrm{alt}}\Outer{\varphi}{\varphi}^{\otimes 2}} 
    = \Tr{\frac{1}{2}(\bbI_{\cH^{\otimes 2}} - \SWAP)\state{\varphi}\otimes \state{\varphi}\cstate{\varphi}\otimes \cstate{\varphi}} 
    = \frac{1}{2}(1 - 1) = 0,
\end{align}where $P_{\mathrm{alt}} = \frac{1}{2}(\bbI_{\cH^{\otimes 2}} - \SWAP)$ denotes the orthogonal projector onto $\Altt{\cH}$. Since $\rho$ and $\cM$ have rank one, $\rho^{\otimes 2}$ and $\cM^{\otimes 2}$ are of this form. Consequently, whatever the distributions $\bbU_\pm$, the operators $\cT_+(\rho)$ and $\cT_-(\cM)$ are positive semidefinite and supported on $\Symt{\cH}$, being averages of operators of this form.

\begin{lemma}[Rank-one identity]
\label{app:lem:rank1}
    Let $\rho$ and $\cM$ be rank one as above, and suppose that at least one of $\bbU_-$ and $\bbU_+$ is a $2$-design with respect to $\compactG$. Then
    \begin{align}
        \var=\frac12\,\mathrm{Re}\,\Tr{\cT_-(\cM)\,(H\otimes\bbI-\bbI\otimes H)^\dagger(H\otimes\bbI-\bbI\otimes H)\,\cT_+(\rho)}. \label{rank:1:identity}
    \end{align}
\end{lemma}

\begin{proof}
    Expanding the double commutator shows that the four traces of \cref{true:var} are
    \begin{align*}
        \var=\Tr{\cT_-(\cM)\commutator{H\otimes\bbI}{\commutator{\bbI\otimes H}{\cT_+(\rho)}}}.
    \end{align*}
    Suppose first that $\bbU_+$ is a $2$-design. Then $\cT_+(\rho)$ is the Haar twirl of $\rho^{\otimes 2}$ and commutes with $U^{\otimes 2}$ for every $U\in\compactG$, in particular with $\expm{tH}^{\otimes 2}=\expm{t(H\otimes\bbI+\bbI\otimes H)}$ for all $t$, and therefore with $H\otimes\bbI+\bbI\otimes H$. We write $\bbI\otimes H=\frac12(H\otimes\bbI+\bbI\otimes H)-\frac12(H\otimes\bbI-\bbI\otimes H)$ and $H\otimes\bbI=\frac12(H\otimes\bbI+\bbI\otimes H)+\frac12(H\otimes\bbI-\bbI\otimes H)$. Since $H\otimes\bbI+\bbI\otimes H$ commutes with $(H\otimes\bbI-\bbI\otimes H)$, as $H\otimes\bbI$ and $\bbI\otimes H$ commute, and with $\cT_+(\rho)$, the terms that contain $H\otimes\bbI+\bbI\otimes H$ drop out, and
    \begin{align*}
        \commutator{H\otimes\bbI}{\commutator{\bbI\otimes H}{\cT_+(\rho)}}=-\frac14\commutator{(H\otimes\bbI-\bbI\otimes H)}{\commutator{(H\otimes\bbI-\bbI\otimes H)}{\cT_+(\rho)}}.
    \end{align*}
    If instead $\bbU_-$ is a $2$-design, the same computation applies to $\cT_-(\cM)$, after the double commutator has been moved onto it with $\iTr{A[B,[C,D]]}=\iTr{[[A,B],C]D}$. In both cases, expanding the double commutator with $(H\otimes\bbI-\bbI\otimes H)^\dagger=-(H\otimes\bbI-\bbI\otimes H)$ gives
    \begin{align*}
        \var=&\,\frac12\,\mathrm{Re}\,\Tr{\cT_-(\cM)\,(H\otimes\bbI-\bbI\otimes H)^\dagger(H\otimes\bbI-\bbI\otimes H)\,\cT_+(\rho)}\\
        &-\frac12\Tr{\cT_-(\cM)\,(H\otimes\bbI-\bbI\otimes H)\,\cT_+(\rho)\,(H\otimes\bbI-\bbI\otimes H)^\dagger},
    \end{align*}
    where we used that $\iTr{\cT_-(\cM)\,\cT_+(\rho)\,(H\otimes\bbI-\bbI\otimes H)^\dagger(H\otimes\bbI-\bbI\otimes H)}$ is the complex conjugate of the trace in the first term, all three operators in it being Hermitian. Up to this point, $\rho$ and $\cM$ are arbitrary. Since they have rank one, $\cT_+(\rho)$ and $\cT_-(\cM)$ are supported on $\Symt{\cH}$. As $\SWAP\,(H\otimes\bbI-\bbI\otimes H)\,\SWAP=-(H\otimes\bbI-\bbI\otimes H)$, the operator $(H\otimes\bbI-\bbI\otimes H)\,\cT_+(\rho)\,(H\otimes\bbI-\bbI\otimes H)^\dagger$ is therefore supported on $\Altt{\cH}$, and the last trace vanishes.
\end{proof}

Following the approach of the previous section, we distinguish three cases. In all of them we use that the Schatten-$\infty$ norm of $(H\otimes\bbI-\bbI\otimes H)^\dagger(H\otimes\bbI-\bbI\otimes H)$ equals $(\infsnorm{H\otimes\bbI-\bbI\otimes H})^2\leq4(\infsnorm{H})^2$.

\paragraph{Case 1:}
Suppose that $U_-$ converges to 2-design with respect to $\compactG$ while $U_+$ does not. Then $\cT_-(\cM)=\sum_{\lambda}\sum_{a,b}\frac{\Tr{\cJ^\dagger_{\lambda, ab}\cM^{\otimes 2}}}{\dim V_\lambda}\cJ_{\lambda, ab}$, and \cref{rank:1:identity} reads
\begin{align}
    \var=\frac12\,\mathrm{Re}\sum_{\lambda}\sum_{a,b}\frac{\Tr{\cJ^\dagger_{\lambda, ab}\cM^{\otimes 2}}}{\dim V_\lambda}\Tr{\cJ_{\lambda, ab}\,(H\otimes\bbI-\bbI\otimes H)^\dagger(H\otimes\bbI-\bbI\otimes H)\,\cT_+(\rho)}. \label{rank:1:case:1:eq:1}
\end{align}
By the complete reducibility of the unitary representation $(\cH, \Pi)$, the Hilbert space decomposes as $\cH = \bigoplus_{\lambda^\prime}\bigoplus_{a^\prime}^{m^\prime_{\lambda^\prime}}\cH_{\lambda^\prime, a^\prime}$, where the $\cH_{\lambda^\prime, a^\prime}$ are mutually orthogonal irreducible invariant subspaces. We fix one such decomposition and let $P_{\lambda^\prime, a^\prime}$ denote the orthogonal projector onto $\cH_{\lambda^\prime, a^\prime}$. Expanding the tensor square $(\bigoplus_{\lambda^\prime}\bigoplus_{a^\prime}^{m^\prime_{\lambda^\prime}}\cH_{\lambda^\prime, a^\prime})^{\otimes 2}$ yields
\begin{align*}
    \cH^{\otimes 2} = \bigoplus_{\lambda^\prime, \nu^\prime} \bigoplus_{a^\prime}^{m^\prime_{\lambda^\prime}}\bigoplus_{b^\prime}^{m^\prime_{\nu^\prime}}\cH_{\lambda^\prime, a^\prime}\otimes \cH_{\nu^\prime, b^\prime} = \bigoplus_{\lambda}\bigoplus_{a}^{m_\lambda} V_{\lambda, a}.
\end{align*}We call the summands $\cH_{\lambda^\prime, a^\prime}\otimes \cH_{\nu^\prime, b^\prime}$ of the first decomposition the \emph{tensor product spaces}. Each tensor product space is an invariant subspace of $\cH^{\otimes 2}$, because every $U\in\compactG$ maps $\cH_{\lambda^\prime, a^\prime}$ into itself and $\cH_{\nu^\prime, b^\prime}$ into itself, and $P_{\lambda^\prime, a^\prime}\otimes P_{\nu^\prime, b^\prime}$ is the orthogonal projector onto it. The second equality is obtained from the Clebsch--Gordan decomposition of the tensor product spaces: each tensor product space is decomposed separately into mutually orthogonal irreducible invariant subspaces, and the copies $V_{\lambda, a}$ are the irreducible subspaces obtained in this way from all the tensor product spaces. Consequently, for every $\lambda$ and every $a$, there exists a unique tensor product space containing the $a$-th copy of $V_{\lambda}$. We denote the indices of this tensor product space by $(\lambda^\prime_{\lambda, a}, a^\prime_{\lambda, a})$ and $(\nu^\prime_{\lambda, a}, b^\prime_{\lambda, a})$, that is, $V_{\lambda, a}\subseteq\cH_{\lambda^\prime_{\lambda, a}, a^\prime_{\lambda, a}}\otimes \cH_{\nu^\prime_{\lambda, a}, b^\prime_{\lambda , a}}$, and we let
\begin{align*}
    Q_{\lambda,a}\defeq P_{\lambda^\prime_{\lambda, a}, a^\prime_{\lambda, a}}\otimes P_{\nu^\prime_{\lambda, a}, b^\prime_{\lambda , a}}
\end{align*}be the orthogonal projector onto it. In formulas that contain two copies $a$ and $b$ of $V_\lambda$, we follow the main text and append the labels $1$ and $2$ to the subscripts of the indices that belong to the first copy and to the second copy, respectively; for instance, $Q_{\lambda,a}=P_{\lambda^\prime_{\lambda, a, 1}, a^\prime_{\lambda, a, 1}}\otimes P_{\nu^\prime_{\lambda, a, 1}, b^\prime_{\lambda , a, 1}}$ and $Q_{\lambda,b}=P_{\lambda^\prime_{\lambda, b, 2}, a^\prime_{\lambda, b, 2}}\otimes P_{\nu^\prime_{\lambda, b, 2}, b^\prime_{\lambda, b, 2}}$. These labels carry no further information, and where the representation $\lambda$ is clear from the context it is omitted from the subscripts, as in $P_{\lambda^\prime_{a, 1}, a^\prime_{a, 1}}$. Because the $a$-th copy of $V_\lambda$ is contained in the range of $Q_{\lambda,a}$ and the $b$-th copy is contained in the range of $Q_{\lambda,b}$, we have $Q_{\lambda,a}\state{\phi_{\lambda, a, m}}=\state{\phi_{\lambda, a, m}}$ and $Q_{\lambda,b}\state{\phi_{\lambda, b, m}}=\state{\phi_{\lambda, b, m}}$ for every $m$, and we find that
\begin{align*}
    Q_{\lambda,b}\,\cJ_{\lambda, ab}\,Q_{\lambda,a} = \sum_{m=1}^{\dim V_\lambda}Q_{\lambda,b}\Outer{\phi_{\lambda, b, m}}{\phi_{\lambda, a, m}}Q_{\lambda,a} = \cJ_{\lambda, ab}.
\end{align*}Moreover, every projector $P_{\lambda^\prime, a^\prime}$ belongs to $\iComm{1, \compactG}$, because $\cH_{\lambda^\prime, a^\prime}$ and its orthogonal complement are invariant subspaces of $\cH$. Hence $P_{\lambda^\prime, a^\prime}$ commutes with every $U\in\compactG$ and with $\pi_*(X)$ for every $X\in\fg$, and therefore $Q_{\lambda,a}$ commutes with $U^{\otimes 2}$, with $\pi_*(X)\otimes\bbI$ and with $\bbI\otimes\pi_*(X)$. In particular, $Q_{\lambda,a}$ and $Q_{\lambda,b}$ commute with $H\otimes\bbI-\bbI\otimes H$. Therefore
\begin{align*}
    &\Tr{\cJ_{\lambda, ab}\,(H\otimes\bbI-\bbI\otimes H)^\dagger(H\otimes\bbI-\bbI\otimes H)\,\cT_+(\rho)}\\
    &\qquad=\EXPD{U\sim \bbU_{+}}{\Tr{\cJ_{\lambda, ab}\,(H\otimes\bbI-\bbI\otimes H)^\dagger(H\otimes\bbI-\bbI\otimes H)\,U^{\otimes 2}\,Q_{\lambda,a}\,\rho^{\otimes 2}\,Q_{\lambda,b}\,(U^\dagger)^{\otimes 2}}}.
\end{align*}
For simplicity, we define the operator $\rho_{\lambda, a, 1 \to \lambda, b, 2}^{\,(\lambda^\prime, a^\prime) \otimes (\nu^\prime, b^\prime)} \defeq Q_{\lambda,a}\, \rho^{\otimes 2}\, Q_{\lambda,b}$, in which the projector on the left belongs to the first index pair and the projector on the right to the second. Applying \cref{fact:holder} with $\infsnorm{\cJ_{\lambda,ab}}=1$, Jensen's inequality and the invariance of the trace norm under unitary conjugation, we obtain
\begin{align}
    \var\leq 2\sum_{\lambda} \sum_{a,b}\frac{\abs{\Tr{\cJ^\dagger_{\lambda, ab}\cM^{\otimes 2}}}\cdot (\infsnorm{H})^2 \cdot \onesnorm{\rho_{\lambda, a, 1 \to \lambda, b, 2}^{\,(\lambda^\prime, a^\prime) \otimes (\nu^\prime, b^\prime)}}}{\dim V_\lambda}.
    \label{rank:1:upper:case:1}
\end{align}
The decompositions used above are not unique: the decomposition of $\cH$ is not unique when some $m^\prime_{\lambda^\prime}>1$, and the Clebsch--Gordan decomposition of a tensor product space is not unique when an irreducible representation occurs in it more than once. The projectors $Q_{\lambda,a}$ are defined only if every copy is contained in a single tensor product space. The bound of this case, the bound of Case 2 below and the lower bounds of \cref{sec:lower}, which contain $Q_{\lambda,a}$, are therefore stated for copies $V_{\lambda, a}$ obtained from the Clebsch--Gordan decomposition of the tensor product spaces as described above; they hold for every such choice, with right-hand sides that may depend on it. We refer to these copies as the copies of the Clebsch--Gordan decomposition.

\paragraph{Case 2:}Suppose that $U_+$ converges to 2-design with respect to $\compactG$ while $U_-$ does not. Exchanging the roles of $\rho$ and $\cM$ in the previous case, \cref{true:var} admits the upper bound
\begin{align}
    2\sum_{\lambda} \sum_{a,b}\frac{\abs{\Tr{\cJ^\dagger_{\lambda, ab}\rho^{\otimes 2}}} \cdot (\infsnorm{H})^2 \cdot \onesnorm{\cM_{\lambda, a, 1 \to \lambda, b, 2}^{\,(\lambda^\prime, a^\prime) \otimes (\nu^\prime, b^\prime)}}}{\dim V_\lambda},
    \label{rank:1:upper:case:2}
\end{align}
with $\cM_{\lambda, a, 1 \to \lambda, b, 2}^{\,(\lambda^\prime, a^\prime) \otimes (\nu^\prime, b^\prime)}\defeq Q_{\lambda,a}\,\cM^{\otimes 2}\,Q_{\lambda,b}$.

\paragraph{Case 3:}Suppose that both $U_{-}$ and $U_{+}$ converge to 2-design with respect to $\compactG$. Applying \cref{fact:holder} to \cref{rank:1:identity} gives
\begin{align}
    \var\leq\frac12\,\infsnorm{(H\otimes\bbI-\bbI\otimes H)^\dagger(H\otimes\bbI-\bbI\otimes H)}\,\onesnorm{\cT_+(\rho)\,\cT_-(\cM)}\leq2\,(\infsnorm{H})^2\,\onesnorm{\cT_+(\rho)\,\cT_-(\cM)}. \label{rank:1:upper:case3:apprx}
\end{align}
The right-hand side does not depend on how the copies are chosen, because $\cT_+(\rho)$ and $\cT_-(\cM)$ do not. To express it through the intertwiners, we insert \cref{eq:upper:1,eq:upper:2} and use $\cJ_{\lambda,ab}\cJ_{\lambda^\prime,cd}=\delta_{\lambda\lambda^\prime}\delta_{ad}\,\cJ_{\lambda,cb}$, which gives
\begin{align}
    \cT_+(\rho)\,\cT_-(\cM)=\sum_\lambda\sum_{a,b}^{m_\lambda}\frac{1}{(\dim V_\lambda)^2}\Big(\sum_{c=1}^{m_\lambda}\Tr{\cJ^\dagger_{\lambda,ac}\cM^{\otimes 2}}\,\Tr{\cJ^\dagger_{\lambda,cb}\rho^{\otimes 2}}\Big)\cJ_{\lambda,ab}. \label{rank:1:product}
\end{align}
The term with the label $\lambda$ is the product of the terms with the label $\lambda$ in \cref{eq:upper:2,eq:upper:1}, and the terms with different labels act on the mutually orthogonal subspaces $m_\lambda V_\lambda$. The trace norm of \cref{rank:1:product} is therefore the sum of the trace norms of its terms, and each of these is at most the product of the Hilbert--Schmidt norms of the two factors. By the orthogonality of the intertwiners, the term with the label $\lambda$ in \cref{eq:upper:2} has the Hilbert--Schmidt norm $\big(\sum_{c,d}\abs{\iTr{\cJ^\dagger_{\lambda,cd}\rho^{\otimes 2}}}^2/\dim V_\lambda\big)^{1/2}$, and likewise for \cref{eq:upper:1}. We obtain
\begin{align}
    \var\leq 2\,(\infsnorm{H})^2\sum_{\lambda}\frac{1}{\dim V_{\lambda}}\Big(\sum_{c,d}^{m_\lambda}\abs{\Tr{\cJ^\dagger_{\lambda,cd}\rho^{\otimes 2}}}^2\Big)^{1/2}\Big(\sum_{a,b}^{m_\lambda}\abs{\Tr{\cJ^\dagger_{\lambda,ab}\cM^{\otimes 2}}}^2\Big)^{1/2}.
    \label{rank:1:upper:case:3}
\end{align}
Each square root in \cref{rank:1:upper:case:3} is at most the corresponding sum of absolute values, and a sum over $\lambda$ of products of non-negative numbers is at most the product of the sums over $\lambda$. The right-hand side of \cref{rank:1:upper:case:3}, and hence that of \cref{rank:1:upper:case3:apprx}, is therefore at most half of the right-hand side of \cref{app:upperbound:case_3}. If $\cT_+(\rho)$ and $\cT_-(\cM)$ commute---for instance when every $V_\lambda$ that carries $\rho^{\otimes 2}$ has multiplicity one, because $\cT_+(\rho)$ is then a linear combination of the projectors $\sum_a\cJ_{\lambda,aa}$, which commute with every intertwiner---then $\cT_+(\rho)\cT_-(\cM)$ is positive semidefinite, because $\cT_+(\rho)$ is positive semidefinite and, since $\cM$ has rank one, so is $\cT_-(\cM)$. Then $\onesnorm{\cT_+(\rho)\cT_-(\cM)}=\iTr{\cT_+(\rho)\cT_-(\cM)}=\iTr{\cT_+(\rho)\cM^{\otimes 2}}$, and \cref{rank:1:upper:case3:apprx} becomes
\begin{align}
    \var\leq 2\,(\infsnorm{H})^2\sum_{\lambda}\sum_{a,b}\frac{\Tr{\cJ^\dagger_{\lambda, ab} \rho^{\otimes 2} }\cdot \Tr{\cJ_{\lambda, ab}\, \cM^{\otimes 2}}}{\dim V_{\lambda}}
    =2\,(\infsnorm{H})^2\,\EXPD{U\sim\mu_{\compactG}}{\Tr{\cM\,U\rho\,U^\dagger}^2}.
    \label{rank:1:upper:case:3:commuting}
\end{align}
Note the second factor $\Tr{\cJ_{\lambda, ab}\cM^{\otimes 2}}=\overline{\Tr{\cJ^\dagger_{\lambda, ab}\cM^{\otimes 2}}}$: the sum $\sum_{a,b}\Tr{\cJ^\dagger_{\lambda,ab}\rho^{\otimes 2}}\Tr{\cJ^\dagger_{\lambda,ab}\cM^{\otimes 2}}$, in which both factors carry $\cJ^\dagger_{\lambda,ab}$, is not invariant under a change of the copies and does not bound the variance in general.

Combining all of the above, the variance is upper bounded by
\begin{align}
    \begin{cases}
        2\sum_{\lambda} \sum_{a,b}\frac{\abs{\Tr{\cJ^\dagger_{\lambda, ab}\cM^{\otimes 2}}}\cdot (\infsnorm{H})^2 \cdot \onesnorm{\rho_{\lambda, a, 1 \to \lambda, b, 2}^{\,(\lambda^\prime, a^\prime) \otimes (\nu^\prime, b^\prime)}}}{\dim V_\lambda} & \bbU_- \text{ converges but }\bbU_{+}\text{ does not}\\
        2\sum_{\lambda} \sum_{a,b}\frac{\abs{\Tr{\cJ^\dagger_{\lambda, ab}\rho^{\otimes 2}}} \cdot (\infsnorm{H})^2 \cdot \onesnorm{\cM_{\lambda, a, 1 \to \lambda, b, 2}^{\,(\lambda^\prime, a^\prime) \otimes (\nu^\prime, b^\prime)}}}{\dim V_\lambda} & \bbU_{+}\text{ converges but }\bbU_{-}\text{ does not}\\
        2\,(\infsnorm{H})^2\,\onesnorm{\cT_+(\rho)\,\cT_-(\cM)} & \text{Both $\bbU_-$ and $\bbU_+$ converge}
    \end{cases}
\end{align}

\subsection{Lower Bound}
\label{sec:lower}
In this section, we derive a lower bound on the variance. To achieve this, we exploit the cyclic property of the trace to express the variance as
\begin{align}
    \var=&\EXPD{U_{-}\sim \bbU_{-}, U_{+}\sim \bbU_{+}}{ \Tr{\cM U_{-}\commutator{H}{U_{+}\rho U_{+}^\dagger } U_{-}^\dagger}^2} \\
    =& \EXPD{U_{-}\sim \bbU_{-}, U_{+}\sim \bbU_{+}}{ \Tr{\tcM^{\otimes 2} \commutator{H}{\trho}^{\otimes 2}} } 
\end{align}where we define the rotated operators $\tcM \defeq U_{-}^\dagger \cM U_{-}$ and $\trho \defeq U_{+}\rho U_{+}^\dagger$. Applying \cref{app:prelim:nested:comm} and incorporating the notation introduced in \cref{sec:upper}, the variance simplifies to
\begin{align}
    \var =& \EXPD{U_{-}\sim \bbU_{-}, U_{+}\sim \bbU_{+}}{ \Tr{\trho^{\otimes 2} \commutator{H \otimes I} { \commutator{ I \otimes H} {\tcM^{\otimes 2}}} } } \\
    =& \Tr{\EXPD{U\sim \bbU_{-}}{\tcM^{\otimes 2}} \commutator{H \otimes I} { \commutator{ I \otimes H} {\EXPD{U \sim \bbU_{+}}{\trho^{\otimes 2}}}} } \\
    =& \Tr{\cT_{-}(\cM) \commutator{H \otimes I} { \commutator{ I \otimes H} {\cT_{+}(\rho)}} } \label{app:var:compact}
\end{align} To evaluate this bound further, we proceed by analyzing the underlying structure case by case.
\paragraph{Case 1:}
Suppose $U_-$ converges but $U_+$ does not. Under these conditions, the variance is approximated by
\begin{align}
    \begin{split}
        \var =& \sum_{\lambda} \sum_{a}^{m_\lambda} \frac{\Tr{\cJ_{\lambda,aa}^\dagger\cM^{\otimes 2}}}{\dim V_\lambda} \Tr{\cJ_{\lambda, aa} \commutator{H\otimes \bbI}{\commutator{\bbI\otimes H}{\cT_{+}(\rho)}}} \\
        &\hspace{1cm}+\sum_{\lambda} \sum_{a\neq b}^{m_\lambda} \frac{\Tr{\cJ_{\lambda,ab}^\dagger\cM^{\otimes 2}}}{\dim V_\lambda} \Tr{\cJ_{\lambda, ab} \commutator{H\otimes \bbI}{\commutator{\bbI\otimes H}{\cT_{+}(\rho)}}}.
    \end{split}
\end{align}Applying the reverse triangle inequality, we can lower bound the variance:
\begin{align}
    \begin{split}
        & \abs{\sum_{\lambda} \sum_{a}^{m_\lambda} \frac{\Tr{\cJ_{\lambda,aa}^\dagger\cM^{\otimes 2}}}{\dim V_\lambda} \Tr{\cJ_{\lambda, aa} \commutator{H\otimes \bbI}{\commutator{\bbI\otimes H}{\cT_{+}(\rho)}}}} \\
        &\hspace{1cm} - \abs{\sum_{\lambda} \sum_{a\neq b}^{m_\lambda} \frac{\Tr{\cJ_{\lambda,ab}^\dagger\cM^{\otimes 2}}}{\dim V_\lambda} \Tr{\cJ_{\lambda, ab} \commutator{H\otimes \bbI}{\commutator{\bbI\otimes H}{\cT_{+}(\rho)}}}}.
    \end{split}
\end{align}For the copies $V_{\lambda,a}$ of the Clebsch--Gordan decomposition (Case 1 of \cref{sec:upper:rank1}), we have the projector relation
\begin{align}
    P_{\lambda^\prime_{b, 2}, a^\prime_{b, 2}}\otimes P_{\nu^\prime_{b, 2}, b^\prime_{b, 2}} \cJ_{\lambda, ab} P_{\lambda^\prime_{a, 1}, a^\prime_{a, 1}}\otimes P_{\nu^\prime_{a, 1}, b^\prime_{a, 1}} = \cJ_{\lambda, ab},
\end{align}and the trace evaluates to
\begin{align}
    &\Tr{\cJ_{\lambda, ab} \commutator{H\otimes \bbI}{\commutator{\bbI\otimes H}{\cT_{+}(\rho)}}} \\
    =& \EXPD{U \sim \bbU_{+}}{\Tr{\cJ_{\lambda, ab}\commutator{H\otimes \bbI}{\commutator{\bbI\otimes H}{U^{\otimes 2} P_{\lambda^\prime_{a, 1}, a^\prime_{a, 1}}\rho P_{\lambda^\prime_{b, 2}, a^\prime_{b, 2}} \otimes P_{\nu^\prime_{a, 1}, b^\prime_{a, 1}}\rho P_{\nu^\prime_{b, 2}, b^\prime_{b, 2}}  (U^\dagger)^{\otimes 2}}}}}
\end{align}Using the intermediate operator $\rho_{\lambda, a, 1 \to \lambda, b, 2}^{\,(\lambda^\prime, a^\prime) \otimes (\nu^\prime, b^\prime)}=Q_{\lambda,a}\,\rho^{\otimes 2}\,Q_{\lambda,b}$ defined in \cref{sec:upper:rank1}, which is the operator conjugated by $U^{\otimes 2}$ in the last line, we apply Jensen's inequality to bound the absolute value of the cross-terms:
\begin{align}
    &\abs{\sum_{\lambda} \sum_{a\neq b}^{m_\lambda} \frac{\Tr{\cJ_{\lambda,ab}^\dagger\cM^{\otimes 2}}}{\dim V_\lambda} \Tr{\cJ_{\lambda, ab} \commutator{H\otimes \bbI}{\commutator{\bbI\otimes H}{\cT_{+}(\rho)}}}} \\
    \leq& \EXPD{U \sim \bbU_{+}}{\abs{\sum_{\lambda} \sum_{a\neq b}^{m_\lambda} \frac{\Tr{\cJ_{\lambda,ab}^\dagger\cM^{\otimes 2}}}{\dim V_\lambda} \Tr{\cJ_{\lambda, ab}\commutator{H\otimes \bbI}{\commutator{\bbI\otimes H}{U^{\otimes 2} \rho_{\lambda, a, 1 \to \lambda, b, 2}^{\,(\lambda^\prime, a^\prime) \otimes (\nu^\prime, b^\prime)} (U^\dagger)^{\otimes 2}}}} }}
\end{align}Combining these results yields the final lower bound on the variance:
\begin{align}
    \begin{split}
        & \abs{\sum_{\lambda} \sum_{a}^{m_\lambda} \frac{\Tr{\cJ_{\lambda,aa}^\dagger\cM^{\otimes 2}}}{\dim V_\lambda} \Tr{\cJ_{\lambda, aa} \commutator{H\otimes \bbI}{\commutator{\bbI\otimes H}{\cT_{+}(\rho)}}}} \\
        &\hspace{0.5cm} - \EXPD{U \sim \bbU_{+}}{\abs{\sum_{\lambda} \sum_{a\neq b}^{m_\lambda} \frac{\Tr{\cJ_{\lambda,ab}^\dagger\cM^{\otimes 2}}}{\dim V_\lambda} \Tr{\cJ_{\lambda, ab}\commutator{H\otimes \bbI}{\commutator{\bbI\otimes H}{U^{\otimes 2} \rho_{\lambda, a, 1 \to \lambda, b, 2}^{\,(\lambda^\prime, a^\prime) \otimes (\nu^\prime, b^\prime)} (U^\dagger)^{\otimes 2}}}} }}
    \end{split} \label{app:lowerbound:case_1}
\end{align}

\paragraph{Case 2:}
Suppose $U_{+}$ converges but $U_-$ does not. By a similar argument, the variance is lower bounded by
\begin{align}
    \begin{split}
        & \abs{\sum_{\lambda} \sum_{a}^{m_\lambda} \frac{\Tr{\cJ_{\lambda,aa}^\dagger\rho^{\otimes 2}}}{\dim V_\lambda} \Tr{\cT_{-}(\cM) \commutator{H\otimes \bbI}{\commutator{\bbI\otimes H}{\cJ_{\lambda, aa}}}}} \\
        &\hspace{0.5cm} -\sum_{\lambda} \sum_{a\neq b}^{m_\lambda} \frac{\snorm{\cJ_{\lambda, bb}\sqrt{{\rho^{\otimes 2}}}}_2\cdot \snorm{\cJ_{\lambda, aa}\sqrt{{\rho^{\otimes 2}}}}_2}{(\dim V_\lambda)}\\
        &\hspace{1cm}\times\EXPD{U\sim \bbU_{-}}{\abs{\Tr{(U^\dagger)^{\otimes 2} \cM_{\lambda, a, 1 \to \lambda, b, 2}^{\,(\lambda^\prime, a^\prime) \otimes (\nu^\prime, b^\prime)} U^{\otimes 2} \commutator{H\otimes \bbI}{\commutator{\bbI\otimes H}{\cJ_{\lambda, ab}}}}}} \\
    \end{split} \label{app:lowerbound:case_2}
\end{align}

\paragraph{Case 3:}
Assuming simultaneous convergence of $U_-$ and $U_+$, the variance is given by \cref{eq:true:var}. Writing the four traces of \cref{eq:true:var} as the double commutator of \cref{app:var:compact}, with $\nu$ in place of $\lambda^\prime$, and separating the terms with $\lambda=\nu$ from those with $\lambda\neq\nu$, we obtain
\begin{align}
    \begin{split}
        \var =& \sum_{\lambda}\sum_{a,b,c,d}^{m_\lambda}\frac{\Tr{\cJ_{\lambda,ab}^\dagger\cM^{\otimes 2}}\cdot \Tr{\cJ_{\lambda,cd}^\dagger\rho^{\otimes 2}}}{(\dim V_\lambda)^2}\Tr{\cJ_{\lambda, ab}\commutator{H\otimes \bbI}{\commutator{\bbI\otimes H}{\cJ_{\lambda, cd}}}} \\
        &\hspace{1cm}+ \sum_{\lambda \neq \nu}\sum_{a,b}^{m_\lambda}\sum_{c,d}^{m_\nu}\frac{\Tr{\cJ_{\lambda,ab}^\dagger\cM^{\otimes 2}}\cdot \Tr{\cJ_{\nu,cd}^\dagger\rho^{\otimes 2}}}{\dim V_\lambda \cdot \dim V_\nu}\Tr{\cJ_{\lambda, ab}\commutator{H\otimes \bbI}{\commutator{\bbI\otimes H}{\cJ_{\nu, cd}}}}.
    \end{split} \label{app:var:case3:sum}
\end{align}Again, applying the reverse triangle inequality, the variance is lower bounded by
\begin{align}
    \begin{split}
        & \abs{\sum_{\lambda}\sum_{a,b,c,d}^{m_\lambda}\frac{\Tr{\cJ_{\lambda,ab}^\dagger\cM^{\otimes 2}}\cdot \Tr{\cJ_{\lambda,cd}^\dagger\rho^{\otimes 2}}}{(\dim V_\lambda)^2}\Tr{\cJ_{\lambda, ab}\commutator{H\otimes \bbI}{\commutator{\bbI\otimes H}{\cJ_{\lambda, cd}}}}} \\
        &\hspace{1cm} - \abs{\sum_{\lambda \neq \nu}\sum_{a,b}^{m_\lambda}\sum_{c,d}^{m_\nu}\frac{\Tr{\cJ_{\lambda,ab}^\dagger\cM^{\otimes 2}}\cdot \Tr{\cJ_{\nu,cd}^\dagger\rho^{\otimes 2}}}{\dim V_\lambda \cdot \dim V_\nu}\Tr{\cJ_{\lambda, ab}\commutator{H\otimes \bbI}{\commutator{\bbI\otimes H}{\cJ_{\nu, cd}}}}}
    \end{split}
\end{align}To proceed, we note that for any Hermitian operator $M$ with spectral decomposition $\sum_{j}\lambda_j \Outer{\lambda_j}{\lambda_j}$, evaluating the absolute value of the trace yields
\begin{align*}
    &\abs{\Tr{\cJ_{\lambda, ab}^\dagger M^{\otimes 2}}} \\
    =&\abs{\sum_{m=1}^{\dim V_\lambda}\Tr{\state{\phi_{m,a}^\lambda}\cstate{\phi_{m,b}^\lambda}M^{\otimes 2}}} \\
    =& \abs{\sum_{m=1}^{\dim V_\lambda} \sum_j \lambda_j \Inner{\phi_{m,b}^\lambda}{\lambda_j}\Inner{\lambda_j}{\phi_{m,a}^\lambda}} \\
    \leq& \sqrt{\sum_{m,j}\left(\sqrt{|\lambda_j|}\Inner{\phi_{m,b}^\lambda}{\lambda_j}\right)^2}\cdot \sqrt{\sum_{m,j}\left(\sqrt{|\lambda_j|}\Inner{\lambda_j}{\phi_{m,a}^\lambda}\right)^2} & \text{(Cauchy-Schwarz)} \\
    \begin{split}
        =& \sqrt{\Tr{\left(\sum_{m}\Outer{\phi_{m,b}^\lambda}{\phi_{m,b}^\lambda}\right)\left(\sum_j |\lambda_j|\Outer{\lambda_j}{\lambda_j}\right)}}\\
        &\hspace{1cm}\times \sqrt{\Tr{\left(\sum_{m}\Outer{\phi_{m,a}^\lambda}{\phi_{m,a}^\lambda}\right)\left(\sum_j |\lambda_j|\Outer{\lambda_j}{\lambda_j}\right)}}
    \end{split}\\
    =& \snorm{\cJ_{\lambda, bb}\sqrt{\abs{M^{\otimes 2}}}}_2\cdot \snorm{\cJ_{\lambda, aa}\sqrt{\abs{M^{\otimes 2}}}}_2
\end{align*}Combining these results, we bound the cross-term as
\begin{align}
    \begin{split}
        &\abs{\sum_{\lambda \neq \nu}\sum_{a,b}^{m_\lambda}\sum_{c,d}^{m_\nu}\frac{\Tr{\cJ_{\lambda,ab}^\dagger\cM^{\otimes 2}}\cdot \Tr{\cJ_{\nu,cd}^\dagger\rho^{\otimes 2}}}{\dim V_\lambda \cdot \dim V_\nu}\Tr{\cJ_{\lambda, ab}\commutator{H\otimes \bbI}{\commutator{\bbI\otimes H}{\cJ_{\nu, cd}}}}}
    \end{split} \\
    \begin{split}
        \leq&\sum_{\nu}\sum_{c,d}^{m_\nu}\frac{\snorm{\cJ_{\nu, dd}\sqrt{{\rho^{\otimes 2}}}}_2\snorm{\cJ_{\nu, cc}\sqrt{{\rho^{\otimes 2}}}}_2}{\dim V_\nu}\abs{\Tr{\left(\sum_{\lambda\neq \nu}\sum_{a,b}^{m_\lambda}\frac{\Tr{\cJ^\dagger_{\lambda, ab}\cM^{\otimes 2}}}{\dim V_\lambda}\cJ_{\lambda, ab}\right)\commutator{H\otimes \bbI}{\commutator{\bbI\otimes H}{\cJ_{\nu, cd}}}}}
    \end{split}
    \label{eq:lb:eq:1}
\end{align}
For the copies of the Clebsch--Gordan decomposition (Case 1 of \cref{sec:upper:rank1}), we have the projector identity
\begin{align*}
    P_{\lambda^\prime_{\nu, d, 2}, a^\prime_{\nu, d, 2}}\otimes P_{\nu^\prime_{\nu, d, 2}, b^\prime_{\nu, d, 2}} \cJ_{\nu, cd} P_{\lambda^\prime_{\nu, c, 1}, a^\prime_{\nu, c, 1}}\otimes P_{\nu^\prime_{\nu, c, 1}, b^\prime_{\nu, c, 1}} = \cJ_{\nu, cd}
\end{align*}Because the tensor factors of these projectors belong to the commutant $\iComm{1, \compactG}$, we find that 
\begin{align}
    \commutator{H\otimes \bbI}{\commutator{\bbI\otimes H}{\cJ_{\nu, cd}}} = P_{\lambda^\prime_{\nu, d, 2}, a^\prime_{\nu, d, 2}}\otimes P_{\nu^\prime_{\nu, d, 2}, b^\prime_{\nu, d, 2}} \commutator{H\otimes \bbI}{\commutator{\bbI\otimes H}{\cJ_{\nu, cd}}} P_{\lambda^\prime_{\nu, c, 1}, a^\prime_{\nu, c, 1}}\otimes P_{\nu^\prime_{\nu, c, 1}, b^\prime_{\nu, c, 1}} \label{eq:lb:blocks}
\end{align}
Additionally, we define the restricted intertwiner
\begin{align*}
    &\cJ_{\lambda, ab}^{(\lambda^\prime, a^\prime, \nu^\prime, b^\prime)_{\nu, c, 1}, \; (\lambda^\prime, a^\prime, \nu^\prime, b^\prime)_{\nu, d, 2}}
    \defeq P_{\lambda^\prime_{\nu, c, 1}, a^\prime_{\nu, c, 1}}\otimes P_{\nu^\prime_{\nu, c, 1}, b^\prime_{\nu, c, 1}} \cJ_{\lambda, ab} P_{\lambda^\prime_{\nu, d, 2}, a^\prime_{\nu, d, 2}}\otimes P_{\nu^\prime_{\nu, d, 2}, b^\prime_{\nu, d, 2}} \\
    =& \begin{cases}
        J_{\lambda, ab}  & V_{\lambda, a} \subseteq  \cH_{\lambda^\prime_{\nu, d, 2}, a^\prime_{\nu, d, 2}}\otimes \cH_{\nu^\prime_{\nu, d, 2}, b^\prime_{\nu, d, 2}} \text{ and } V_{\lambda, b} \subseteq \cH_{\lambda^\prime_{\nu, c, 1}, a^\prime_{\nu, c, 1}}\otimes \cH_{\nu^\prime_{\nu, c, 1}, b^\prime_{\nu, c, 1}} \\
        0 & o.w.
    \end{cases}
\end{align*}where $V_{\lambda, a}, V_{\lambda, b}$ denote the $a$-th and $b$-th copy of $V_{\lambda}$, respectively. By \cref{eq:lb:blocks} and the cyclic property of the trace, every $\cJ_{\lambda,ab}$ in the weighted sum inside the trace of \cref{eq:lb:eq:1} can be replaced by its restricted intertwiner without changing the trace. The sum obtained in this way evaluates to
\begin{align*}
    & \sum_{a,b}^{m_\lambda}\frac{\Tr{\cJ^\dagger_{\lambda, ab}\cM^{\otimes 2}}}{\dim V_\lambda}P_{\lambda^\prime_{\nu, c, 1}, a^\prime_{\nu, c, 1}}\otimes P_{\nu^\prime_{\nu, c, 1}, b^\prime_{\nu, c, 1}} \cJ_{\lambda, ab} P_{\lambda^\prime_{\nu, d, 2}, a^\prime_{\nu, d, 2}}\otimes P_{\nu^\prime_{\nu, d, 2}, b^\prime_{\nu, d, 2}} \\
    =& \sum_{a,b}^{m_\lambda}\frac{\Tr{\cJ^\dagger_{\lambda, ab}(P_{\lambda^\prime_{\nu, c, 1}, a^\prime_{\nu, c, 1}}\otimes P_{\nu^\prime_{\nu, c, 1}, b^\prime_{\nu, c, 1}}) \cM^{\otimes 2}(P_{\lambda^\prime_{\nu, d, 2}, a^\prime_{\nu, d, 2}}\otimes P_{\nu^\prime_{\nu, d, 2}, b^\prime_{\nu, d, 2}})}}{\dim V_\lambda} \cJ_{\lambda, ab}^{(\lambda^\prime, a^\prime, \nu^\prime, b^\prime)_{\nu, c, 1}, \; (\lambda^\prime, a^\prime, \nu^\prime, b^\prime)_{\nu, d, 2}} \\
    =& \sum_{a,b}^{m_\lambda}\frac{\Tr{\cJ^\dagger_{\lambda, ab}(P_{\lambda^\prime_{\nu, c, 1}, a^\prime_{\nu, c, 1}}\cM P_{\lambda^\prime_{\nu, d, 2}, a^\prime_{\nu, d, 2}}) \otimes (P_{\nu^\prime_{\nu, c, 1}, b^\prime_{\nu, c, 1}}\cM P_{\nu^\prime_{\nu, d, 2}, b^\prime_{\nu, d, 2}})}}{\dim V_\lambda} \cJ_{\lambda, ab}^{(\lambda^\prime, a^\prime, \nu^\prime, b^\prime)_{\nu, c, 1}, \; (\lambda^\prime, a^\prime, \nu^\prime, b^\prime)_{\nu, d, 2}} \\
    =& \sum_{a,b}^{m_\lambda}\frac{\Tr{\cJ^\dagger_{\lambda, ab}\cM^{(\lambda^\prime, a^\prime)_{\nu, c, 1}, \; (\lambda^\prime, a^\prime)_{\nu, d, 2}} \otimes \mathcal{M}^{(\nu^\prime, b^\prime)_{\nu, c, 1}, \; (\nu^\prime, b^\prime)_{\nu, d, 2}}}}{\dim V_\lambda} \cJ_{\lambda, ab}^{(\lambda^\prime, a^\prime, \nu^\prime, b^\prime)_{\nu, c, 1}, \; (\lambda^\prime, a^\prime, \nu^\prime, b^\prime)_{\nu, d, 2}}
\end{align*}Expanding the trace $\Tr{\cJ_{\lambda, ab}^{(\lambda^\prime, a^\prime, \nu^\prime, b^\prime)_{\nu, c, 1}, \; (\lambda^\prime, a^\prime, \nu^\prime, b^\prime)_{\nu, d, 2}}\commutator{H\otimes \bbI}{\commutator{\bbI\otimes H}{\cJ_{\nu, cd}}}}$ and applying the orthogonality condition $\cJ_{\lambda, ab}\cJ_{\nu, cd} = 0$ for $\lambda \neq \nu$, alongside $\commutator{H \otimes \bbI + \bbI \otimes H}{\cJ_{\nu, cd}} = 0$, the expression simplifies to
\begin{align}
    2\Tr{\cJ_{\lambda, ab}^{(\lambda^\prime, a^\prime, \nu^\prime, b^\prime)_{\nu, c, 1}, \; (\lambda^\prime, a^\prime, \nu^\prime, b^\prime)_{\nu, d, 2}} (H \otimes \bbI) \cJ_{\nu, cd} (H\otimes \bbI)}
\end{align}
Combining these results, the final variance is lower bounded by
\begin{align}
    \begin{split}
        &\abs{\sum_{\lambda}\sum_{a,b,c,d}^{m_\lambda}\frac{\Tr{\cJ_{\lambda,ab}^\dagger\cM^{\otimes 2}}\cdot \Tr{\cJ_{\lambda,cd}^\dagger\rho^{\otimes 2}}}{(\dim V_\lambda)^2}\Tr{\cJ_{\lambda, ab}\commutator{H\otimes \bbI}{\commutator{\bbI\otimes H}{\cJ_{\lambda, cd}}}}} \\
        &\hspace{0.5cm}-2\sum_{\nu}\sum_{c,d}^{m_\nu}\frac{\snorm{\cJ_{\nu, dd}\sqrt{{\rho^{\otimes 2}}}}_2\snorm{\cJ_{\nu, cc}\sqrt{{\rho^{\otimes 2}}}}_2}{\dim V_\nu}\\
        &\hspace{1cm}\times\abs{\Tr{\left(\sum_{\lambda\neq \nu}\sum_{a,b}^{m_\lambda}\frac{\Tr{\cJ^\dagger_{\lambda, ab} \cM_{\nu, c, 1 \to \nu, d, 2}^{\,(\lambda^\prime, a^\prime) \otimes (\nu^\prime, b^\prime)}}}{\dim V_\lambda}\cJ_{\lambda, ab}^{(\lambda^\prime, a^\prime, \nu^\prime, b^\prime)_{\nu, c, 1}, \; (\lambda^\prime, a^\prime, \nu^\prime, b^\prime)_{\nu, d, 2}}\right) (H \otimes \bbI) \cJ_{\nu, cd} (H\otimes \bbI)}},
    \end{split} \label{app:lowerbound:case_3}
\end{align}where $\cM_{\nu, c, 1 \to \nu, d, 2}^{\,(\lambda^\prime, a^\prime) \otimes (\nu^\prime, b^\prime)} \defeq\cM^{(\lambda^\prime, a^\prime)_{\nu, c, 1}, \; (\lambda^\prime, a^\prime)_{\nu, d, 2}} \otimes \mathcal{M}^{(\nu^\prime, b^\prime)_{\nu, c, 1}, \; (\nu^\prime, b^\prime)_{\nu, d, 2}}$

\subsection{Trainability Criteria}
\label{sec:criteria}

In this section, we derive the trainability criteria of \cref{main:text:theory:sec:discussion} of the main text from the bounds of Case~3, \cref{app:upperbound:case_3,app:lowerbound:case_3}. Throughout, both $\bbU_-$ and $\bbU_+$ are $2$-designs with respect to $\compactG$, so that $\cT_-(\cM)$ and $\cT_+(\rho)$ are given by \cref{eq:upper:1,eq:upper:2}, and the copies $V_{\lambda,a}$ are those of the Clebsch--Gordan decomposition, for which \cref{app:lowerbound:case_3} is stated. By the definition of the intertwiners after \cref{eq:35}, $\cJ_{\lambda,ab}\cJ_{\nu,cd}=\delta_{\lambda\nu}\delta_{ad}\cJ_{\lambda,cb}$ and $\iTr{\cJ_{\lambda,ab}\cJ^\dagger_{\nu,cd}}=\delta_{\lambda\nu}\delta_{ac}\delta_{bd}\dim V_\lambda$. In particular, $\sum_a\cJ_{\lambda,aa}$ is the orthogonal projector onto $m_\lambda V_\lambda$, and $\big(\sum_a\cJ_{\lambda,aa}\big)\cJ_{\nu,cd}=\cJ_{\nu,cd}\sum_a\cJ_{\lambda,aa}=\delta_{\lambda\nu}\cJ_{\nu,cd}$. Hence $\sum_a\cJ_{\lambda,aa}$ commutes with $\cT_-(\cM)$ and $\cT_+(\rho)$, and $\cT_-(\cM)\sum_a\cJ_{\lambda,aa}$ and $\cT_+(\rho)\sum_a\cJ_{\lambda,aa}$ are the terms with the label $\lambda$ in \cref{eq:upper:1,eq:upper:2}. For complex numbers $z_{ab}$, the operator $\sum_{a,b}z_{ab}\cJ_{\lambda,ab}$ vanishes on the orthogonal complement of $m_\lambda V_\lambda$ and maps $\state{\phi_{\lambda,a,m}}$ to $\sum_bz_{ab}\state{\phi_{\lambda,b,m}}$. It therefore acts in the same way on each of the $\dim V_\lambda$ mutually orthogonal subspaces spanned by $\state{\phi_{\lambda,1,m}},\dots,\state{\phi_{\lambda,m_\lambda,m}}$, so that each of its non-zero singular values, and each of its non-zero eigenvalues, occurs a multiple of $\dim V_\lambda$ times. The intertwiners commute with $U^{\otimes 2}$ for $U\in\compactG$, and by \cref{eq:35} every element of $\iComm{2,\compactG}$, being its own twirl, is a linear combination of them. They are therefore an orthogonal basis of $\iComm{2,\compactG}$, the commutant of $\Pi(G)^{\otimes 2}$, and $\dim\iComm{2,\compactG}=\sum_\lambda m_\lambda^2$.

\subsubsection{Implications of the Upper Bound}
\label{sec:criteria:upper}

Since $\rho$ and $\cM$ are Hermitian, the absolute values in the two sums of \cref{app:upperbound:case_3} are $\iabs{\iTr{\cJ^\dagger_{\lambda,cd}\rho^{\otimes 2}}}$ and $\iabs{\iTr{\cJ^\dagger_{\lambda,ab}\cM^{\otimes 2}}}$. By \cref{eq:upper:2} and the orthogonality of the intertwiners, 
\begin{align*}
    (\snorm{\cT_+(\rho)}_2)^2=\sum_\lambda\sum_{c,d}\abs{\iTr{\cJ^\dagger_{\lambda,cd}\rho^{\otimes 2}}}^2/\dim V_\lambda.
\end{align*}As the sum with $\rho^{\otimes 2}$ in \cref{app:upperbound:case_3} has $\dim\iComm{2,\compactG}$ terms, the Cauchy--Schwarz inequality bounds it by $\big(\dim\iComm{2,\compactG}\big)^{1/2}\snorm{\cT_+(\rho)}_2$; likewise, the sum with $\cM^{\otimes 2}$ is at most $\big(\dim\iComm{2,\compactG}\big)^{1/2}\snorm{\cT_-(\cM)}_2$. Hence the right-hand side of \cref{app:upperbound:case_3} is at most
\begin{align}
    4\dim\iComm{2,\compactG}\,(\infsnorm{H})^2\,\snorm{\cT_+(\rho)}_2\,\snorm{\cT_-(\cM)}_2. \label{app:crit:factors}
\end{align}
Here $\snorm{\cT_+(\rho)}_2$ is the state overlap and $\snorm{\cT_-(\cM)}_2$ the observable overlap of \cref{main:text:theory:sec:discussion}. When the commutant of $\Pi(G)^{\otimes 2}$ has polynomial dimension, the upper bound can therefore be at least inverse-polynomial only if the product $(\infsnorm{H})^2\snorm{\cT_+(\rho)}_2\snorm{\cT_-(\cM)}_2$ is.

All terms of the two sums are non-negative. We keep in the sum with $\rho^{\otimes 2}$ only the terms with the label $\lambda^\prime$ and $c=d$, which are $\iTr{\cJ_{\lambda^\prime,cc}\rho^{\otimes 2}}/\sqrt{\dim V_{\lambda^\prime}}$ because $\cJ_{\lambda^\prime,cc}$ is an orthogonal projector, and in the sum with $\cM^{\otimes 2}$ only those with the label $\lambda$ and $a=b$, whose sum is at least $\abs{\iTr{\sum_a\cJ_{\lambda,aa}\cM^{\otimes 2}}}/\sqrt{\dim V_\lambda}$. Hence, for every pair $\lambda,\lambda^\prime$, the right-hand side of \cref{app:upperbound:case_3} is at least
\begin{align}
    4(\infsnorm{H})^2\,\frac{\iTr{\sum_c\cJ_{\lambda^\prime,cc}\rho^{\otimes 2}}\,\abs{\iTr{\sum_a\cJ_{\lambda,aa}\cM^{\otimes 2}}}}{\sqrt{\dim V_\lambda\dim V_{\lambda^\prime}}}. \label{app:crit:sufficient}
\end{align}
The upper bound is therefore at least inverse-polynomial if $\infsnorm{H}$ is and if there are irreducible representations $V_\lambda$ and $V_{\lambda^\prime}$ of polynomial dimension on which the Hilbert--Schmidt overlaps $\abs{\iTr{\sum_a\cJ_{\lambda,aa}\cM^{\otimes 2}}}$ and $\iTr{\sum_a\cJ_{\lambda^\prime,aa}\rho^{\otimes 2}}$ are at least inverse-polynomial. This is the sufficient condition illustrated in \cref{fig:upper-bound}. An upper bound that is at least inverse-polynomial does not, however, imply that the variance is (\cref{sec:S72}).

The weight of $\rho^{\otimes 2}$ on $m_\nu V_\nu$ is $\iTr{\sum_c\cJ_{\nu,cc}\rho^{\otimes 2}}$, and these weights add up to $\iTr{\rho^{\otimes 2}}=1$. The state overlap is small when $\rho^{\otimes 2}$ has small weight on the irreducible representations of small dimension: for every $d_0>0$,
\begin{align}
    (\snorm{\cT_+(\rho)}_2)^2\leq\sum_\nu\frac{\big(\iTr{\sum_c\cJ_{\nu,cc}\rho^{\otimes 2}}\big)^2}{\dim V_\nu}\leq\sum_{\nu:\,\dim V_\nu<d_0}\iTr{\textstyle\sum_c\cJ_{\nu,cc}\rho^{\otimes 2}}+\frac{1}{d_0}. \label{app:crit:state}
\end{align}
Indeed, by \cref{eq:upper:2}, $\cT_+(\rho)$ is the sum over $\nu$ of the mutually orthogonal operators $\cT_+(\rho)\sum_c\cJ_{\nu,cc}$, so that $(\snorm{\cT_+(\rho)}_2)^2$ is the sum of their squared Hilbert--Schmidt norms. Each of them is positive semidefinite, with trace equal to the weight of $\rho^{\otimes 2}$ on $m_\nu V_\nu$, since $\cT_+(\rho)$ is an average of the positive semidefinite operators $(U^\dagger)^{\otimes 2}\rho^{\otimes 2}U^{\otimes 2}$ and the projector $\sum_c\cJ_{\nu,cc}$ commutes with $\cT_+(\rho)$ and with $U^{\otimes 2}$; and by the statement on $\sum_{a,b}z_{ab}\cJ_{\lambda,ab}$ above, each of its non-zero eigenvalues occurs a multiple of $\dim V_\nu$ times. The sum of the squares of its eigenvalues is therefore at most the square of its trace divided by $\dim V_\nu$, which gives the first inequality; the second holds because every weight is at most $1$ and the weights add up to $1$. Hence, if $\rho^{\otimes 2}$ has only exponentially small weight on the irreducible representations of dimension below some exponentially large $d_0$, the state overlap is exponentially small, and by \cref{app:crit:factors} the upper bound is exponentially small unless $(\infsnorm{H})^2\dim\iComm{2,\compactG}\,\snorm{\cT_-(\cM)}_2$ grows exponentially with $n$.

\subsubsection{Implications of the Lower Bound}
\label{sec:criteria:lower}

This section proves the sufficient condition of \cref{main:text:theory:sec:discussion}, \cref{main:text:thm:sufficient} of the main text (\cref{app:thm:concentration}), and its extension to states with weight outside $m_{\lambda^\ast}V_{\lambda^\ast}$, \cref{main:text:cor:leakage} of the main text. We call the summand of $\lambda$ in the sum inside the absolute value of \cref{app:lowerbound:case_3} the isolated term of $\lambda$, and the summand of $\nu$ in the subtracted sum, including the factor $2$, the cross term of $\nu$. If $\rho^{\otimes 2}$ is supported on $m_{\lambda^\ast}V_{\lambda^\ast}$, then $\cT_+(\rho)\sum_c\cJ_{\nu,cc}=0$ and $\cJ_{\nu,cc}\sqrt{\rho^{\otimes 2}}=0$ for every $\nu\neq\lambda^\ast$, so that the isolated terms and the cross terms of all $\nu\neq\lambda^\ast$ vanish, and \cref{app:lowerbound:case_3} is the absolute value of the isolated term of $\lambda^\ast$, the term $\lambda=\lambda^\ast$ of \cref{main:text:theory:isolated:terms}, minus the cross term of $\lambda^\ast$, the residual cross term built from \cref{eq:conc-crossnorm,eq:conc-crosstrace}.

\paragraph{Isolated terms.} Inserting the terms with the label $\lambda$ of \cref{eq:upper:1,eq:upper:2}, we obtain by linearity that the isolated term of $\lambda$ is
\begin{align}
    \begin{split}
        &\sum_{a,b,c,d}^{m_\lambda}\frac{\Tr{\cJ_{\lambda,ab}^\dagger\cM^{\otimes 2}}\cdot\Tr{\cJ_{\lambda,cd}^\dagger\rho^{\otimes 2}}}{(\dim V_\lambda)^2}\Tr{\cJ_{\lambda,ab}\commutator{H\otimes\bbI}{\commutator{\bbI\otimes H}{\cJ_{\lambda,cd}}}}\\
        &\qquad=\Tr{\cT_-(\cM)\Big({\textstyle\sum_a}\cJ_{\lambda,aa}\Big)\commutator{H\otimes\bbI}{\commutator{\bbI\otimes H}{\cT_+(\rho){\textstyle\sum_a}\cJ_{\lambda,aa}}}}.
    \end{split} \label{app:crit:terms}
\end{align}
The operator $\cT_+(\rho)\sum_a\cJ_{\lambda,aa}$ is positive semidefinite, and its trace is the weight of $\rho^{\otimes 2}$ on $m_\lambda V_\lambda$ (\cref{sec:criteria:upper}), so that this weight is its trace norm. By \cref{eq:upper:1} and the statement on $\sum_{a,b}z_{ab}\cJ_{\lambda,ab}$ above, each non-zero singular value of $\cT_-(\cM)\sum_a\cJ_{\lambda,aa}$ occurs a multiple of $\dim V_\lambda$ times, so that its Schatten-$\infty$ norm is at most its trace norm divided by $\dim V_\lambda$; and as a compression of $\cT_-(\cM)$, an average of unitary conjugates of $\cM^{\otimes 2}$, it has Schatten-$\infty$ norm at most $(\infsnorm{\cM})^2$ and trace norm at most $(\onesnorm{\cM})^2$. By \cref{fact:holder} and $\onesnorm{\commutator{H\otimes\bbI}{\commutator{\bbI\otimes H}{A}}}\leq4(\infsnorm{H})^2\onesnorm{A}$ for every operator $A$, the absolute value of the isolated term of $\lambda$ is therefore at most
\begin{align}
    4(\infsnorm{H})^2\,\iTr{\textstyle\sum_a\cJ_{\lambda,aa}\rho^{\otimes 2}}\,\min\CurB{(\infsnorm{\cM})^2,\,\frac{(\onesnorm{\cM})^2}{\dim V_\lambda}}. \label{app:crit:isolated}
\end{align}
Since the weight is at most $1$, the second entry shows that the absolute value of the isolated term of $\lambda$ is at most $4(\infsnorm{H})^2(\onesnorm{\cM})^2/\dim V_\lambda$, so that, when $\infsnorm{H}$ and the trace norm of the observable are polynomially bounded, as for a projector onto a pure state, it can be at least inverse-polynomial only if $\dim V_\lambda$ is at most polynomial in $n$.

The isolated terms depend on how the generator acts on the copies through the traces $\iTr{\cJ_{\lambda,ab}\commutator{H\otimes\bbI}{\commutator{\bbI\otimes H}{\cJ_{\lambda,cd}}}}$. The projector $\cJ_{\lambda,aa}$ onto the $a$-th copy of $V_\lambda$ commutes with $U^{\otimes 2}$ for every $U\in\compactG$, in particular with $\expm{tH}^{\otimes 2}=\expm{t(H\otimes\bbI+\bbI\otimes H)}$ for every $t$ (as $\expm{tH}\in\compactG$ by \cref{reduction:thm}), and hence with $H\otimes\bbI+\bbI\otimes H$. Therefore $\commutator{\bbI\otimes H}{\cJ_{\lambda,aa}}=-\commutator{H\otimes\bbI}{\cJ_{\lambda,aa}}$ and $\iTr{\cJ_{\lambda,aa}\commutator{H\otimes\bbI}{\commutator{\bbI\otimes H}{\cJ_{\lambda,aa}}}}=-\iTr{\cJ_{\lambda,aa}\commutator{H\otimes\bbI}{\commutator{H\otimes\bbI}{\cJ_{\lambda,aa}}}}=\iTr{\commutator{\cJ_{\lambda,aa}}{H\otimes\bbI}^2}$. Inserting $\commutator{\cJ_{\lambda,aa}}{H\otimes\bbI}=\cJ_{\lambda,aa}(H\otimes\bbI)(\bbI-\cJ_{\lambda,aa})-(\bbI-\cJ_{\lambda,aa})(H\otimes\bbI)\cJ_{\lambda,aa}$ and using $\cJ_{\lambda,aa}(\bbI-\cJ_{\lambda,aa})=0$ and $(H\otimes\bbI)^\dagger=-H\otimes\bbI$, we obtain
\begin{align}
    \Tr{\cJ_{\lambda,aa}\commutator{H\otimes\bbI}{\commutator{\bbI\otimes H}{\cJ_{\lambda,aa}}}}=2\big(\snorm{(\bbI-\cJ_{\lambda,aa})(H\otimes\bbI)\cJ_{\lambda,aa}}_2\big)^2. \label{app:crit:copy}
\end{align}
This trace vanishes if and only if $H\otimes\bbI$ maps the $a$-th copy into itself. If $m_\lambda=1$, the sum in \cref{app:crit:terms} has the single term $a=b=c=d=1$, and by \cref{app:crit:copy} the isolated term of $\lambda$ is
\begin{align*}
    \frac{2\,\iTr{\cJ_{\lambda,11}\cM^{\otimes 2}}\,\iTr{\cJ_{\lambda,11}\rho^{\otimes 2}}}{(\dim V_\lambda)^2}\,\big(\snorm{(\bbI-\cJ_{\lambda,11})(H\otimes\bbI)\cJ_{\lambda,11}}_2\big)^2,
\end{align*}
the product of the trace overlaps of the observable and the state with $V_\lambda$ and of the squared Hilbert--Schmidt norm of the part of $(H\otimes\bbI)\cJ_{\lambda,11}$ outside $V_\lambda$, which measures how far $H\otimes\bbI$ moves $V_\lambda$. It vanishes when $H\otimes\bbI$ maps $V_\lambda$ into itself, however large the two trace overlaps are; this is why the generator must move $V_{\lambda^\ast}$.

\paragraph{Cross terms.} The restricted operators can be removed from the trace in the summand $(c,d)$ of the cross term of $\nu$:
\begin{align*}
    &\Tr{\left(\sum_{\lambda\neq \nu}\sum_{a,b}^{m_\lambda}\frac{\Tr{\cJ^\dagger_{\lambda, ab} \cM_{\nu, c, 1 \to \nu, d, 2}^{\,(\lambda^\prime, a^\prime) \otimes (\nu^\prime, b^\prime)}}}{\dim V_\lambda}\cJ_{\lambda, ab}^{(\lambda^\prime, a^\prime, \nu^\prime, b^\prime)_{\nu, c, 1}, \; (\lambda^\prime, a^\prime, \nu^\prime, b^\prime)_{\nu, d, 2}}\right) (H \otimes \bbI) \cJ_{\nu, cd} (H\otimes \bbI)}\\
    &\qquad=\Tr{\left(\sum_{\lambda\neq\nu}\sum_{a,b}^{m_\lambda}\frac{\Tr{\cJ^\dagger_{\lambda,ab}Q_{\nu,c}\cM^{\otimes 2}Q_{\nu,d}}}{\dim V_\lambda}\,Q_{\nu,c}\cJ_{\lambda,ab}Q_{\nu,d}\right)(H\otimes\bbI)\,\cJ_{\nu,cd}\,(H\otimes\bbI)}\\
    &\qquad=\Tr{Q_{\nu,c}\Big(\bbI-{\textstyle\sum_a}\cJ_{\nu,aa}\Big)\cT_-(\cM)\,Q_{\nu,d}\,(H\otimes\bbI)\,\cJ_{\nu,cd}\,(H\otimes\bbI)}\\
    &\qquad=\Tr{\Big(\bbI-{\textstyle\sum_a}\cJ_{\nu,aa}\Big)\cT_-(\cM)\,(H\otimes\bbI)\,\cJ_{\nu,cd}\,(H\otimes\bbI)}.
\end{align*}
The first equality holds by the definitions of the restricted intertwiner and of $\cM_{\nu,c,1\to\nu,d,2}^{\,(\lambda^\prime,a^\prime)\otimes(\nu^\prime,b^\prime)}=Q_{\nu,c}\cM^{\otimes 2}Q_{\nu,d}$ (\cref{sec:lower}; Case~2 of \cref{sec:upper:rank1}), where $Q_{\nu,c}$ is the orthogonal projector onto the tensor product space that contains the $c$-th copy of $V_\nu$ (Case~1 of \cref{sec:upper:rank1}). By \cref{eq:upper:1}, the operator $Q_{\nu,c}\big(\bbI-\sum_a\cJ_{\nu,aa}\big)\cT_-(\cM)Q_{\nu,d}$ in the third line is the sum in the second line with $\iTr{\cJ^\dagger_{\lambda,ab}\cM^{\otimes 2}}$ in place of $\iTr{\cJ^\dagger_{\lambda,ab}Q_{\nu,c}\cM^{\otimes 2}Q_{\nu,d}}$. As every copy lies in exactly one tensor product space, $Q_{\nu,c}\cJ_{\lambda,ab}Q_{\nu,d}$ is either zero or $\cJ_{\lambda,ab}$, and in the second case the two traces are equal, because $\iTr{\cJ^\dagger_{\lambda,ab}Q_{\nu,c}\cM^{\otimes 2}Q_{\nu,d}}=\iTr{(Q_{\nu,c}\cJ_{\lambda,ab}Q_{\nu,d})^\dagger\cM^{\otimes 2}}$; this gives the second equality. The third holds because $Q_{\nu,c}$ and $Q_{\nu,d}$ commute with $H\otimes\bbI$, $Q_{\nu,d}\cJ_{\nu,cd}Q_{\nu,c}=\cJ_{\nu,cd}$ (Case~1 of \cref{sec:upper:rank1}) and the trace is cyclic. For $\nu=\lambda^\ast$, the absolute value of the left-hand side is \cref{eq:conc-crosstrace}, and the cross term of $\lambda^\ast$ is the residual cross term of \cref{main:text:theory:sec:discussion}, built from \cref{eq:conc-crossnorm,eq:conc-crosstrace}.

The observable enters the right-hand side only through $\big(\bbI-\sum_a\cJ_{\nu,aa}\big)\cT_-(\cM)$. As $\bbI-\sum_a\cJ_{\nu,aa}$ commutes with $\cT_-(\cM)$ and with $U^{\otimes 2}$, this operator is the average of the unitary conjugates of the part $\big(\bbI-\sum_a\cJ_{\nu,aa}\big)\cM^{\otimes 2}\big(\bbI-\sum_a\cJ_{\nu,aa}\big)$ of $\cM^{\otimes 2}$ outside $m_\nu V_\nu$, so that its Schatten-$\infty$ norm is at most that of this part. By \cref{fact:holder} with $\infsnorm{H\otimes\bbI}=\infsnorm{H}$ and $\onesnorm{\cJ_{\nu,cd}}=\iTr{\cJ_{\nu,cc}}=\dim V_\nu$, the absolute value of the right-hand side is at most $(\infsnorm{H})^2\dim V_\nu$ times this norm. Since $\big(\sum_c\snorm{\cJ_{\nu,cc}\sqrt{\rho^{\otimes 2}}}_2\big)^2\leq m_\nu\iTr{\sum_c\cJ_{\nu,cc}\rho^{\otimes 2}}$ by Cauchy--Schwarz, the cross term of $\nu$ is at most
\begin{align}
    2m_\nu(\infsnorm{H})^2\,\snorm{\Big(\bbI-{\textstyle\sum_a}\cJ_{\nu,aa}\Big)\cM^{\otimes 2}\Big(\bbI-{\textstyle\sum_a}\cJ_{\nu,aa}\Big)}_\infty\Tr{{\textstyle\sum_c}\cJ_{\nu,cc}\rho^{\otimes 2}}, \label{app:crit:crossbound}
\end{align}
which is small when $m_\nu$ times the weight of $\rho^{\otimes 2}$ on $m_\nu V_\nu$ is small, or when $m_\nu$ times the Schatten-$\infty$ norm of the part of $\cM^{\otimes 2}$ outside $m_\nu V_\nu$ is small. If $\rho^{\otimes 2}$ is supported on $m_{\lambda^\ast}V_{\lambda^\ast}$, its weight there is $1$, and the residual cross term is at most $2m_{\lambda^\ast}(\infsnorm{H})^2$ times the Schatten-$\infty$ norm of the part of $\cM^{\otimes 2}$ outside $m_{\lambda^\ast}V_{\lambda^\ast}$. If $\cM^{\otimes 2}$ is supported on $m_{\lambda^\ast}V_{\lambda^\ast}$ as well, the residual cross term vanishes, and since $\cT_-(\cM)$ and $\cT_+(\rho)$ then consist of the terms with the label $\lambda^\ast$ in \cref{eq:upper:1,eq:upper:2}, \cref{app:var:compact,app:crit:terms} show that the variance equals the isolated term of $\lambda^\ast$.

With \cref{app:lowerbound:case_3}, these statements give the following corollary, which is \cref{main:text:thm:sufficient} of the main text.

\begin{corollary}[Lower bound under concentration]
\label{app:thm:concentration}
    Suppose that $\bbU_\pm$ are $2$-designs with respect to $\compactG$, let the copies $V_{\lambda,a}$ be those of the Clebsch--Gordan decomposition, and let $\lambda^\ast$ be a label such that $\rho^{\otimes 2}$ is supported on $m_{\lambda^\ast}V_{\lambda^\ast}$. Then
    \begin{align}
        \begin{split}
            \var\geq&\,\abs{\sum_{a,b,c,d}^{m_{\lambda^\ast}}\frac{\Tr{\cJ_{\lambda^\ast,ab}^\dagger\cM^{\otimes 2}}\cdot\Tr{\cJ_{\lambda^\ast,cd}^\dagger\rho^{\otimes 2}}}{(\dim V_{\lambda^\ast})^2}\Tr{\cJ_{\lambda^\ast,ab}\commutator{H\otimes\bbI}{\commutator{\bbI\otimes H}{\cJ_{\lambda^\ast,cd}}}}}\\
            &-\frac{2}{\dim V_{\lambda^\ast}}\sum_{c,d}^{m_{\lambda^\ast}}\snorm{\cJ_{\lambda^\ast,cc}\sqrt{\rho^{\otimes 2}}}_2\,\snorm{\cJ_{\lambda^\ast,dd}\sqrt{\rho^{\otimes 2}}}_2\\
            &\qquad\times\abs{\Tr{\Big(\bbI-{\textstyle\sum_a}\cJ_{\lambda^\ast,aa}\Big)\cT_-(\cM)\,(H\otimes\bbI)\,\cJ_{\lambda^\ast,cd}\,(H\otimes\bbI)}}.
        \end{split} \label{app:crit:concentration}
    \end{align}
    The second term is at most $2(\infsnorm{H})^2m_{\lambda^\ast}\isnorm{\big(\bbI-\sum_a\cJ_{\lambda^\ast,aa}\big)\cM^{\otimes 2}\big(\bbI-\sum_a\cJ_{\lambda^\ast,aa}\big)}_\infty$; if $\cM^{\otimes 2}$ is supported on $m_{\lambda^\ast}V_{\lambda^\ast}$ as well, it vanishes, and the variance equals the isolated term of $\lambda^\ast$. In particular, if the second term is at most half of the first and the first term is $\Omega(1/\poly{n})$, then the variance is at least half of the first term, hence $\Omega(1/\poly{n})$, and there is no barren plateau.
\end{corollary}

In \cref{main:text:thm:sufficient}, where $H=\pi_*(g_j)$, $\Theta=\cT_-(\cM)$ and $P_{\lambda^\ast}=\sum_a\cJ_{\lambda^\ast,aa}$, the first term of \cref{app:crit:concentration} is written $\abs{I_{\lambda^\ast}}$ and the norms $\isnorm{\cJ_{\lambda^\ast,cc}\sqrt{\rho^{\otimes 2}}}_2$ are written $s_c$. We now prove \cref{main:text:cor:leakage} of the main text, in which $\rho^{\otimes 2}$ may also have weight outside $m_{\lambda^\ast}V_{\lambda^\ast}$. For every label $\lambda$, let $I_\lambda$ be the isolated term of $\lambda$; for $\lambda=\lambda^\ast$, it is the quantity $I_{\lambda^\ast}$ of \cref{main:text:thm:sufficient}. With the traces of the cross terms written without the restricted operators, as above, \cref{app:lowerbound:case_3} reads
\begin{align*}
    \var\geq&\,\abs{\sum_\lambda I_\lambda}-\sum_\nu\frac{2}{\dim V_\nu}\sum_{c,d}^{m_\nu}\snorm{\cJ_{\nu,cc}\sqrt{\rho^{\otimes 2}}}_2\,\snorm{\cJ_{\nu,dd}\sqrt{\rho^{\otimes 2}}}_2\\*
    &\qquad\times\abs{\Tr{\Big(\bbI-{\textstyle\sum_a}\cJ_{\nu,aa}\Big)\cT_-(\cM)\,(H\otimes\bbI)\,\cJ_{\nu,cd}\,(H\otimes\bbI)}},
\end{align*}
where the summand of $\nu$ in the second term is the cross term of $\nu$. A label $\nu\neq\lambda^\ast$ enters this bound in two places, and both can be non-zero once $\rho^{\otimes 2}$ has weight on $m_\nu V_\nu$. First, its isolated term $I_\nu$ lies inside the absolute value. Its sign is not known, so that it may cancel part of $I_{\lambda^\ast}$, and we separate it by the triangle inequality, $\iabs{\sum_\lambda I_\lambda}\geq\iabs{I_{\lambda^\ast}}-\sum_{\nu\neq\lambda^\ast}\iabs{I_\nu}$. Second, its cross term is already subtracted in \cref{app:lowerbound:case_3}. Keeping the label $\lambda^\ast$ apart in both sums, and writing its isolated term and its cross term in the notation of \cref{main:text:thm:sufficient}, we obtain
\begin{align*}
    \var\geq&\,\abs{I_{\lambda^\ast}}-\frac{2}{\dim V_{\lambda^\ast}}\sum_{c,d}^{m_{\lambda^\ast}}s_cs_d\,\abs{\Tr{(\bbI-P_{\lambda^\ast})\Theta\,(H\otimes\bbI)\,\cJ_{\lambda^\ast,cd}\,(H\otimes\bbI)}}\\
    &-\sum_{\nu\neq\lambda^\ast}\abs{I_\nu}\\
    &-\sum_{\nu\neq\lambda^\ast}\frac{2}{\dim V_\nu}\sum_{c,d}^{m_\nu}\snorm{\cJ_{\nu,cc}\sqrt{\rho^{\otimes 2}}}_2\,\snorm{\cJ_{\nu,dd}\sqrt{\rho^{\otimes 2}}}_2\\*
    &\qquad\times\abs{\Tr{\Big(\bbI-{\textstyle\sum_a}\cJ_{\nu,aa}\Big)\cT_-(\cM)\,(H\otimes\bbI)\,\cJ_{\nu,cd}\,(H\otimes\bbI)}}.
\end{align*}
The first two terms are the right-hand side of \cref{main:text:eq:concentration}. The third term is the sum of the isolated terms of the labels $\nu\neq\lambda^\ast$, and the fourth term is the sum of their cross terms. We bound both by the weights $\iTr{\sum_c\cJ_{\nu,cc}\rho^{\otimes 2}}$ of $\rho^{\otimes 2}$ on $m_\nu V_\nu$. For the third term, \cref{app:crit:isolated} with the first entry of the minimum gives
\begin{align*}
    \sum_{\nu\neq\lambda^\ast}\abs{I_\nu}\leq\sum_{\nu\neq\lambda^\ast}4(\infsnorm{H})^2(\infsnorm{\cM})^2\,\iTr{\textstyle\sum_c\cJ_{\nu,cc}\rho^{\otimes 2}},
\end{align*}
where the factor $4$ comes from the bound used for \cref{app:crit:isolated}: for every operator $A$, the trace norm of the double commutator $\commutator{H\otimes\bbI}{\commutator{\bbI\otimes H}{A}}$ is at most $4(\infsnorm{H})^2\onesnorm{A}$, since each of its four terms has trace norm at most $(\infsnorm{H})^2\onesnorm{A}$. For the fourth term, the part of $\cM^{\otimes 2}$ outside $m_\nu V_\nu$ in \cref{app:crit:crossbound} has Schatten-$\infty$ norm at most that of $\cM^{\otimes 2}$, which is $(\infsnorm{\cM})^2$, because $\bbI-\sum_a\cJ_{\nu,aa}$ is an orthogonal projector. By \cref{app:crit:crossbound}, the fourth term is therefore at most
\begin{align*}
    \sum_{\nu\neq\lambda^\ast}2m_\nu(\infsnorm{H})^2(\infsnorm{\cM})^2\,\iTr{\textstyle\sum_c\cJ_{\nu,cc}\rho^{\otimes 2}},
\end{align*}
where the factor $2$ is that of the cross terms in \cref{app:lowerbound:case_3}, and the factor $m_\nu$ comes from the Cauchy--Schwarz inequality $\big(\sum_c\snorm{\cJ_{\nu,cc}\sqrt{\rho^{\otimes 2}}}_2\big)^2\leq m_\nu\iTr{\sum_c\cJ_{\nu,cc}\rho^{\otimes 2}}$ used for \cref{app:crit:crossbound}. Adding the two bounds, the third and the fourth term are together at most
\begin{align*}
    (\infsnorm{H})^2(\infsnorm{\cM})^2\sum_{\nu\neq\lambda^\ast}(4+2m_\nu)\,\iTr{\textstyle\sum_c\cJ_{\nu,cc}\rho^{\otimes 2}},
\end{align*}
which is \cref{main:text:eq:leakage}. Hence \cref{main:text:eq:concentration} holds with \cref{main:text:eq:leakage} subtracted from its right-hand side. If the second term of \cref{main:text:eq:concentration} and \cref{main:text:eq:leakage} are together at most $\iabs{I_{\lambda^\ast}}/2$, the variance is at least $\iabs{I_{\lambda^\ast}}/2$. This proves \cref{main:text:cor:leakage} of the main text.

\subsection{Comparison with Lie-Algebraic Frameworks} 
\label{sec:comparison_prior_works}

In this section, we compare our results with those of \cite{fontana2024characterizing,ragone2024lie}. Their framework provides a Lie-algebraic description of the variance, which is exact when the observable is generated by the algebra; ours is organized by the invariant subspaces of $\cH^{\otimes 2}$ and does not need that assumption. In particular, \cite{fontana2024characterizing} express the variance in terms of the simple-ideal decomposition of the compact Lie algebra,
\begin{align}
    \fg = \fc \oplus \bigoplus_\alpha \fg_\alpha ,
\end{align}
with contributions of the form
\begin{align}
    \sum_\alpha
    \frac{
        \|H_{\fg_\alpha}\|_{K}^{2}
        \|O_{\fg_\alpha}\|_{F}^{2}
        \|\rho_{\fg_\alpha}\|_{F}^{2}
    }{
        d_{\fg_\alpha}^{2}
    } . \label{eq:comparison:lie}
\end{align}
This gives a powerful, algebra-centric perspective where the scaling is organized by the dimensions $d_{\fg_\alpha}$ of the simple ideals. Therefore, it effectively captures how the variance depends on the decomposition of $\fg$. Furthermore, the exact compact-group variance formula (Theorem~2.9 in \cite{fontana2024characterizing}) assumes that the observable is generated directly by the Lie algebra, such that
\begin{align}
    i\cM=d\phi(\fm),
    \qquad
    \fm\in\fg .
\end{align}
This is a major structural restriction. In many variational quantum algorithms, the observable is chosen independently of the compact Lie algebra $\fg$ associated with the ansatz. For a general observable, since the observable $\cM$ and the state $\rho$ act on the state space $\cH$ and do not fully live within the Lie algebra representation (i.e., $i d\phi(\fg)$), one must instead use their generalized lower bound (e.g., Theorem~F.3). While this generalized bound covers more cases, its evaluation is complicated by the presence of a covariance term. As a result, it is not obvious from the statement alone when the lower bound is strictly positive or at least inverse-polynomial in $n$. In particular, because the covariance term contains competing contributions, determining when it remains strictly positive or sufficiently small in magnitude—and thus verifying a non-vanishing variance—often requires a more careful, case-specific analysis.

When the observable does lie in the algebra, the two descriptions agree, since both then give the variance exactly. For a simple algebra the sum in \cref{eq:comparison:lie} has a single term, which we write as $(\norm{H}_K)^2(\snorm{\cM}_2)^2(\snorm{\rho_{\fg}}_2)^2/D^2$. Here $D\defeq\dim\pi_*(\fg)$, $\rho_{\fg}$ is the orthogonal projection of $\rho$ onto the space $i\pi_*(\fg)$ of Hermitian operators, and $(\norm{H}_K)^2\defeq\sum_{i}(\snorm{[H,B_i]}_2)^2$ for a basis $\CurB{B_i}_{i=1}^{D}$ of $i\pi_*(\fg)$ that is orthonormal for the Hilbert--Schmidt inner product; the subscript $K$ refers to the Killing form. Three of the examples of \cref{sec:examples}---the standard representation of $\SU{2^n}$ with a traceless observable (\cref{app:example:SUN}), the collective-spin representation with $\cM=S_X+S_Y+S_Z$ (\cref{app:example:SU2}) and free fermions with $\cM=Z^{(1)}$ (\cref{app:example:ff:local})---are of this type: the values obtained there from \cref{eq:true:var}, for instance $\frac23(\lambda^*)^2$ and $4(n-1)/\big(n(2n-1)^2\big)$, are exactly what this formula gives, as is checked in each of these sections. The representation-theoretic formulas go beyond it in two situations: when the observable has a component outside the algebra, for which \cref{eq:true:var} gives the variance without a covariance term (\cref{app:example:compound}), and when one asks which invariant subspaces make a given state and observable trainable (\cref{sec:criteria}).

\section{Sampling-Based Variance Estimator}
\label{sec:tester}
In this section, we propose a sampling-based algorithm to estimate the variance of the partial derivative $\partial_{\theta_{j,\ell}}C$. As established in the preceding section, the direct computation of this variance conventionally requires decomposing the unitary representation $\cH^{\otimes 2}$ and explicitly constructing the associated projectors $\cJ_{\lambda, ab}$. Although this decomposition is theoretically accessible via standard Lie algebraic methods \cite{hall2013lie} and Weingarten calculus \cite{collins2022weingarten}, such procedures are computationally intractable for large system sizes $n$. Our sampling approach bypasses this decomposition entirely: it runs the ansatz itself, and it decides whether the variance exceeds a tolerance $\tau$ with a number of runs polynomial in $1/\tau$ (\cref{app:tester:prop}).

Following the notation established in the preceding section, we denote the generator of the unitary operation $\Pi(\expm{\theta_{\ell, j}g_j}) = \expm{\theta_{\ell, j}\pi_*(g_j)}$ by $H$.

\subsection{Derivation}
We first restructure the variance into a form that accommodates random sampling. Rather than relying on the explicit algebraic decomposition, we approach the problem through a finite-difference approximation of the exact variance. Following \cref{lemma:expectation}, we assume that the first moment of the partial derivative $\partial_{\theta_{j,\ell}}C$ vanishes. Under this condition, the variance can be written as
\begin{align}
    \begin{split}
        \var =& \EXPD{U_{-} \sim \bbU_{-}}{\Tr{\cM^{\otimes 2}(U_{-})^{\otimes 2} \commutator{H \otimes I} { \commutator{ I \otimes H} {\EXPD{U_{+}\sim \bbU_{+}}{(U_{+})^{\otimes 2}\rho^{\otimes 2}(U_{+}^\dagger)^{\otimes 2}}}} (U_{-}^\dagger)^{\otimes 2} }}
    \end{split}
\end{align}Next, recalling the generator identity for any operators $H$ and $A$,
\begin{align*}
    \frac{d \expm{t H} A \expm{-t H}}{dt}\bigg|_{t = 0} = [H, A]
\end{align*}we can approximate the nested commutators using a central finite-difference method with a small step size $x$, i.e.,
\begin{align*}
    \commutator{H\otimes \bbI}{A}\approx \frac{(\expm{xH}\otimes I) A (\expm{-xH}\otimes I) - (\expm{-xH}\otimes I) A (\expm{xH}\otimes I)}{2x}.
\end{align*}and
\begin{align*}
    \commutator{\bbI \otimes H}{A} \approx \frac{(\bbI\otimes \expm{xH}) A (\bbI\otimes \expm{-xH}) - (\bbI\otimes \expm{-xH}) A (\bbI\otimes \expm{xH})}{2x}
\end{align*}For notational simplicity, we define the superoperators
\begin{align}
    f(x;A,H) \defeq \frac{(\expm{xH}\otimes I) A (\expm{-xH}\otimes I) - (\expm{-xH}\otimes I) A (\expm{xH}\otimes I)}{2x} \\
    g(x;A,H) \defeq \frac{(I\otimes \expm{xH}) A (I\otimes \expm{-xH}) - (I\otimes \expm{-xH}) A (I\otimes \expm{xH})}{2x}
\end{align}Applying this finite-difference approximation, we define the approximate variance $\bbV_{\mathrm{FD}}$ via the composition $f \circ g$:
\begin{align}
    \bbV_{\mathrm{FD}} \defeq \EXPD{U_{-} \sim \bbU_{-}}{\Tr{\cM^{\otimes 2}(U_{-})^{\otimes 2} f\circ g \left(\EXPD{U_{+}\sim \bbU_{+}}{(U_{+})^{\otimes 2}\rho^{\otimes 2}(U_{+}^\dagger)^{\otimes 2}}\right) (U_{-}^\dagger)^{\otimes 2} }} \label{app:tester:emp:var}
\end{align}By introducing the shifted unitary operators $U_{\pm \varepsilon} \equiv \exp(\pm \varepsilon H) \otimes \bbI$ and $V_{\pm \varepsilon} \equiv \bbI \otimes \exp(\pm \varepsilon H)$, expanding this composition yields
\begin{align*}
    \begin{split}
        &\frac{1}{4\varepsilon^2} \left(\EXPD{U_{-} \sim \bbU_{-}}{ \Tr{ \cM^{\otimes 2}(U_{-})^{\otimes 2} U_\varepsilon V_\varepsilon \left(\EXPD{U_{+}\sim \bbU_{+}}{(U_{+})^{\otimes 2}\rho^{\otimes 2}(U_{+}^\dagger)^{\otimes 2}}\right) V_{-\varepsilon} U_{-\varepsilon} (U_{-}^\dagger)^{\otimes 2} }}\right. \\
        &\hspace{1cm}- \EXPD{U_{-} \sim \bbU_{-}}{ \Tr{ \cM^{\otimes 2}(U_{-})^{\otimes 2} U_\varepsilon V_{-\varepsilon} \left(\EXPD{U_{+}\sim \bbU_{+}}{(U_{+})^{\otimes 2}\rho^{\otimes 2}(U_{+}^\dagger)^{\otimes 2}}\right) V_\varepsilon U_{-\varepsilon}(U_{-}^\dagger)^{\otimes 2} }} \\
        &\hspace{1cm}- \EXPD{U_{-} \sim \bbU_{-}}{ \Tr{ \cM^{\otimes 2}(U_{-})^{\otimes 2}U_{-\varepsilon} V_\varepsilon \left(\EXPD{U_{+}\sim \bbU_{+}}{(U_{+})^{\otimes 2}\rho^{\otimes 2}(U_{+}^\dagger)^{\otimes 2}}\right) V_{-\varepsilon} U_\varepsilon(U_{-}^\dagger)^{\otimes 2} }} \\
        &\hspace{1cm}+ \left.\EXPD{U_{-} \sim \bbU_{-}}{ \Tr{ \cM^{\otimes 2}(U_{-})^{\otimes 2}U_{-\varepsilon} V_{-\varepsilon} \left(\EXPD{U_{+}\sim \bbU_{+}}{(U_{+})^{\otimes 2}\rho^{\otimes 2}(U_{+}^\dagger)^{\otimes 2}}\right) V_\varepsilon U_\varepsilon(U_{-}^\dagger)^{\otimes 2} }} \right).
    \end{split}
\end{align*}
Every operator in this expansion acts on one copy at a time: for $s, t\in\{\pm\varepsilon\}$,
\begin{align*}
    (U_{-})^{\otimes 2} U_{s} V_{t} (U_{+})^{\otimes 2} = \left(U_{-}\expm{sH}U_{+}\right)\otimes\left(U_{-}\expm{tH}U_{+}\right).
\end{align*}
Writing
\begin{align}
    C_{U_\pm}(s) \defeq \Tr{\cM\, U_{-}\expm{sH}U_{+}\,\rho\,U_{+}^\dagger\expm{-sH}U_{-}^\dagger}
    \label{app:tester:cost}
\end{align}
for the cost along the differentiated gate, and using $\Tr{X\otimes Y}=\Tr{X}\Tr{Y}$ together with the linearity of the expectation over $U_{+}$, each term of the expansion becomes a product of two single-copy expectation values:
\begin{align}
    \Tr{\cM^{\otimes 2}(U_{-})^{\otimes 2} U_{s} V_{t} (U_{+})^{\otimes 2}\rho^{\otimes 2}(U_{+}^\dagger)^{\otimes 2} V_{t}^\dagger U_{s}^\dagger (U_{-}^\dagger)^{\otimes 2}} = C_{U_\pm}(s)\, C_{U_\pm}(t).
    \label{app:tester:product}
\end{align}
Hence
\begin{align}
    \bbV_{\mathrm{FD}} = \EXPD{U_{\pm}\sim\bbU_{\pm}}{\left(\frac{C_{U_\pm}(\varepsilon)-C_{U_\pm}(-\varepsilon)}{2\varepsilon}\right)^{2}},
    \qquad
    \var = \EXPD{U_{\pm}\sim\bbU_{\pm}}{\big(\partial_s C_{U_\pm}(0)\big)^{2}},
    \label{app:tester:factorized}
\end{align}
where the second identity follows in the same way from $\commutator{H\otimes\bbI}{\commutator{\bbI\otimes H}{X\otimes X}} = \commutator{H}{X}\otimes\commutator{H}{X}$. Both quantities can therefore be sampled with ordinary runs of the $n$-qubit circuit; no operation acting jointly on two copies is required.

\subsection{Algorithm}

Building on the finite-difference approximation derived above, we now formalize a sampling procedure to evaluate the approximate variance $\bbV_{\mathrm{FD}}$. 

\begin{algorithm} [H]
\caption{Sampling-Based Variance Estimator}\label{app:new:alg}

\Input{Input quantum state $\rho$, Measurement $\cM$, $\varepsilon\in (0,1)$, distribution $\bbU_{+}$, distribution $\bbU_{-}$}

Prepare $\rho^{\otimes 2}$.

$b_1, b_2 \sim \Bernoulli{1/2}$.

Sample $U_{+}$ from $\bbU_{+}$ and $U_{-}$ from $\bbU_{-}$.

$\rho_{out} \leftarrow (U_{-})^{\otimes 2} U_{(-1)^{b_2}\cdot\varepsilon}V_{(-1)^{b_1}\cdot\varepsilon}(U_{+})^{\otimes 2} \rho^{\otimes 2} (U_{+}^\dagger)^{\otimes 2} V_{(-1)^{b_1}\cdot\varepsilon}^\dagger U_{(-1)^{b_2}\cdot\varepsilon}^\dagger (U_{-}^\dagger)^{\otimes 2}$.

Measure $\rho_{out}$ with $\cM^{\otimes 2}$ and store the outcome to $O$.

\Return{$\frac{1}{\varepsilon^2}\cdot (-1)^{b_1 + b_2}\cdot O$}

\end{algorithm}

Every operator in the step that defines $\rho_{out}$ acts on one copy at a time, so that $\rho_{out}$ is the product of the two states $U_{-}\expm{sH}U_{+}\,\rho\,U_{+}^\dagger\expm{-sH}U_{-}^\dagger$ with $s=(-1)^{b_2}\varepsilon$ and $s=(-1)^{b_1}\varepsilon$, and the outcome of $\cM^{\otimes 2}$ is distributed as the product of the outcomes of $\cM$ measured on these two states. \Cref{app:new:alg} can therefore be run as two executions of the $n$-qubit circuit whose outcomes are multiplied; written in this way, it is \Cref{main:text:new:alg} of the main text.

The following lemma establishes that this sampling-based algorithm provides an unbiased estimator for the finite-difference expansion $\bbV_{\mathrm{FD}}$.

\begin{lemma}
    \Cref{app:new:alg} provides an unbiased estimate of the approximate variance $\bbV_{\mathrm{FD}}$. \label{app:tester:unbiased}
\end{lemma}

\begin{proof} [Proof of \cref{app:tester:unbiased}]
    Conditioned on the binary variables $b_1, b_2$ and the sampled unitaries $U_{-}, U_{+}$, the expected value of the measurement outcome is
    \begin{align*}
        &\EXP{\frac{1}{\varepsilon^2}\cdot (-1)^{b_1 + b_2}\cdot O\middle| b_1, b_2, U_{-}, U_{+}} \\
        =& \frac{1}{\varepsilon^2}\cdot (-1)^{b_1 + b_2}\cdot\Tr{\cM^{\otimes 2}(U_{-})^{\otimes 2} U_{(-1)^{b_2}\varepsilon}V_{(-1)^{b_1}\varepsilon}(U_{+})^{\otimes 2} \rho^{\otimes 2} (U_{+}^\dagger)^{\otimes 2} V_{(-1)^{b_1}\varepsilon}^\dagger U_{(-1)^{b_2}\varepsilon}^\dagger (U_{-}^\dagger)^{\otimes 2}}
    \end{align*}Taking the expectation over the unitary distributions $\bbU_{-}$ and $\bbU_{+}$ yields
    \begin{align*}
        \frac{(-1)^{b_1 + b_2}}{\varepsilon^2}\cdot\Tr{\EXPD{U_{-}\sim \bbU_{-}}{(U_{-}^\dagger)^{\otimes 2} \cM^{\otimes 2}(U_{-})^{\otimes 2}} U_{(-1)^{b_2}\varepsilon}V_{(-1)^{b_1}\varepsilon}\EXPD{U_{+}\sim \bbU_{+}}{(U_{+})^{\otimes 2} \rho^{\otimes 2} (U_{+}^\dagger)^{\otimes 2}} V_{(-1)^{b_1}\varepsilon}^\dagger U_{(-1)^{b_2}\varepsilon}^\dagger }
    \end{align*}
    Finally, taking the total expectation over the independent coin flips $b_1, b_2 \sim \Bernoulli{1/2}$, and applying the linearity of expectation alongside the cyclic property of the trace, the terms expand to
    \begin{align*}
        &\frac{1}{4\varepsilon^2}\left(\EXPD{U_{-} \sim \bbU_{-}}{ \Tr{ \cM^{\otimes 2}(U_{-})^{\otimes 2} U_\varepsilon V_\varepsilon \left(\EXPD{U_{+}\sim \bbU_{+}}{(U_{+})^{\otimes 2}\rho^{\otimes 2}(U_{+}^\dagger)^{\otimes 2}}\right) V_{-\varepsilon} U_{-\varepsilon} (U_{-}^\dagger)^{\otimes 2} }}\right. \\
        &\hspace{1cm}- \EXPD{U_{-} \sim \bbU_{-}}{ \Tr{ \cM^{\otimes 2}(U_{-})^{\otimes 2} U_\varepsilon V_{-\varepsilon} \left(\EXPD{U_{+}\sim \bbU_{+}}{(U_{+})^{\otimes 2}\rho^{\otimes 2}(U_{+}^\dagger)^{\otimes 2}}\right) V_\varepsilon U_{-\varepsilon}(U_{-}^\dagger)^{\otimes 2} }} \\
        &\hspace{1cm}- \EXPD{U_{-} \sim \bbU_{-}}{ \Tr{ \cM^{\otimes 2}(U_{-})^{\otimes 2}U_{-\varepsilon} V_\varepsilon \left(\EXPD{U_{+}\sim \bbU_{+}}{(U_{+})^{\otimes 2}\rho^{\otimes 2}(U_{+}^\dagger)^{\otimes 2}}\right) V_{-\varepsilon} U_\varepsilon(U_{-}^\dagger)^{\otimes 2} }} \\
        &\hspace{1cm}+ \left.\EXPD{U_{-} \sim \bbU_{-}}{ \Tr{ \cM^{\otimes 2}(U_{-})^{\otimes 2}U_{-\varepsilon} V_{-\varepsilon} \left(\EXPD{U_{+}\sim \bbU_{+}}{(U_{+})^{\otimes 2}\rho^{\otimes 2}(U_{+}^\dagger)^{\otimes 2}}\right) V_\varepsilon U_\varepsilon(U_{-}^\dagger)^{\otimes 2} }} \right) 
    \end{align*}
    This matches the definition of $\bbV_{\mathrm{FD}}$, confirming the estimator is unbiased.
\end{proof}

\subsection{Analysis}
In this section, we analyze the error rate and sample complexity of \Cref{app:new:alg}. We begin by bounding the finite-difference approximation error.

\begin{lemma}
    For any anti-Hermitian $H$, Hermitian operator $A$, and $\varepsilon\in (0,1)$, the finite-difference approximation error is bounded by
    \begin{align*}
        \onesnorm{f\circ g(A) - \commutator{H\otimes \bbI}{\commutator{\bbI \otimes H}{A}}} \leq \frac{16}{3} \varepsilon^2 (\infsnorm{H})^4 \onesnorm{A} + \cO(\varepsilon^4)
    \end{align*} \label{app:tester:lemma:1}
\end{lemma}

\begin{proof} [Proof of \cref{app:tester:lemma:1}]
    For simplicity, we write $\ad_{H_1}(A) \defeq [H \otimes \bbI, A]$ and $\ad_{H_2}(A) = [\bbI \otimes H, A]$. Then, using the identity $e^{X} A e^{-X} = e^{\ad_X}(A)$ and the Taylor expansions of $e^{\ad_X}$ and $\sinh{\ad_X}$, we can rewrite the functions $f$ and $g$ following 
    \begin{align*}
        f(\varepsilon; A, H) = \frac{e^{\varepsilon \ad_{H_1}}(A) - e^{-\varepsilon \ad_{H_1}}(A)}{2\varepsilon} = \frac{\sinh(\varepsilon \ad_{H_1})}{\varepsilon}(A)    
    \end{align*}
    Similarly, 
    \begin{align*}
        g(\varepsilon; A, H) = \frac{\sinh(\varepsilon \ad_{H_2})}{\varepsilon}(A)
    \end{align*}
    
    The difference between the finite-difference composition and the true double commutator is
    \begin{align*}
        f \circ g(A) - \ad_{H_1} \circ \ad_{H_2}(A) = \left( \frac{\sinh(\varepsilon \ad_{H_1})\circ \sinh(\varepsilon \ad_{H_2})}{\varepsilon^2} - \ad_{H_1} \circ \ad_{H_2} \right)(A)    
    \end{align*}
    Expanding to fourth order in $\varepsilon$, we get
    \begin{align*}
        &\frac{1}{\varepsilon^2}\left(\varepsilon^2(\ad_{H_1} \circ \ad_{H_2}) + \frac{\varepsilon^4}{6} \left( \ad_{H_1} \circ \ad_{H_2}^{\circ 3} + \ad_{H_1}^{\circ 3} \circ \ad_{H_2} + \cO(\varepsilon^6) \right)\right) - \ad_{H_1} \circ \ad_{H_2} \\
        =&\frac{\varepsilon^2}{6} \left( \ad_{H_1}^{\circ 3} \circ \ad_{H_2} + \ad_{H_1} \circ \ad_{H_2}^{\circ 3} \right)(A) + \cO(\varepsilon^4)
    \end{align*}To bound the superoperators acting on $A$ in the trace norm, we use Hölder's inequality for Schatten norms (see \cref{fact:holder}), which states that for any operators $X$ and $Y$:
    \begin{align*}
        \onesnorm{XY} \leq \infsnorm{X}\onesnorm{Y}, \quad \onesnorm{YX} \leq \infsnorm{X}\onesnorm{Y}
    \end{align*}
    Applying this to a commutator $[X, A]$ for any operator $X$ and $A$, we get
    \begin{align*}
        \onesnorm{[X, A]} = \onesnorm{ XA - AX } \leq \onesnorm{ XA } + \onesnorm{ AX } \leq 2 \infsnorm{ X } \onesnorm{ A }
    \end{align*}
    Using the property above, the induced norms of adjoint operator $\ad_{H_1}$ and $\ad_{H_2}$ mapping from the trace-class space to itself are bounded by the operator norm of the Hamiltonian:
    \begin{align*}
        \onesnorm{ [H \otimes \bbI, A] }, \onesnorm{ [\bbI \otimes H, A] } \leq 2 \infsnorm{ H } \infsnorm{ \bbI } \onesnorm{ A } = 2 \infsnorm{ H } \onesnorm{ A }
    \end{align*}and 
    \begin{align*}
         \onesnorm{\ad_{H_1}\circ \ad_{H_2}(A)} \leq 4 (\infsnorm{ H })^2 \onesnorm{ A }
    \end{align*}
    Combining all of the above together with properties of norms, we have 
    \begin{align*}
        \onesnorm{f \circ g(A) - \ad_{H_1} \circ \ad_{H_2}(A)} \leq& \onesnorm{\frac{\varepsilon^2}{6} \left( \ad_{H_1}^{\circ 3} \circ \ad_{H_2} + \ad_{H_1} \circ \ad_{H_2}^{\circ 3} \right)(A) + \cO(\varepsilon^4)} \\
        \leq& \frac{\varepsilon^2}{6} \Big( (2\infsnorm{H})^3 (2\infsnorm{H}) + (2\infsnorm{H}) (2\infsnorm{H})^3 \Big) \onesnorm{A} \\
        \leq& \frac{16}{3} \varepsilon^2 (\infsnorm{H})^4 \onesnorm{A} + \cO(\varepsilon^4)
    \end{align*} 
\end{proof}

Next, we determine the number of experimental runs necessary to achieve an additive error $\teps$ with high probability. We define the empirical estimator
\begin{align}
    \bbO = \frac{1}{\cN}\sum_{j=1}^{\cN} \bbO^{(j)}
\end{align}where $\bbO^{(j)}$ denotes the outcome of the $j$-th run of \Cref{app:new:alg} and $\cN$ is the number of runs.

\begin{lemma}
    Let $\varepsilon\in (0,1)$ be a constant. For any desired precision $\teps > 0$ and failure probability $\delta \in (0, 1)$, if the number of samples satisfies
    \begin{align*}
        \cN \geq \frac{2 (\infsnorm{\cM})^4}{\teps^2 \varepsilon^4} \ln\left(\frac{2}{\delta}\right)
    \end{align*}
    then the empirical estimator $\bbO$ approximates the approximate variance $\bbV_{\mathrm{FD}}$ up to an additive error of $\teps$ with probability at least $1 - \delta$.  \label{app:tester:lemma:2}
\end{lemma}

\begin{proof} [Proof of \cref{app:tester:lemma:2}]
    Since $O$ in \Cref{app:new:alg} is an outcome of measuring $\cM^{\otimes 2}$,
    \begin{align*}
        \abs{\frac{1}{\varepsilon^2}\cdot (-1)^{b_1 + b_2}\cdot O} = \frac{1}{\varepsilon^2}\abs{O} \leq \frac{1}{\varepsilon^2}(\infsnorm{\cM})^2,
    \end{align*}so each output lies in an interval of length $2(\infsnorm{\cM})^2/\varepsilon^2$. Applying Hoeffding's inequality to the empirical mean of $\cN$ independent bounded variables, we have
    \begin{align*}
        \Prob{\abs{\bbO - \bbV_{\mathrm{FD}}}> \teps} \leq  2 \expm{-\frac{2\cN^2\teps^2}{\frac{4\cN}{\varepsilon^4}(\infsnorm{\cM})^4}} = 2 \expm{-\frac{\cN\teps^2\varepsilon^4}{2(\infsnorm{\cM})^4}}
    \end{align*}Solving $2 \expm{-\frac{\cN\teps^2\varepsilon^4}{2(\infsnorm{\cM})^4}} \leq \delta$ for $\cN$, we have
    \begin{align*}
        \cN \geq \frac{2 (\infsnorm{\cM})^4}{\teps^2 \varepsilon^4} \ln\left(\frac{2}{\delta}\right)
    \end{align*}
\end{proof}

Finally, synthesizing \cref{app:tester:lemma:1,app:tester:lemma:2} alongside Hölder's inequality (\cref{fact:holder}), we apply the triangle inequality to bound the total absolute error between the empirical estimator $\bbO$ and the true variance $\var$. With probability at least $1-\delta$, the total error is bounded by:
\begin{align*}
    &\abs{\bbO - \var} \\
    =&\abs{\bbO - \bbV_{\mathrm{FD}} + \bbV_{\mathrm{FD}} - \var}\\
    \leq& \abs{\bbO - \bbV_{\mathrm{FD}}} + \abs{\bbV_{\mathrm{FD}} - \var} \\
    \leq& \teps + \infsnorm{\left(\EXPD{U_{-}\sim \bbU_{-}}{(U_{-})^{\otimes 2}\cM^{\otimes 2}(U_{-}^\dagger)^{\otimes 2}}\right)}\cdot \onesnorm{(f\circ g - \ad_{H\otimes \bbI}\circ \ad_{\bbI \otimes H})\left(\EXPD{U_{+}\sim \bbU_{+}}{(U_{+})^{\otimes 2}\rho^{\otimes 2}(U_{+}^\dagger)^{\otimes 2}}\right)} \\
    \leq& \teps + \frac{16}{3} \varepsilon^2 (\infsnorm{H})^4 \infsnorm{\cM^{\otimes 2}} \onesnorm{\rho^{\otimes 2}} + \cO(\varepsilon^4)
\end{align*}

\begin{proposition}[Deciding trainability; restatement of \cref{main:text:prop:tolerance}]
    Let $\tau>0$ and $\delta\in(0,1)$, and let $\bbO$ be the mean of $\cN$ independent outputs of \Cref{app:new:alg}, where either (i) the exact form is used and $\cN\geq 8\gamma^4(\infsnorm{\cM})^4\ln(2/\delta)/\tau^2$, or (ii) the finite-difference form is used with $\varepsilon$ small enough that $\abs{\bbV_{\mathrm{FD}}-\var}\leq\tau/4$, and $\cN$ satisfies \cref{app:tester:lemma:2} with $\teps=\tau/4$. Then $\abs{\bbO-\var}\leq\tau/2$ with probability at least $1-\delta$. In particular, the rule ``declare the ansatz trainable at tolerance $\tau$ if $\bbO>\tau$'' is correct with probability at least $1-\delta$ whenever $\var>3\tau/2$ or $\var<\tau/2$. \label{app:tester:prop}
\end{proposition}

\begin{proof}[Proof of \cref{app:tester:prop}]
    In case (i), \cref{app:tester:lemma:exact} with $\teps=\tau/2$ gives $\abs{\bbO-\var}\leq\tau/2$ with probability at least $1-\delta$. In case (ii), \cref{app:tester:lemma:2} with $\teps=\tau/4$ gives $\abs{\bbO-\bbV_{\mathrm{FD}}}\leq\tau/4$ with probability at least $1-\delta$, and $\abs{\bbV_{\mathrm{FD}}-\var}\leq\tau/4$ by the choice of $\varepsilon$, so again $\abs{\bbO-\var}\leq\tau/2$. On this event, $\var>3\tau/2$ implies $\bbO\geq\var-\tau/2>\tau$, and $\var<\tau/2$ implies $\bbO\leq\var+\tau/2<\tau$.

    For the cost of case (ii), the total-error bound above (with $\infsnorm{\cM^{\otimes 2}}=(\infsnorm{\cM})^2$ and $\onesnorm{\rho^{\otimes 2}}=1$) shows that the bias is at most $\tau/4$ once $\varepsilon^2\leq 3\tau/(64(\infsnorm{H})^4(\infsnorm{\cM})^2)$ up to higher-order terms, so \cref{app:tester:lemma:2} requires $\cN=\cO\big((\infsnorm{H})^8(\infsnorm{\cM})^8\ln(2/\delta)/\tau^4\big)$ runs. In case (i), $\cN=\cO\big(\gamma^4(\infsnorm{\cM})^4\ln(2/\delta)/\tau^2\big)$. Both are polynomial in $n$ whenever $\tau=1/\poly{n}$ and $\gamma$, $\infsnorm{H}$ and $\infsnorm{\cM}$ are at most polynomial in $n$.
\end{proof}

\subsection{Discussion}
\subsubsection{Time Complexity}
The computational overhead required to execute a single run of \Cref{app:new:alg} is remarkably efficient. Specifically, the time complexity per run scales strictly as $\cO(L \cdot k)$, where $L$ denotes the total number of parameterized layers within the quantum circuit, and $k$ represents the number of basis generators defining the underlying Lie algebra $\Lie{G}$.

The total number of runs is set by the tolerance. By \cref{app:tester:prop}, deciding whether the variance exceeds $\tau$ takes $\cO(\tau^{-2}\log\delta^{-1})$ runs with an exact shift rule and $\cO(\tau^{-4}\log\delta^{-1})$ with finite differences, which is polynomial whenever $\tau = 1/\poly{n}$. Only resolving an exponentially small variance to relative precision requires exponentially many runs. That cost is a property of the quantity being estimated and is shared by every sampling method, and it is not needed to decide trainability at a polynomial tolerance.

\subsubsection{Non-zero expected gradient}

Notably, the algorithm's validity does not depend on whether the unitary ensembles $U_\pm$ converge to a 2-design with respect to $\Pi(G)$. The methodology generalizes naturally to any parameterized quantum circuit of the form $\prod_{l=1}^L \expm{i\theta_l H_l}$, where $H_l$ are Hermitian generators. It can be deployed provided that the first moment of the partial derivative with respect to $\theta_l$ vanishes. This holds at any depth when $\expm{2\pi iH_l}$ is a multiple of the identity (\cref{lemma:expectation}). For other generators, this condition cannot be universally guaranteed, but regimes exhibiting a non-zero expected gradient possess sufficient global features in their optimization landscape to drive parameter updates. Under such conditions, standard first-order gradient estimation techniques (such as the parameter-shift rule) are entirely adequate to guide the optimization trajectory, rendering the computationally demanding task of bounding higher-order moments unnecessary.

Therefore, for such generators, decisively identifying whether the expected gradient vanishes is a critical prerequisite. To determine if the expectation of $\partial_{\theta_l}C$ is exactly zero, we first observe that it can be expressed as:
\begin{align*}
    \EXP{\partial_{\theta_l}C} = \EXPD{U_-}{\Tr{\cM U_- \commutator{H_{\theta_l}}{\EXPD{U_+}{U_+ \rho U_+^\dagger}}U_-^\dagger}}.
\end{align*}
We can utilize the same finite-difference principle to approximate this value. Defining the adjoint action as $\ad_{A}(\cdot) \defeq \expm{A}(\cdot) \expm{A}^\dagger$, we approximate the commutator via the linear map $\widetilde{f}$:
\begin{align*}
    [H_{\theta_l},\cdot] \approx \widetilde{f}(\cdot) \defeq \frac{\ad_{\varepsilon_1 H_l}(\cdot) - \ad_{-\varepsilon_1 H_l}(\cdot)}{2\varepsilon_1}.
\end{align*}
Substituting this approximation back into the trace yields a computationally accessible form for the expected gradient for a sufficiently small $\varepsilon_1$:
\begin{align*}
    \EXP{\partial_{\theta_l}C} \approx \EXPD{U_-}{\Tr{\cM U_- \widetilde{f}\left(\EXPD{U_+}{U_+ \rho U_+^\dagger}\right)U_-^\dagger}}.
\end{align*}
Following arguments analogous to those presented above, the estimation error is bounded by 
\begin{align*}
    \teps_1 + \frac{4\varepsilon_1^2}{3} \infsnorm{\cM}(\infsnorm{H_l})^3 \onesnorm{\rho}
\end{align*}with probability at least $1-\delta_1$. Achieving this bound requires a sample complexity of $\cO\left(\frac{(\infsnorm{\cM})^4}{\teps_1^2 \varepsilon_1^4} \ln\left(\frac{2}{\delta_1}\right)\right)$, where $\varepsilon_1, \teps_1, \delta_1 \in (0,1)$.

\subsubsection{Comparison with the deterministic cross-product estimator}
The approximate variance in \cref{app:tester:emp:var} can also be estimated with a deterministic cross-product estimator. By \cref{app:tester:factorized},
\begin{align*}
    \bbV_{\mathrm{FD}} = \bbE_{U_\pm} \left[ \left( \frac{\Delta C}{2\varepsilon} \right)^2 \right],
\end{align*}where $\Delta C = \mathrm{tr}\left(\cM U_-(e^{\varepsilon H}\trho e^{-\varepsilon H} - e^{-\varepsilon H}\trho e^{\varepsilon H})U_-^\dagger\right),$ and $\trho = U_+\rho U_+^\dagger$. Estimating the square of this expected value without bias requires two independent estimates. For a single sample step $j$, one draws $U_+$ and $U_-$ from $\bbU_+$ and $\bbU_-$ and forms
\begin{align}
    \hat{X}_1^{(j)} = \frac{O_{+\varepsilon, 1}^{(j)} - O_{-\varepsilon, 1}^{(j)}}{2\varepsilon},\quad \hat{X}_2^{(j)} = \frac{O_{+\varepsilon, 2}^{(j)} - O_{-\varepsilon, 2}^{(j)}}{2\varepsilon}
\end{align}from four single-shot executions. By independence, $\bbE_{\mathrm{shots}}[\hat{X}_1 \hat{X}_2] = (\frac{\Delta C}{2\varepsilon})^2$, so $Z^{(j)} = \hat{X}_1^{(j)} \hat{X}_2^{(j)}$ is unbiased for $\bbV_{\mathrm{FD}}$. Run as two executions of the $n$-qubit circuit, \Cref{app:new:alg} is the randomized version of this estimator: each factor is replaced by a single shot at a randomly signed shift, multiplied by this sign and divided by $\varepsilon$, which has the same conditional mean $\Delta C/(2\varepsilon)$.

Both estimators are bounded by $(\infsnorm{\cM})^2/\varepsilon^2$, so Hoeffding's inequality requires the same number of samples, $\cN \geq 2(\infsnorm{\cM})^4\ln(2/\delta)/(\teps^2\varepsilon^4)$, for an additive precision $\teps$ with failure probability at most $\delta$. Since a sample of $Z$ costs four executions and a run of \Cref{app:new:alg} costs two, the randomized version needs half as many circuit executions for the same worst-case guarantee.

The variances differ in the other direction. For an observable with eigenvalues $\pm 1$, every output of \Cref{app:new:alg} equals $\pm 1/\varepsilon^2$, so its second moment is $1/\varepsilon^4$, whereas $\bbE[Z^2 \mid U_\pm] = \big((1-m_{+}m_{-})/(2\varepsilon^2)\big)^2$ with $m_{\pm} = C_{U_\pm}(\pm\varepsilon)$. Since $m_{+}m_{-}\geq -(m_{+}-m_{-})^2/4$ and $m_{+}-m_{-}=\cO(\varepsilon)$, the latter is at most $(1+\cO(\varepsilon^2))/(4\varepsilon^4)$. For a fixed number of circuit executions, the deterministic estimator therefore has a variance smaller by at least a factor of about $2$ for small $\varepsilon$; in the setting of \cref{fig:placeholder} at $n=4$, the factor is about $2.5$. When the number of runs is fixed in advance from the Hoeffding bound, \Cref{app:new:alg} is the cheaper choice; when sampling stops on the observed spread, the deterministic estimator is. Both are products of two independent bounded factors, so neither has heavy tails, and \cref{app:tester:prop} applies to both.

\subsubsection{Exact Variance Estimation via the Parameter-Shift Rule}
\label{app:tester:exact}
The finite-difference approach naturally introduces an analytical bias $\cO(\varepsilon^2)$ and a corresponding $1/\varepsilon^4$ blow-up in sample complexity. However, if the spectrum of the generator $H$ is known and has exactly two eigenvalues, an exact parameter-shift rule removes this bias completely, turning the finite-difference approximation into an analytically exact algorithm. Let the generator of the unitary operation be $H = -i \frac{P}{2}$, where $P$ is an involutory operator (e.g., a Pauli string) such that $P^2 = \bbI$. Using Euler's formula for involutions ($e^{i \theta P} = \cos(\theta)\bbI + i \sin(\theta)P$), we apply a shift $s = \pi/2$:
\begin{align}
    \expm{s H} &= \expm{-i \frac{\pi}{4} P} = \frac{1}{\sqrt{2}} (\bbI - i P), \\ \expm{-s H} &= \expm{i \frac{\pi}{4} P} = \frac{1}{\sqrt{2}} (\bbI + i P).
\end{align}Substituting these into the shifted operator difference yields:
\begin{align}
    \expm{s H} A \expm{-s H} &= \frac{1}{2}(\bbI - iP)A(\bbI + iP) = \frac{1}{2} \left( A - i\commutator{P}{A} + PAP \right), \\ \expm{-s H} A \expm{s H} &= \frac{1}{2}(\bbI + iP)A(\bbI - iP) = \frac{1}{2} \left( A + i\commutator{P}{A} + PAP \right).
\end{align}Taking the difference between these two shifted operators directly recovers the exact commutator:
\begin{align}
    \expm{s H} A \expm{-s H} - \expm{-s H} A \expm{s H} &= -i\commutator{P}{A} = 2 \commutator{-i \frac{P}{2}}{A} = 2\commutator{H}{A}.
\end{align}In the two-copy notation of the derivation, we define the parameter-shift superoperators $f_{\text{PSR}}$ and $g_{\text{PSR}}$ acting on the tensor space. By replacing the infinitesimal step $\varepsilon$ with $s = \pi/2$ and adjusting the normalizer to $\frac{1}{2}$, we obtain:
\begin{align}
    f_{\text{PSR}}(A) &\defeq \frac{1}{2} \left( (\expm{sH}\otimes \bbI) A (\expm{-sH}\otimes \bbI) - (\expm{-sH}\otimes \bbI) A (\expm{sH}\otimes \bbI) \right), \\
    g_{\text{PSR}}(A) &\defeq \frac{1}{2} \left( (\bbI\otimes \expm{sH}) A (\bbI\otimes \expm{-sH}) - (\bbI\otimes \expm{-sH}) A (\bbI\otimes \expm{sH}) \right).
\end{align}Because this shift is analytically exact, the composition of these superoperators identically equals the exact nested commutator:
\begin{align}
    f_{\text{PSR}} \circ g_{\text{PSR}}(A) \equiv \commutator{H\otimes \bbI}{\commutator{\bbI \otimes H}{A}}.
\end{align}Consequently, the bias vanishes. Running \Cref{app:new:alg} with the shifts $\pm\pi/2$ in place of $\pm\varepsilon$ and the prefactor $1$ in place of $1/\varepsilon^2$ outputs $(-1)^{b_1+b_2}O$, an unbiased estimate of $\var$ itself, bounded by $(\infsnorm{\cM})^2$; this is the exact form with $\gamma=1$, and the $1/\varepsilon^4$ factor disappears from the sample complexity.

\paragraph{General generators.} The same holds for any generator with a known spectrum~\cite{wierichs2022general}. If $\Omega_1,\dots,\Omega_R$ are the distinct positive differences between eigenvalues of $iH$, then $s\mapsto C_{U_\pm}(s)$ in \cref{app:tester:cost} is a trigonometric polynomial with these frequencies, and there are $2R$ shifts $x_k$ and coefficients $c_k$, depending only on the frequencies, with $\partial_s C_{U_\pm}(0) = \sum_k c_k C_{U_\pm}(x_k)$ for all $U_{\pm}$. \Cref{app:tester:lemma:exact} then applies with $\gamma=\sum_k\abs{c_k}$. For equidistant frequencies $\Omega_l = l$, $l=1,\dots,R$, one choice is $x_k = (2k-1)\pi/(2R)$ and $c_k = (-1)^{k-1}/\big(4R\sin^2(x_k/2)\big)$ for $k=1,\dots,2R$~\cite{wierichs2022general}. 

\begin{lemma}
    The exact form of \Cref{app:new:alg} is an unbiased estimator of $\var$ with outputs bounded by $\gamma^2(\infsnorm{\cM})^2$. Consequently, for any $\teps>0$ and $\delta\in(0,1)$, if $\cN \geq 2\gamma^4(\infsnorm{\cM})^4\ln(2/\delta)/\teps^2$, then $\abs{\bbO-\var}\leq\teps$ with probability at least $1-\delta$. \label{app:tester:lemma:exact}
\end{lemma}

\begin{proof}[Proof of \cref{app:tester:lemma:exact}]
    Conditioned on $U_{\pm}$ and on $k_1,k_2$, the expected outcome is 
    \begin{align*}
        \EXP{O\middle| k_1,k_2,U_{-},U_{+}}=C_{U_\pm}(x_{k_2})\,C_{U_\pm}(x_{k_1})
    \end{align*}by \cref{app:tester:product}, so that the conditional mean of the output is
    \begin{align*}
        \sum_{k_1,k_2}\frac{\abs{c_{k_1}}\abs{c_{k_2}}}{\gamma^2}\,\gamma^2\,\mathrm{sgn}(c_{k_1})\,\mathrm{sgn}(c_{k_2})\,C_{U_\pm}(x_{k_1})\,C_{U_\pm}(x_{k_2}) = \Big(\sum_k c_k\,C_{U_\pm}(x_k)\Big)^2 = \big(\partial_s C_{U_\pm}(0)\big)^2,
    \end{align*}and its expectation over $U_{\pm}$ is $\var$ by \cref{app:tester:factorized}. The bound follows from $\abs{O}\leq(\infsnorm{\cM})^2$. Each output then lies in an interval of length $2\gamma^2(\infsnorm{\cM})^2$, and Hoeffding's inequality gives $\Prob{\abs{\bbO-\var}>\teps}\leq 2\expm{-\cN\teps^2/(2\gamma^4(\infsnorm{\cM})^4)}\leq\delta$ for the stated $\cN$.
\end{proof}
For the gates of \cref{fig:placeholder}, $H = -\tfrac{i}{2}(X_1X_2+Y_1Y_2)$ has spectrum $\{0,\pm i\}$, so $R=2$, and the rule
\begin{align*}
    \partial_s C_{U_\pm}(0) = \frac{C_{U_\pm}(\tfrac{\pi}{4}) - C_{U_\pm}(-\tfrac{\pi}{4})}{8\sin^2\frac{\pi}{8}} - \frac{C_{U_\pm}(\tfrac{3\pi}{4}) - C_{U_\pm}(-\tfrac{3\pi}{4})}{8\sin^2\frac{3\pi}{8}}
\end{align*}
is exact, with $\gamma = 2$. Every output of the exact form is then $\pm 4$ for $\cM = X_1X_2$, and \cref{app:tester:lemma:exact} with $\teps = 0.999/n^2$ and $\delta = 10^{-3}$ gives the run counts quoted in \cref{fig:placeholder}.

\section{Extension to Generating Sets of the Lie Algebra}
\label{sec:generating_set}

In this section, we consider the case where we only have access to a generating set of the Lie algebra $\fg$ associated with a compact and connected Lie group $G$. Previously, we assumed access to a basis (spanning set) of $\fg$. Here, we relax this: we assume there exists a set $\cG = \{x_1, \dots, x_k\} \subseteq \fg$ such that every $x \in \fg$ can be expressed as a linear combination of these elements and their nested commutators, i.e., for all $x \in \fg$, $x = \sum_{j_1=1}^k c_{j_1} x_{j_1} + \sum_{j_1,j_2} c_{j_1j_2} [x_{j_1}, x_{j_2}] + \sum_{j_1,j_2,j_3} c_{j_1j_2j_3} [x_{j_1}, [x_{j_2}, x_{j_3}]] + \dots$, where coefficients are from the underlying fields. Since $\fg$ is finite-dimensional, the summation above can be taken to be finite. Given a unitary representation $(\cH, \Pi)$ of $G$, the ansatz is defined by 
\begin{align}
    \absU  = \prod_{l=1}^L \prod_{q = 1}^{|\cG|} \Pi\!\left(\exp(\theta_{l,q} x_q)\!\right). \label{app:generating:ansetz}
\end{align}

Next, we evaluate the expressibility of the aforementioned ansatz. Following the arguments presented in \cref{sec:reduction}, it is straightforward to verify that $\Pi(G)$ is a Lie group associated with the Lie algebra $\pi_*(\mathfrak{g})$, which is generated by the set $\{\pi_*(x_1), \dots, \pi_*(x_k)\}$. Consequently, the ansatz can be represented in the product form:
\begin{align}
    \prod_{l=1}^L \prod_{q = 1}^{|\cG|} \expm{\theta_{l,q} \pi_*(x_q)}.
\end{align}

Adopting the methodology from \cref{sec:express}, we first establish that the $t$-th fold commutant of the set defined by $\prod_{q = 1}^{|\cG|} \expm{\theta_{l,q} \pi_*(x_q)}$ is identical to the $t$-th fold commutant of $\Pi(G)$.

\begin{lemma}
    Let $\compactG$ be a compact and connected matrix Lie group associated with Lie algebra $\tfg$ \textbf{generated by} $\tx_1, \cdots, \tx_\cN$. Define 
    \begin{align}
        S = \CurB{\prod_{j=1}^{\cN}\expm{\theta_j \tx_j}: \theta_j \in [-\pi,\pi]}. 
    \end{align}Then, 
    \begin{align}
        \Comm{t, S} = \Comm{t, \compactG}
    \end{align}for all $t\in \bbN_{\geq 1}$ \label{app:generating:lemma:1}
\end{lemma}

\begin{proof} [Proof of \cref{app:generating:lemma:1}]
    Since $S\subseteq \compactG$, it follows immediately that $\iComm{t, \compactG}\subseteq \iComm{t, S}$. Again, it suffices to show $\iComm{t, S} \subseteq \iComm{t, \compactG}$. To show the reverse inclusion, let $A\in \iComm{t, S}$ be arbitrary. By definition, $A$ commutes with $\big(\prod_{j=1}^\cN \exp(\theta_j \tx_j)\big)^{\otimes t}$ for all values of the parameters. By setting all $\theta_l = 0$ except for $\theta_j$, we find $[\exp(\theta_j \tx_j)^{\otimes t}, A] = 0$ for each $j=1,\cdots,\cN$. Since $\exp(\theta_j\tx_j)^{\otimes t}=\exp\big(\theta_j\sum_{i=1}^{t}\bbI^{\otimes(i-1)}\otimes\tx_j\otimes\bbI^{\otimes(t-i)}\big)$, differentiating with respect to $\theta_j$ at $\theta_j = 0$ yields:
    \begin{align}
        \frac{d}{d\theta_j} \expm{\theta_j \tx_j}^{\otimes t} A \expm{-\theta_j \tx_j}^{\otimes t} \Big|_{\theta_j=0} = \commutator{\textstyle\sum_{i=1}^{t}\bbI^{\otimes(i-1)}\otimes\tx_j\otimes\bbI^{\otimes(t-i)}}{A} = 0.
    \end{align}Thus, $A$ commutes with the $t$-fold actions $\sum_{i=1}^{t}\bbI^{\otimes(i-1)}\otimes\tx_j\otimes\bbI^{\otimes(t-i)}$ of all generators $\{\tx_j\}_{j=1}^{\cN}$. Since these generate $\tfg$, the map that sends $B$ to its $t$-fold action preserves commutators, and the operators that commute with $A$ are closed under commutators, $A$ commutes with the $t$-fold action of every $B \in \tfg$. Using the Baker–Campbell–Hausdorff formula \cite{hall2013lie}:
    \begin{align}
        \exp(B)^{\otimes t} A \exp(-B)^{\otimes t} = A + \commutator{\textstyle\sum_{i}\bbI^{\otimes(i-1)}\otimes B\otimes\bbI^{\otimes(t-i)}}{A} + \dots = A.
    \end{align}Because $\compactG$ is connected (\cref{def:connected:lie:subgp}), any element $U \in \compactG$ can be written as a finite product of exponentials $U = \prod_l \exp(B_l)$ for $B_l \in \tfg$. Consequently, $[U^{\otimes t}, A] = 0$ for all $U \in \compactG$, implying $A \in \iComm{t, \compactG}$.
\end{proof}

The \cref{app:generating:lemma:1} ensures that if an operator commutes with the set of generators, it commutes with the entire group. With the commutant equivalence established, we bound the spectral radius of the difference between the layer integral and the Haar integral.

\begin{lemma}
    Let $\compactG$ be the compact, connected Lie group associated with the Lie algebra $\tfg$ \textbf{generated} by anti-Hermitian matrices $H_1, \cdots, H_\cN$. For $t\in\mathbb N_{\geq 1}$, define
    \begin{align}
        \cV(\theta_1,\ldots,\theta_\cN) := \prod_{j=1}^\cN \exp\!\bigl(\theta_j H_j\bigr),\qquad(\theta_1,\ldots,\theta_\cN)\in[-\pi,\pi]^\cN .
    \end{align}Then, the operator $\EXPD{\theta_1, \cdots, \theta_\cN\sim [-\pi, \pi]}{(\cV)^{\otimes t}\otimes (\cV^*)^{\otimes t}} - \EXPD{U \sim \mu_{\compactG}}{(U)^{\otimes t}\otimes (U^*)^{\otimes t}}$ has spectral radius strictly smaller than 1, i.e., 
         \begin{align}
         \rho\left(\EXPD{\theta_1, \cdots, \theta_\cN\sim [-\pi, \pi]}{(\cV)^{\otimes t}\otimes (\cV^*)^{\otimes t}} - \EXPD{U \sim \mu_{\compactG}}{(U)^{\otimes t}\otimes (U^*)^{\otimes t}}\right) < 1
     \end{align}Here $\mu_{\compactG}$ denotes the Haar probability measure on $\compactG$, and $(\cdot)^*$ denotes complex conjugation.\label{generating:small:spectral:radius}
\end{lemma}

\begin{proof} [Proof of \cref{generating:small:spectral:radius}]
    Invoking \cref{app:generating:lemma:1} and following the same arguments presented in proof of \cref{express:small:spectral:radius}, one can show this.
\end{proof}

Finally, we state the expressibility of the multi-layer ansatz given in \cref{app:generating:ansetz} with the following theorem.

\begin{theorem}
    Let $\compactG$ be the compact, connected matrix Lie group associated with the Lie algebra $\fg$ \textbf{generated} by anti-Hermitian matrices $\cH_1, \cdots, \cH_\cN$. Let $t\in\mathbb N_{\geq 1}$. Define
    \begin{align}
        \cV_l(\vtheta_l) := \prod_{j=1}^\cN \exp\!\bigl((\vtheta_l)_j H_j\bigr),\qquad\vtheta_l\in[-\pi,\pi]^\cN.
    \end{align}Then, 
    \begin{align}
        \lim_{L \to \infty} \norm{\EXPD{\vtheta_1,\cdots,\vtheta_L \sim [-\pi, \pi]^\cN}{\prod_{l=1}^L\left((\cV_l)^{\otimes t}\otimes (\cV_l^*)^{\otimes t}\right)} - \EXPD{U \sim \mu_{\compactG}}{(U)^{\otimes t}\otimes (U^*)^{\otimes t}}} = 0
    \end{align}for any sub-multiplicative matrix norm.
    \label{generating:express:thm}
\end{theorem}

Following the same arguments used in the proof of \cref{express:thm}, we can show \cref{generating:express:thm}. This theorem implies that for any positive integer $t$ and any precision $\varepsilon \in (0,1)$, there exists a depth $L^*$ such that the ansatz defined in \cref{app:generating:ansetz} forms an $\varepsilon$-approximate $t$-design with respect to the Lie group $\Pi(G)$.

Given this observation, and by applying the techniques presented in \cref{sec:upper} and \cref{sec:lower}, it follows that the upper and lower variance bounds are identical to those in \cref{app:upperbound:case_1,app:upperbound:case_2,app:upperbound:case_3} and \cref{app:lowerbound:case_1,app:lowerbound:case_2,app:lowerbound:case_3}. Consequently, the algorithms developed in \cref{sec:tester} remain valid and applicable without further adaptation.

\definecolor{chred}{RGB}{195,10,20}
\newcommand{\new}[1]{{\color{blue}#1}}
\section{Examples}
\label{sec:examples}
To clearly illustrate how these theoretical upper and lower bounds behave in practice, we will now walk through following concrete examples. For the sake of clarity and to keep the underlying mathematics manageable, we will look specifically at the case where both unitary, $\bbU_-$ and $\bbU_+$, form 2-designs with respect to the group $\compactG$.

\subsection{Standard Representation of SU(N)}
\label{app:example:SUN}
Consider the special unitary group $\SU{d}$ with $d=2^n$ and its associated Lie algebra $\su{d}$ generated by $H_1, \dots, H_k$. Under the standard representation acting on $\bbC^{d}$, the tensor square space $(\bbC^{d})^{\otimes 2}$ decomposes into exactly two irreducible invariant subspaces: the symmetric subspace $\iSymt{\bbC^{d}}$ and the alternating subspace $\iAltt{\bbC^{d}}$, of dimensions $d(d+1)/2$ and $d(d-1)/2$. Both subspaces have dimensions scaling asymptotically as $\cO(2^{2n})$.

\paragraph{Upper bound.} Let $\rho$ be a pure state and let $\cM$ be traceless, i.e., reduced in the sense of \cref{main:text:rem:reduced}. With $P_{sym}=\frac12(\bbI+\SWAP)$, $P_{alt}=\frac12(\bbI-\SWAP)$ and $\Tr{\SWAP\,(A\otimes A)}=\Tr{A^2}$ we obtain
\begin{align*}
    \Tr{\rho^{\otimes 2} P_{sym} }=1,\qquad \Tr{\rho^{\otimes 2} P_{alt} }=0,\qquad \Tr{\cM^{\otimes 2} P_{sym} }=-\Tr{\cM^{\otimes 2} P_{alt} }=\frac12\big(\snorm{\cM}_2\big)^2 .
\end{align*}
Consequently, applying the result from \cref{sec:upper}, the variance is upper bounded by
\begin{align}
    \var &\leq 4\big(\infsnorm{H}\big)^2\cdot\frac{\sqrt2}{\sqrt{d(d+1)}}\cdot\frac{(\snorm{\cM}_2)^2}{2}\left(\frac{\sqrt2}{\sqrt{d(d+1)}}+\frac{\sqrt2}{\sqrt{d(d-1)}}\right) \nonumber\\
    &=4\big(\infsnorm{H}\big)^2\big(\snorm{\cM}_2\big)^2\left(\frac{1}{d(d+1)}+\frac{1}{d\sqrt{d^2-1}}\right). \label{eq:app:sun:upper}
\end{align}
Since $(\snorm{\cM}_2)^2\leq d\,(\infsnorm{\cM})^2$, the right-hand side is $\cO\left(2^{-n}\right)$ whenever $\infsnorm{\cM}$ and $\infsnorm{H}$ are $\cO(1)$.

\paragraph{Exact variance.} For a pure state, $\rho^{\otimes 2}$ lies in $\iSymt{\bbC^{d}}$, so that $\cT_+(\rho)=P_{sym}/\dim\iSymt{\bbC^{d}}$ by \cref{eq:35}, and $\cT_-(\cM)=\theta_{sym}P_{sym}+\theta_{alt}P_{alt}$ with $\theta_{sym}=\frac{(\snorm{\cM}_2)^2}{d(d+1)}$ and $\theta_{alt}=-\frac{(\snorm{\cM}_2)^2}{d(d-1)}$. Write $K=H\otimes\bbI$, $K^\prime=\bbI\otimes H$ and $\Delta\defeq K-K^\prime$. Since $P_{sym}+P_{alt}=\bbI$ and the trace of a commutator vanishes, \cref{app:var:compact} gives $\var=(\theta_{sym}-\theta_{alt})\iTr{P_{sym}\commutator{K}{\commutator{K^\prime}{P_{sym}}}}/\dim\iSymt{\bbC^{d}}$, and by \cref{app:crit:copy}, with $\cJ_{\lambda,aa}=P_{sym}$, the trace equals $2(\snorm{P_{alt}KP_{sym}}_2)^2$. Here $P_{alt}KP_{sym}=\frac12\Delta P_{sym}$, because $K+K^\prime$ maps $\iSymt{\bbC^{d}}$ into itself and $\Delta$, which changes sign under $\SWAP$, maps it into $\iAltt{\bbC^{d}}$. For a traceless generator, $\Delta^\dagger\Delta=-H^2\otimes\bbI-\bbI\otimes H^2+2H\otimes H$ gives $(\snorm{\frac12\Delta P_{sym}}_2)^2=\frac14\Tr{P_{sym}\Delta^\dagger\Delta}=\frac{d}{4}(\snorm{H}_2)^2$. Hence
\begin{align}
    \var=\frac{d\,(\snorm{H}_2)^2}{2\dim\iSymt{\bbC^{d}}}\big(\theta_{sym}-\theta_{alt}\big)=\frac{2\,(\snorm{H}_2)^2\,(\snorm{\cM}_2)^2}{(d^2-1)(d+1)}. \label{eq:app:sun:exact}
\end{align}
Every traceless observable lies in $i\,\su{d}$, so \cref{eq:app:sun:exact} is also the value given by the Lie-algebraic formula of \cref{sec:comparison_prior_works}, with $D=d^2-1$, $(\norm{H}_K)^2=2d(\snorm{H}_2)^2$ and $(\snorm{\rho_\fg}_2)^2=\Tr{\rho^2}-1/d$. For a state that is not pure, the same formula gives \cref{eq:app:sun:exact} multiplied by $(d\Tr{\rho^2}-1)/(d-1)$.

\paragraph{Comparison.} For a Pauli observable and a Pauli generator, $\infsnorm{\cM}=\infsnorm{H}=1$ and $(\snorm{\cM}_2)^2=(\snorm{H}_2)^2=d$, so that
\begin{align}
    \var=\frac{2d^2}{(d^2-1)(d+1)}\approx\frac{2}{2^n},\qquad\text{while \cref{eq:app:sun:upper} equals}\quad\frac{4}{d+1}+\frac{4}{\sqrt{d^2-1}}\approx\frac{8}{2^n}.
\end{align}
The upper bound has the correct exponent and exceeds the exact variance by a factor that tends to four ($4.11$ at $n=4$).

\paragraph{The role of the trace.} Whether the bound is evaluated for $\cM$ or for its traceless part matters. For $\cM=\frac12(\bbI+Z^{(1)})$ one has $\Tr{\cM^{\otimes 2} P_{sym} }=\frac{d^2}{8}+\frac{d}{4}$ and $\Tr{\cM^{\otimes 2} P_{alt} }=\frac{d^2}{8}-\frac{d}{4}$, so that the right-hand side of \cref{app:upperbound:case_3} becomes
\begin{align*}
    8\big(\infsnorm{H}\big)^2\left(\frac{d^2/8+d/4}{d(d+1)}+\frac{d^2/8-d/4}{d\sqrt{d^2-1}}\right)\longrightarrow2\big(\infsnorm{H}\big)^2\qquad(n\to\infty),
\end{align*}
whereas the variance is that of the traceless part $\frac12Z^{(1)}$, namely $\frac{d^2}{2(d^2-1)(d+1)}\approx\frac12\,2^{-n}$ for a Pauli generator.

\subsection{Inverse-Polynomially Large Upper Bound with Vanishing Variance}\label{sec:S72}
We give two examples in which the upper bound \cref{app:upperbound:case_3} is only polynomially small while the variance is exactly zero.

\paragraph{Direct sum.} Let $d,d^\prime\geq2$ with $d+d^\prime\leq2^n$, put $r\defeq2^n-d-d^\prime$, and let the direct product group $\SU{d} \times \SU{d^\prime}$ act on $\cH=\bbC^{d}\oplus \bbC^{d^\prime}\oplus\bbC^{r}$ by $(g_1,g_2)\mapsto g_1\oplus g_2\oplus\bbI_r$. Let $H$ be an element within the generating set of its associated Lie algebra $\Lie{\SU{d} \times \SU{d^\prime}}$. The tensor square decomposes as
\begin{align*}
    \cH^{\otimes 2} = \Symt{\bbC^{d}} \oplus \Altt{\bbC^{d}} \oplus \Symt{\bbC^{d^\prime}} \oplus \Altt{\bbC^{d^\prime}} \oplus 2\left(\bbC^{d} \otimes \bbC^{d^\prime}\right)\oplus 2r\,\bbC^{d}\oplus2r\,\bbC^{d^\prime}\oplus r^2\,\bbC .
\end{align*}Suppose $\rho$ is a rank-1 density operator supported exclusively on $\bbC^{d}$, and $\cM$ is a rank-1 projector supported entirely on $\bbC^{d^\prime}$. Consequently, $\rho^{\otimes 2}$ and $\cM^{\otimes 2}$ have support strictly within the symmetric subspaces $\iSymt{\bbC^{d}}$ and $\iSymt{\bbC^{d^\prime}}$, respectively; both are irreducible and occur once.

The derivative vanishes for every value of the parameters: $U_+\rho U_+^\dagger$, and with it $\commutator{H}{U_+\rho U_+^\dagger}$, is supported on $\bbC^{d}$, whereas $U_-^\dagger\cM U_-$ is supported on $\bbC^{d^\prime}$, so that $\Tr{U_-^\dagger\cM U_-\commutator{H}{U_+\rho U_+^\dagger}}=0$. Therefore the exact variance is $0$.

Following \cref{sec:upper}, the variance is upper bounded by
\begin{align}
    4\big(\infsnorm{H}\big)^2\left(\frac{\Tr{\rho^{\otimes 2} P_{sym, d} }}{\sqrt{d(d+1)/2}}\right)\left(\frac{\Tr{\cM^{\otimes 2} P_{sym, d^\prime} }}{\sqrt{d^\prime(d^\prime+1)/2}}\right) = \frac{8\big(\infsnorm{H}\big)^2}{\sqrt{d(d+1)\,d^\prime(d^\prime+1)}}=\Theta\!\left(\frac{1}{d\,d^\prime}\right), \label{eq:app:witness:directsum}
\end{align}and the lower bound \cref{app:lowerbound:case_3} is $0$. For the reduced observable $\cM-P_{d^\prime}/d^\prime$ of \cref{main:text:rem:reduced}(i), where $P_{d^\prime}$ projects onto $\bbC^{d^\prime}$, the upper bound becomes $4(\infsnorm{H})^2\big(1-\frac{1}{d^\prime}\big)\big[\big(d(d+1)d^\prime(d^\prime+1)\big)^{-1/2}+\big(d(d+1)d^\prime(d^\prime-1)\big)^{-1/2}\big]$, which is again $\Theta(1/(dd^\prime))$. The upper bound is thus polynomially small in $n$ if and only if \emph{both} $d$ and $d^\prime$ are polynomial in $n$; if either $d$ or $d^\prime$ is exponentially large, for instance $d^\prime=2^n-d$, the bound is itself exponentially small and the example shows nothing. Finally, this overestimate disappears under the compression of \cref{main:text:rem:reduced}(ii), since $P_A\cM P_A=0$ for $\cH_A=\bbC^{d}$.

\paragraph{Tensor product.} The second example is unaffected by both reductions. Let $d_1d_2=2^n$, and let $\SU{d_1}\times\SU{d_2}$ act on $\cH=\bbC^{d_1}\otimes\bbC^{d_2}$ by $(g_1,g_2)\mapsto g_1\otimes g_2$; this representation is irreducible. Take $\cM=\cM_1\otimes\bbI$ with $\cM_1$ traceless, and let the differentiated generator be $H=\bbI\otimes H_2$. Every $U_-$ is of the form $V_1\otimes V_2$, so $U_-^\dagger\cM U_-=(V_1^\dagger\cM_1V_1)\otimes\bbI$ commutes with $H$ and
\begin{align*}
    \partial_{\theta_{\ell,j}}C=\Tr{\commutator{U_-^\dagger\cM U_-}{H}\,U_+\rho U_+^\dagger}=0
\end{align*}
for every state and every value of the parameters. On the other hand, $\cH^{\otimes 2}$ is the direct sum of the four inequivalent irreducible subspaces $W_{ss^\prime}$, $s,s^\prime\in\CurB{sym,alt}$, obtained from the symmetric or alternating square of each tensor factor. For a product state, $\rho^{\otimes 2}$ lies in $W_{sym,sym}$, and $\abs{\Tr{\cM^{\otimes 2}P_{ss^\prime}}}=\frac12(\snorm{\cM_1}_2)^2\dim(s^\prime)$, where $\dim(sym)=d_2(d_2+1)/2$ and $\dim(alt)=d_2(d_2-1)/2$. Thus \cref{app:upperbound:case_3} equals
\begin{align}
    \frac{8\big(\infsnorm{H}\big)^2\big(\snorm{\cM_1}_2\big)^2}{\sqrt{d_1(d_1+1)\,d_2(d_2+1)}}\left(\sqrt{\frac{d_2(d_2+1)}{4}}+\sqrt{\frac{d_2(d_2-1)}{4}}\right)\left(\frac{1}{\sqrt{d_1(d_1+1)}}+\frac{1}{\sqrt{d_1(d_1-1)}}\right), \label{eq:app:witness:tensor}
\end{align}
which for $(\snorm{\cM_1}_2)^2=d_1$---a Pauli operator on the first register---tends to $16(\infsnorm{H})^2/d_1$ when $d_1$ and $d_2$ are large. With $d_1=\poly{n}$ the upper bound is polynomially small although the variance vanishes identically. No reduction of the observable repairs this, because \cref{app:upperbound:case_3} depends on the generator only through its norm.

\subsection{\texorpdfstring{$n-$}{}Qubit Collective-Spin Representation of SU(2)}
\label{app:example:SU2}

Consider the special unitary group $\SU{2}$ and its associated Lie algebra $\su{2}$, spanned by the generators $L_x = -iX, L_y = -iY, L_z = -iZ$. We evaluate the $n$-qubit collective-spin representation acting on $(\bbC^{2^n}, \Pi)$. The ansatz operators are defined as
\begin{align*}
    \Pi\left(\expm{\theta L_x}\right) = \expm{{\theta\pi_*(L_x)}} =& \expm{-i \theta\sum_{j=1}^n X^{(j)}}, \\
    \Pi\left(\expm{\theta L_y}\right) = \expm{{\theta\pi_*(L_y)}} =& \expm{-i \theta\sum_{j=1}^n Y^{(j)}}, \\
    \Pi\left(\expm{\theta L_z}\right) = \expm{{\theta\pi_*(L_z)}} =& \expm{-i \theta\sum_{j=1}^n Z^{(j)}},
\end{align*}where $X^{(j)}, Y^{(j)}$, and $Z^{(j)}$ denote the standard Pauli matrices. For notational simplicity, we define the collective spin operators
\begin{align}
    S_M = {\frac{1}{2}}\sum_{j=1}^n M^{(j)}
\end{align}for $M\in \{X, Y, Z\}${, which satisfy $\commutator{S_X}{S_Y}=iS_Z$ (and cyclic permutations); in particular, $\Pi\left(\expm{\theta L_x}\right)=\expm{-2i\theta S_X}$}. The tensor square space $(\bbC^{2^n})^{\otimes 2}$ decomposes as:
\begin{align}
    \left(\bbC^{2^n}\right)^{\otimes 2} = \bigoplus_{\lambda=0}^n m_\lambda V_\lambda
\end{align}where $V_\lambda$ are the irreducible invariant subspaces of dimension $2\lambda + 1$, and $m_\lambda = \binom{2n}{n-\lambda} - \binom{2n}{n-\lambda - 1}$ denotes the multiplicity.
 
Let $\rho = \Outer{\lambda^*}{\lambda^*}$, where $\state{\lambda^*}$ is {a} highest weight vector. By definition, this vector is an eigenvector of the $z$-axis generator $S_Z$ with eigenvalue $\lambda^*$, and it is annihilated by the raising operator $S_+${, where $S_\pm\defeq S_X\pm iS_Y$. We denote by $\cH_{\opl}\subseteq\bbC^{2^n}$ the irreducible subspace generated by $\state{\lambda^*}$, which carries the spin-$\lambda^*$ representation}. For an $n$-qubit system, the allowed highest weights are $\lambda^* \in \{0, 1, 2, \dots, n/2\}$ for even $n$, and $\lambda^* \in \{1/2, 3/2, 5/2, \dots, n/2\}$ for odd $n$.
 
We consider the measurement observable $\cM = S_X + S_Y + S_Z$ and the generator $H = -iS_Z$. Because $\state{\lambda^*}$ is a highest weight vector, {$\state{\lambda^*}\otimes\state{\lambda^*}$ is a highest weight vector of weight $2\lambda^*$ for the tensor-square representation and generates a single copy of} the irreducible subspace $V_{2\lambda^*}${, namely the copy contained in $\cH_{\opl}\otimes\cH_{\opl}$}. To establish a clear reference frame, we {choose the multiplicity basis such that} $\rho^{\otimes 2}$ is supported exclusively on the {$\opc$}-th copy of $V_{2\lambda^*}$. Evaluating the first term in \cref{app:lowerbound:case_3} yields:
\begin{align*}
    &\sum_{\lambda}\sum_{a,b,c,d}^{m_{\lambda}}\frac{\Tr{\cJ_{\lambda,ab}^\dagger\cM^{\otimes 2}}\cdot \Tr{\cJ_{\lambda,cd}^\dagger\rho^{\otimes 2}}}{(\dim V_\lambda)^2}\Tr{\cJ_{\lambda, ab}\commutator{H\otimes \bbI}{\commutator{\bbI\otimes H}{\cJ_{\lambda, cd}}}} \\
    =& \sum_{a,b}^{m_{{2\opl}}} \frac{\Tr{\cJ_{2\opl,ab}^\dagger\cM^{\otimes 2}}}{(\dim V_{2\opl})^2}\Tr{\cJ_{2\opl, ab}\commutator{H\otimes \bbI}{\commutator{\bbI\otimes H}{\cJ_{2\opl, \opc\opc}}}}
\end{align*}Expanding the nested commutator trace {with $H=-iS_Z$, and rotating the $z$-axis into the $x$-axis (which leaves every $\cJ_{\lambda,ab}$ invariant)}, we have
\begin{align*}
    T_{ab} \defeq &{\Tr{\cJ_{2\opl, ab}\commutator{H\otimes \bbI}{\commutator{\bbI\otimes H}{\cJ_{2\opl, \opc\opc}}}}} \\
    {=}&\mathrm{tr}\left\{\cJ_{2\opl, ab} \bigg[ (S_X \otimes \bbI) \cJ_{2\opl, \opc\opc} (\bbI \otimes S_X) + (\bbI \otimes S_X) \cJ_{2\opl, \opc\opc} (S_X \otimes \bbI) \right. \\ &\hspace{3cm}\left.- \cJ_{2\opl, \opc\opc} (S_X \otimes S_X) - (S_X \otimes S_X) \cJ_{2\opl, \opc\opc} \bigg] \right\}
\end{align*}By spatial symmetry, the trace reduces to
\begin{align}
    &\Tr{\cJ_{2\opl, \opc\opc} (S_X \otimes S_X)\cJ_{2\opl, ab}} \\
    =& \frac{1}{3}\Tr{\cJ_{2\opl, \opc\opc} (S_X \otimes S_X + S_Y \otimes S_Y + S_Z \otimes S_Z)\cJ_{2\opl, ab}} \\
    =& \delta_{{b}\opc}\frac{1}{3} \sum_{K=-2\opl}^{2\opl} \cstate{2\opl, {\opc}, K } (S_X \otimes S_X + S_Y \otimes S_Y + S_Z \otimes S_Z) \state{ 2\opl, {a}, K },
\end{align}where $\CurB{\state{2\opl, a, K}}$ denotes the orthonormal basis of the $a$-th copy of $V_{2\opl}$, in which 
\begin{align*}
    \cJ_{2\opl,ab}=\sum_{K}\state{2\opl,a,K}\cstate{2\opl,b,K}.
\end{align*}
Here $S_X \otimes S_X + S_Y \otimes S_Y + S_Z \otimes S_Z$ is the spin--spin coupling operator. It equals $\frac{1}{2}\big(\widetilde{S}^2 - S^2\otimes\bbI - \bbI\otimes S^2\big)$, where $\widetilde{S}_M\defeq S_M\otimes\bbI+\bbI\otimes S_M$, $\widetilde{S}^2\defeq\widetilde{S}_X^2+\widetilde{S}_Y^2+\widetilde{S}_Z^2$ and $S^2\defeq S_X^2+S_Y^2+S_Z^2$, and therefore acts as the scalar $\frac{1}{2}\big(\lambda(\lambda+1)-2\opl(\opl+1)\big)$ on every copy of $V_\lambda$ contained in $\cH_{\opl}\otimes\cH_{\opl}$. Because the $\opc$-th copy lies in $\cH_{\opl}\otimes\cH_{\opl}$, $\state{ 2\opl, \opc, K }$ is an eigenvector of the {spin--spin coupling} operator with eigenvalue $(\opl)^2$ for all $K${. Hence} this expression evaluates to $\frac{1}{3}(\opl)^2(4\opl+1) \delta_{a\opc}\delta_{b\opc}$. A symmetric argument yields $\Tr{\cJ_{2\opl, ab}(S_X \otimes S_X) \cJ_{2\lambda^*, \opc\opc}} = \frac{1}{3}(\opl)^2(4\opl+1) \delta_{a\opc}\delta_{b\opc}$.
 
{
The cross terms in $T_{ab}$ involve matrix elements of $S_X\otimes\bbI$ and $\bbI\otimes S_X$ between the $\opc$-th copy of $V_{2\opl}$ and an arbitrary copy of $V_{2\opl}$.
Both $S_X\otimes\bbI$ and $\bbI\otimes S_X$ leave $\cH_{\opl}\otimes\cH_{\opl}$ invariant, and by the Clebsch--Gordan rule, $\cH_{\opl}\otimes\cH_{\opl}=\bigoplus_{\mu=0}^{2\opl}V_\mu$ contains exactly one copy of $V_{2\opl}$, namely the $\opc$-th copy.
Every other copy $a\neq\opc$ of $V_{2\opl}$ is orthogonal to $\cH_{\opl}\otimes\cH_{\opl}$: it is orthogonal to the $\opc$-th copy by construction, and to the components $V_\mu$ with $\mu\neq 2\opl$ because these carry inequivalent irreducible representations.
Since $(S_X\otimes\bbI)\state{2\opl,\opc,K^\prime}\in\cH_{\opl}\otimes\cH_{\opl}$, we therefore have
\begin{align}
    \cstate{2\opl,a,K}(S_X\otimes\bbI)\state{2\opl,\opc,K^\prime} = \delta_{a\opc}\,\cstate{2\opl,\opc,K}(S_X\otimes\bbI)\state{2\opl,\opc,K^\prime},
\end{align}
and likewise for $\bbI\otimes S_X$.
It remains to evaluate the matrix elements within the $\opc$-th copy.
This copy is spanned by the vectors obtained from the swap-symmetric vector $\state{\opl}\otimes\state{\opl}$ by repeated application of the swap-symmetric lowering operator $S_-\otimes\bbI+\bbI\otimes S_-$.
Hence it lies in $\iSymt{\bbC^{2^n}}$, i.e., $\SWAP\state{2\opl,\opc,K}=\state{2\opl,\opc,K}$ for all $K$, and using $\SWAP(S_X\otimes\bbI)\SWAP=\bbI\otimes S_X$ we obtain
\begin{align*}
    \cstate{2\opl,\opc,K}(S_X\otimes\bbI)\state{2\opl,\opc,K^\prime}
    =& \cstate{2\opl,\opc,K}\SWAP(S_X\otimes\bbI)\SWAP\state{2\opl,\opc,K^\prime} \\
    =& \cstate{2\opl,\opc,K}(\bbI\otimes S_X)\state{2\opl,\opc,K^\prime} \\
    =& \frac{1}{2}\cstate{2\opl,\opc,K}(S_X\otimes\bbI+\bbI\otimes S_X)\state{2\opl,\opc,K^\prime}.
\end{align*}
The operator $\widetilde{S}_X=S_X\otimes\bbI+\bbI\otimes S_X$ generates the diagonal action of $\SU{2}$; it therefore maps every copy of $V_{2\opl}$ to itself and acts on each copy by the same matrix $(\widetilde{S}_X)_{K,K^\prime}$ in the bases $\CurB{\state{2\opl,a,K}}_K$.
Combining the above, we find for every copy $a$
\begin{align}
    \cstate{2\opl,a,K}(S_X\otimes\bbI)\state{2\opl,\opc,K^\prime} = \cstate{2\opl,a,K}(\bbI\otimes S_X)\state{2\opl,\opc,K^\prime} = \frac{1}{2}\delta_{a\opc}(\widetilde{S}_X)_{K,K^\prime},
\end{align}
and, taking adjoints, $\cstate{2\opl,\opc,K}(S_X\otimes\bbI)\state{2\opl,b,K^\prime} = \cstate{2\opl,\opc,K}(\bbI\otimes S_X)\state{2\opl,b,K^\prime} = \frac{1}{2}\delta_{\opc b}(\widetilde{S}_X)_{K,K^\prime}$.
This identity relies on one of the two copies being the $\opc$-th copy: for two arbitrary copies of $V_{2\opl}$ with opposite swap parity, the matrix elements of $S_X\otimes\bbI$ need not vanish.
} {Hence} the cross-term evaluates to:
\begin{align}
    \Tr{\cJ_{2\opl, ab} (S_X \otimes \bbI) \cJ_{2\opl, \opc\opc} (\bbI \otimes S_X)} =& \frac{1}{4} \delta_{b\opc} \delta_{\opc a} \sum_{K, K'} ({\widetilde{S}_X})_{K, K^\prime} ({\widetilde{S}_X})_{K^\prime, K} \\
    =& \frac{1}{4} \delta_{b\opc} \delta_{\opc a} \sum_{K} ({\widetilde{S}_X^2})_{K, K}.
\end{align}
Exploiting spatial symmetry once more,
\begin{align*}
    {\sum_K(\widetilde{S}_X^2)_{K,K} = \frac{1}{3}\sum_K\big(\widetilde{S}_X^2 + \widetilde{S}_Y^2 + \widetilde{S}_Z^2\big)_{K,K}} = \frac{1}{3} (2\opl)(2\opl+1)(4\opl+1).
\end{align*}
Thus, we conclude:
\begin{align}
    \Tr{\cJ_{2\opl, ab} (S_X \otimes \bbI) \cJ_{2\opl, \opc\opc} (\bbI \otimes S_X)} = \frac{\opl(2\opl+1)(4\opl+1)}{{6}} \delta_{\opc a} \delta_{b\opc}.
\end{align}
Similarly, we have $\Tr{\cJ_{2\opl, ab} (\bbI \otimes S_X) \cJ_{2\opl, \opc\opc} (S_X \otimes \bbI)} = \frac{\opl(2\opl+1)(4\opl+1)}{{6}} \delta_{\opc a} \delta_{b\opc}$. Combining all of the above {with the two terms evaluated previously}, we note that
\begin{align}
    T_{ab} = {2\cdot\frac{\opl(2\opl+1)(4\opl+1)}{6}\,\delta_{\opc a}\delta_{b\opc} - 2\cdot\frac{(\opl)^2(4\opl+1)}{3}\,\delta_{\opc a}\delta_{b\opc} =} \frac{\opl(4\opl+1)}{3} \delta_{\opc a} \delta_{b\opc}.
\end{align}
This identity directly simplifies the nested commutator trace:
\begin{align}
    &\sum_{a,b}^{m_{{2\opl}}} \frac{\Tr{\cJ_{2\opl,ab}^\dagger\cM^{\otimes 2}}}{(\dim V_{2\opl})^2}\Tr{\cJ_{2\opl, ab}\commutator{H\otimes \bbI}{\commutator{\bbI\otimes H}{{\cJ_{2\opl,\opc\opc}}}}} = \frac{\Tr{\cJ_{2\opl,{\opc\opc}}^\dagger\cM^{\otimes 2}}}{(\dim V_{2\opl})^2}\frac{\opl(4\opl+1)}{3}.
\end{align}

Finally, we expand $\cM^{\otimes 2} = \sum_{M,M^\prime \in \{X,Y,Z\}} S_M \otimes S_{M^\prime}$. For $M\neq M^\prime$, consider the $\pi$-rotation $R_M\defeq(-i)^nM^{\otimes n}$, e.g., $R_X=\Pi\big(e^{\frac{\pi}{2}L_x}\big)$. It satisfies $R_MS_MR_M^\dagger=S_M$ and $R_MS_{M^\prime}R_M^\dagger=-S_{M^\prime}$, so that 
\begin{align*}
    (R_M\otimes R_M)(S_M\otimes S_{M^\prime})(R_M\otimes R_M)^\dagger=-S_M\otimes S_{M^\prime}.
\end{align*}
Since every $\cJ_{\lambda,ab}$ commutes with $R_M\otimes R_M$, it follows that $\Tr{\cJ_{\lambda,ab}^\dagger(S_M\otimes S_{M^\prime})}=0$ for $M\neq M^\prime$, and we evaluate the observable trace as:

\begin{align*}
    \Tr{\cJ_{2\opl,{\opc\opc}}^\dagger\cM^{\otimes 2}} &= \sum_{K=-2\opl}^{2\opl} \cstate{2\opl, {\opc}, K } (S_X \otimes S_X + S_Y \otimes S_Y + S_Z \otimes S_Z) \state{ 2\opl, \opc,K} \\
    &= (\opl)^2(4\opl+1).
\end{align*}
Substituting this back into the expansion, the target summation evaluates to:
\begin{align}
    \sum_{\lambda}\sum_{a,b,c,d}^{m_{\lambda}}\frac{\Tr{\cJ_{\lambda,ab}^\dagger\cM^{\otimes 2}}\cdot \Tr{\cJ_{\lambda,cd}^\dagger\rho^{\otimes 2}}}{(\dim V_\lambda)^2}\Tr{\cJ_{\lambda, ab}\commutator{H\otimes \bbI}{\commutator{\bbI\otimes H}{\cJ_{\lambda, cd}}}} = \frac{(\opl)^3}{3}.
\end{align}
 
Next, we evaluate the second term in \cref{app:lowerbound:case_3}:
\begin{align*}
    \begin{split}
        &2\sum_{\nu}\sum_{c,d}^{m_\nu}\frac{\snorm{\cJ_{{\nu}, dd}\sqrt{{\rho^{\otimes 2}}}}_2\snorm{\cJ_{{\nu}, cc}\sqrt{{\rho^{\otimes 2}}}}_2}{\dim V_\nu}\\
        &\hspace{0.5cm}\cdot\abs{\Tr{\left(\sum_{\lambda\neq \nu}\sum_{a,b}^{m_\lambda}\frac{\Tr{\cJ^\dagger_{\lambda, ab} \cM_{c, 1 \to d, 2}^{\,(\lambda^\prime, a^\prime) \otimes (\nu^\prime, b^\prime)}}}{\dim V_\lambda}\cJ_{\lambda, ab}^{(\lambda^\prime, a^\prime, \nu^\prime, b^\prime)_{c, 1}, \; (\lambda^\prime, a^\prime, \nu^\prime, b^\prime)_{d, 2}}\right) (H \otimes \bbI) \cJ_{\nu, cd} (H\otimes \bbI)}}
    \end{split}
\end{align*}
Because $\rho^{\otimes 2}$ is supported on the $\opc$-th copy of $V_{2\opl}$, {only $\nu=2\opl$ and $c=d=\opc$ contribute, and for this term $(\snorm{\cJ_{2\opl,\opc\opc}\sqrt{\rho^{\otimes 2}}}_2)^2=\Tr{\cJ_{2\opl,\opc\opc}\,\rho^{\otimes 2}}=1$. The second term therefore equals}
\begin{align*}
    \frac{2}{\dim V_{2\opl}}&\Biggl|\tr\Biggl\{\biggl(\sum_{\lambda\neq {2\opl}}\sum_{a,b}^{m_\lambda}\frac{\Tr{\cJ^\dagger_{\lambda, ab} \cM_{\opc, 1 \to \opc, 2}^{\,(\lambda^\prime, a^\prime) \otimes (\nu^\prime, b^\prime)}}}{\dim V_\lambda}\cJ_{\lambda, ab}^{(\lambda^\prime, a^\prime, \nu^\prime, b^\prime)_{\opc, 1}, \; (\lambda^\prime, a^\prime, \nu^\prime, b^\prime)_{\opc, 2}}\biggr) \\ &\hspace{2cm}
    \times(H \otimes \bbI) \cJ_{2\opl, \opc\opc} (H\otimes \bbI)\Biggr\}\Biggr|,
\end{align*}
where $\dim V_{2\opl}=4\opl+1$. Recall that $V_{2\opl}\subseteq \cH_{\opl} \otimes \cH_{\opl} = \bigoplus_{{\mu} = |\opl - \opl|}^{\opl + \opl} V_{{\mu}} =  \bigoplus_{{\mu} = 0}^{2\opl} V_{{\mu}}$. For each allowable $\lambda${, i.e., $0\leq\lambda\leq 2\opl$}, the tensor product $\cH_{\opl} \otimes \cH_{\opl}$ contains exactly one copy of $V_{\lambda}$, which means that
\begin{align}
    \sum_{a,b}^{m_\lambda}\frac{\Tr{\cJ^\dagger_{\lambda, ab} \cM_{\opc, 1 \to \opc, 2}^{\,(\lambda^\prime, a^\prime) \otimes (\nu^\prime, b^\prime)}}}{\dim V_\lambda}\cJ_{\lambda, ab}^{(\lambda^\prime, a^\prime, \nu^\prime, b^\prime)_{\opc, 1}, \; (\lambda^\prime, a^\prime, \nu^\prime, b^\prime)_{\opc, 2}} = \frac{\Tr{\cJ^\dagger_{\lambda, k_{{\lambda}}k_{{\lambda}}} \cM^{\otimes 2}}}{\dim V_\lambda}\cJ_{\lambda, k_{{\lambda}}k_{{\lambda}}},
\end{align}
{where $k_\lambda\in[m_\lambda]$ labels the copy of $V_\lambda$ contained in $\cH_{\opl}\otimes\cH_{\opl}$ (in particular, $k_{2\opl}=\opc$). Copies of $V_\lambda$ with $\lambda>2\opl$ do not occur in $\cH_{\opl}\otimes\cH_{\opl}$ and do not contribute, since $(H\otimes\bbI)\cJ_{2\opl,\opc\opc}(H\otimes\bbI)$ is supported on $\cH_{\opl}\otimes\cH_{\opl}$.}
Following the arguments established previously, {the spin--spin coupling operator acts as the scalar $\frac{1}{2}\big(\lambda(\lambda+1)-2\opl(\opl+1)\big)$ on the $k_\lambda$-th copy of $V_\lambda$, since this copy lies in $\cH_{\opl}\otimes\cH_{\opl}$. Together with the vanishing of the terms $S_M\otimes S_{M^\prime}$ for $M\neq M^\prime$, this yields}:
\begin{align}
    \frac{\Tr{\cJ^\dagger_{\lambda, k_{{\lambda}}k_{{\lambda}}} \cM^{\otimes 2}}}{\dim V_\lambda}\cJ_{\lambda, k_{{\lambda}}k_{{\lambda}}} =& \frac{\Tr{\cJ^\dagger_{\lambda, k_{{\lambda}}k_{{\lambda}}} ({S_X \otimes S_X + S_Y \otimes S_Y + S_Z \otimes S_Z})}}{\dim V_\lambda}\cJ_{\lambda, k_{{\lambda}}k_{{\lambda}}} \\
    =&  \frac{1}{2}(\lambda(\lambda+1) - 2\opl(\opl+1))\cJ_{\lambda, k_{{\lambda}}k_{{\lambda}}}.
\end{align}
Summing over the relevant irreducible representations {and letting $Q$ denote the orthogonal projector onto $\cH_{\opl}\otimes\cH_{\opl}$, so that $Q=\sum_{\lambda=0}^{2\opl}\cJ_{\lambda,k_\lambda k_\lambda}$,} gives:
\begin{align}
    \sum_{{\lambda=0}}^{{2\opl-1}} \frac{\Tr{\cJ^\dagger_{\lambda, k_{{\lambda}}k_{{\lambda}}} \cM^{\otimes 2}}}{\dim V_\lambda}\cJ_{\lambda, k_{{\lambda}}k_{{\lambda}}} =& \sum_{{\lambda=0}}^{{2\opl-1}}\frac{1}{2}(\lambda(\lambda+1) - 2\opl(\opl+1))\cJ_{\lambda, k_{{\lambda}}k_{{\lambda}}} \\
    =& {Q}({S_X \otimes S_X + S_Y \otimes S_Y + S_Z \otimes S_Z}){Q} - {(\opl)^2\,}\cJ_{2\opl, \opc\opc}.
\end{align}
{Consequently, since $(H\otimes\bbI)\cJ_{2\opl,\opc\opc}(H\otimes\bbI)$ is supported on $\cH_{\opl}\otimes\cH_{\opl}$, the projectors $Q$ can be dropped, and the quantity inside the absolute value in the second term evaluates to}
\begin{align}
    \begin{split}
        &{\frac{2}{4\opl+1}\bigg[}\Tr{({S_X \otimes S_X + S_Y \otimes S_Y + S_Z \otimes S_Z}) (H\otimes \bbI) \cJ_{2\opl, \opc\opc} (H\otimes \bbI)} \\
        &\hspace{1cm}- {(\opl)^2\,}\Tr{\cJ_{2\opl, \opc\opc} (H\otimes \bbI) \cJ_{2\opl, \opc\opc} (H\otimes \bbI)}{\bigg]}
    \end{split} \\
    =& 2 \left( -\frac{1}{3}(\opl)^3(\opl+1) + \frac{1}{3}(\opl)^2 - (\opl)^2\left[-\frac{1}{6}\opl(2\opl+1)\right] \right)\\
    =& -\frac{1}{3}(\opl)^3 + \frac{2}{3}(\opl)^2{,}
\end{align}
{
where we used
\begin{align*}
    \Tr{(S_X \otimes S_X + S_Y \otimes S_Y + S_Z \otimes S_Z) (H\otimes \bbI) \cJ_{2\opl, \opc\opc} (H\otimes \bbI)} &= -\frac{1}{3}(4\opl+1)\big((\opl)^3(\opl+1)-(\opl)^2\big),\\
    \Tr{\cJ_{2\opl, \opc\opc} (H\otimes \bbI) \cJ_{2\opl, \opc\opc} (H\otimes \bbI)} &= -\frac{1}{6}\opl(2\opl+1)(4\opl+1).
\end{align*}
}
Combining these evaluations, the variance is lower-bounded by:
\begin{align}
    \frac{1}{3}(\opl)^3 - \abs{-\frac{1}{3}(\opl)^3 + \frac{2}{3}(\opl)^2} = {\frac{1}{3}(\opl)^3 - \frac{1}{3}(\opl)^2\abs{\opl-2} = \begin{cases} \frac{2}{3}(\opl)^2, & \opl\geq 2,\\[2pt] \frac{2}{3}(\opl)^2(\opl-1), & \opl<2. \end{cases}}
\end{align}
{In particular, for $\opl\geq 2$ the variance is lower-bounded by $\frac{2}{3}(\opl)^2$, whereas for $\opl<2$ the bound is weaker and becomes vacuous for $\opl\in\{\frac{1}{2},1\}$.}

\paragraph{Exact value.} The two quantities evaluated above also give the variance exactly. Since $\rho^{\otimes 2}$ lies in the $\opc$-th copy of $V_{2\opl}$ alone, the only non-zero trace of $\rho^{\otimes 2}$ in the exact expression \cref{app:var:case3:sum} is $\Tr{\cJ^\dagger_{2\opl,\opc\opc}\rho^{\otimes 2}}=(\snorm{\cJ_{2\opl,\opc\opc}\sqrt{\rho^{\otimes 2}}}_2)^2=1$. Hence \cref{eq:lb:eq:1} holds with equality, the steps from there to \cref{app:lowerbound:case_3} are identities, and the two sums of \cref{app:var:case3:sum} are the two quantities evaluated above, with their signs: the first is $\frac13(\opl)^3$, the first term of \cref{app:lowerbound:case_3}, and the second is $-\frac13(\opl)^3+\frac23(\opl)^2$, the second term of \cref{app:lowerbound:case_3} before its absolute value is taken, i.e., minus the contribution $\frac13(\opl)^3-\frac23(\opl)^2$ of the cross terms quoted in \cref{main:text:theory:sec:discussion} of the main text. Hence
\begin{align}
    \var=\frac13(\opl)^3-\frac13(\opl)^3+\frac23(\opl)^2=\frac23(\opl)^2\qquad\text{for all }\opl\geq\tfrac12. \label{eq:app:su2:exact}
\end{align}
The lower bound subtracts the absolute value of the second sum, which is positive for $\opl<2$; it therefore equals the exact variance precisely for $\opl\geq2$. Because $i\cM\in\pi_*(\su{2})$, the same value follows from the Lie-algebraic formula of \cref{sec:comparison_prior_works}: with $N\defeq(\snorm{S_Z}_2)^2$ one has $D=3$, $(\snorm{\cM}_2)^2=3N$, $(\snorm{\rho_\fg}_2)^2=(\opl)^2/N$ and $(\norm{H}_K)^2=2$, so that $\var=2\cdot3N\cdot\frac{(\opl)^2}{N}\cdot\frac19=\frac23(\opl)^2$.

\paragraph{Upper bound after compression.} \Cref{fig:variance-bounds}(b) shows \cref{app:upperbound:case_3} for $\opl=n/2$ after the compression of \cref{main:text:rem:reduced}(ii) to the sub-representation $\cH_{\opl}$, on which $\cM$ is traceless. There $\cH_{\opl}\otimes\cH_{\opl}=\bigoplus_{\lambda=0}^{n}V_\lambda$ without multiplicities, $\rho^{\otimes 2}$ lies in $V_n$, and $\cT_-(\cM)$ is the spin--spin coupling operator, which equals $\frac12\big(\lambda(\lambda+1)-\frac{n(n+2)}{2}\big)$ on $V_\lambda$. With $\dim V_\lambda=2\lambda+1$ and $\infsnorm{H}=n/2$, \cref{app:upperbound:case_3} becomes
\begin{align}
    4\Big(\frac n2\Big)^2\cdot\frac{1}{\sqrt{2n+1}}\cdot\sum_{\lambda=0}^{n}\frac{\sqrt{2\lambda+1}}{2}\,\abs{\lambda(\lambda+1)-\frac{n(n+2)}{2}}=\frac{n^2}{2\sqrt{2n+1}}\sum_{\lambda=0}^{n}\sqrt{2\lambda+1}\,\abs{\lambda(\lambda+1)-\frac{n(n+2)}{2}}. \label{eq:app:su2:upper:compressed}
\end{align}
Since $\infsnorm{\cM}=\sqrt3\,n/2$ on $\cH_{\opl}$, the trivial bound $4(\infsnorm{H})^2(\infsnorm{\cM})^2$ of \cref{sec:upper} equals $\frac34n^4$, and \cref{eq:app:su2:upper:compressed} exceeds it for every $n\geq5$, by a factor $1.07$ at $n=5$ and $2.8$ at $n=20$.

\subsection{Free-Fermion}
\label{app:example:free:fermine}

In this section, we consider the fermionic Fock space $\cF \defeq \SPAN{\state{b_1, \dots, b_n} : b_j \in \CurB{0,1}} \simeq \bbC^{2^n}$, equipped with the spinor representation $\Pi: \Spin{2n} \to \SU{V}$. For simplicity, we assume $n \geq 8$. Under this representation, the full Fock space canonically decomposes into its even- and odd-parity invariant subspaces, defined respectively as $S_+ \defeq \cF_{\mathrm{even}} \defeq \ispan{\state{b_1, \dots, b_n} : \sum_j b_j \equiv 0 \pmod 2}$ and $S_- \defeq \cF_{\mathrm{odd}} \defeq \ispan{\state{b_1, \dots, b_n} : \sum_j b_j \equiv 1 \pmod 2}$ with $\dim S_\pm = 2^{n-1}$.

To describe this representation, we first map the $n$-qubit spin operators to fermionic creation and annihilation operators using the Jordan--Wigner transformation. Define
\begin{align*}
    c_j=\left(\prod_{k=1}^{j-1}Z^{(k)}\right)\sigma_-^{(j)},\qquad c_j^\dagger=\left(\prod_{k=1}^{j-1}Z^{(k)}\right)\sigma_+^{(j)},
\end{align*}
where $\sigma_-^{(j)}=\frac{1}{2}\left(X^{(j)}+iY^{(j)}\right),\enspace\sigma_+^{(j)}=\frac{1}{2}\left(X^{(j)}-iY^{(j)}\right).$ 

\begin{remark}
    Here $\sigma_-^{(j)} = \Outer{0}{1}_j$ lowers the fermionic \emph{occupation} of mode $j$; in the spin conventions in which $\state{0}$ denotes the empty mode, this operator is the spin-raising operator, so the apparent mismatch of signs with the convention $\sigma_\pm = \frac{1}{2}(X\pm iY)$ is only one of nomenclature.  In particular $c_j\state{0^{\otimes n}}=0$, so $\state{0^{\otimes n}}$ is the Fock vacuum, and since the $Z$-string acts only on modes $k<j$ it commutes with $\sigma_\pm^{(j)}$, so that $c_j^\dagger$ really is the adjoint of $c_j$.
\end{remark}

These operators satisfy the canonical anticommutation relations
\begin{align}
    \CurB{c_j,c_k^\dagger}=\delta_{jk}\bbI,\qquad \CurB{c_j,c_k}=\CurB{c_j^\dagger,c_k^\dagger}=0. \label{app:example:cl:rel}
\end{align}
To formalize the unitary representation of these free fermions, it is algebraically convenient to pass to the Majorana basis by defining $2n$ Hermitian operators
\begin{align*}
    \gamma_{2j-1}=c_j+c_j^\dagger,\qquad \gamma_{2j}=-i(c_j-c_j^\dagger),\quad 1\leq j\leq n.
\end{align*}
Under the Jordan--Wigner transformation, these operators take the explicit form
\begin{align}
    \gamma_{2j-1}=\left(\prod_{k=1}^{j-1}Z^{(k)}\right)X^{(j)},\qquad\gamma_{2j}=\left(\prod_{k=1}^{j-1}Z^{(k)}\right)Y^{(j)}. \label{app:example:ff:gamma}
\end{align}
They satisfy the Clifford relations $\CurB{\gamma_\mu,\gamma_\nu}=2\delta_{\mu\nu}\bbI$ for $1\leq\mu,\nu\leq2n$.

The Lie algebra $\mathfrak{spin}(2n)$ is realized on the Fock space by quadratic Majorana operators. More precisely, the anti-Hermitian operators $\CurB{\frac{1}{2}\gamma_\mu\gamma_\nu \middle|1\leq\mu<\nu\leq2n}$ form a real basis for the image of $\mathfrak{spin}(2n)$ under the derived representation $\pi_*$.; there are $\binom{2n}{2}=n(2n-1)=\dim\mathfrak{so}(2n)$
of them, as required. Equivalently, the corresponding Hermitian quadratic Hamiltonians are proportional to $i\gamma_\mu\gamma_\nu$. Let $W\equiv\bbC^{2n}$ denote the vector representation of $\Spin{2n}$. This representation can be defined through the conjugation action of $\Spin{2n}$ on the linear span of the Majorana operators. In particular, for each $g\in\Spin{2n}$, there exists a linear transformation $\rho_W(g):W\to W$ such that 
\begin{align*}
    \Pi(g)\gamma_\mu\Pi(g)^\dagger = \sum_{\nu=1}^{2n} \rho_W(g)_{\nu\mu}\gamma_\nu. 
\end{align*}
Note that in the preceding equation the $2n$ Majorana operators are treated as basis vectors of a $2n$-dimensional vector space: they are linearly independent, so their complex span is exactly $2n$-dimensional, and the Clifford relations \cref{app:example:cl:rel} play no role in this count.  What the Clifford relations do ensure is that this span is a \emph{proper, conjugation-invariant} subspace of the $4^n$-dimensional algebra that the $\gamma_\mu$ generate.  Thus, the linear span of the Majorana operators transforms according to the $2n$-dimensional representation $(W,\rho_W)$. We reserve the symbol $\rho_W$ for this representation; the symbol $\rho$ is used exclusively for density matrices in \cref{app:example:ff:local}.

We next consider the tensor representation $\left(V^{\otimes2},\Pi^{\otimes2}\right)$.
A standard Clifford-algebra {isomorphism} gives the canonical
$\Spin{2n}$-equivariant identification
\begin{align}
    V\otimes V^* \cong\operatorname{End}(V) \cong \mathrm{Cl}_{2n}(\bbC)\cong \bigoplus_{k=0}^{2n}\Lambda^k(W),
    \label{eq:app:clifford:iso}
\end{align}
where $\mathrm{Cl}_{2n}(\bbC)$ is the Clifford algebra. Since the full Dirac spinor
representation $V$ is self-dual
choosing a $\Spin{2n}$-equivariant isomorphism $V\cong V^*$ yields a generally
noncanonical isomorphism
\begin{align}
    V^{\otimes2} \cong\bigoplus_{k=0}^{2n}\Lambda^k(W).
    \label{eq:app:tensor:square}
\end{align}
{We fix this isomorphism once and choose it unitary, which is possible by compactness; all orthogonal projectors and Hilbert–Schmidt norms below refer to that choice.}
The spaces $\Lambda^k(W)$ and $\Lambda^{2n-k}(W)$ are equivalent $\Spin{2n}$-representations.
Hence, for every $0\leq k<n$, the irreducible representation $\Lambda^k(W)$ occurs
with multiplicity two {in $V^{\otimes 2}$}.

{Three structural facts about \cref{eq:app:clifford:iso} will be used repeatedly. First, the identification of
$\mathrm{Cl}_{2n}(\bbC)$ with $\bigoplus_k\Lambda^k(W)$ is the \emph{symbol map}
\begin{align}
    \gamma_{i_1}\gamma_{i_2}\cdots\gamma_{i_k} \longmapsto e_{i_1}\wedge e_{i_2}\wedge\cdots\wedge e_{i_k},
    \qquad i_1<i_2<\cdots<i_k,
    \label{eq:app:symbol:map}
\end{align}
which is an isomorphism of $\Spin{2n}$-representations but not of algebras.
Second, under $V^{\otimes 2}\cong\End{V}$ an operator $A\otimes\bbI$ acts as
\emph{left} multiplication by $A$. Combining this with the standard Clifford action
$\gamma_\mu\cdot\omega=e_\mu\wedge\omega+\iota_{e_\mu}\omega$, where $\iota$ denotes
contraction, left multiplication by a single Majorana operator changes the exterior
degree by exactly $\pm1$, and hence left multiplication by a quadratic Majorana
operator $\gamma_\mu\gamma_\nu$ satisfies
\begin{align}
    \gamma_\mu\gamma_\nu\,\Lambda^k(W)\subseteq\Lambda^{k-2}(W)\oplus\Lambda^{k}(W)\oplus\Lambda^{k+2}(W).
    \label{eq:app:degree:shift}
\end{align}
Third, because $V\cong V^*$ was chosen $\Spin{2n}$-equivariant, the tensor-square action on $V^{\otimes 2}\cong\End{V}$ becomes conjugation, $\Pi(g)^{\otimes 2}=\Ad\big(\Pi(g)\big)$ with $\Ad{g}X\defeq gXg^{-1}$. \footnote{%
Explicitly, let $\phi:V\to V^*$ be the chosen isomorphism, so that equivariance reads
$\phi\,\Pi(g)=\Pi^*(g)\,\phi$ with $\Pi^*(g)f\defeq f\circ\Pi(g)^{-1}$. Then
\begin{align*}
    \underbrace{\Pi(g)\otimes\Pi(g)}_{\text{on } V\otimes V}
    \;\xrightarrow{\;\bbI\otimes\phi\;}\;
    \underbrace{\Pi(g)\otimes\Pi^*(g)}_{\text{on } V\otimes V^*}
    \;\xrightarrow{\;v\otimes f\,\mapsto\,\lvert v\rangle\langle f\rvert\;}\;
    \underbrace{X\mapsto\Pi(g)\,X\,\Pi(g)^{-1}}_{\text{on } \End{V}},
\end{align*}
where $\lvert v\rangle\langle f\rvert$ denotes the operator $u\mapsto f(u)\,v$ and the
second arrow is the canonical isomorphism $V\otimes V^*\cong\End{V}$.} For $\mu\neq\nu$, conjugation by $\gamma_\mu\gamma_\nu$ fixes every $\gamma_\kappa$ with $\kappa\notin\CurB{\mu,\nu}$ and reverses the signs of $\gamma_\mu$ and $\gamma_\nu$; since conjugation is an algebra automorphism, these signs multiply over products, so the symbol map \cref{eq:app:symbol:map} carries $\Ad(\gamma_\mu\gamma_\nu)$ to the diagonal operator $e_I\mapsto(-1)^{\abs{I\cap\CurB{\mu,\nu}}}e_I$.}

To describe the middle exterior power, let $W_{\bbR}$ denote the real vector space
spanned by the Majorana basis, equipped with its standard Euclidean inner product
and orientation. The corresponding Hodge-star operator extends complex linearly to
$\star:\Lambda^k(W)\longrightarrow\Lambda^{2n-k}(W)$. On the middle exterior power
$\Lambda^n(W)$, it satisfies $\star^2=(-1)^n\bbI$. Define the normalized Hodge-star
operator $\cD\defeq i^n\star$. Then $\cD^2=\bbI$ on $\Lambda^n(W)$. We define
\begin{align*}
    \Lambda^n_\pm(W)\defeq\CurB{\omega\in\Lambda^n(W)\middle|\mathcal D\omega=\pm\omega}.
\end{align*}
The middle exterior power therefore decomposes as
$\Lambda^n(W)=\Lambda^n_+(W)\oplus\Lambda^n_-(W),$ where $\Lambda^n_+(W)$ and
$\Lambda^n_-(W)$ are irreducible $\Spin{2n}$-representations satisfying
$\dim\Lambda^n_\pm(W)=\frac{1}{2}\binom{2n}{n}$. Therefore, the decomposition of the tensor-square
representation into irreducible $\Spin{2n}$-representations, including their
multiplicities, is
\begin{align}
    V^{\otimes2}\cong\bigoplus_{k=0}^{n-1}\left(2\Lambda^k(W)\right)\oplus\Lambda^n_+(W)\oplus\Lambda^n_-(W).
    \label{eq:app:full:decomposition}
\end{align}
The calculation in \cref{app:example:ff:local,app:example:ff:non-local} takes place in the even-parity subspace $S_+\otimes S_+$, whose decomposition differs from \cref{eq:app:full:decomposition}. Under the $\bbZ_2$-grading of $\mathrm{Cl}_{2n}(\bbC)$, the block of $\End{V}$ containing $S_+\otimes S_+$ receives exactly one copy of each $\Lambda^k(W)$ with $k<n$ and $k\equiv n \pmod 2$, together with one of the two chiralities $\Lambda^n_\pm(W)$, so that
\begin{align}
    S_+\otimes S_+ \;\cong\; \Lambda^n_\varepsilon(W)\;\oplus\;\bigoplus_{j=1}^{\lfloor n/2\rfloor}\Lambda^{n-2j}(W),
    \label{eq:app:even:decomposition}
\end{align}
for a single chirality $\varepsilon$, \emph{every summand occurring with
multiplicity one}; the dimensions sum to $4^{n-1}=\dim(S_+\otimes S_+)$.  In
particular, of the two copies of $\Lambda^{n-2}(W)$ in
\cref{eq:app:full:decomposition} exactly one lies in $S_+\otimes S_+$, and it is a
Hodge-paired combination of the degree-$(n-2)$ and degree-$(n+2)$ subspaces rather
than either one alone.

For completeness, given a basis $\CurB{e_1,e_2,\dots,e_{2n}}$ of $W$, the $k$-th
exterior power is defined by
\begin{align*}
    \Lambda^k(W)\defeq\operatorname{span}\CurB{e_{i_1}\wedge\cdots\wedge e_{i_k}\middle| 1\leq i_1<\cdots<i_k\leq2n}.
\end{align*}
The wedge products are given by
\begin{align*}
    e_{i_1}\wedge\cdots\wedge e_{i_k}\defeq\frac{1}{k!}\sum_{\sigma\in S_k}\sgn(\sigma)e_{i_{\sigma(1)}}\otimes\cdots\otimes e_{i_{\sigma(k)}}.
\end{align*}
We also set $\Lambda^0(W)\defeq\bbC$. {We equip $\Lambda^k(W)$ with the $\Spin{2n}$-invariant inner product for which the exterior basis $\CurB{e_I}_{\abs{I}=k}$ is \emph{orthonormal}.} The dimension of each exterior power is $\dim\Lambda^k(W)=\binom{2n}{k}$. Consequently,
\begin{align*}
    \sum_{k=0}^{2n}\dim\Lambda^k(W)=\sum_{k=0}^{2n}\binom{2n}{k}=2^{2n} =\dim V^{\otimes2},
\end{align*}
as required.

To demonstrate the effectiveness of our upper and lower bounds, we provide two
choices for $\rho$ and $\cM$. One is known to have non-vanishing variance and is
used to verify our lower bound; the other is known to have exponentially vanishing
variance, demonstrating that our upper bound can successfully capture it.

\subsubsection{Local Observable} 
\label{app:example:ff:local}
First, we consider 
\begin{align} 
    \rho = \Outer{0^{\otimes n}}{0^{\otimes n}}, \qquad \cM = Z^{(1)} = -i\gamma_1\gamma_2, \qquad H = \frac{1}{2}\gamma_1\gamma_2. \label{eq:free:fermion:rho:M:H}
\end{align}
{Recall from assumptions above that $\frac12\gamma_1\gamma_2$ is an anti-Hermitian element of the image of $\mathfrak{spin}(2n)$, so $H$ is anti-Hermitian; we retain this convention rather than the Hermitian normalization $iH$, and note that the sign matters, since $\commutator{iA}{\commutator{iA}{\cdot}}=-\commutator{A}{\commutator{A}{\cdot}}$.} We evaluate the second term of the lower bound
\begin{align}
    \begin{split}
        &2\sum_{\nu}\sum_{c,d=1}^{m_\nu}
        \frac{\snorm{\cJ_{\nu,dd}\sqrt{\rho^{\otimes 2}}}_{2} \snorm{\cJ_{\nu,cc}\sqrt{\rho^{\otimes 2}}}_{2}}{\dim V_\nu} \\
        &\quad\times \abs{\Tr{\left(\sum_{\lambda\neq\nu} \sum_{a,b=1}^{m_\lambda} \frac{\Tr{\cJ_{\lambda,ab}^{\dagger} \cM_{\nu,c,1\to\nu,d,2}^{(\lambda^\prime,a^\prime) \otimes (\nu^\prime,b^\prime)}}}{\dim V_\lambda} \cJ_{\lambda,ab}^{(\lambda^\prime,a^\prime,\nu^\prime,b^\prime)_{\nu,c,1}, (\lambda^\prime,a^\prime,\nu^\prime,b^\prime)_{\nu,d,2}}\right) (H\otimes\bbI) \cJ_{\nu,cd} (H\otimes\bbI)}}.
    \end{split}
    \label{eq:free:fermion:cross:def}
\end{align}
We use the standard spinor--exterior-algebra realization of the tensor-square
representation. In particular, writing
\begin{align*}
    W &\defeq \bbC^{2n},
\end{align*}
the tensor product of two spinor representations decomposes into exterior-form
representations of $\Spin{2n}$. The two middle-form representations are the chiral
components
\begin{align*}
    \Lambda^n(W) &= \Lambda_+^n(W) \oplus \Lambda_-^n(W).
\end{align*}
We denote by $V_{\nu_0}$ the unique chiral middle-form representation containing the
tensor-product vacuum. Thus,
\begin{align*}
    V_{\nu_0} &\simeq \Lambda_\varepsilon^n(W)
\end{align*}
for some chirality $\varepsilon\in\CurB{+,-}$, and
\begin{align}
    \dim V_{\nu_0} &= \frac{1}{2}\binom{2n}{n}.
    \label{eq:free:fermion:middle:dimension}
\end{align}
Since $\rho$ is a rank-one {projector, it is idempotent, and hence}
$\sqrt{\rho^{\otimes 2}} = \rho^{\otimes 2}$.
The vector ${\state{0^{\otimes n}}^{\otimes 2}}$ is a highest-weight vector of
the {unique component of $S_+\otimes S_+$ isomorphic to $\Lambda^n_\varepsilon(W)$,
which we label $V_{\nu_0,c_0}$; uniqueness is guaranteed by the multiplicity-one
statement in \cref{eq:app:even:decomposition} together with the inequivalence of the
two chiralities}. Therefore,
\begin{align}
    \cJ_{\nu,cc}{\state{0^{\otimes n}}^{\otimes 2}} &=
    \begin{cases}
        {\state{0^{\otimes n}}^{\otimes 2}}, & \nu=\nu_0 \text{ and } c=c_0, \\
        0, & \text{otherwise}.
    \end{cases}
    \label{eq:free:fermion:vacuum:projector}
\end{align}
Consequently,
\begin{align}
    \begin{split}
        \left(\snorm{\cJ_{\nu,cc}\sqrt{\rho^{\otimes 2}}}_{2}\right)^2
        &= \Tr{\rho^{\otimes 2} \cJ_{\nu,cc}^{\dagger} \cJ_{\nu,cc} \rho^{\otimes 2}} \\
        &= {\cstate{0^{\otimes n}}^{\otimes 2}}\cJ_{\nu,cc}{\state{0^{\otimes n}}^{\otimes 2}} \\
        &= \delta_{\nu,\nu_0} \delta_{c,c_0}.
    \end{split}
    \label{eq:free:fermion:vacuum:norm}
\end{align}
Hence,
\begin{align*}
    \snorm{\cJ_{\nu,cc} \sqrt{\rho^{\otimes 2}}}_{2} &= \delta_{\nu,\nu_0} \delta_{c,c_0}.
\end{align*}
The same identity holds with $c$ replaced by $d$. It follows that every term in
\cref{eq:free:fermion:cross:def} vanishes except $\nu = \nu_0, c = d = c_0$.
{Moreover, by \cref{eq:app:even:decomposition} the multiplicity of $V_{\nu_0}$ in
$S_+\otimes S_+$ is one, so $m_{\nu_0}=1$ and $c_0$ is the only available value of
the multiplicity index.} Therefore, the expression can be written as
\begin{align}
    \begin{split}
        \frac{2}{\dim V_{\nu_0}} \abs{\Tr{\left(\sum_{\lambda\neq\nu_0} \sum_{a,b=1}^{m_\lambda} \frac{\Tr{\cJ_{\lambda,ab}^{\dagger} \cM_{\nu_0,c_0,1\to\nu_0,c_0,2}^{(\lambda^\prime,a^\prime) \otimes (\nu^\prime,b^\prime)}}}{\dim V_\lambda} \cJ_{\lambda,ab}^{\mathrm{res}}\right) (H\otimes\bbI) \cJ_{\nu_0,c_0c_0} (H\otimes\bbI)}},
    \end{split}
    \label{eq:free:fermion:cross:vacuum:reduction}
\end{align}
where, for brevity, we write
\begin{align*}
    \cJ_{\lambda,ab}^{\mathrm{res}} &\defeq \cJ_{\lambda,ab}^{(\lambda^\prime,a^\prime,\nu^\prime,b^\prime)_{\nu_0,c_0,1}, (\lambda^\prime,a^\prime,\nu^\prime,b^\prime)_{\nu_0,c_0,2}}.
\end{align*}

The vacuum state $\state{0^{\otimes n}}$ has even fermion parity because it contains
zero fermionic excitations, meaning it is an eigenstate of the total parity operator
with an eigenvalue of $+1$. Hence, the relevant parent tensor-product space is
\begin{align*}
    S_+\otimes S_+,
\end{align*}
where $S_+$ is spanned by all computational-basis states containing an even number of
occupied fermionic modes. Because $\cM=-i\gamma_1\gamma_2$ is an even product of
Majorana operators, it preserves fermion parity $\commutator{\cM}{P_+} = 0$, where
$P_+$ is the projector onto $S_+$. Therefore, $P_+\cM P_+ = P_+\cM = \cM P_+$. For
the surviving indices $c=d=c_0$, the restricted observable becomes
\begin{align}
    \begin{split}
        \cM_{\nu_0,c_0,1\to\nu_0,c_0,2}^{(+,1)\otimes(+,1)} = \left(P_+\cM P_+\right) \otimes \left(P_+\cM P_+\right) = (P_+\otimes P_+) \cM^{\otimes 2} (P_+\otimes P_+),
    \end{split}
    \label{eq:free:fermion:restricted:M}
\end{align}
For the surviving terms, both $V_{\lambda,a}$ and $V_{\lambda,b}$ are contained in
$S_+\otimes S_+$. Let $Q_+\defeq P_+\otimes P_+$. Since
\begin{align*}
    \cM_{\nu_0,c_0,1\to\nu_0,c_0,2}^{(+,1)\otimes(+,1)} = Q_+\cM^{\otimes 2}Q_+,
\end{align*}
cyclicity of the trace gives
\begin{align*}
    \Tr{\cJ_{\lambda,ab}^\dagger\cM_{\nu_0,c_0,1\to\nu_0,c_0,2}^{(+,1)\otimes(+,1)}}= \Tr{\left(Q_+\cJ_{\lambda,ab}Q_+\right)^\dagger\cM^{\otimes 2}}.
\end{align*}
Because $\cJ_{\lambda,ab}$ maps between irreducible copies contained entirely in
$S_+\otimes S_+$, we have $Q_+\cJ_{\lambda,ab}Q_+=\cJ_{\lambda,ab}$. Therefore,
\begin{align*}
    \Tr{\cJ_{\lambda,ab}^\dagger\cM_{\nu_0,c_0,1\to\nu_0,c_0,2}^{(+,1)\otimes(+,1)}}=\Tr{\cJ_{\lambda,ab}^\dagger\cM^{\otimes 2}}.
\end{align*}

Set $K \defeq H\otimes\bbI$. {By \cref{eq:app:degree:shift}}, left Clifford
multiplication by $H = \frac{1}{2}\gamma_1\gamma_2$ can change the exterior degree by
$-2$, $0$, or $+2$. Therefore,
\begin{align}
    K\Lambda^k(W) &\subseteq \Lambda^{k-2}(W) \oplus \Lambda^k(W) \oplus \Lambda^{k+2}(W).
    \label{eq:free:fermion:H:degree}
\end{align}
{Since $\gamma_1\gamma_2$ is an even product of Majorana operators, $K$ preserves the parity decomposition and in particular maps $S_+\otimes S_+$ into itself.} Starting from {the irreducible invariant subspace $V_{\nu_0}\simeq \Lambda_\varepsilon^n(W)$}, the degree-preserving component remains inside $V_{\nu_0}$ and is excluded by the condition $\lambda\neq\nu_0$. The degree-lowering and degree-raising components together generate the irreducible representation
\begin{align}
    V_{\lambda_0} &\simeq \Lambda^{n-2}(W),
    \label{eq:free:fermion:lambda0}
\end{align}
where the degree-$(n+2)$ realization is identified with the degree-$(n-2)$ realization by Hodge duality. This representation occurs with multiplicity one in $S_+\otimes S_+$, by eq. (8). Consequently, after insertion between $K\cJ_{\nu_0,c_0c_0}K$, the weighted intertwiner sum reduces to
\begin{align}
    \begin{split}
        \Tr{\left(\sum_{\lambda\neq\nu_0} \sum_{a,b=1}^{m_\lambda} \frac{\Tr{\cJ_{\lambda,ab}^{\dagger} \cM^{\otimes 2}}}{\dim V_\lambda} \cJ_{\lambda,ab}^{\mathrm{res}}\right) K \cJ_{\nu_0,c_0c_0} K}
        &= \frac{\Tr{\cJ_{\lambda_0,11}^{\dagger} \cM^{\otimes 2}}}{\dim V_{\lambda_0}} \Tr{\cJ_{\lambda_0,11} K \cJ_{\nu_0,c_0c_0} K}.
    \end{split}
    \label{eq:free:fermion:weighted:sum:reduction}
\end{align}
By \cref{eq:free:fermion:H:degree}, $K\cJ_{\nu_0,c_0c_0}K$ occupies exterior degrees $n-2$, $n$ and $n+2$, while by \cref{eq:app:even:decomposition} the copy of $\Lambda^{n-2j}(W)$ in $S_+\otimes S_+$ occupies degrees $n\mp2j$; only $j=1$ survives, $j=0$ being excluded by $\lambda\neq\nu_0$.

Let
\begin{align*}
    e_I &= e_{i_1}\wedge\cdots\wedge e_{i_k} \in \Lambda^k(W)
\end{align*}
be an exterior-basis vector, where $I = \{i_1, \cdots, i_k\}$ is a multi-index such
that $1\leq i_1,\cdots, i_k\leq 2n$. By \cref{eq:app:symbol:map} and the
adjoint-action, we note that the operator
$\cM = -i\gamma_1\gamma_2$ induces a sign determined by the number of indices from
$\CurB{1,2}$ contained in $I$:
\begin{align}
    \cM^{\otimes 2}e_I &= (-1)^{\abs{I\cap\CurB{1,2}}+1}e_I =
    \begin{cases}
        -e_I, & \abs{I\cap\CurB{1,2}}=0, \\
        +e_I, & \abs{I\cap\CurB{1,2}}=1, \\
        -e_I, & \abs{I\cap\CurB{1,2}}=2.
    \end{cases}
    \label{eq:free:fermion:M:action}
\end{align}
{Explicitly, $(\gamma_1\gamma_2)^2=-\bbI$, so $\exp\big(\pi\cdot\tfrac12\gamma_1\gamma_2\big)=\gamma_1\gamma_2$ lies in $\Pi(\Spin{2n})$, and therefore $\cM^{\otimes2}=(-i)^2\Ad(\gamma_1\gamma_2)=-\Ad(\gamma_1\gamma_2)$.  The overall $+1$ in the exponent of \cref{eq:free:fermion:M:action} comes from the prefactor $(-i)^2=-1$ and \emph{not} from the adjoint action, which contributes only $(-1)^{\abs{I\cap\CurB{1,2}}}$.} The number of degree-$k$ basis vectors satisfying these three conditions is, respectively,
\begin{align}
    \begin{cases}
        \binom{2n-2}{k} & \abs{I\cap\CurB{1,2}}=0, \\
        2\binom{2n-2}{k-1} & \abs{I\cap\CurB{1,2}}=1, \\
        \binom{2n-2}{k-2} & \abs{I\cap\CurB{1,2}}=2.
    \end{cases}\label{eq:free:fermion:number:of:basis}
\end{align}
For notational simplicity, we denote by $\cP_k$ the orthogonal projector onto the
entire degree-$k$ subspace $\Lambda^k(\bbC^{2n})$. Then, the trace contains
contributions from all possible values of $\abs{I\cap\CurB{1,2}}$. To make this
explicit, define {the index sets}
\begin{align*}
    \cI_{k,r} \defeq \CurB{{I}: I\subseteq\CurB{1,\ldots,2n}, \abs{I}=k, \abs{I\cap\CurB{1,2}}=r}, \qquad r\in\CurB{0,1,2}.
\end{align*}
The exterior basis of $\Lambda^k(\bbC^{2n})$ then admits the disjoint decomposition
\begin{align*}
    \CurB{e_I:I\subseteq\CurB{1,\ldots,2n}, \abs{I}=k}=\bigsqcup_{r=0}^{2}{\CurB{e_I : I\in\cI_{k,r}}}.
\end{align*}
To express this decomposition at the level of orthogonal projectors, define
\begin{align*}
    \cP_{k,r} \defeq \sum_{{I\in\cI_{k,r}}} \Outer{e_I}{e_I}.
\end{align*}
Then,
\begin{align*}
    \cP_k = \sum_{r=0}^{2}\cP_{k,r} \qquad\cP_{k,r}\cP_{k,s} =\delta_{rs}\cP_{k,r}.
\end{align*}
Therefore,
\begin{align}
    \begin{split}
        \Tr{\cP_k^\dagger\cM^{\otimes2}}
        &= \sum_{r=0}^{2} \sum_{I\in\mathcal I_{k,r}} \cstate{e_I}\cM^{\otimes2}\state{e_I} = -\sum_{I\in\mathcal I_{k,0}}1 + \sum_{I\in\mathcal I_{k,1}}1 - \sum_{I\in\mathcal I_{k,2}}1.
    \end{split} \label{app:example:tr:proj:Lambda:k}
\end{align}
Using \cref{eq:free:fermion:number:of:basis}, we obtain
\begin{align*}
    \Tr{\cP_k^\dagger\cM^{\otimes2}} = -\binom{2n-2}{k} + 2\binom{2n-2}{k-1} - \binom{2n-2}{k-2}.
\end{align*}
Since $\dim\Lambda^k(W) = \binom{2n}{k}$, we obtain
\begin{align}
    \begin{split}
        \frac{\Tr{{\cP_k}^{\dagger} \cM^{\otimes 2}}}{\dim\Lambda^k(W)} &= \frac{2\binom{2n-2}{k-1} - \binom{2n-2}{k} - \binom{2n-2}{k-2}}{\binom{2n}{k}} = \frac{2n-(2n-2k)^2}{2n(2n-1)}.
    \end{split}
    \label{eq:free:fermion:M:normalized:trace:binomial}
\end{align}

For the representation $V_{\lambda_0} \simeq \Lambda^{n-2}(W)$, we set $k=n-2$ in \cref{eq:free:fermion:M:normalized:trace:binomial}. Although $\cJ_{\lambda_0,11}\neq\cP_{n-2}$, this is legitimate. The range of $\cJ_{\lambda_0,11}$ lies in $\Lambda^{n-2}(W)\oplus\Lambda^{n+2}(W)$ and, being contained in $S_+\otimes S_+$, consists of vectors $X$ with $P_+X=X$, i.e. $\Gamma X=X$, where $P_+=\tfrac12(\bbI+\Gamma)$ acts by left multiplication and $\Gamma\defeq\prod_jZ^{(j)}=(-i)^n\gamma_1\cdots\gamma_{2n}$ is the parity operator. Since the indices in $J$ cancel in $\gamma_1\cdots\gamma_{2n}\gamma_J$, left multiplication by $\Gamma$ sends $e_J$ to $\eta_J\,e_{J^c}$ with $\abs{\eta_J}=1$. Thus $\Gamma e_J\neq e_J$, and the solutions of $\Gamma X=X$ in $\Lambda^{n-2}(W)\oplus\Lambda^{n+2}(W)$ are spanned by the $\binom{2n}{n-2}=\dim V_{\lambda_0}$ orthonormal vectors $\tfrac{1}{\sqrt2}\left(e_J+\eta_J\,e_{J^c}\right)$, $\abs{J}=n-2$, which therefore form a basis of the range. Moreover, $e_J$ and $e_{J^c}$ share an $\cM^{\otimes2}$-eigenvalue, because $\abs{J^c\cap\CurB{1,2}}=2-\abs{J\cap\CurB{1,2}}$ and the sign in \cref{eq:free:fermion:M:action} is symmetric under $r\mapsto2-r$. Each basis vector therefore has the same expectation value as $e_J$, so $\Tr{\cJ_{\lambda_0,11}\cM^{\otimes2}}=\Tr{\cP_{n-2}\cM^{\otimes2}}$, while $\dim V_{\lambda_0}=\dim\Lambda^{n-2}(W)$. We obtain
\begin{align}
    \begin{split}
        \frac{\Tr{\cJ_{\lambda_0,11}^{\dagger} \cM^{\otimes 2}}}{\dim V_{\lambda_0}}= \frac{n-8}{n(2n-1)}.
    \end{split}
    \label{eq:free:fermion:M:normalized:trace:n-2}
\end{align}
{Since $n\geq8$, this quantity is nonnegative, and therefore}
\begin{align}
    \abs{\frac{\Tr{\cJ_{\lambda_0,11}^{\dagger} \cM^{\otimes 2}}}{\dim V_{\lambda_0}}} &= \frac{{n-8}}{n(2n-1)}.
    \label{eq:free:fermion:M:normalized:trace:abs}
\end{align}

We next compute
$\Tr{\cJ_{\lambda_0,11} (H\otimes\bbI) \cJ_{\nu_0,c_0c_0} (H\otimes\bbI)}$.
{By assumption, $H$ is anti-Hermitian, and
therefore so is $K = H\otimes\bbI$.} Because $\cJ_{\lambda_0,11}$ and
$\cJ_{\nu_0,c_0c_0}$ are mutually orthogonal projectors,
\begin{align}
    \begin{split}
        \Tr{\cJ_{\lambda_0,11} K \cJ_{\nu_0,c_0c_0} K}
        &= - \Tr{\left(\cJ_{\lambda_0,11} K \cJ_{\nu_0,c_0c_0}\right)^\dagger \left(\cJ_{\lambda_0,11} K \cJ_{\nu_0,c_0c_0}\right)} \\
        &= - \left(\snorm{\cJ_{\lambda_0,11} K \cJ_{\nu_0,c_0c_0}}_{2}\right)^{2}.
    \end{split}
    \label{eq:free:fermion:transition:trace:norm}
\end{align}
To compute this Hilbert--Schmidt norm, consider an $n$-element multi-index
$I \subseteq \CurB{1,\ldots,2n}$ such that $\CurB{1,2} \subseteq I$. Then, we write
$J \defeq I\setminus\CurB{1,2}$, so that $\abs{J} = n-2$. Let $I^c$ be the
complement of $I$. Then, $J^c=I^c\cup\CurB{1,2}$. After fixing the Hodge-star
phases, a normalized vector in {the irreducible invariant subspace $V_{\nu_0}\simeq \Lambda_\varepsilon^n(W)$} can be written
as
\begin{align*}
    \phi_I^\varepsilon = \frac{1}{\sqrt{2}} \left(e_I+\varepsilon\alpha_Ie_{I^c}\right), \qquad \abs{\alpha_I}=1, \quad \varepsilon\in \CurB{\pm 1}.
\end{align*}
Under left Clifford multiplication, $\gamma_1\gamma_2$ removes the indices $1$ and
$2$ from $e_I$ and inserts them into $e_{I^c}$. Therefore, for some signs
$\sigma_I,\tau_I\in\CurB{+1,-1}$,
\begin{align*}
    \gamma_1\gamma_2 e_I = \sigma_I e_J, \qquad \gamma_1\gamma_2 e_{I^c} = \tau_I e_{J^c}.
\end{align*}
Since $H=\frac{1}{2}\gamma_1\gamma_2$, we obtain
\begin{align}
    \begin{split}
        (H\otimes\bbI)\phi_I^\varepsilon &= \frac{1}{2\sqrt{2}} \left(\sigma_Ie_J + \varepsilon\alpha_I\tau_Ie_{J^c}\right) = \frac{1}{2}\psi_J^\varepsilon,
    \end{split}
\end{align}
where $\psi_J^\varepsilon \defeq ({1}/{\sqrt{2}})\cdot \left(\sigma_Ie_J + \varepsilon\alpha_I\tau_Ie_{J^c}\right)$
is a normalized vector in the copy of the $\Lambda^{n-2}(W)$ representation obtained
by pairing the degree-$(n-2)$ and degree-$(n+2)$ components through Hodge duality.
{Because $K=H\otimes\bbI$ preserves $S_+\otimes S_+$, the vector
$\psi_J^\varepsilon$ lies in that subspace, and hence --- by the multiplicity-one
statement of \cref{eq:app:even:decomposition} --- in the \emph{unique} copy of
$\Lambda^{n-2}(W)$ contained in $S_+\otimes S_+$, which is exactly the range of
$\cJ_{\lambda_0,11}$.}
Hence,
\begin{align*}
    \left(\snorm{(H\otimes\bbI)\phi_I^\varepsilon}_{2}\right)^2 = \frac{1}{4}.
\end{align*}
Let $\CurB{\phi_I^\varepsilon}$ be an orthonormal basis of {$\Lambda^n_\varepsilon(W)$,
indexed by one representative $I$ from each complementary pair $\CurB{I,I^c}$}. The
Hilbert--Schmidt norm can then be expanded as
\begin{align*}
    \left(\snorm{\cJ_{\lambda_0,11} (H\otimes\bbI) \cJ_{\nu_0,c_0c_0}}_{2}\right)^2 = \sum_I \left(\snorm{\cJ_{\lambda_0,11} (H\otimes\bbI)\phi_I^\varepsilon}_{2}\right)^2.
\end{align*}
Only the complementary pairs for which the representative $I$ contains both $1$ and
$2$ contribute to the degree-$(n-2)$ and degree-$(n+2)$ components. There are
$\binom{2n-2}{n-2}$ such pairs, since the remaining $n-2$ elements of $I$ are chosen
from $\CurB{3,\ldots,2n}$. Moreover,
\begin{align*}
    \cJ_{\lambda_0,11} (H\otimes\bbI)\phi_I^\varepsilon = \frac{1}{2}\psi_J^\varepsilon
\end{align*}
for a normalized vector $\psi_J^\varepsilon$, and hence each contributing basis
vector satisfies
\begin{align*}
    \left(\snorm{\cJ_{\lambda_0,11} (H\otimes\bbI)\phi_I^\varepsilon}_{2}\right)^2 = \frac{1}{4}.
\end{align*}
Combining these observations yields
\begin{align}
    \abs{\Tr{\cJ_{\lambda_0,11} (H\otimes\bbI) \cJ_{\nu_0,c_0c_0} (H\otimes\bbI)}} &= \frac{1}{4} \binom{2n-2}{n-2}.
    \label{eq:free:fermion:transition:trace:abs}
\end{align}
Substituting \eqref{eq:free:fermion:middle:dimension},
\eqref{eq:free:fermion:M:normalized:trace:abs}, and
\eqref{eq:free:fermion:transition:trace:abs} into
\eqref{eq:free:fermion:cross:vacuum:reduction} gives
\begin{align}
    \begin{split}
        \frac{2}{\frac{1}{2}\binom{2n}{n}} \frac{{n-8}}{n(2n-1)} \frac{1}{4} \binom{2n-2}{n-2} &= \frac{{n-8}}{n(2n-1)} \frac{\binom{2n-2}{n-2}}{\binom{2n}{n}} \\ &= \frac{{(n-8)}(n-1)}{2n(2n-1)^2},
    \end{split}
    \label{eq:free:fermion:cross:before:simplification}
\end{align}
{where we used $\binom{2n-2}{n-2}\big/\binom{2n}{n} = (n-1)\big/\big(2(2n-1)\big)$.}

Next, we evaluate the diagonal term. Using \cref{app:example:tr:proj:Lambda:k}, we find that
\begin{align*}
    \frac{\Tr{\cJ_{\nu_0,c_0c_0}\cM^{\otimes 2}}}{\dim V_{\nu_0}} = \frac{2\binom{2n-2}{n-1} - 2\binom{2n-2}{n-2}}{\binom{2n}{n}} = \frac{1}{2n-1}.
\end{align*}
{Here we note that $\binom{2n-2}{n}=\binom{2n-2}{n-2}$.}
Moreover, because $\cJ_{\nu_0,c_0c_0}$ is invariant under the diagonal action
generated by $H\otimes\bbI+\bbI\otimes H$, we have
$\commutator{H\otimes\bbI+\bbI\otimes H}{\cJ_{\nu_0,c_0c_0}} = 0$. Equivalently,
$\commutator{\bbI\otimes H}{\cJ_{\nu_0,c_0c_0}} = - \commutator{H\otimes\bbI}{\cJ_{\nu_0,c_0c_0}}$.
It follows that
\begin{align*}
    \Tr{\cJ_{\nu_0,c_0c_0} \commutator{H\otimes\bbI}{\commutator{\bbI\otimes H}{\cJ_{\nu_0,c_0c_0}}}}
    &= - \Tr{\cJ_{\nu_0,c_0c_0} \commutator{K}{\commutator{K}{{\cJ_{\nu_0,c_0c_0}}}}} \\
    &= 2 \left(\snorm{(\bbI-\cJ_{\nu_0,c_0c_0})(H\otimes\bbI)\cJ_{\nu_0,c_0c_0}}_{2}\right)^2 \\
    &= \frac{1}{2} \binom{2n-2}{{n-2}}.
\end{align*}
The second equality uses $K^\dagger=-K$, as in \cref{eq:free:fermion:transition:trace:norm}. For the third, $KV_{\nu_0}\subseteq V_{\nu_0}\oplus V_{\lambda_0}$ gives $(\bbI-\cJ_{\nu_0,c_0c_0})K\cJ_{\nu_0,c_0c_0}=\cJ_{\lambda_0,11}K\cJ_{\nu_0,c_0c_0}$, whose norm was computed in \cref{eq:free:fermion:transition:trace:abs}. Since $\rho^{\otimes 2} = \Outer{0^{\otimes n}}{0^{\otimes n}}^{\otimes 2}$ is
supported entirely on $V_{\nu_0,c_0}$, we have
\begin{align*}
    \Tr{\cJ_{\lambda,cd}^{\dagger}\rho^{\otimes 2}} = \delta_{\lambda,\nu_0} \delta_{c,c_0} \delta_{d,c_0}.
\end{align*}
{Together with $m_{\nu_0}=1$, which forces $a=b=c_0$ as well,} the diagonal
contribution reduces, using $\dim V_{\nu_0} = \frac{1}{2} \binom{2n}{n}$, to
\begin{align}
    &\abs{\sum_{\lambda} \sum_{a,b,c,d=1}^{m_\lambda} \frac{\Tr{\cJ_{\lambda,ab}^{\dagger}\cM^{\otimes 2}} \Tr{\cJ_{\lambda,cd}^{\dagger}\rho^{\otimes 2}}}{\left(\dim V_\lambda\right)^2} \Tr{\cJ_{\lambda,ab} \commutator{H\otimes\bbI}{\commutator{\bbI\otimes H}{\cJ_{\lambda,cd}}}}}  \label{eq:free:fermion:diagonal}\\
    =& \frac{\Tr{\cJ_{\nu_0,c_0c_0}\cM^{\otimes 2}}}{\dim V_{\nu_0}} \frac{1}{\dim V_{\nu_0}} \Tr{\cJ_{\nu_0,c_0c_0} \commutator{H\otimes\bbI}{\commutator{\bbI\otimes H}{{\cJ_{\nu_0,c_0c_0}}}}} \nonumber\\
    =& \frac{1}{2n-1} \frac{1}{\dim V_{\nu_0}} \frac{1}{2} \binom{2n-2}{{n-2}} \nonumber \\
    =& \frac{n-1}{2(2n-1)^2} \nonumber.
\end{align}

Combining the diagonal contribution with the cross-irrep contribution, the lower
bound becomes
\begin{align*}
     \frac{n-1}{2(2n-1)^2} - \frac{{(n-8)}(n-1)}{2n(2n-1)^2} = \frac{n-1}{2(2n-1)^2} \left(1 - \frac{{n-8}}{n}\right)=\frac{4(n-1)}{n(2n-1)^2},
\end{align*}
since $n\geq 8$ by assumption.

\paragraph{Exact value.} As in \cref{app:example:SU2}, the two quantities evaluated above give the variance exactly, and without the restriction $n\geq8$, which they use only to remove absolute values. Since $\rho^{\otimes 2}$ lies in $V_{\nu_0,c_0}$ alone, \cref{eq:lb:eq:1} holds with equality, and the two sums of \cref{app:var:case3:sum} are the diagonal contribution \cref{eq:free:fermion:diagonal}, $\frac{n-1}{2(2n-1)^2}$, and the cross-irrep contribution with its sign. By \cref{eq:free:fermion:M:normalized:trace:n-2,eq:free:fermion:transition:trace:norm,eq:free:fermion:transition:trace:abs}, the latter equals
\begin{align*}
    \frac{2}{\frac12\binom{2n}{n}}\cdot\frac{n-8}{n(2n-1)}\cdot\Big(-\frac14\binom{2n-2}{n-2}\Big)=-\frac{(n-8)(n-1)}{2n(2n-1)^2}.
\end{align*}
Hence
\begin{align}
    \var=\frac{n-1}{2(2n-1)^2}-\frac{(n-8)(n-1)}{2n(2n-1)^2}=\frac{4(n-1)}{n(2n-1)^2}\qquad\text{for all }n\geq2. \label{eq:app:ff:exact}
\end{align}
For $n\geq8$ the cross-irrep contribution is not positive, and the lower bound above coincides with the exact variance. For $n<8$ it is positive, and the lower bound, $\frac{(n-1)(n-4)}{n(2n-1)^2}$, is weaker; it is not positive for $n\leq4$. Since $i\cM=\gamma_1\gamma_2\in\pi_*(\mathfrak{spin}(2n))$, the same value follows from the Lie-algebraic formula of \cref{sec:comparison_prior_works} for $n\geq3$, where the algebra is simple: with the orthonormal basis $\CurB{i\gamma_\mu\gamma_\nu/\sqrt{2^n}}_{\mu<\nu}$ of $i\pi_*(\mathfrak{spin}(2n))$ one finds $D=n(2n-1)$, $(\snorm{\cM}_2)^2=2^n$, $(\snorm{\rho_\fg}_2)^2=n/2^n$ and $(\norm{H}_K)^2=4(n-1)$, so that $\var=4(n-1)\cdot2^n\cdot\frac{n}{2^n}\cdot\frac{1}{n^2(2n-1)^2}$. For $n=2$ the algebra $\mathfrak{spin}(4)$ has two simple ideals, of which only one acts non-trivially on $S_+$; the formula applied to the sub-representation $S_+\cong\bbC^2$ again gives $\frac29$.

\paragraph{Upper bound after compression.} \Cref{fig:variance-bounds}(a) shows \cref{app:upperbound:case_3} after the compression of \cref{main:text:rem:reduced}(ii) to $S_+$, on which $Z^{(1)}$ is traceless. By \cref{eq:app:even:decomposition}, $S_+\otimes S_+$ consists of $V_{\nu_0}$, of dimension $\frac12\binom{2n}{n}$, and of one copy of $\Lambda^{n-2j}(\bbC^{2n})$, of dimension $\binom{2n}{n-2j}$, for each $1\leq j\leq\lfloor n/2\rfloor$. The argument that led to \cref{eq:free:fermion:M:normalized:trace:n-2} applies to each of these copies, so that by \cref{eq:free:fermion:M:normalized:trace:binomial} with $k=n-2j$ the operator $\cT_-(\cM)$ equals $\frac{n-8j^2}{n(2n-1)}$ on the copy of $\Lambda^{n-2j}(\bbC^{2n})$, and $\frac{1}{2n-1}$ on $V_{\nu_0}$. Since $\rho^{\otimes 2}$ lies in $V_{\nu_0}$ and $\infsnorm{H}=\frac12$, \cref{app:upperbound:case_3} becomes
\begin{align}
    \frac{1}{\sqrt{\frac12\binom{2n}{n}}}\left[\frac{\sqrt{\frac12\binom{2n}{n}}}{2n-1}+\sum_{j=1}^{\lfloor n/2\rfloor}\sqrt{\binom{2n}{n-2j}}\,\frac{\abs{n-8j^2}}{n(2n-1)}\right]
    =\frac{1}{2n-1}\left[1+\frac{\sqrt2}{n}\sum_{j=1}^{\lfloor n/2\rfloor}\abs{8j^2-n}\sqrt{\binom{2n}{n-2j}\Big/\binom{2n}{n}}\,\right]. \label{eq:app:ff:upper:compressed}
\end{align}
It decays as $n^{-1/2}$ and stays below the trivial bound $4(\infsnorm{H})^2(\infsnorm{\cM})^2=1$. Without the compression the observable factor also collects the copies in $S_-\otimes S_-$, $S_+\otimes S_-$ and $S_-\otimes S_+$, and the bound is larger: evaluated numerically it is $1.54$ at $n=4$, against $0.44$ from \cref{eq:app:ff:upper:compressed}.

\subsubsection{Non-Local Observable}
\label{app:example:ff:non-local}
Let
\begin{align*}
    \rho = \cM = \Outer{0^{\otimes n}}{0^{\otimes n}}, \qquad H = \frac{1}{2}\gamma_1\gamma_2,
\end{align*}
and again assume that $G=\Spin{2n}$ is the full fermionic Gaussian unitary group generated by quadratic Majorana operators. Then
\begin{align*}
    \rho^{\otimes 2} = \cM^{\otimes 2} = \Outer{0^{\otimes n}}{0^{\otimes n}}^{\otimes 2}.
\end{align*}

For each $j\in\{1,\ldots,n\}$, define the commuting Cartan elements
\begin{align*}
    K_j = -\frac{i}{2}\gamma_{2j-1}\gamma_{2j} = \frac{1}{2}Z^{(j)}.
\end{align*}
Let $\fh=\operatorname{span}\CurB{K_1,\ldots,K_n}$ be the chosen Cartan subalgebra, and let $\CurB{\epsilon_1,\ldots,\epsilon_n}\subseteq\fh^*$ denote its dual basis, defined by $\epsilon_j(K_k)=\delta_{jk}$. Accordingly, if a vector $v$ satisfies $K_jv=\mu_jv$ for all $j$, its weight is given by $\sum_{j=1}^n\mu_j\epsilon_j$. Since $K_j\state{0^{\otimes n}} = \frac{1}{2}\state{0^{\otimes n}}$, the vacuum $\state{0^{\otimes n}}$ is a highest-weight vector of a half-spin representation with highest weight $\omega_{\mathrm{vac}} = \frac{1}{2} \left(\epsilon_1+\cdots+\epsilon_n\right)$. Consequently, the tensor product $\state{0^{\otimes n}}^{\otimes 2}$ is a highest-weight vector with weight $2\omega_{\mathrm{vac}} = \epsilon_1+\cdots+\epsilon_n$. Therefore, $\state{0^{\otimes n}}^{\otimes 2}$ lies entirely in the irreducible representation $V_{\lambda_\star} = V_{2\omega_{\mathrm{vac}}}$.

Under the Clifford--exterior-algebra correspondence, this representation is one of the two irreducible middle-degree components, $V_{\lambda_\star} \simeq \Lambda_{\pm}^{n}\left(\bbC^{2n}\right)$. Its dimension is $\dim V_{\lambda_\star} = \frac{1}{2}\binom{2n}{n}$. Moreover, $V_{\lambda_\star}$ appears with multiplicity one. Hence,
\begin{align*}
    m_{\lambda_\star}=1, \qquad \cJ_{\lambda_\star,11} = P_{\lambda_\star},
\end{align*}
where $P_{\lambda_\star}$ is the orthogonal projector onto $V_{\lambda_\star}$. Since $\state{0^{\otimes 2n}}\in V_{\lambda_\star}$, $P_{\lambda_\star}\state{0^{\otimes 2n}} = \state{0^{\otimes 2n}}$. It follows that
\begin{align}
    \Tr{\rho^{\otimes 2} \cJ_{\lambda_\star,11}} &= \cstate{0^{\otimes 2n}} P_{\lambda_\star} \state{0^{\otimes 2n}} = 1, \\
    \Tr{\cM^{\otimes 2} \cJ_{\lambda_\star,11}} &= \cstate{0^{\otimes 2n}} P_{\lambda_\star} \state{0^{\otimes 2n}} = 1. \label{eq:free_fermion:M:overlap}
\end{align}
For every other irreducible invariant subspace,
\begin{align*}
    \Tr{\rho^{\otimes 2} \cJ_{\lambda,ab}} = \Tr{\cM^{\otimes 2} \cJ_{\lambda,ab}} = 0.
\end{align*}
Therefore,
\begin{align}
    \sum_{\lambda^\prime} \sum_{c,d} \frac{\Tr{\rho^{\otimes 2} \cJ_{\lambda^\prime,cd}}}{\sqrt{\dim V_{\lambda^\prime}}} &= \frac{1}{\sqrt{\dim V_{\lambda_\star}}}, \label{eq:free_fermion:rho:sum} \\
    \sum_{\lambda} \sum_{a,b} \frac{\abs{\Tr{\cM^{\otimes 2} \cJ_{\lambda,ab}}}}{\sqrt{\dim V_{\lambda}}} &= \frac{1}{\sqrt{\dim V_{\lambda_\star}}}. \label{eq:free_fermion:M:sum}
\end{align}

Next, using the Majorana relation, we note that $H = \frac{i}{2}Z^{(1)}$. Since $Z^{(1)}$ is unitary,
\begin{align}
    \snorm{H}_{\infty} = \frac{1}{2}, \qquad \left(\snorm{H}_{\infty}\right)^2 = \frac{1}{4}. \label{eq:free_fermion:H:norm}
\end{align}

Substituting the expressions for the relevant sums and the norm into the variance upper bound, we obtain the upper bound for this case:
\begin{align}
    \begin{split}
        \mathrm{Var} &\leq 4 \left(\frac{1}{\sqrt{\dim V_{\lambda_\star}}}\right) \left(\frac{1}{\sqrt{\dim V_{\lambda_\star}}}\right) \left(\frac{1}{4}\right) = \frac{1}{\dim V_{\lambda_\star}} = \frac{2}{\binom{2n}{n}}.
    \end{split}
\end{align}
Finally, using Stirling's approximation $\binom{2n}{n} \sim \frac{4^n}{\sqrt{\pi n}}$, we find $\mathrm{Var} \lesssim \frac{2\sqrt{\pi n}}{4^n}$. Hence, the variance upper bound decreases exponentially with the number of fermionic modes.

\paragraph{Exact value.} Here $\cM=\rho$, so $\cM^{\otimes 2}$ is supported on the irreducible subspace $V_{\lambda_\star}$, and $\cT_-(\cM)=\cT_+(\rho)=P_{\lambda_\star}/\dim V_{\lambda_\star}$ by Schur's lemma. The subspace $V_{\lambda_\star}$ is the subspace $V_{\nu_0}$ of \cref{app:example:ff:local}. By \cref{app:var:compact,app:crit:copy}, with $\cJ_{\lambda_\star,11}=P_{\lambda_\star}$,
\begin{align}
    \mathrm{Var}=\frac{\Tr{P_{\lambda_\star}\commutator{H\otimes\bbI}{\commutator{\bbI\otimes H}{P_{\lambda_\star}}}}}{(\dim V_{\lambda_\star})^2}=\frac{2\big(\snorm{(\bbI-P_{\lambda_\star})(H\otimes\bbI)P_{\lambda_\star}}_2\big)^2}{(\dim V_{\lambda_\star})^2}=\frac{2\binom{2n-2}{n-2}}{\binom{2n}{n}^2}=\frac{n-1}{(2n-1)\binom{2n}{n}}, \label{eq:app:ff:nonlocal:exact}
\end{align}
since $(\bbI-P_{\lambda_\star})(H\otimes\bbI)P_{\lambda_\star}=\cJ_{\lambda_0,11}(H\otimes\bbI)P_{\lambda_\star}$, as shown in \cref{app:example:ff:local} before \cref{eq:free:fermion:diagonal}, and its squared norm is $\frac14\binom{2n-2}{n-2}$ by \cref{eq:free:fermion:transition:trace:abs}.
The upper bound $2/\binom{2n}{n}$ obtained above, which is evaluated for the projector itself and not for the reduced observable, exceeds the exact variance by the factor $2(2n-1)/(n-1)$, which tends to four.

\subsection{Number-Conserving Fermions: an Observable Outside the Algebra}
\label{app:example:compound}

In the examples of \cref{app:example:SU2,app:example:ff:local} the observable is generated by the algebra. We now derive, from \cref{app:var:compact}, the variance shown in \cref{fig:placeholder}, for which it is not, and for which $\rho^{\otimes 2}$ does not lie in a single irreducible subspace.

\paragraph{Setting.} Let $a_a\defeq Z^{(1)}\cdots Z^{(a-1)}\,\Outer{0}{1}_a$, $a=1,\dots,n$, be Jordan--Wigner annihilation operators on $\cH=(\bbC^2)^{\otimes n}$; these are the operators $c_a$ of \cref{app:example:free:fermine}, written $a_a$ as in \cref{fig:placeholder}, and they satisfy the canonical anticommutation relations of \cref{app:example:cl:rel}. The group $G=\SU{n}$ acts on $\cH$ through the representation $\Pi$ with
\begin{align}
    \pi_*(X)=\sum_{a,b=1}^nX_{ab}\,a_a^\dagger a_b\quad(X\in\su{n}),\qquad \Pi(V)\,a_c^\dagger\,\Pi(V)^\dagger=\sum_{e=1}^nV_{ec}\,a_e^\dagger\quad(V\in\SU{n}). \label{eq:app:compound:rep}
\end{align}
It conserves the particle number $\sum_aa_a^\dagger a_a$; on the $m$-particle sector, which is isomorphic to $\Lambda^m(\bbC^n)$, $\Pi(V)$ is the $m$-th compound matrix of $V$. We take
\begin{align}
    \rho=\Outer{+}{+}^{\otimes n},\qquad\cM=X^{(1)}X^{(2)},\qquad H=\pi_*(h)=-\tfrac{i}{2}\big(X^{(1)}X^{(2)}+Y^{(1)}Y^{(2)}\big),\qquad h\defeq-i(E_{12}+E_{21}),
\end{align}
where $E_{ab}$ are the matrix units, so that $h\in\su{n}$, and assume that $\bbU_\pm$ form $2$-designs with respect to $\Pi(\SU{n})$.

\paragraph{Splitting the observable.} On the first two qubits $a_1^\dagger a_2=\Outer{1}{0}\otimes\Outer{0}{1}$ and $a_1^\dagger a_2^\dagger=\Outer{1}{0}\otimes\Outer{1}{0}$, so that
\begin{align}
    X^{(1)}X^{(2)}=\cM_h+\cM_p,\qquad
    \begin{aligned}
        \cM_h&\defeq a_1^\dagger a_2+a_2^\dagger a_1=\tfrac12\big(X^{(1)}X^{(2)}+Y^{(1)}Y^{(2)}\big),\\
        \cM_p&\defeq a_1^\dagger a_2^\dagger+a_2a_1=\tfrac12\big(X^{(1)}X^{(2)}-Y^{(1)}Y^{(2)}\big).
    \end{aligned} \label{eq:app:compound:split}
\end{align}
The hopping part $\cM_h=iH$ lies in $i\pi_*(\su{n})$. The pairing part $\cM_p$ changes the particle number by two and is orthogonal to $\pi_*(\su{n})$. By \cref{eq:app:compound:rep}, conjugation by $\Pi(V)$ acts on the span of the operators $a_c^\dagger a_d^\dagger$, $c<d$, as on $\Lambda^2(\bbC^n)$, on the span of their adjoints as on its dual, and on $i\pi_*(\su{n})$ by the adjoint representation. For $n\geq3$ these three representations are irreducible, and the adjoint representation, of dimension $n^2-1$, is equivalent neither to $\Lambda^2(\bbC^n)$ nor to its dual, of dimension $n(n-1)/2$. We use the orthogonal projector $\cP_2$ onto $\iComm{2,\compactG}$ of \cref{prop:cP:is:orth:proj}, with $\cU=\Pi$; since $\bbU_-$ is a $2$-design and the Haar measure is invariant under $U\mapsto U^\dagger$, $\cT_-(X)=\cP_2(X^{\otimes 2})$ for every observable $X$. Since $\cP_2$ averages $A\otimes B$ over the simultaneous conjugation of both factors, $\cP_2(\cM_h\otimes\cM_p)$ is an invariant element of the tensor product of the representation on $i\pi_*(\su{n})$ with that on the span of the $a_c^\dagger a_d^\dagger$ and their adjoints; by Schur's lemma such an element is zero unless the second representation contains the dual of the first, the adjoint representation. Hence $\cP_2(\cM_h\otimes\cM_p)=\cP_2(\cM_p\otimes\cM_h)=0$, so that $\cT_-(\cM)=\cT_-(\cM_h)+\cT_-(\cM_p)$, and \cref{app:var:compact}, which is linear in $\cT_-(\cM)$, gives
\begin{align}
    \var\big[X^{(1)}X^{(2)}\big]=\var\big[\cM_h\big]+\var\big[\cM_p\big]. \label{eq:app:compound:additive}
\end{align}
For $n=2$ the operator $\cM_p$ commutes with $\Pi(\SU{2})$, so that $\var[\cM_p]=0$ and \cref{eq:app:compound:additive} holds by \cref{main:text:rem:reduced}(i).

\paragraph{Hopping part.} Since $i\cM_h\in\pi_*(\su{n})$, we apply the Lie-algebraic formula of \cref{sec:comparison_prior_works}. The representation rescales the Hilbert--Schmidt inner product: $\Tr{a_a^\dagger a_b\,a_d^\dagger a_c}=2^{n-2}(\delta_{ac}\delta_{bd}+\delta_{ab}\delta_{cd})$, and therefore $\Tr{\pi_*(X)^\dagger\pi_*(Y)}=2^{n-2}\Tr{X^\dagger Y}$ for traceless $X,Y$. Hence $(\snorm{\cM_h}_2)^2=(\snorm{H}_2)^2=2^{n-2}\Tr{h^\dagger h}=2^{n-1}$, and $(\norm{H}_K)^2=2n\Tr{h^\dagger h}=4n$, because the Casimir operator of the adjoint representation of $\su{n}$ with respect to the trace form equals $2n$. For the state, $\Tr{\pi_*(X)\rho}=\sum_{a,b}X_{ab}\Gamma_{ba}$ with the one-body density matrix
\begin{align}
    \Gamma_{ba}\defeq\cstate{+^{\otimes n}}a_a^\dagger a_b\state{+^{\otimes n}}=\tfrac12\delta_{ab}+\tfrac14\big(\delta_{a,b+1}+\delta_{a+1,b}\big), \label{eq:app:compound:gamma}
\end{align}
since for $a<b$ the operator $a_a^\dagger a_b=\Outer{1}{0}_a\,Z^{(a+1)}\cdots Z^{(b-1)}\Outer{0}{1}_b$ contains a string of $Z$ operators unless $b=a+1$, and $\cstate{+}Z\state{+}=0$. The traceless part of $\Gamma$ has squared Hilbert--Schmidt norm $2(n-1)/16$, so that $(\snorm{\rho_\fg}_2)^2=2^{-(n-2)}(n-1)/8=(n-1)/2^{n+1}$. With $D=\dim\su{n}=n^2-1$, the formula yields
\begin{align}
    \var\big[\cM_h\big]=\frac{4n\cdot2^{n-1}\cdot(n-1)\,2^{-(n+1)}}{(n^2-1)^2}=\frac{n}{(n-1)(n+1)^2}. \label{eq:app:compound:hopping}
\end{align}

\paragraph{Pairing part.} Let $n\geq3$, $N_2\defeq\binom{n}{2}$ and $p_{cd}\defeq a_c^\dagger a_d^\dagger$ for $c<d$, so that $\cM_p=p_{12}+p_{12}^\dagger$. The operators $p_{cd}$ are mutually orthogonal with $(\snorm{p_{cd}}_2)^2=2^{n-2}$, and conjugation by $\Pi(V)$ acts on their span unitarily and irreducibly. By the Schur orthogonality relations,
\begin{align}
    \cP_2\big(p_{12}\otimes p_{12}^\dagger\big)=\frac{1}{N_2}\sum_{c<d}p_{cd}\otimes p_{cd}^\dagger, \label{eq:app:compound:schur}
\end{align}
and likewise with the two tensor factors exchanged, whereas $\cP_2(p_{12}\otimes p_{12})=0$: for $n\neq4$ because $\Lambda^2(\bbC^n)$ is not self-dual, and for $n=4$ because the invariant bilinear form $(\omega,\omega^\prime)\mapsto\omega\wedge\omega^\prime$ on $\Lambda^2(\bbC^4)$ vanishes on the pair $(e_1\wedge e_2,e_1\wedge e_2)$. Inserting $\cT_-(\cM_p)$ into \cref{app:var:compact} and using \cref{app:prelim:nested:comm}, we obtain
\begin{align}
    \var\big[\cM_p\big]=\frac{2}{N_2}\sum_{c<d}\EXPD{U_+}{\abs{\Tr{p_{cd}\commutator{H}{\trho}}}^2},\qquad\trho=U_+\rho U_+^\dagger. \label{eq:app:compound:pairing:1}
\end{align}
Since $\commutator{\pi_*(X)}{a_c^\dagger}=\sum_eX_{ec}a_e^\dagger$, we have $\commutator{H}{p_{cd}}=\sum_{e<f}\big[d\Lambda^2(h)\big]_{ef,cd}\,p_{ef}$, where $d\Lambda^2(h)$ is the matrix by which $h$ acts on $\Lambda^2(\bbC^n)$ in the basis $\CurB{e_c\wedge e_d}_{c<d}$. Writing $y_{ef}(\sigma)\defeq\Tr{p_{ef}\,\sigma}$, this gives $\Tr{p_{cd}\commutator{H}{\trho}}=-\sum_{e<f}\big[d\Lambda^2(h)\big]_{ef,cd}\,y_{ef}(\trho)$. Under $\rho\mapsto\Pi(V)\rho\Pi(V)^\dagger$ the vector $y$ transforms by the complex conjugate of the representation $\Lambda^2$, which is unitary and irreducible, so that $\EXPD{U_+}{y(\trho)\,y(\trho)^\dagger}=\frac{\norm{y(\rho)}^2}{N_2}\bbI$ by Schur's lemma and
\begin{align}
    \sum_{c<d}\EXPD{U_+}{\abs{\Tr{p_{cd}\commutator{H}{\trho}}}^2}=\frac{(\snorm{d\Lambda^2(h)}_2)^2\,\norm{y(\rho)}^2}{N_2}. \label{eq:app:compound:pairing:2}
\end{align}
The two remaining quantities are elementary. For traceless $X$ one has $\Tr{d\Lambda^2(X)^\dagger d\Lambda^2(X)}=(n-2)\Tr{X^\dagger X}$, hence $(\snorm{d\Lambda^2(h)}_2)^2=2(n-2)$. And $y_{ef}(\rho)=\cstate{+^{\otimes n}}a_e^\dagger a_f^\dagger\state{+^{\otimes n}}=\frac14\delta_{f,e+1}$ by the argument used for \cref{eq:app:compound:gamma}, so that $\norm{y(\rho)}^2=(n-1)/16$. Altogether,
\begin{align}
    \var\big[\cM_p\big]=\frac{2}{N_2^2}\cdot2(n-2)\cdot\frac{n-1}{16}=\frac{n-2}{n^2(n-1)}. \label{eq:app:compound:pairing}
\end{align}

\paragraph{Result.} Adding \cref{eq:app:compound:hopping,eq:app:compound:pairing},
\begin{align}
    \var\big[X^{(1)}X^{(2)}\big]=\frac{n}{(n-1)(n+1)^2}+\frac{n-2}{n^2(n-1)}=\frac{2n^3-3n-2}{n^2(n-1)(n+1)^2}\qquad\text{for all }n\geq2, \label{eq:app:compound:result}
\end{align}
which is the solid line of \cref{fig:placeholder}. The variance decays as $2/n^2$, so there is no barren plateau. The Lie-algebraic formula applied to the component of the observable inside the algebra gives the first term only, which is $56\%$ of the variance at $n=4$ and tends to one half of it as $n\to\infty$; the remainder comes from the pairing component, which lies outside $i\pi_*(\su{n})$. The lower bound of \cref{app:lowerbound:case_3}, shown as squares in \cref{fig:placeholder}, is positive there but captures only part of \cref{eq:app:compound:result}: at $n=4$ it is approximately $0.0075$, about $8\%$ of the variance.

\paragraph{Simulation of the runs of \cref{fig:placeholder}.} A run of \Cref{main:text:new:alg} needs the value of the cost function $C=\cstate{+^{\otimes n}}\Pi(V)^\dagger X^{(1)}X^{(2)}\,\Pi(V)\state{+^{\otimes n}}$ for the matrix $V\in\SU{n}$ with $\Pi(V)=U_-\expm{x_kH}U_+$, where $x_k$ is the shift of the execution. By \cref{eq:app:compound:rep}, $\Pi(V)^\dagger a_c\,\Pi(V)=\sum_{e}V_{ce}a_e$, and $X^{(1)}X^{(2)}=(a_1^\dagger-a_1)(a_2+a_2^\dagger)$, so that $C$ is computed, in a time polynomial in $n$, from the first two rows of $V$ and the two-point functions $\Gamma$ of \cref{eq:app:compound:gamma} and $y(\rho)$. For each run, $V_-$ and $V_+$ are drawn from the Haar measure on $\SU{n}$, so that $U_\pm=\Pi(V_\pm)$, the matrix $V=V_-\expm{x_kh}V_+$ is formed for the shifts $x_{k_1}$ and $x_{k_2}$ of the two executions of the run, and the outcome $\pm1$ of $X^{(1)}X^{(2)}$ in each execution is drawn with probability $(1\pm C)/2$. This evaluation agrees with a state-vector simulation of the circuits for $n=4,5,6$.

\paragraph{Overlap factors of the upper bound.} The state overlap $\snorm{\cT_+(\rho)}_2$ and the observable overlap $\snorm{\cT_-(\cM)}_2$ of \cref{sec:criteria:upper} behave in opposite ways in this example. For the state, by \cref{prop:cP:is:orth:proj}, $(\snorm{\cT_+(\rho)}_2)^2=\iTr{\cT_+(\rho)\,\rho^{\otimes 2}}=\EXPD{U}{\abs{\cstate{+^{\otimes n}}U\state{+^{\otimes n}}}^4}\leq\EXPD{U}{\abs{\cstate{+^{\otimes n}}U\state{+^{\otimes n}}}^2}=\iTr{\rho\,\bar\rho}$, where the expectation is over the Haar measure of $\Pi(\SU{n})$ and $\bar\rho\defeq\EXPD{U}{U\rho U^\dagger}$ is the orthogonal projection of $\rho$ onto $\iComm{1,\compactG}$ (\cref{prop:cP:is:orth:proj} with $k=1$). The $m$-particle sectors with $1\leq m\leq n-1$ carry the irreducible representations $\Lambda^m(\bbC^n)$, which are pairwise non-isomorphic and non-trivial, and the sectors $m=0$ and $m=n$ carry the trivial representation. Hence $\iComm{1,\compactG}$ is spanned by the orthogonal projectors onto the sectors with $1\leq m\leq n-1$ and by the operators on the span of $\state{0^{\otimes n}}$ and $\state{1^{\otimes n}}$, and the weight $\binom{n}{m}/2^n$ of $\state{+^{\otimes n}}$ on the $m$-particle sector, of dimension $\binom{n}{m}$, gives
\begin{align*}
    \iTr{\rho\,\bar\rho}=\sum_{m=1}^{n-1}\frac{1}{\binom{n}{m}}\left(\frac{\binom{n}{m}}{2^n}\right)^2+\left(\frac{2}{2^n}\right)^2=\frac{2^n+2}{4^n}.
\end{align*}
For the observable, let $n\geq3$. By \cref{eq:app:compound:schur} and $\cP_2(p_{12}\otimes p_{12})=0$, $\cT_-(\cM_p)=\frac{1}{N_2}\sum_{c<d}\big(p_{cd}\otimes p_{cd}^\dagger+p_{cd}^\dagger\otimes p_{cd}\big)$, and $\cT_-(\cM)=\cT_-(\cM_h)+\cT_-(\cM_p)$, as in the derivation of \cref{eq:app:compound:additive}. By this and \cref{prop:cP:is:orth:proj}, $(\snorm{\cT_-(\cM)}_2)^2=\iTr{\cT_-(\cM_h)\cM_h^{\otimes 2}}+\iTr{\cT_-(\cM_p)\cM_p^{\otimes 2}}+2\iTr{\cT_-(\cM_p)\cM_h^{\otimes 2}}$, and the last trace vanishes, because $\cM_h$ conserves the particle number and every $p_{cd}$ changes it by two. By \cref{eq:app:compound:rep}, $\Pi(V)\cM_h\Pi(V)^\dagger=\sum_{a,b}(VxV^\dagger)_{ab}\,a_a^\dagger a_b$ with $x\defeq E_{12}+E_{21}$, and the trace formula used for \cref{eq:app:compound:hopping} gives $\iTr{\cM_h\Pi(V)\cM_h\Pi(V)^\dagger}=2^{n-2}\Tr{xVxV^\dagger}$. By the Schur orthogonality relations for the adjoint representation, which is irreducible on the traceless $n\times n$ matrices, $\EXPD{V}{\Tr{xVxV^\dagger}^2}=(\Tr{x^2})^2/(n^2-1)=4/(n^2-1)$. For the pairing part, $\iTr{p_{cd}\cM_p}=\iTr{p_{cd}^\dagger\cM_p}=2^{n-2}$ if $(c,d)=(1,2)$ and $0$ otherwise. Altogether,
\begin{align}
    (\snorm{\cT_+(\rho)}_2)^2\leq\frac{2^n+2}{4^n},\qquad(\snorm{\cT_-(\cM)}_2)^2=\frac{4\cdot4^{n-2}}{n^2-1}+\frac{2\cdot4^{n-2}}{N_2}=4^{n-1}\left(\frac{1}{n^2-1}+\frac{1}{n(n-1)}\right). \label{app:compound:overlaps}
\end{align}
The state overlap is exponentially small and the observable overlap exponentially large, whereas the variance \cref{eq:app:compound:result} decays only as $2/n^2$. The upper bound \cref{app:upperbound:case_3}, which cannot be smaller than the variance, is therefore not exponentially small, although the state overlap is.

\end{document}